\documentclass[11pt]{article}
\usepackage{eurosym}
\usepackage{amsfonts}
\usepackage{amssymb}
\usepackage{graphicx}
\usepackage{amsmath}
\usepackage{makeidx}
\usepackage{indentfirst}
\usepackage[T1]{fontenc}
\usepackage[utf8]{inputenc}

\newcounter{resultnum}[section]
\newcounter{conclusionnum}[section]
\newcounter{conditionnum}[section]
\newcounter{conjecturenum}[section]
\newcounter{examplenum}[section]
\newcounter{exercisenum}[section]
\newcounter{lemmanum}[section]
\newcounter{notationnum}[section]
\newcounter{theoremnum}[section]
\newcounter{definitionnum}[section]
\newcounter{corollarynum}[section]
\newcounter{remarknum}[section]
\newcounter{propositionnum}[section]
\newcounter{acknowledgementnum}[section]
\newcounter{algorithmnum}[section]
\newcounter{axiomnum}[section]
\newcounter{casenum}[section]
\newcounter{claimnum}[section]
\newcounter{summarynum}[section]
\newcounter{problemnum}[section]
\begin{document}

\title{Nonassociative Ricci flows, star product and R-flux deformed black
holes, and swampland conjectures}
\date{April 30, 2023}
\author{{\textbf{Lauren\c{t}iu Bubuianu}\thanks{%
email: laurentiu.bubuianu@tvr.ro and laurfb@gmail.com}} \and {\small \textit{%
SRTV - Studioul TVR Ia\c{s}i} and \textit{University Appolonia}, 2 Muzicii
street, Ia\c{s}i, 700399, Romania} \vspace{.1 in} \and \textbf{Sergiu I.
Vacaru} \thanks{%
emails: sergiu.vacaru@fulbrightmail.org ; sergiu.vacaru@gmail.com ; \textit{%
Address for post correspondence in 2022-2023 as a visitor senior researcher
at YF CNU Ukraine:\ } Yu. Gagarin street, 37-3, Chernivtsi, Ukraine, 58008}
\and {\small \textit{Department of Physics, California State University at
Fresno, Fresno, CA 93740, USA; }} \and {\small \textit{Institute of
Applied-Physics and Computer Sciences, Kotsyubinsky 2, Chernivtsi, 58012,
Ukraine; }} \and {\small \textit{Faculty of Physics, Babes-Bolyai
University, 1 Mihail Kogalniceanu Street, Cluj-Napoca, 400084, Romania}} 
\vspace{.1 in} \and {\textbf{El\c{s}en Veli Veliev}} \thanks{%
email: elsen@kocaeli.edu.tr and elsenveli@hotmail.com} \\
{\small \textit{\ Department of Physics,\ Kocaeli University, 41380, Izmit,
Turkey }}}
\maketitle

\begin{abstract}
We extend to a theory of nonassociative geometric flows a string-inspired model of nonassociative gravity determined by star product and R-flux deformations. The nonassociative Ricci tensor and curvature scalar defined
by (non) symmetric metric structures and generalized (non) linear connections are used for defining nonassociative versions of Grigori Perelman F- and W-functionals for Ricci flows and computing associated thermodynamic variables. We develop and apply the anholonomic frame and connection deformation method, AFCDM, which allows us to construct exact and parametric solutions describing nonassociative geometric flow evolution
scenarios and modified Ricci soliton configurations with quasi-stationary generic off-diagonal metrics. There are provided explicit examples of solutions modelling geometric and statistical thermodynamic evolution on a
temperature-like parameter of modified black hole configurations encoding nonassociative star-product and R-flux deformation data. Further perspectives of the paper are motivated by nonassociative off-diagonal geometric flow extensions of the swampland program, related conjectures and claims on geometric and physical properties of new classes of quasi-stationary Ricci flow and black hole solutions.

\vskip3pt

\textbf{Keywords:}\ nonassociative geometric flows; nonassociative gravity;
nonholonomic star product; R-flux deformations; Perelman F- and
W-functionals; geometric flow thermodynamics; exact and parametric
solutions; nonassociative black holes; modified swampland program; swampland
conjectures.
\end{abstract}

\tableofcontents

\section{Introduction, preliminaries, and motivations}

\label{sec1}

\subsection{On nonassociative geometric flows, gravity, new methods of
constructing exact/ para\-metric \newline solutions, and the swampland program}

One of the most important results in modern mathematics consists from the
proof of the Poincar\'{e}-Thurston conjecture due to Grigori Perelman \cite%
{perelman1,perelman2,perelman3}. The approach involved a study of geometric
flow evolution equations of Riemannian metrics introduced independently by
R. Hamilton \cite{hamilton82}, in mathematics, and D. Friedan \cite%
{friedan80}), in physics. We note that a new concept of W-entropy and a
respective statistical thermodynamic model for geometric flows \cite%
{perelman1} were elaborated, when the thermodynamic variables are defined in
terms of the Riemannian metric volume forms with a normalizing function and
using the Ricci tensor and scalar curvature. Such ideas and geometric
methods are very important for elaborating new directions and applications
in modern physics, cosmology and astrophysics, and quantum information
theory; see new results, research programs and references from \cite%
{vacaru20,ib19,bubuianu19}. Comprehensive reviews of advanced topological
and geometric analysis methods involved in the Ricci flow theory can be
found in monographs \cite{kleiner06,morgan06,cao06}.\footnote{%
Readers may familiarize themselves with the outstanding scientific and
social impact of such results presented, respectively, in a magazine article
and a YouTube clip:\ S. Nasar and D. Gruber, Manifold Destiny -- A legendary
problem and the battle over who solved it. The New Yorker, Annals of
Mathematics, August 28, 2006, \newline
https://www.newyorker.com/magazine/2006/08/28/manifold-destiny ; and
"Grigory Perelman documentary" [Russian with English subtitles]
https://www.youtube.com/watch?v=Ng1W2KUHI2s}

\vskip5pt A challenging problem in modern particle physics and gravity is to
develop the fundamental mathematical results on geometric flows and G.
Perelman's thermodynamics in a relativistic form, for metrics with Lorentz
signature and/or for non-Riemannian geometric objects derived, for instance,
for string modifications of Einstein gravity. There is a substantial
difference between the original mathematical methods with Riemannian metrics
and the directions for elaborating new methods and research on relativistic
geometric flows, supersymmetric and (non) commutative generalizations, and
recent applications \cite{v07,v08,ib19}, see also references therein.%
\footnote{%
Some hundred of works devoted to possible implications in modern physics of
the Ricci flow theory were published during the last 15 years and it is not
possible to analyse all new and original results in this research article.
We cite and discuss only a series of most relevant papers which motivate our
research program on "nonassociative geometric flows and applications in
physics and information theory" stated in \cite{partner01,partner02} and
allow to develop the geometric and analytic methods which are used for
constructing new classes of exact and parametric solutions of geometric
evolution and modified gravity field equations.} Even, at present, certain
variants of the Poincar\'{e}-Thurston conjecture for pseudo-Riemannian
metrics and/or generalized connections were not formulated and proven, we
can elaborate on self-consistent causal geometric flow evolution models in
physics and information theory if we apply the anholonomic frame and
connection deformation methods, AFCDM, and construct/select more general
classes of solutions with well-defined and verifiable classical and quantum
properties \cite{v00,lb18}.

\vskip5pt Geometric flow theories are closely related to the RG flow models
and underlying nonlinear sigma-models with beta-functions computed in a
framework of string gravity theory, or a modified/quantum gravity model with
ultraviolet, UV, completion and UV/ IR correspondence (IR, from infrared) 
\cite{gomez19}. In connection to this, we note the swampland program \cite%
{vaf05,oog06,pal19} which main goal is to elaborate rigorous criteria how to
distinguish the low-energy effective field theories that can be completed in
the UV from those that cannot. The swampland hypothesis in quantum gravity,
QG, the infinite distance conjecture and various other someway related
conjectures, were revisited recently in a series of works \cite%
{luest19,kehagias19,biasio20,biasio21,lueben21,biasio22} using (non)
commutative geometric flow, exact solutions in gravity theories, and quantum
field methods. Here we emphasize that to elaborate on rigorous algebraic and
geometric approaches to mathematical particle physics and QG, we have to
consider models of nonassociative quantum mechanics, QM, \cite%
{jordan32,jordan34} and further developments with nonassicative and
noncommutative algebras \cite%
{okubo,castro2,v08,mylonas13,kupriyanov15,gunaydin}. We cite \cite%
{szabo19,blumenhagen16,aschieri17,partner01,partner02,partner03,partner04}
for reviews and recent results on nonassociative and noncommutative geometry
and physics. An important task in string and M-theory, and modern quantum
field theory, QFT, and QG, is to extend the swampland program in a form
incorporating geometric and physical models with nonassociative/
noncommutative structures. \textit{A general scope of this work is to
investigate how exact and parametric solutions in nonassociative gravity can
be correlated to the infinite distance conjecture and corresponding
criteria/conjectures/ claims involving nontrivial running cosmological
constants, non-Riemannian and pseudo-Riemannian geometric flows, and
modified gravity theories, MGTs. } From a plethora of above mentioned
nonassociative and noncommutative geometric and field theories, we study an
explicit class of models with nonassociative twist deformations defined by 
R-flux deformations in string theory. In such an approach, we are able to
elaborate on physically well-defined geometric flow evolution and gravity
theories encoding nonassociative data, when the results are verifiable in
linear order on a small deformation parameter. Various classes of new
parametric solutions and applications in modern cosmology, astrophysics,
information thermodynamics etc. can be also considered. At the end of
subsection 7.2, we discuss the validity of our methods and claims and
further perspectives for general nonassociative theories. We argue that
fundamental geometric flow equations and important statistical thermodynamic
functionals can be always postulated in abstract geometric form and then all
order decompositions on deformation parameter can be performed. In the
linear approximation, certain additional variational principles can be
formulated and then recurrently extended to higher order decompositions on a
small deformation parameter.

\vskip5pt In \cite{blumenhagen16,aschieri17}, two quite similar and
self-consistent approaches to nonassociative gravity (defined by star
product R-flux deformations in string theory) were elaborated up to levels
of definition and parametric computation of the nonassociative Ricci tensor $%
\mathcal{R}ic^{\star }[\nabla ^{\star }]$ and corresponding curvature scalar 
$\mathcal{R}s^{\star }[\nabla ^{\star }].$\footnote{%
Formulas of type $...[\nabla ^{\star }]$ state a functional dependence
involving possible (star) products, partial derivations etc. We follow the
notations and conventions from \cite{partner01,partner02} generalizing for
arbitrary nonholonomic frames the formulas from \cite%
{blumenhagen16,aschieri17}. This allows us to use in the Introduction
section certain abstract nonassociative formulas if they are analogs of
associative and commutative ones. In result, the motivations and purposes of
the work can be formulated in a more compact form. Of course, in index and
coordinate/frame forms, which are necessary for finding solutions of
physically important systems of PDEs, the formulas for nonassociative
geometry/gravity are more sophisticate because they involve terms from star
product and R-flux deformations. Such computations will be considered in
next sections. In Appendix \ref{appendixb}, there are provided necessary
details and explanations. We recommend readers to study the mentioned works
and summaries of previous results before reading the main part of the
article.} Such nonassociative geometric objects are determined by a
nonassociative Levi-Civita, LC, connection $\nabla ^{\star }[\mathfrak{g}%
^{\star }]$, constructed as a nonlinear functional using nonassociative
symmetric, $g^{\star }$, and nonsymmetric, $\check{g}^{\star },$ components
of a star-metric, $\star $-metric, $\mathfrak{g}^{\star }=(g^{\star },\check{%
g}^{\star }).$ As in string gravity and M-theory \cite%
{mylonas13,kupriyanov15,gunaydin}, the nonassociative geometric
constructions and (vacuum) gravity theories involve nonassociative star, $%
\star $, product deformations computed for a prescribed Moyal--Weyl tensor
product and determined by non-geometric fluxes (R-fluxes). Such a $\star $%
--product allows us to define and compute nonassociative deformations of the
LC-connection in (pseudo) Riemannian geometry, $\nabla \rightarrow \nabla
^{\star }$, and (for more general constructions) to elaborate on various
models of nonassociative non-Riemannian geometry with nontrivial torsion and
non-metricity involving nonsymmetric metric tensors.

\vskip5pt

We follow in nonassociative geometry and gravity a symbolic abstract
geometric formalism \cite{aschieri17,partner01,partner02} which is similar
to that for GR \cite{misner} but (in our approach) is more formalized and
adapted to nonassociative and noncommutative geometric structures. It
simplifies the geometric constructions which in many cases are formal
re-definitions and more sophisticate transforms/ deformations of certain
fundamental geometric objects and formulas into similar ones with star
labels. The nonassociative vacuum gravitations equations can be postulated/
derived, and computed in abstract geometric form, or using cumbersome
nonholonomic frame / coordinate formulas, as respective star product R-flux
deformations of some standard commutative ones. In abstract (non)
associative geometric form, we can postulate the $\star$--deformed vacuum
Einstein equations: 
\begin{equation}
\ ^{\shortmid }\mathcal{R}ic^{\star }[\ ^{\shortmid }\nabla ^{\star }]=\
^{\shortmid }\Lambda _{0}\ ^{\shortmid }\mathfrak{g}^{\star },
\label{nonassocveq}
\end{equation}%
where $\ ^{\shortmid }\Lambda _{0}$ is a conventional cosmological constant.
The nonassociative geometric objects in these equations (i.e. $\ ^{\shortmid
}\mathfrak{g}^{\star },\ ^{\shortmid }\nabla ^{\star },$ and $\ ^{\shortmid }%
\mathcal{R}ic^{\star }$) are defined on a $\star $-product deformed phase
space $\ ^{\shortmid }\mathcal{M}=T^{\ast }V\rightarrow \ ^{\shortmid }%
\mathcal{M}^{\star },$ where $T^{\ast }V$ is the cotangent bundle of a
Lorentz manifold $V$.\footnote{%
To elaborate on nonassociative star product and R-flux deformations of the
Einstein equations in general relativity, GR, we consider a basic
associative and commutative spacetime manifold $V,$ of dimension $\dim V=4,$
for instance, enabled with metrics of signature $(+++-).$ We suppose that
readers are familiar with the basic concepts from the mathematical
relativity, and methods of constructing exact solutions, as well with the
formalism of (non) linear connections in (co) vector/ tangent bundle
geometry \cite{misner,hawking73,wald82,kramer03,partner01,partner02}. Here
we note that the label "$\ ^{\shortmid }$" is used in order to emphasize
that the geometric objects are defined on cotangent bundles and not on usual
tangent bundles $\ \mathcal{M}=TV,$ when respective geometric/physical
objects are written $\mathcal{R}ic^{\star }[\nabla ^{\star }]=\Lambda _{0}\ 
\mathfrak{g}^{\star }.$}

\vskip5pt

Similar assumptions and a respective abstract geometric formalism for star
product R-flux deformations can be used in order to postulate nonassociative
generalizations of the R. Hamilton equations \cite%
{hamilton82,friedan80,perelman1},%
\begin{equation}
\frac{\partial \ ^{\shortmid }\mathfrak{g}^{\star }}{\partial \tau }=-2\
^{\shortmid }\mathcal{R}ic^{\star }[\ ^{\shortmid }\nabla ^{\star }]+...,
\label{nonassocheq}
\end{equation}%
describing the flow evolution on $\ ^{\shortmid }\mathcal{M}$ of a family of
nonassociative metrics $\ ^{\shortmid }\mathfrak{g}^{\star }(\tau )$
parameterized by a positive parameter $0\leq \tau \leq \tau _{0}.$ In these
formulas, dots are used for additional terms which can be defined/computed
for respective star product R-flux deformations of a corresponding
variational calculus \cite{perelman1,kleiner06,morgan06,cao06,ib19}, see
references therein and details in section \ref{ssngfeq}.\footnote{%
Any point $\ ^{\shortmid }u=(x,p)=\{\ ^{\shortmid }u^{\alpha
}=(x^{i},p_{a})\}$ in a phase space $\ ^{\shortmid }\mathcal{M}$ is
parameterized by spacetime coordinates $x=(x^{i})$ and cofiber (momentum
like) coordinates $p=(p_{a}),$ which are dual to conventional velocities $%
v=(v^{a}),$ for indices $i,j,k,...=1,2,3,4$ and $a,b,c,...=5,6,7,8.$ In a
similar form, we label a point $u=(x,v)=\{u^{\alpha}=(x^{i},v^{a})\}$ in $%
\mathcal{M}$. In this work, small Greek indices run values $\alpha ,\beta
,...=1,2,...8,$ but they may take values up to 10, 11, ... for a
corresponding modified gravity theory, MGT, (super) string/ gravity models
etc.}

\vskip5pt The coefficients of geometric objects, effective and matter field
sources, and respective physically important geometric evolution/ dynamical
equations in such theories can be decomposed into real and complex terms and
parametric forms using decompositions on $\hbar ,$ the Planck constant, and $%
\kappa :=\ell _{s}^{3}/6\hbar ,$ the string constant, were $\ell _{s}$ is a
length parameter \cite{aschieri17}.\footnote{%
Nonassociative geometric and physical theories can be formulated in certain
forms encoding quasi-Hopf \cite{drinf,aschieri17,partner01} and/or
exceptional algebraic structures, for instance, with octonionic and Clifford
configurations \cite{okubo,castro2,gunaydin,szabo19}. For different
nonassociative algebraic and geometric configurations and respective
generalized nonassocitative and noncommutative differential and integral
calculi, we can elaborate on different types of classical and quantum
physical theories. In our works on nonassociative geometry and physics, we
follow an explicit approach with quasi-Hopf nonholonomic geometric
structures determined by nonassociative star products and R-fluxes \cite%
{blumenhagen16,aschieri17,partner01,partner02,partner03,partner04}.} In \cite%
{partner01,partner02}, the AFCDM was generalized for finding exact and
parametric solutions of systems of nonlinear partial differential equations,
PDEs, of type (\ref{nonassocveq}). We proved a general splitting and
integration property for quasi-stationary configurations (with Killing
symmetry on a time like coordinate) of such nonassociative nonlinear
dynamical systems and shown how to construct nonassociative four
dimensional, 4-d, and 8-d, black hole, BH, and black ellipsoid, BE,
solutions in \cite{partner02,partner03,partner04}. Generic off-diagonal
quasi-stationary solutions can be generated for $\kappa $-linear parametric
decompositions transforming the real part of (\ref{nonassocveq}) into
associative and commutative modified Einstein equations (see appendix \ref%
{appendixaa}), 
\begin{equation}
\ ^{\shortmid }\widehat{\mathbf{R}}_{\ \ \gamma _{s}}^{\beta _{s}}={\delta }%
_{\ \ \gamma _{s}}^{\beta _{s}}\ _{s}^{\shortmid }\mathcal{K}.
\label{modeinst}
\end{equation}%
In this formula, the canonical Ricci s-tensor $\ ^{\shortmid }\widehat{%
\mathcal{R}}ic=\{\ ^{\shortmid }\widehat{\mathbf{R}}_{\ \ \gamma
_{s}}^{\beta _{s}}\}$ is a tensor adapted to a so-called nonholonomic shell
structure, s-structure, with a conventional (2+2)+(2+2)-splitting containing
2-d shells labeled in abstract form by $s=1,2,3,4.$ It is defined by a
respective canonical s-connection%
\begin{equation}
\ ^{\shortmid }\widehat{\mathbf{D}}_{\gamma _{s}}=\ ^{\shortmid }\nabla
_{\gamma _{s}}+\ ^{\shortmid }\widehat{\mathbf{Z}}_{\gamma _{s}},
\label{distort}
\end{equation}%
where the distortion s-tensor $\ ^{\shortmid }\widehat{\mathbf{Z}}_{\gamma
_{s}}$ is such way chosen that the system of nonlinear PDEs (\ref{modeinst})
can be decoupled and integrated in general form with respect to a
correspondingly defined nonholonomic s-frames $\ ^{\shortmid }\mathbf{e}%
_{\gamma _{s}}$, see details in \cite{partner02}.\footnote{\label{fncoord}In
s-adapted form, we follow such a convention of indices and local
coordinates: For instance, $\beta _{2}=(j_{1},b_{2}),$ where $%
j_{1}=1,2;b_{2}=3,4,$ for the shell $s=2$ and the coordinate $u^{4}=y^{4}=t$
considered as a time like one, $t,$ but $u^{\beta
_{2}}=(x^{i_{1}},y^{3},y^{4}),$ for $(x^{i_{1}},y^{3})$ being space like
coordinates. In a similar form, we split the indices and coordinates, for
instance, on the shall $s=4,$ when $\beta _{4}=(j_{3},b_{4}),$ for $%
j_{3}=1,2,...6$ and $b_{4}7,8;$ and coordinates $\ _{4}^{\shortmid }u=\{\
^{\shortmid }u^{\beta
_{4}}=(x^{i_{1}},y^{3},y^{4}=t,p_{5},p_{6},p_{7},p_{8}=E)\},$ with $E$ being
a conventional energy type coordinate for a relativistic phase space $\ \
^{\shortmid }\mathcal{M}$. Here we note also that nonholonomic frames $\
^{\shortmid }\mathbf{e}_{\gamma _{s}}=\ ^{\shortmid }\mathbf{e}_{\ \gamma
_{s}}^{\gamma _{s}^{\prime }}\ ^{\shortmid }\mathbf{\partial }_{\gamma
_{s}^{\prime }}$ can be related via frame transforms, using matrices $\
^{\shortmid }\mathbf{e}_{\ \gamma _{s}}^{\gamma ;_{s}},$ to local coordinate
bases $\ ^{\shortmid }\mathbf{\partial }_{\gamma _{s}^{\prime }},$ when the
Einstein convention on repeating "up-low" indices is applied. In a similar
form, we can consider s-splitting of dual frames $\ ^{\shortmid }\mathbf{e}%
^{\beta _{s}},$ when $\ ^{\shortmid }\mathbf{e}^{\beta _{s}}$ $\ ^{\shortmid
}\mathbf{e}_{\gamma _{s}}=\ \delta _{\ \ \gamma _{s}}^{\beta _{s}},$ with $%
\delta _{\ \ \gamma _{s}}^{\beta _{s}}$ being the Kronecker symbol.} The
effective sources in (\ref{modeinst}) encodes contributions from
nonassociative $\star $-product with nontrivial R-flux terms, $\mathcal{R}%
_{\quad \cdot }^{\cdot \cdot }\sim $ $\mathcal{R}_{\quad \alpha _{2}}^{\tau
_{s}\xi _{s}},$ being parameterized in s-adapted form as $\ ^{\shortmid }%
\mathbf{K}_{\ \ \gamma _{s}}^{\beta _{s}}=\delta _{\ \ \gamma _{s}}^{\beta
_{s}}\ \ _{s}^{\shortmid }\mathcal{K}(\kappa ,\hbar ,\mathcal{R}_{\quad
\cdot }^{\cdot \cdot },\ _{s}^{\shortmid }u)$. General classes of solutions
(they are generic off-diagonal because such solutions can't be diagonalized
in a general form in a finite phase space region via coordinate transforms)
are determined by s-metrics 
\begin{eqnarray}
g=\ _{s}^{\shortmid }\mathbf{g} &=&(h_{1}\ ^{\shortmid }\mathbf{g},~v_{2}\
^{\shortmid }\mathbf{g},\ c_{3}\ ^{\shortmid }\mathbf{g,}c_{4}\ ^{\shortmid }%
\mathbf{g})\in T\mathbf{T}^{\ast }\mathbf{V}\otimes T\mathbf{T}^{\ast }%
\mathbf{V}  \label{sdm} \\
&=&\ ^{\shortmid }\mathbf{g}_{\alpha _{s}\beta _{s}}(\ _{s}^{\shortmid }u)\
\ ^{\shortmid }\mathbf{e}^{\alpha _{s}}\otimes _{s}\ ^{\shortmid }\mathbf{e}%
^{\beta _{s}}=\{\ \ ^{\shortmid }\mathbf{g}_{\alpha _{s}\beta _{s}}=(\ \
^{\shortmid }\mathbf{g}_{i_{1}j_{1}},\ \ ^{\shortmid }\mathbf{g}%
_{a_{2}b_{2}},\ \ ^{\shortmid }\mathbf{g}^{a_{3}b_{3}},\ \ ^{\shortmid }%
\mathbf{g}^{a_{4}b_{4}})\}.  \notag
\end{eqnarray}%
adapted to a corresponding nonlinear connection, N-connection, structure and
for respective s-adapted tensor products $\otimes _{s},$ see geometric
preliminaries in next section.

\vskip5pt The AFCDM allows us to construct/find various classes of
quasi-stationary solutions of (\ref{modeinst}), and of (\ref{nonassocveq}),
when the coefficients of (\ref{sdm}) are computed as functionals 
\begin{equation}
\ ^{\shortmid }\mathbf{g}_{\alpha _{s}\beta _{s}}(\ _{s}^{\shortmid }u)\ =\
^{\shortmid }\mathbf{g}_{\alpha _{s}\beta _{s}}[\ _{s}^{\shortmid }\mathcal{K%
},\ _{s}^{\shortmid }\Psi \ ]=\ ^{\shortmid }\mathbf{g}_{\alpha _{s}\beta
_{s}}[\ _{s}^{\shortmid }\mathcal{K},\ _{s}^{\shortmid }\Phi ,\
_{s}^{\shortmid }\Lambda _{0}\ ]  \label{smfunct}
\end{equation}%
determined by effective sources $\ _{s}^{\shortmid }\mathcal{K}$ and
generating functions $\ _{s}^{\shortmid }\Psi (\kappa ,\hbar ,\mathcal{R}%
_{\quad \cdot }^{\cdot \cdot },\ _{s}^{\shortmid }u)$, or $\ _{s}^{\shortmid
}\Phi (\kappa ,\hbar ,\mathcal{R}_{\quad \cdot }^{\cdot \cdot },\
_{s}^{\shortmid }u)$.\footnote{\label{fnreal}we note that in this work there
are used for explicit computations certain real co-fiber coordinates with
labels "$\ _{s}^{\shortmid }$" even in nonassociative gravity \cite%
{aschieri17,partner01,partner02} it is necessary to consider for various
purposes complex coordinates with labels $~$"$_{s}^{\shortparallel }"$} In
general, such solutions depend also on integration functions and constants,
effective sources, broken (non) linear symmetries etc., which should be
prescribed/defined from certain experimental/ observational data, boundary
conditions etc. A very important property of the functionals for s-metrics (%
\ref{smfunct}) is that they posses certain general nonlinear symmetries for
generating functions and effective sources and give possibilities to
introduce into consideration some effective cosmological constants $\
_{s}^{\shortmid }\Lambda _{0}$ on any shell $s=1,2,3,4.$ For nonassociative
vacuum gravitational equations, such a proof is provided in section 5.4 of 
\cite{partner02}. Here, we present, for simplicity, only the formulas for
the spacetime shell $s=2,$ for changing the generating data, $\ (\ _{2}\Psi
(\hbar ,\kappa ,x^{i_{1}},y^{3}),\ ~_{2}^{\shortmid }\mathcal{K}(\hbar
,\kappa ,x^{i_{1}},y^{3}))\leftrightarrow (\ _{2}\Phi (\hbar ,\kappa
,x^{i_{1}},y^{3}),\ _{2}\Lambda _{0}),$ (see appendix \ref{bassnsym} with
formulas for all shells $s=1,2,3,4$), 
\begin{eqnarray}
\frac{\partial _{3}[(\ _{2}\Psi )^{2}]}{\ ~_{2}^{\shortmid }\mathcal{K}} &=&%
\frac{\partial _{3}[(\ _{2}\Phi )^{2}]}{\ _{2}\Lambda _{0}},%
\mbox{ which can be
integrated as  }  \notag \\
(_{2}\Phi )^{2} &=&\ _{2}\Lambda _{0}\int dy^{3}(~_{2}^{\shortmid }\mathcal{K%
})^{-1}\partial _{3}[(\ _{2}\Psi )^{2}]\mbox{ and/or }(\ _{2}\Psi )^{2}=(\
_{2}\Lambda _{0})^{-1}\int dy^{3}(~_{2}^{\shortmid }\mathcal{K})\partial
_{3}[(\ _{2}\Phi )^{2}].  \label{ntransf2}
\end{eqnarray}%
We can chose in such formulas $\ _{s}^{\shortmid }\Lambda _{0}=\Lambda
_{0}=const,$ or to study running cosmological constants $\ _{s}^{\shortmid
}\Lambda (\tau )$, for instance, on a geometric flow parameter $\tau $; and
phase space polarizations via $\ _{s}^{\shortmid }\Lambda _{0}\rightarrow \
_{s}^{\shortmid }\Lambda (\tau ,\ _{s}^{\shortmid }u)$ for more general
considerations. Such nonlinear transforms allow us to rewrite equivalently
the nonassociative modified vacuum gravitational equations (\ref{modeinst})
in a form with effective constants in the right part and redefined
functional dependence of the s-metrics in the Ricci s-tensor%
\begin{equation}
\ ^{\shortmid }\widehat{\mathbf{R}}_{\ \ \gamma _{s}}^{\beta
_{s}}~[~_{s}^{\shortmid }\mathcal{K}(\kappa ,\hbar ,\mathcal{R}_{\quad \cdot
}^{\cdot \cdot },\ _{s}^{\shortmid }u),\ _{s}^{\shortmid }\Phi (\kappa
,\hbar ,\mathcal{R}_{\quad \cdot }^{\cdot \cdot },\ _{s}^{\shortmid }u)]=\
\delta _{\ \ \gamma _{s}}^{\beta _{s}}\ \ _{s}^{\shortmid }\Lambda _{0}.
\label{nonassocrsol}
\end{equation}%
These equations define a particular class of nonholonomic Ricci solitons
encoding nonassociative data, see section \ref{ssngfeq}.

\vskip5pt We emphasize that using a class of quasi-stationary solutions $\
^{\shortmid }\mathbf{g}_{\alpha _{s}\beta _{s}}[\ _{s}^{\shortmid }\mathcal{K%
},\ _{s}^{\shortmid }\Phi ,\ _{s}^{\shortmid }\Lambda _{0}\ ]$ (\ref{smfunct}%
) we can compute in general functional form respective components of a
nonassociative $\star $-metric, $\mathfrak{g}^{\star }[...]=(g^{\star }[...],%
\check{g}^{\star }[...])$ as it is explained in \cite{partner02} and, for
respective 4-d and 8-d BH/BE solutions, in \cite{partner03,partner04}. In
result, we generate parametric solutions for nonassociative vacuum Einstein
equations (\ref{nonassocveq}) if we restrict the geometric constructions for
a subclass of generating functions when the distortion s-tensor $\
^{\shortmid }\widehat{\mathbf{Z}}_{\gamma _{s}}$ (\ref{distort}) is
constrained to zero and $\ ^{\shortmid }\widehat{\mathbf{D}}_{\gamma
_{s}}\rightarrow \ ^{\shortmid }\nabla _{\gamma _{s}}.$\footnote{\label%
{fndual}Here we note that in a similar (in certain sense) dual form one can
be constructed nonassociative locally anisotropic and inhomogeneous
cosmological solutions with Killing symmetry, for instance, on $\partial
_{3},$ when the s-metrics in certain s-adapted frame do not depend on
coordinate $y^{3}$ but depend on $y^{4}=t$ and other space and (co) fiber
coordinates. In associative and commutative forms, such solutions are
discussed in \cite{lb18,bubuianu19,vacaru20}, see references therein.
Cosmological models encoding nonassociative star product R-flux deformed
data will be studied in our future works as it was stated in \cite{partner01}%
, see also query Q5 at the end of conclusion section.}

\subsection{The structure, aims, and the main hypothesis of the paper}

\label{ssint12}This work is a natural and logical development of a series of
articles on nonassociative geometry and physics \cite%
{szabo19,blumenhagen16,aschieri17} and a new research program on
nonassociative geometric and quantum information flows and gravity \cite%
{partner01,partner02,partner03,partner04}. It is related to the swampland
program \cite{vaf05,oog06,pal19} and revised conjectures \cite%
{kehagias19,biasio20,biasio21,lueben21,biasio22} following such five aims:

\vskip5pt \textbf{The first} aim stated for section \ref{sec2} is to prove
that using nonassociative star products determined by R-flux deformations we
can formulate and provide a physical motivation for the generalized
nonassociative R. Hamilton equations (\ref{nonassocheq}). In general, such
nonassociative geometric flow equations can be derived for certain canonical
data $\left[ \ _{s}^{\shortmid }\mathfrak{g}^{\star }(\tau ),\
_{s}^{\shortmid }\widehat{\mathbf{D}}^{\star }(\tau )\right] $ stated for
generalized nonassociative G. Perelman F- and W-functionals, $\
_{s}^{\shortmid }\widehat{\mathcal{F}}^{\star }(\tau )$ and $\
_{s}^{\shortmid }\widehat{\mathcal{W}}^{\star}(\tau )$. Formulas (54) in 
\cite{partner04} present $\kappa $-linear parametric versions of such
functionals encoding string star product R-flux data which, in this work,
are generalized to describe nonassociative geometric flow evolution and/or
nonholonomic Ricci solitons (\ref{nonassocrsol}) (in particular, they
include nonassociative modified vacuum equations).

\vskip5pt \textbf{The second} aim, for section \ref{sec3}, is to elaborate
on statistical thermodynamic models for nonassociative geometric flows with
thermodynamic variables derived from the W-entropy $\ ^{\shortmid }\widehat{%
\mathcal{W}}^{\star }(\tau ).$ We extend the constructions from \cite%
{partner04} and provide general formulas and respective $\kappa $-linear
parametric versions of effective canonical energy, $\ _{s}^{\shortmid }%
\widehat{\mathcal{E}}^{\star }(\tau )\rightarrow \ _{s}^{\shortmid }\widehat{%
\mathcal{E}}(\tau );$ entropy, $\ _{s}^{\shortmid }\widehat{\mathcal{S}}%
^{\star }(\tau )\rightarrow \ _{s}^{\shortmid }\widehat{\mathcal{S}}(\tau );$
and quadratic fluctuations, $\ _{s}^{\shortmid }\widehat{\sigma }^{\star
}(\tau )\rightarrow \ _{s}^{\shortmid }\widehat{\sigma }(\tau ),$ all
encoding nonassociative star and R-flux deformed data. We emphasize that the
off-diagonal quasi-stationary and BH/\ BE solutions in nonassociative
gravity are not described, in general, in the framework of the
Bekenstein-Hawking thermodynamics paradigm for BHs \cite%
{bek1,bek2,haw1,haw2,misner,hawking73,wald82} if there are not defined
certain hypersurface horizons, duality and holography conditions. Contrary
to this, modified versions of G. Perelman W-entropy $\ ^{\shortmid }\widehat{%
\mathcal{W}}(\tau)$ and geometric-statistic thermodynamic entropy $\
_{s}^{\shortmid }\widehat{\mathcal{S}}(\tau )$ can be defined and computed
for any class of exact/ parametric solutions in GR and MGTs.

\vskip5pt Then,\textbf{\ the third} aim, in section \ref{sec4}, is to extend
the AFCDM \cite{v00,lb18,partner02,partner03,partner04} in such forms which
allows us to decouple and integrate the nonassociative R. Hamilton equations
in certain generalized forms describing the geometric evolution of
quasi-stationary Ricci soliton and vacuum gravitational structures (with
effective sources encoding star product and R-flux deformations). We analyze
a class of nonlinear symmetries relating different types of generating
functions, generating (effective) sources and cosmological constants running
on temperature like geometric flow parameter. There are provided explicit
formulas for quadratic linear elements for nonassociative geometric
evolution in $\kappa $-linear parametric forms of quasi-stationary generic
off-diagonal metrics and gravitational polarization functions.

\vskip5pt \textbf{The forth} aim, in section \ref{sec5}, is to elaborate on
parametric geometric flows and related thermodynamics models of
quasi-stationary solutions describing nonassociative evolution of star
R-flux deformed BHs. Additionally to the classes of nonassociative BH
solutions of Tangherlini and double BH phase space solutions studied in \cite%
{partner04}, we construct two other types nonassociative generic
off-diagonal $\kappa $-linear parametric BHs subjected to geometric
evolution. There are analyzed nonassociative phase space double
Schwarzschild--AdS black ellipsoid, BE, configurations and nonassociative
flows of phase space Reisner-Nordstr\"{o}m BHs. We show also how to compute
G. Perelman thermodynamic variables for generic off-diagonal solutions
related via nonlinear symmetries (\ref{ntransf2}) to effective running on
temperature $\tau $ cosmological constants $\ _{s}^{\shortmid
}\Lambda(\tau). $ Then, there are stated the conditions when the concept
Bekenstein-Hawking entropy can be applied for certain particular examples of
such nonassociative BHs and their deformations which may have different
physical interpretations.

\vskip5pt \textbf{The fifth} aim, in section \ref{sec6}, is to study how the
swampland program should be generalized/modified in order to include
nonassociative geometric flows and related exact/ parametric solutions. We
note that the main goal of the swampland program is to elaborate certain
criteria which allow to distinguish low-energy effective field theories
which can be completed into QG in the UV from those theories that cannot. In
fact, for the AdS spaces, the Ricci flow swampland conjecture is equivalent
to the anti- de Sitter distance conjecture (ADS) \cite%
{oog06,kehagias19,biasio20}. Swampland conjectures were studied recently in
connection to BH physics, extra dimensions and geometric flow conjectures,
when the concept of Bekenstein-Hawking entropy was applied \cite%
{biasio21,lueben21,biasio22}. A modern approach to QG and string and
M-theory involves models with nonassocitative structures for QM, QFT and
MGTs. This modifies substantially the mathematical formalism and methods for
constructing exact/parametric solutions and quantization and providing a
physical interpretation of nonassociative geometric flows and gravity. The
Bekenstein--Hawking thermodynamic paradigm should be completed with a more
general one which is based on the concept of G. Perelman W-entropy and
derived (non) associative/ commutative geometric thermodynamics.

\vskip5pt In section \ref{sec7}, we discuss and conclude the main results
based on: \newline
\textbf{The Main Hypothesis, MH,} \emph{of this work is that the Swampland
Program and related conjectures have to be generalized and modified
following above Aims 1-5 (objectives of this work) with the purpose to
formulate well-defined criteria how to include nonassociative and
noncommutative geometric flows, QM, QFTs, and MGTs in elaborating QG
theories related to M-theory and string gravity. Self-consistent geometric
and physical models and solutions should encode at least in parametric form
certain nonassociative star product and R-flux data in low-energy limits of
corresponding effective geometric flow evolution and field theories which
can be completed into QG in the UV forms and distinguished from another
classes of theories which do not have such properties.}

\vskip5pt In section \ref{appendixa}, we outline the main concepts and most
important formulas on nonassociative differential geometry with symmetric
and nonsymmetric metrics and (non) linear connections \cite%
{blumenhagen16,aschieri17,partner01,partner02,partner03,partner04}. Finally,
in Appendix \ref{appendixb}, we summarize the main ideas and steps for
constructing generic off-diagonal quasi-stationary and BH solutions using
the AFCDM \cite{v00,lb18,partner02,partner03,partner04}. Such formulas
describe as particular cases exact/parametric solutions for nonassociative
Ricci solitons and quasi-stationary gravitational polarizations for $\kappa $%
-linear geometric flows studied in the main text of the paper.

\section{A model of nonassociative Ricci flows with star product R-flux
deformations}

\label{sec2}

G. Perelman elaborated his approach to the Ricci flow theory \cite{perelman1}
by introducing (postulating) the concepts of F- and W-functionals for a
familiy of geometric flows of of Riemannian metrics $\ ^{n}g(\tau
)=\{g_{ij}(\tau )\simeq g_{ij}(\tau ,x^{k})\}$ on a closed manifold $\
^{n}V, $ $\dim \ ^{n}V=n$ (in this work, we can consider $n=3$),%
\begin{equation}
F(\tau )\simeq F[\tau ,\ ^{n}Rsc(\tau ),\ ^{n}\nabla (\tau ),\ ^{n}g(\tau
),f(\tau )]\mbox{ and }W(\tau )=W[\tau ,\ ^{n}Rsc(\tau ),\ ^{n}\nabla (\tau
),\ ^{n}g(\tau ),f(\tau )].  \label{perelmfst}
\end{equation}%
In such formulas, it is used a flow parameter $\tau ,0\leq \tau \leq \tau
_{0},$ which can be treated as \textbf{temperature}; the scalar curvatures $%
\ ^{n}Rsc(\tau )=\ ^{n}Rsc(\tau ,x^{k})$ are determined by a corresponding
family of LC--connections $\ ^{n}\nabla (\tau )=\ ^{n}\nabla (\tau ,x^{k})$.
A normalizing function $f(\tau )\simeq f(\tau ,x^{k})$ \ is used for
defining integration measures $\left( 4\pi \tau \right) ^{-n/2}e^{-\ f}d%
\mathcal{V}ol(\tau )$ with volume elements $d\mathcal{V}ol(\tau )=\sqrt{|\
^{n}g(\tau )|}\ d^{n}\ x^{i}.$ It should be emphasized that a type of $%
f(\tau )$ may have different implications and interpretations in topology
and/or geometric analysis, and differential geometry theories.\footnote{%
\label{fnnormf}We can fix it, for instance, in order to elaborate certain
geometric/ thermodynamic models for some prescribed topological
configurations, or to simplify the procedure of finding exact/parametric
solutions of respective geometric flow/ Ricci soliton equations and
associated thermodynamic values \cite%
{lb18,ib19,bubuianu19,vacaru20,partner04}. In a series of recent works \cite%
{kehagias19,biasio20,biasio21,lueben21,biasio22}, respective physical
applications are elaborated for $f(\tau ,x^{k})$ considered as
scalar/Higgs/moduli fields.}

\vskip5pt Using a (3+1) splitting on 4-d, Lorentz manifolds, the functionals
(\ref{perelmfst}) can be generalized in relativistic form, which allows us
to prove (using standard variational procedures, or abstract geometric
methods) respective generalizations of the R. Hamilton equations \cite%
{hamilton82} and elaborate on relativistic geometric thermodynamic models.
This does not results in a formulation and proof of a relativistic version
of the Poincar\'{e}--Thorston conjecture. Nevertheless, we can elaborate on
important physical models for certain classes of solutions of relativistic
flow equations with well-defined causal evolution and describing important
physical processed. Such solutions can be found in nonassociative geometric
(flow) and gravity theories using the AFCDM \cite%
{partner02,partner03,partner04}.

\vskip5pt In \cite{blumenhagen16,aschieri17}, the nonassociative geometry
and gravity with star product and R-flux deformations were formulated up to
defining and computing nonassociative versions and for $\kappa $-parametric
decompositions of respective Ricci tensors and scalar curvatures defined by
nonassociative Levi Civita, LC, connections. In principle, those
constructions allow to elaborated on nonassociative versions of G. Perelmans
functionals (\ref{perelmfst}). Applying rigorous geometric methods and
respective variational procedures generalizing the geometric analysis
formalism from \cite{perelman1,kleiner06,morgan06,cao06,ib19}, we can derive
respective nonassociative geometric flow equaitons. Such star deformed R.
Hamilton equations and, in particular nonassociative Ricci solitons and/or
nonassociative vacuum Einstein equations consists very sophisticate systems
of nonlinear partial differential equations, PDEs. It is a very difficult
technical problem to solve and analyse possible physical implications of
such nonassociative geometric evolution and/or star deformed dynamical
gravitational field equations. In \cite{partner02}, we proved that
nonassociative vacuum Einstein equations can be decoupled and intergrated in
very general forms for the so-called canonical s-connection structure. The
same geometric methods can be generalized for generating solutions of
nonassociative geometric flow equations. We shall study such applications of
the AFCDM in section \ref{sec4}. The goal of this section is to outline
necessary geometric methods and formulate a nonassociative generalization of
F- and W-functionals (\ref{perelmfst}) is some forms which allow to derive
nonassociative versions of R.\ Hamilton equations which can be solved in
certain general off-diagonal forms of nonassociative metrics and (non)
linear connections which can be nonholonomically constrained to
LC-configurations.

\subsection{Nonassociative differential geometry with (non) linear
connections}

\label{appendixa}The nonassociative differential geometry on phase spaces
enabled with a nonholonomic dyadic shell adapted (s-adapted) star product
and R-flux deformations and respective symmetric and nonsymmetric metric and
(non) linear connection structures \cite{partner01,partner02} is reviewed.
Some necessary definitions and constructions from \cite%
{blumenhagen16,aschieri17} are also considered but in a form which will
allow extensions to so-called exactly and $\kappa $-parametric solvable
nonassociative geometric flow models. We follow both an abstract (index and
coordinate free) description of nonassociative geometric objects and
formulas and present certain s-adapted frame (co) bases and index formulas
which are used for providing exact/ parametric solutions and swampland
conjectures in the sections \ref{sec4}-\ref{sec6}. It is supposed that
readers are familiar with the main concepts on mathematical relativity and
(non) linear connection formalism described in \cite%
{misner,hawking73,wald82,lb18}. The notations and definitions were stated in
partner works \cite{partner01,partner02,partner03,partner04}.\footnote{%
Unfortunately, it is not possible to simplify such notations because we have
to distinguish various abstract and frame constructions in associative/
commutative nonholonomic geometry, their nonassociative and nonassociative
generalizations, $\kappa $-parametric \ decompostions and generating
solutions with dependences on spacetime and momentum type coordinates (which
in nonassociative geometry can be complex, or almost complex), star product
and R-flux deformations of certain prime metrics into noncommutative ones
with symmetric and nonsymmetric components, consider nonlinear symmetries
etc. Here we also note that there are different styles/ traditions for
notations in geometric and statistical thermodynamics, MGTs, nonassociative
geometry and coventions for constructing solutions in the theory of
nonlinear PDEs. We have to use various abstract left/write, up/low labels,
boldface symbols and shell coordinates in orde to show how the AFCDM can be
applied and physically important thermodynamic variables can be computed.}

\subsubsection{Associative and commutative dyadic and nonlinear connection
formalism}

\label{appendixaa} The associative and commutative geometric arena consists
from a phase space modeled as a cotangent Lorentz bundle $\mathcal{M}%
=T^{\ast }V$ on a spacetime manifold $V$ of signature $(+++-)$. Such a phase
space can be enabled with conventional (2+2)+(2+2) splitting determined by a
nonholonomic (equivalently, anholonomic/non-integrable) dyadic, 2-d,
decomposition into four oriented shells $s=1,2,3,4$ (in brief,
s-decomposition). A s-splitting is defined by a nonlinear connection,
N-connection (equivalently, s-connection), structure: 
\begin{eqnarray}
\ _{s}^{\shortmid }\mathbf{N}:\ \ _{s}T\mathbf{T}^{\ast }\mathbf{V} &=&\
^{1}hT^{\ast }V\oplus \ ^{2}vT^{\ast }V\oplus \ ^{3}cT^{\ast }V\oplus \
^{4}cT^{\ast }V,\mbox{ which is dual to }  \notag \\
\ _{s}\mathbf{N}:\ \ _{s}T\mathbf{TV} &=&\ ^{1}hTV\oplus \ ^{2}vTV\oplus \
^{3}vTV\oplus \ ^{4}vTV,\mbox{  for }s=1,2,3,4.  \label{ncon}
\end{eqnarray}%
We use $\ ^{1}h$ for a conventional 2-d shell (dyadic) splitting on
cotangent bundle, with $x^{i_{1}\text{ }}$local coordinates; then $\ ^{2}v$
for a 2-d vertical like splitting with $y^{a_{2}}$ coordinates on the shell $%
s=2;$ at the next shell $s=3,$ the splitting is convetional co-vertical, we
write $\ ^{3}c$ and use local coordinates $p_{a_{3}};$ for the 4th shell $%
s=4,$ the respective symbols are $\ ^{4}c$ and $p_{a_{4}}.$ Such s-splitting
will allow to decouple and integrate in general off-diagonal form
nonassociative geometric and physically important systems of nonlinear PDEs.
In a local coordinate basis (see conventions from footnotes \ref{fncoord}, %
\ref{fnreal}, and \ref{fndual}), a nonlinear s-connection defined by a
Whitney sum $\oplus $ as in (\ref{ncon}), for instance, is characterized by
coefficients $\ _{s}^{\shortmid }\mathbf{N}=\{\ ^{\shortmid }N_{\
i_{s}a_{s}}(\ ^{\shortmid }u)\},$ for $u=(x,p)=\ ^{\shortmid }u=(\ _{1}x,\
_{2}y,\ _{3}p,\ _{4}p)$. Such coefficients allow us to construct N-elongated
bases (N-/ s-adapted bases) as linear N-operators: 
\begin{equation}
\ ^{\shortmid }\mathbf{e}_{\alpha _{s}}[\ ^{\shortmid }N_{\ i_{s}a_{s}}]=(\
^{\shortmid }\mathbf{e}_{i_{s}}=\ \frac{\partial }{\partial x^{i_{s}}}-\
^{\shortmid }N_{\ i_{s}a_{s}}\frac{\partial }{\partial p_{a_{s}}},\ \
^{\shortmid }e^{b_{s}}=\frac{\partial }{\partial p_{b_{s}}})\mbox{ on }\
_{s}T\mathbf{T}_{\shortmid }^{\ast }\mathbf{V,}  \label{nadapbdsc}
\end{equation}%
and, dual s-adapted bases, s-cobases,%
\begin{equation}
\ ^{\shortmid }\mathbf{e}^{\alpha _{s}}[\ ^{\shortmid }N_{\ i_{s}a_{s}}]=(\
^{\shortmid }\mathbf{e}^{i_{s}}=dx^{i_{s}},\ ^{\shortmid }\mathbf{e}%
_{a_{s}}=d\ p_{a_{s}}+\ ^{\shortmid }N_{\ i_{s}a_{s}}dx^{i_{s}})\mbox{ on }\
\ _{s}T^{\ast }\mathbf{T}_{\shortmid }^{\ast }\mathbf{V.}  \label{nadapbdss}
\end{equation}%
Such s-frames are not integrable, i.e. nonholonomic (equivalently,
anholonomic) because, in general, they satisfy certain anholonomy
conditions, $\ ^{\shortmid }\mathbf{e}_{\beta _{s}}\ ^{\shortmid }\mathbf{e}%
_{\gamma _{s}}-\ ^{\shortmid }\mathbf{e}_{\gamma _{s}}\ ^{\shortmid }\mathbf{%
e}_{\beta _{s}}=\ ^{\shortmid }w_{\beta _{s}\gamma _{s}}^{\tau _{s}}\
^{\shortmid }\mathbf{e}_{\tau _{s}}$, see details in \cite%
{partner01,partner02}. For a 4+4 splitting, we write, for instance, $\
^{\shortmid }\mathbf{N}=\{\ ^{\shortmid }N_{\ ia}(x^{j},p_{b})\},$ and use
the term N-connection. We shall put a left label $s$ for corresponding
spaces and geometric objects (labeled with bold letters if they are written
in a N-adapted form) if, for instance, a phase space is enabled with a
s-adapted dyadic structure, $\ _{s}\mathcal{M},$ and use terms like
s-tensor, s-metric, s-connection etc.

The geometric s-objects and respective formulas (\ref{ncon})-(\ref{nadapbdss}%
) can be generalized for additional running on a geometric flow evolution
parameter $\tau ,$ for instance, writting $\ ^{\shortmid }\mathbf{N(}\tau 
\mathbf{)\simeq }\ ^{\shortmid }\mathbf{N(}\tau ,\ ^{\shortmid }u\mathbf{)}%
=\{\ ^{\shortmid }N_{\ ia}(\tau )\mathbf{\simeq }\ ^{\shortmid }N_{\
ia}(\tau ,x^{j},p_{b})\}$ and, respectively, $\ ^{\shortmid }\mathbf{e}%
_{\alpha _{s}}(\tau ),\ ^{\shortmid }\mathbf{e}^{\alpha _{s}}(\tau ),$ etc.
For running of geometric/ physical objects, we shall write only the $\tau $%
-dependence if that will not result in ambiguities. Here, we note that in a
similar form we can introduce and write formulas for geometric objects on $\
\ _{s}T\mathbf{TV,}$ i.e. when the total space coordinates are of
spacetime-velocity type. In such case, we omit the labels "$\ ^{\shortmid }"$
and write, for instance, $\ \mathbf{e}_{\alpha _{s}}(\tau )$ and$\ \mathbf{e}%
^{\alpha _{s}}(\tau ),$ for local coordinates $u=(x,v)=\ ^{\shortmid }u=(\
_{1}x,\ _{2}y,\ _{3}v,\ _{4}v).$ In general, such coordinates are not just
dual like fibe and co-fiber ones but may include certain Legendre transforms 
\cite{vbv18,bubuianu19}. In this work, we shall work on nonassociative phase
spaces as in \cite{blumenhagen16,aschieri17} and \cite%
{partner01,partner02,partner03,partner04} using labels "$\ ^{\shortmid }"$
in order to follow an unified system of notations which will allow in furthe
partner works to elaborate on nonassociative models of Finsler-Lagrange
spaces, which are important in quantum information theory.

A metric field in a phase space $\mathcal{M}$ is a second rank symmetric
tensor $\ ^{\shortmid }g=\{\ ^{\shortmid }g_{\alpha \beta }\}\in TT^{\ast
}V\otimes TT^{\ast }V$ of local signature $(+,+,+,-;+,+,+,-).$ It can be
written in equivalent form as a s-metric $\ _{s}^{\shortmid }\mathbf{g}=\{\
^{\shortmid }\mathbf{g}_{\alpha _{s}\beta _{s}}\}$ (\ref{sdm}). For $\tau $%
-families of pahse space metric and s-metrics , $\ $we shall use notations
of type $\ ^{\shortmid }g(\tau )=\{\ ^{\shortmid }g_{\alpha \beta }(\tau )\} 
$ and, respectively, $\ _{s}^{\shortmid }\mathbf{g}(\tau )=\{\ ^{\shortmid }%
\mathbf{g}_{\alpha _{s}\beta _{s}}(\tau )\}$

Another important geometric concept is that of s-connection with a
(2+2)+(2+2) splitting (we use the term distinguished connection,
d-connection, for a (4+4)-splitting), which is a linear connection
preserving under parallel transports a respective N-connection splitting (%
\ref{ncon}):%
\begin{equation}
\ _{s}^{\shortmid }\mathbf{D}=(h_{1}\ ^{\shortmid }\mathbf{D},\ v_{2}\
^{\shortmid }\mathbf{D},\ c_{3}\ ^{\shortmid }\mathbf{D},\ c_{4}\
^{\shortmid }\mathbf{D})=\{\ ^{\shortmid }\Gamma _{\ \ \beta _{s}\gamma
_{s}}^{\alpha _{s}}\},  \label{sdc}
\end{equation}%
where indices split into respective dyadic components of a respective $%
h_{1},v_{2},c_{3},c_{4}$ decomposition. Using standard definitions from
differential geometry, we can introduce and compute in standard form%
\footnote{%
see details, proofs, and references in \cite{lb18,ib19,bubuianu19,vacaru20},
when the coefficient formulas are provided in s-adapted forms with respect
to s-frames (\ref{nadapbdsc}) and (\ref{nadapbdss}); a number of important
abstract and coefficient formulas with nonassociative generalizations are
contained in \cite{partner01,partner02,partner03,partner04}} as for any
linear connection but for s-adapted $\ _{s}^{\shortmid }\mathbf{D}$ such
fundamental geometric s-objects:%
\begin{eqnarray}
\ _{s}^{\shortmid }\mathcal{T} &=&\{\ ^{\shortmid }\mathbf{T}_{\ \beta
_{s}\gamma _{s}}^{\alpha _{s}}\},\mbox{ the s-torsion };\   \label{mafgeomob}
\\
\ _{s}^{\shortmid }\mathcal{R} &=&\{\ ^{\shortmid }\mathbf{R}_{\ \beta
_{s}\gamma _{s}\delta _{s}}^{\alpha _{s}}\},%
\mbox{ the Riemannian
s-curvature };  \notag \\
\ _{s}^{\shortmid }\mathcal{R}ic &=&\{\ ^{\shortmid }\mathbf{R}_{\ \beta
_{s}\gamma _{s}}:=\ ^{\shortmid }\mathbf{R}_{\ \beta _{s}\gamma _{s}\alpha
_{s}}^{\alpha _{s}}\neq \ ^{\shortmid }\mathbf{R}_{\ \gamma _{s}\beta
_{s}}\},\mbox{ the Ricci s-tensor};  \notag \\
\ _{s}^{\shortmid }\mathcal{R}sc &=&\{\ ^{\shortmid }\mathbf{g}^{\beta
_{s}\gamma _{s}}\ ^{\shortmid }\mathbf{R}_{\ \beta _{s}\gamma _{s}}\},%
\mbox{the Riemannian scalar }.  \notag
\end{eqnarray}%
Geometric data $(\ \ _{s}^{\shortmid }\mathbf{g,}\ _{s}^{\shortmid }\mathbf{%
D)}$ of type (\ref{sdm}) and (\ref{sdc}) enable a $\ _{s}\mathcal{M}$ with a
dyadic metric-affine s-structure which is a N-adapted phase space version of
metric-affine geometry \cite{misner,kramer03}. Additionally to geometric
s-objects (\ref{mafgeomob}), such spaces are characterized by a nonmetricity
s-tensor, $\ _{s}^{\shortmid }\mathcal{Q}=\{\ ^{\shortmid }\mathbf{Q}%
_{\gamma _{s}\alpha _{s}\beta _{s}\ }=\ ^{\shortmid }\mathbf{D}_{\gamma
_{s}}\ ^{\shortmid }\mathbf{g}_{\alpha _{s}\beta _{s}}\}.$ Above formulas
for d-/ s-connections and respective geometric s-objects, can be defined and
computed for geometric flows, for insance, in the forms $\ _{s}^{\shortmid }%
\mathbf{D}(\tau )=\{\ ^{\shortmid }\Gamma _{\ \ \beta _{s}\gamma
_{s}}^{\alpha _{s}}(\tau )\},\ _{s}^{\shortmid }\mathcal{R}ic(\tau )$ etc.
We have to keep a s-label for indices or abstract geometric s-objects in
order to emphasize that the goemetric constructions are performed for a
nonholonomic dyadic formalism. For splitting of type (4+4), the nonholonomic
geometry is different and such decompositions do not allow general
decoupling and integration of fundamenta geometric and physical systems of
PDEs.

Using a s-metric $\ ^{\shortmid }g=$ $\ _{s}^{\shortmid }\mathbf{g}$ (\ref%
{sdm}), we can define and compute in abstract and component forms two
important linear connection structures (the Levi-Civita, LC, connection and
the canonical s-connection): 
\begin{equation}
(\ _{s}^{\shortmid }\mathbf{g,\ _{s}^{\shortmid }N})\rightarrow \left\{ 
\begin{array}{cc}
\ ^{\shortmid }\mathbf{\nabla :} & \ ^{\shortmid }\mathbf{\nabla }\ \
_{s}^{\shortmid }\mathbf{g}=0;\ _{\nabla }^{\shortmid }\mathcal{T}=0,\ %
\mbox{\  LC--connection }; \\ 
\ _{s}^{\shortmid }\widehat{\mathbf{D}}: & 
\begin{array}{c}
\ _{s}^{\shortmid }\widehat{\mathbf{Q}}=0;\ h_{1}\ ^{\shortmid }\widehat{%
\mathcal{T}}=0,v_{2}\ ^{\shortmid }\widehat{\mathcal{T}}=0,c_{3}\
^{\shortmid }\widehat{\mathcal{T}}=0,c_{4}\ ^{\shortmid }\widehat{\mathcal{T}%
}=0, \\ 
h_{1}v_{2}\ ^{\shortmid }\widehat{\mathcal{T}}\neq 0,h_{1}c_{s}\ ^{\shortmid
}\widehat{\mathcal{T}}\neq 0,v_{2}c_{s}\ ^{\shortmid }\widehat{\mathcal{T}}%
\neq 0,c_{3}c_{4}\ ^{\shortmid }\widehat{\mathcal{T}}\neq 0,%
\end{array}%
\begin{array}{c}
\mbox{ canonical } \\ 
\mbox{ s-connection  }.%
\end{array}%
\end{array}%
\right.  \label{twocon}
\end{equation}%
In this work, "hat" labels are used for geometric s-objects written in
canonical form, for instance, $\ _{s}^{\shortmid }\widehat{\mathbf{D}},$ $\
_{s}^{\shortmid }\widehat{\mathcal{R}}=\{\ ^{\shortmid }\widehat{\mathbf{R}}%
_{\ \beta _{s}\gamma _{s}\delta _{s}}^{\alpha _{s}}\}$ etc. There are
canonical distortion relations for linear connections (of type (\ref{distort}%
)) which allow to compute canonical distortions of fundamental geometric
objects (\ref{mafgeomob}) and relate, for instance, two different curvature
tensors, for instance, $\ _{\nabla }^{\shortmid }\mathcal{R}=\{\ _{\nabla
}^{\shortmid }R_{\ \beta _{s}\gamma _{s}\delta _{s}}^{\alpha _{s}}\}$ and $\
_{s}^{\shortmid }\widehat{\mathcal{R}}=\{\ ^{\shortmid }\widehat{\mathbf{R}}%
_{\ \beta _{s}\gamma _{s}\delta _{s}}^{\alpha _{s}}\};\ _{\nabla
}^{\shortmid }\mathcal{R}ic$ and $\ _{s}^{\shortmid }\widehat{\mathcal{R}}ic$
etc. For $\tau $-families such formulas can written, for instance, $\
^{\shortmid }\mathbf{\nabla (}\tau \mathbf{),}\ _{s}^{\shortmid }\widehat{%
\mathbf{D}}\mathbf{(}\tau \mathbf{)},\ _{s}^{\shortmid }\widehat{\mathcal{R}}%
\mathbf{(}\tau \mathbf{)}=\{\ ^{\shortmid }\widehat{\mathbf{R}}_{\ \beta
_{s}\gamma _{s}\delta _{s}}^{\alpha _{s}}\mathbf{(}\tau \mathbf{)}\},\
_{\nabla }^{\shortmid }\mathcal{R}ic\mathbf{(}\tau \mathbf{)},$ etc.

The modified Einstein equations for $\ _{s}^{\shortmid }\widehat{\mathbf{D}}$
(\ref{twocon}) can be derived in abstract geometric form as in GR \cite%
{misner} but on phase space $\ _{s}\mathcal{M},$%
\begin{equation}
\ ^{\shortmid }\widehat{\mathbf{R}}ic_{\alpha _{s}\beta _{s}}=\ ^{\shortmid
}\Upsilon _{\alpha _{s}\beta _{s}},  \label{seinsta}
\end{equation}%
where the s-tensor for effective and/or matter field sources can be
postulated in the forms 
\begin{equation}
\ ^{\shortmid }\Upsilon _{_{\beta _{s}\gamma _{s}}}=\left\{ 
\begin{array}{c}
\ _{s}^{\shortmid }\Lambda _{0}\ ^{\shortmid }\mathbf{g}_{\alpha _{s}\beta
_{s}}=\frac{1}{2}\ ^{\shortmid }\mathbf{g}_{\alpha _{s}\beta _{s}}\
_{s}^{\shortmid }\widehat{\mathbf{R}}sc+\ _{s}^{\shortmid }\lambda \ \
^{\shortmid }\mathbf{g}_{\alpha _{s}\beta _{s}},%
\mbox{vacuum with shell
cosmological constants}\ _{s}^{\shortmid }\Lambda _{0}\mbox{ or  }\
_{s}^{\shortmid }\lambda ; \\ 
\ _{s}^{\shortmid }\Lambda (\tau ,\ ^{\shortmid }u)\ ^{\shortmid }\mathbf{g}%
_{\alpha _{s}\beta _{s}},%
\mbox{ for  polarized constants from geometric
flow/ string / quantum theories}; \\ 
\ ^{\shortmid }\mathbf{Y}_{_{\beta _{s}\gamma _{s}}},%
\mbox{ from
variational/ geometric  principles of interactions on }\ _{s}\mathcal{M}; \\ 
\ ^{\shortmid }\mathbf{K}_{_{\beta _{s}\gamma _{s}}}\left\lceil \hbar
,\kappa \right\rceil ,%
\mbox{ for effective parametric star R-flux
corrections, in this work and \cite{partner02,partner03,partner04} }.%
\end{array}%
\right.  \label{sourca}
\end{equation}%
The gravitational field equations (\ref{seinsta}) can be written in terms of
the LC-connection $\nabla _{\alpha }$\ if we consider distortion relations (%
\ref{distort}). Imposing additional zero s-torsion conditions, 
\begin{equation}
\ _{s}^{\shortmid }\widehat{\mathbf{Z}}=0,\mbox{ which is equivalent to }\
_{s}^{\shortmid }\widehat{\mathbf{D}}_{\mid \ \ _{s}^{\shortmid }\widehat{%
\mathbf{T}}=0}=\ ^{\shortmid }\nabla ,  \label{lccond}
\end{equation}%
we can extract LC-configurations from various classes of solutions of
nonholonomic phase space generalized Einstein equations. Conservation laws
can be formulated as in GR using $\ ^{\shortmid }\nabla $ on $\mathcal{M},$
for instance, 
\begin{equation*}
\ ^{\shortmid }\nabla (\ _{\nabla }^{\shortmid }\mathcal{R}ic_{\alpha
_{s}\beta _{s}}-\frac{1}{2}\ ^{\shortmid }\mathbf{g}_{\alpha _{s}\beta
_{s}}\ \ _{\nabla }^{\shortmid }Rsc)=0,
\end{equation*}%
but such laws are written in a more cumbersome forms if we distort the
geometrical objects and this equations in terms of $\ _{s}^{\shortmid }%
\widehat{\mathbf{D}}$ using formulas (\ref{distort}). This is a typical
property of nonholonomic systems in geometric mechanics and gravity
theories. Here we note that notations for nonholonomic constraints of type $%
\ _{s}^{\shortmid }\widehat{\mathbf{D}}_{\mid \ \ _{s}^{\shortmid }\widehat{%
\mathbf{T}}=0}$ (\ref{lccond}) are used in our parner works \cite%
{partner01,partner02,partner03} even the editors of some journals request a
symplified version for notations like $\ _{s}^{\shortmid }\widehat{\mathbf{D}%
}=\ ^{\shortmid }\nabla $ for $\ _{s}^{\shortmid }\widehat{\mathbf{T}}=0,$
when the zero s-torsion conditions are considered as certain nonholonomic
constraints on a class of some generic off-diagonal soluions.

The main motivation to use the canonical s-connection $\ _{s}^{\shortmid }%
\widehat{\mathbf{D}}$ and respective phase space equations (\ref{seinsta})
with nonholonomic 2+2+2+2 splitting is that in such geometric variables we
can decouple and integrate in very general forms various classes of
(modified) geometric flow and gravitational field equations. Using the
AFCDM, this is proven in \cite{bubuianu19,partner02} and references therein,
see a summary of results in Appendix \ref{appendixb}. Here we note that it
is not possible to decouple such systems of nonlinear PDEs written in terms
of $\nabla _{\alpha }.$ The main idea is to use $\ _{s}^{\shortmid }\widehat{%
\mathbf{D}}$ in order to find explicit exact/parametric solutions and then
to impose additional constraints of type (\ref{lccond}) in order to extract
LC-configurations if it is important for elaborating certain physical
models. We can model $\tau $-evolution of families of equations of type (\ref%
{seinsta})- (\ref{lccond}) for so-called geometric evolution of nonholonomic
Einstein systems, NES, studied in \cite{ib19}.

\subsubsection{Nonassociative vacuum Einstein equations for the canonical
s-connection}

\label{appendixab}

The geometric constructions performed in this work are based on the concept
of star product $\star _{s}$ defined in s-adapted form in our works \cite%
{partner01,partner02} and using the previous constructions from \cite%
{blumenhagen16,aschieri17}: 
\begin{eqnarray}
f\star _{s}q:= &&\cdot \lbrack \mathcal{F}_{s}^{-1}(f,q)]  \label{starpn} \\
&=&\cdot \lbrack \exp (-\frac{1}{2}i\hbar (\ ^{\shortmid }\mathbf{e}%
_{i_{s}}\otimes \ ^{\shortmid }e^{i_{s}}-\ ^{\shortmid }e^{i_{s}}\otimes \
^{\shortmid }\mathbf{e}_{i_{s}})+\frac{i\mathit{\ell }_{s}^{4}}{12\hbar }%
R^{i_{s}j_{s}a_{s}}(p_{a_{s}}\ ^{\shortmid }\mathbf{e}_{i_{s}}\otimes \
^{\shortmid }\mathbf{e}_{j_{a}}-\ ^{\shortmid }\mathbf{e}_{j_{s}}\otimes
p_{a_{s}}\ ^{\shortmid }\mathbf{e}_{i_{s}}))]f\otimes q  \notag \\
&=&f\cdot q-\frac{i}{2}\hbar \lbrack (\ ^{\shortmid }\mathbf{e}_{i_{s}}f)(\
^{\shortmid }e^{i_{s}}q)-(\ ^{\shortmid }e^{i_{s}}f)(\ ^{\shortmid }\mathbf{e%
}_{i_{s}}q)]+\frac{i\mathit{\ell }_{s}^{4}}{6\hbar }%
R^{i_{s}j_{s}a_{s}}p_{a_{s}}(\ ^{\shortmid }\mathbf{e}_{i_{s}}f)(\
^{\shortmid }\mathbf{e}_{j_{s}}q)+\ldots .  \notag
\end{eqnarray}%
In this formula, there are considered actions of $\ ^{\shortmid }\mathbf{e}%
_{i_{s}}$ on some functions $\ f(x,p)$ and $\ q(x,p),$ see formulas for
N-elongated derivatives and differentials (\ref{nadapbdsc}) and (\ref%
{nadapbdss}); a constant $\mathit{\ell }$ characterizes the R-flux
contributions determined by an antisymmetric $R^{i_{s}j_{s}a_{s}}$
background in string theory, when the tensor product $\otimes $ can be
written also in a s-adapted form $\otimes _{s}.$ For explicit computations
and small parametric decompositions on $\hbar $ and $\kappa =\mathit{\ell }%
_{s}^{3}/6\hbar ,$ the tensor products turn into usual multiplications as in
the third line of above formula. A phase space $\ _{s}\mathcal{M}$ enabled
with a star product (\ref{starpn}) transforms into a nonassociative one, $\
_{s}^{\star }\mathcal{M}$, when the s-adapted geometric objects and
(physical) equations are star-deformed.

Considering geometric flows on a parameter $\tau $ of s-frames $\
^{\shortmid }\mathbf{e}_{i_{s}}\mathbf{(}\tau \mathbf{)}$ (\ref{nadapbdsc}),
we obtain define respective flow families s-adapted star product $\star _{s}%
\mathbf{(}\tau \mathbf{)}$ even the functions $\ f$ and $q$ may not depend
on evolution parameter. Similar $\tau $-dependences of geometric/ physical
s-objects and structures have to be defined for evolution on nonassociative
and associative geometric models.

For $_{s}\mathcal{M}\rightarrow \ _{s}^{\star }\mathcal{M},$ a $\star _{s}$%
-structure transforms any symmetric metric $\ _{s}^{\shortmid }\mathbf{g}$
into a general nonsymmetric one with respective symmetric, $\ _{\star
s}^{\shortparallel }\mathbf{g,}$ and nonsymmetric, $\ _{\star
s}^{\shortparallel }\mathfrak{g,}$ components. We use labels $\ _{\star
s}^{\shortparallel }$ instead of $\ _{s}^{\shortmid }\,\ $ because such
metrics may contain complex terms. On (co) tangent bundles, it is always
possible to elaborate almost complex models (with real basic manifolds and
real (co) fibers), or certain decompositions into pure real and imaginary
components (the last ones are not considered for geometric constructions
with real variables). So, labels $\ _{s}^{\shortmid }\,$\ involve a
procedure of transforming geometric constructions into certain s-objects on
real manifolds and bundle spaces. We studied in \cite{partner01,partner02}
the nonassociative (non) symmetric and generalized connection s-structures
on $\ _{s}^{\star }\mathcal{M}$ endowed also with quasi-Hopf s-structure
determined by a nonassociative algebra $\mathcal{A}_{s}^{\star }$
(generalizing the constructions from \cite{aschieri17,drinf}). A
nonassociative symmetric, $\ _{\star s}^{\shortmid }\mathbf{g}$, and
nonsymmetric metric, $\ _{\star }^{\shortmid }\mathfrak{g}_{\alpha _{s}\beta
_{s}},$ s-tensor on a phase space $\ _{s}\mathcal{M}$ with star and R-flux
induced terms on a Lorentz base spacetime manifold can be represented in the
forms 
\begin{eqnarray}
\ _{\star s}^{\shortmid }\mathbf{g} &=&\ _{\star }^{\shortmid }\mathbf{g}%
_{\alpha _{s}\beta _{s}}\star _{s}(\ ^{\shortmid }\mathbf{e}^{\alpha
_{s}}\otimes _{\star s}\ ^{\shortmid }\mathbf{e}^{\beta _{s}}),\mbox{ where }%
\ _{\star }^{\shortmid }\mathbf{g}(\ ^{\shortmid }\mathbf{e}_{\alpha _{s}},\
^{\shortmid }\mathbf{e}_{\beta _{s}})=\ _{\star }^{\shortmid }\mathbf{g}%
_{\alpha _{s}\beta _{s}}=\ _{\star }^{\shortmid }\mathbf{g}_{\beta
_{s}\alpha _{s}}\in \mathcal{A}_{s}^{\star }  \notag \\
\ \ _{\star }^{\shortmid }\mathfrak{g}_{\alpha _{s}\beta _{s}} &=&\ \
_{\star }^{\shortmid }\mathbf{g}_{\alpha _{s}\beta _{s}}-\kappa \mathcal{R}%
_{\quad \alpha _{s}}^{\tau _{s}\xi _{s}}\ \ ^{\shortmid }\mathbf{e}_{\xi
_{s}}\ \ _{\star }^{\shortmid }\mathbf{g}_{\beta _{s}\tau _{s}}=\ _{\star
}^{\shortmid }\mathfrak{g}_{\alpha _{s}\beta _{s}}^{[0]}+\ \ _{\star
}^{\shortmid }\mathfrak{g}_{\alpha _{s}\beta _{s}}^{[1]}(\kappa )=\ \
_{\star }^{\shortmid }\mathfrak{\check{g}}_{\alpha _{s}\beta _{s}}+\ \
_{\star }^{\shortmid }\mathfrak{a}_{\alpha _{s}\beta _{s}},  \label{aux40a}
\end{eqnarray}%
where $\mathcal{R}_{\quad \alpha _{s}}^{\tau _{s}\xi _{s}}$ are related to $%
R^{i_{s}j_{s}a_{s}}$ from\ (\ref{starpn}) via certain frame transforms and
multiplications on some real/complex coefficients. In (\ref{aux40a}), we
consider that $\ _{\star }^{\shortmid }\mathfrak{\check{g}}_{\alpha
_{s}\beta _{s}}$ is the symmetric part and $\ _{\star }^{\shortmid }%
\mathfrak{a}_{\alpha _{s}\beta _{s}}$ is the anti-symmetric part computed, 
\begin{eqnarray}
\ _{\star }^{\shortmid }\mathfrak{\check{g}}_{\alpha _{s}\beta _{s}}:= &&%
\frac{1}{2}(\ _{\star }^{\shortmid }\mathfrak{g}_{\alpha _{s}\beta _{s}}+\
_{\star }^{\shortmid }\mathfrak{g}_{\beta _{s}\alpha _{s}})=\ _{\star
}^{\shortmid }\mathbf{g}_{\alpha _{s}\beta _{s}}-\frac{\kappa }{2}\left( 
\mathcal{R}_{\quad \beta _{s}}^{\tau _{s}\xi _{s}}\ \ ^{\shortmid }\mathbf{e}%
_{\xi _{s}}\ _{\star }^{\shortmid }\mathbf{g}_{\tau _{s}\alpha _{s}}+%
\mathcal{R}_{\quad \alpha _{s}}^{\tau _{s}\xi _{s}}\ \ ^{\shortmid }\mathbf{e%
}_{\xi _{s}}\ _{\star }^{\shortmid }\mathbf{g}_{\beta _{s}\tau _{s}}\right)
\label{aux40b} \\
&=&\ _{\star }^{\shortmid }\mathfrak{\check{g}}_{\alpha _{s}\beta
_{s}}^{[0]}+\ _{\star }^{\shortmid }\mathfrak{\check{g}}_{\alpha _{s}\beta
_{s}}^{[1]}(\kappa ),  \notag \\
&&\mbox{ for }\ _{\star }^{\shortmid }\mathfrak{\check{g}}_{\alpha _{s}\beta
_{s}}^{[0]}=\ \ _{\star }^{\shortmid }\mathbf{g}_{\alpha _{s}\beta _{s}}%
\mbox{ and }\ \ _{\star }^{\shortmid }\mathfrak{\check{g}}_{\alpha _{s}\beta
_{s}}^{[1]}(\kappa )=-\frac{\kappa }{2}\left( \mathcal{R}_{\quad \beta
_{s}}^{\tau _{s}\xi _{s}}\ \ ^{\shortmid }\mathbf{e}_{\xi _{s}}\ \ _{\star
}^{\shortmid }\mathbf{g}_{\tau _{s}\alpha _{s}}+\mathcal{R}_{\quad \alpha
_{s}}^{\tau _{s}\xi _{s}}\ ^{\shortmid }\mathbf{e}_{\xi _{s}}\ \ _{\star
}^{\shortmid }\mathbf{g}_{\beta _{s}\tau _{s}}\right) ;  \notag \\
\ _{\star }^{\shortmid }\mathfrak{a}_{\alpha _{s}\beta _{s}}:= &&\frac{1}{2}%
(\ _{\star }^{\shortmid }\mathfrak{g}_{\alpha _{s}\beta _{s}}-\ _{\star
}^{\shortmid }\mathfrak{g}_{\beta _{s}\alpha _{s}})=\frac{\kappa }{2}\left( 
\mathcal{R}_{\quad \beta _{s}}^{\tau _{s}\xi _{s}}\ ^{\shortmid }\mathbf{e}%
_{\xi _{s}}\ \ _{\star }^{\shortmid }\mathbf{g}_{\tau _{s}\alpha _{s}}-%
\mathcal{R}_{\quad \alpha _{s}}^{\tau _{s}\xi _{s}}\ ^{\shortmid }\mathbf{e}%
_{\xi _{s}}\ _{\star }^{\shortmid }\mathbf{g}_{\beta _{s}\tau _{s}}\right) 
\notag \\
&=&\ _{\star }^{\shortmid }\mathfrak{a}_{\alpha _{s}\beta _{s}}^{[1]}(\kappa
)=\frac{1}{2}(\ _{\star }^{\shortmid }\mathfrak{g}_{\alpha _{s}\beta
_{s}}^{[1]}(\kappa )-\ _{\star }^{\shortmid }\mathfrak{g}_{\beta _{s}\alpha
_{s}}^{[1]}(\kappa )).  \label{aux40aa}
\end{eqnarray}%
Respective nonsymmetric inverse s-metrics can be parameterized in the form $%
\ _{\star }^{\shortmid }\mathfrak{g}^{\alpha _{s}\beta _{s}}=\ _{\star
}^{\shortmid }\mathfrak{\check{g}}^{\alpha _{s}\beta _{s}}+\ _{\star
}^{\shortmid }\mathfrak{a}^{\alpha _{s}\beta _{s}}$, when $\ _{\star
}^{\shortmid }\mathfrak{\check{g}}^{\alpha _{s}\beta _{s}}$ is not the
inverse to $\ _{\star }^{\shortmid }\mathfrak{\check{g}}_{\alpha _{s}\beta
_{s}}$ and $\ _{\star }^{\shortmid }\mathfrak{a}^{\alpha _{s}\beta _{s}}$ is
not inverse to $\ _{\star }^{\shortmid }\mathfrak{a}_{\alpha _{s}\beta
_{s}}. $ We emphasize that to compute inverse metrics and s-metrics, define
s-adapted geometric objects using commutators and anti-commutator, and
contractions with s-tensors and s-metrics on $\ _{s}^{\star }\mathcal{M}$
for such nonassociative geometric models, we have to apply more sophisticate
procedures, see details in \cite{aschieri17,partner01,partner02}. For
modelling geometric flow evoluton of symmetric and nonsymmetric components
of star product deformed metrics, we have to consider respective families of
s-objects and their s-adapted components, for instance, $\ _{\star
s}^{\shortmid }\mathbf{g(}\tau \mathbf{),}\ _{\star }^{\shortmid }\mathbf{g}%
_{\beta _{s}\alpha _{s}}\mathbf{(}\tau \mathbf{),}\ \ _{\star }^{\shortmid }%
\mathfrak{g}_{\alpha _{s}\beta _{s}}\mathbf{(}\tau \mathbf{)}=\ \ _{\star
}^{\shortmid }\mathfrak{\check{g}}_{\alpha _{s}\beta _{s}}\mathbf{(}\tau 
\mathbf{)}+\ \ _{\star }^{\shortmid }\mathfrak{a}_{\alpha _{s}\beta _{s}}%
\mathbf{(}\tau \mathbf{)}$ etc.

Nonassociative star deformations $\star _{s}$ of respective LC- and
canonical s-connections from (\ref{twocon}), adapted to a nonlinear
s--connection structures $\ _{s}^{\shortmid }\mathbf{N,}$ also involve a
canonical s-splitting for nonassociative LC-connection and canonical
s-connection 
\begin{equation}
\ _{s}^{\shortmid }\widehat{\mathbf{D}}\rightarrow \ _{s}^{\shortmid }%
\widehat{\mathbf{D}}^{\star }=(h_{1}\ ^{\shortmid }\widehat{\mathbf{D}}%
^{\star },\ v_{2}\ ^{\shortmid }\widehat{\mathbf{D}}^{\star },\ c_{3}\
^{\shortmid }\widehat{\mathbf{D}}^{\star },\ c_{4}\ ^{\shortmid }\widehat{%
\mathbf{D}}^{\star })=\ ^{\shortmid }\nabla ^{\star }+\ _{\star
s}^{\shortmid }\widehat{\mathbf{Z}},  \label{candistrnas}
\end{equation}%
where {\small 
\begin{equation}
(\ _{\star s}^{\shortmid }\mathbf{g,\ _{s}^{\shortmid }N})\rightarrow
\left\{ 
\begin{array}{cc}
\ _{\star }^{\shortmid }\mathbf{\nabla :}\  & 
\begin{array}{c}
\fbox{\ $\ \ _{\star }^{\shortmid }\mathbf{\nabla }\ _{\star s}^{\ \shortmid
}\mathbf{g}=0$;\ $_{\nabla }^{\shortmid }\mathcal{T}^{\star }=0$}%
\mbox{\ star
LC-connection}; \\ 
\end{array}
\\ 
\ _{s}^{\shortmid }\widehat{\mathbf{D}}^{\star }: & \fbox{$%
\begin{array}{c}
\ _{s}^{\shortmid }\widehat{\mathbf{D}}^{\star }\ _{\star s}^{\ \shortmid }%
\mathbf{g}=0;\ h_{1}\ ^{\shortmid }\widehat{\mathcal{T}}^{\star }=0,v_{2}\
^{\shortmid }\widehat{\mathcal{T}}^{\star }=0,c_{3}\ ^{\shortmid }\widehat{%
\mathcal{T}}^{\star }=0,c_{4}\ ^{\shortmid }\widehat{\mathcal{T}}^{\star }=0,
\\ 
h_{1}v_{2}\ ^{\shortmid }\widehat{\mathcal{T}}^{\star }\neq 0,h_{1}c_{s}\
^{\shortmid }\widehat{\mathcal{T}}^{\star }\neq 0,v_{2}c_{s}\ ^{\shortmid }%
\widehat{\mathcal{T}}^{\star }\neq 0,c_{3}c_{4}\ ^{\shortmid }\widehat{%
\mathcal{T}}^{\star }\neq 0,%
\end{array}%
$}\mbox{ canonical  s-connection }.%
\end{array}%
\right.  \label{twoconsstar}
\end{equation}%
}We note that in the definition of $\ _{s}^{\shortmid }\widehat{\mathbf{D}}%
^{\star }$ we use the s-tensor $_{\star s}^{\ \shortmid }\mathbf{g.}$ There
are alternative possibilities, for instance, to involve directly a
nonsymmetric metric $\ _{\star }^{\shortmid }\mathfrak{g}_{\alpha _{s}\beta
_{s}}$, which makes the procedure of constructing parametric solutions more
sophisticate than the variant with $_{\star s}^{\ \shortmid }\mathbf{g}$.
Working only up to $\kappa $-linear terms, such canonical s-connections are
equivalent for those configurations when $\ _{\star }^{\shortmid }\mathfrak{g%
}_{\alpha _{s}\beta _{s}}=\ _{\star }^{\shortmid }\mathfrak{\check{g}}%
_{\alpha _{s}\beta _{s}}+\ \ _{\star }^{\shortmid }\mathfrak{a}_{\alpha
_{s}\beta _{s}},$ with $\ _{\star }^{\shortmid }\mathfrak{\check{g}}_{\alpha
_{s}\beta _{s}}=\ _{\star }^{\ \shortmid }\mathbf{g}\ _{\alpha _{s}\beta
_{s}}$ and $\ _{\star }^{\shortmid }\mathfrak{a}_{\alpha _{s}\beta
_{s}|\kappa \rightarrow 0}\rightarrow 0,$ but there are non-vanishing terms
for $\ _{\star }^{\shortmid }\mathfrak{a}_{\alpha _{s}\beta _{s}}(\kappa
\neq 0).$ Such conditions can be always stated for certain commutative
nonholonomic configurations on which the star product deformations are
applied to keep such conditions. After certain classes of physically
important $\kappa $-parametric solutions are constructed in explicitly form,
we can consider general nonassociative frame and coordinate transforms.
Families of nonassociative canonical s-connections $\ _{s}^{\shortmid }%
\widehat{\mathbf{D}}^{\star }\mathbf{(}\tau \mathbf{)}=\ ^{\shortmid }\nabla
^{\star }\mathbf{(}\tau \mathbf{)}+\ _{\star s}^{\shortmid }\widehat{\mathbf{%
Z}}\mathbf{(}\tau \mathbf{)}$ have to be considered for elaborating
nonassociative geometric flow models, when all formulas from (\ref%
{candistrnas}) and (\ref{twoconsstar}) are re-defined with $\tau $%
-parametric dependence.

Nonassociative LC-configurations can be extracted similarly to (\ref{lccond}%
) if we impose additional zero s-torsion conditions,%
\begin{equation}
\ _{\star s}^{\shortmid }\widehat{\mathbf{Z}}=0,%
\mbox{ which is
equivalent to }\ _{s}^{\shortmid }\widehat{\mathbf{D}}_{\mid \
_{s}^{\shortmid }\widehat{\mathbf{T}}=0}^{\star }=\ \ _{\star }^{\shortmid
}\nabla .  \label{lccondnonass}
\end{equation}%
In general, all type of metrics on $\ _{s}^{\star }\mathcal{M}$, and related
s-metrics $\ _{\star s}^{\ \shortmid }\mathbf{g}$, subjected/ or not to some
conditions of type (\ref{lccondnonass}) contain certain nonzero anholonomy
coefficients of frame structures. In such cases, respective symmetric and
nonsymmetric s-metrics can be written in local coordinate forms as generic
off-diagonal matrices. For $\tau $-families, \ such conditions for
extracting and flow evolution of LC-connections can be written $\ _{\star
s}^{\shortmid }\widehat{\mathbf{Z}}\mathbf{(}\tau \mathbf{)}=0$ and $\
_{s}^{\shortmid }\widehat{\mathbf{D}}^{\star }\mathbf{(}\tau \mathbf{)}=\ \
_{\star }^{\shortmid }\nabla \mathbf{(}\tau \mathbf{)}$ for $\
_{s}^{\shortmid }\widehat{\mathbf{T}}\mathbf{(}\tau \mathbf{)}=0.$ Here we
note that we do not obtain equalities of some linear connections (by
definition, two different linear connections have different transformation
laws under frame/coordinate transforms) but certain equalities of
coefficients in certain s-adapted fromes.

To define and compute geometric and physical objects on a familiy
nonassociative phase spaces $\ _{s}^{\star }\mathcal{M}\mathbf{(}\tau 
\mathbf{)}$ defined by star product R-flux deformations, we follow:

\textbf{Convention 2 } (see details in \cite%
{partner01,partner02,partner03,partner04}; we can consider that in those
works all definitions and formulas were stated for a fixed value $\tau _{0}$
and in this work all results are extended for arbitrary $\tau $):\ The
commutative and nonassociative geometric data derived for a star product $%
\star _{s}$ (\ref{starpn}), are related by such s-adapted transforms:%
{\scriptsize 
\begin{equation}
\begin{array}{ccc}
\fbox{$(\star _{N}\mathbf{(}\tau \mathbf{)},\ \ \mathcal{A}_{N}^{\star }%
\mathbf{(}\tau \mathbf{)},\ _{\star }^{\shortmid }\mathbf{g}(\tau )\mathbf{%
,\ _{\star }^{\shortmid }\mathfrak{g}}(\tau )\mathbf{\mathfrak{,}\
^{\shortmid }N(}\tau \mathbf{)},\mathbf{\ ^{\shortmid }e}_{\alpha }\mathbf{(}%
\tau \mathbf{),\ ^{\shortmid }D}^{\star }\mathbf{(}\tau \mathbf{)})$} & 
\Leftrightarrow & \fbox{$(\star _{s}(\tau ),\ \ \mathcal{A}_{s}^{\star
}(\tau ),\ _{\star s}^{\shortmid }\mathbf{g}(\tau )\mathbf{,}\ _{\star
s}^{\shortmid }\mathbf{\mathfrak{g}}(\tau )\mathbf{\mathfrak{,}}\ \
_{s}^{\shortmid }\mathbf{N}(\tau ),\mathbf{\ ^{\shortmid }e}_{\alpha
_{s}}(\tau )\mathbf{,}\ _{s}^{\shortmid }\mathbf{D}^{\star }(\tau ))$} \\ 
& \Uparrow &  \\ 
\fbox{$(\mathbf{\ ^{\shortmid }g}(\tau )\mathbf{,\ ^{\shortmid }N(}\tau 
\mathbf{)},\mathbf{\ ^{\shortmid }e}_{\alpha }\mathbf{(}\tau \mathbf{),\
^{\shortmid }}\widehat{\mathbf{D}}\mathbf{(}\tau \mathbf{)})$} & 
\Leftrightarrow & \fbox{$(\ \ _{s}^{\shortmid }\mathbf{g}(\tau )\mathbf{,\ }%
\ _{s}^{\shortmid }\mathbf{N}(\tau ),\mathbf{\ \ ^{\shortmid }e}_{\alpha
_{s}}(\tau )\mathbf{,}\ _{s}^{\shortmid }\widehat{\mathbf{D}}(\tau )),$}%
\end{array}
\label{conv2s}
\end{equation}%
} for certain canonical distortions$\mathbf{\ ^{\shortmid }D}^{\star }(\tau
)=\ _{\star }^{\shortmid }\nabla (\tau )+\ ^{\shortmid }\widehat{\mathbf{Z}}%
^{\star }(\tau ),$ for respective nonholonomic splitting 4+4, and $\
_{s}^{\shortmid }\mathbf{D}^{\star }(\tau )=\ _{\star }^{\shortmid }\nabla
(\tau )+\ _{s}^{\shortmid }\widehat{\mathbf{Z}}^{\star }(\tau ),$ for
corresponding nonholonomic s-splitting. For simplicity, hereafter we shall
not write $\tau $-dependencies of geometric objects and structures if that
will not result in ambiguities.

Applying the rule of Convention 2, we can define and compute star-product
deformations of fundamental geometric s-objects (\ref{mafgeomob}),%
\begin{eqnarray}
\ _{s}^{\shortmid }\mathcal{T} &\rightarrow &\ _{s}^{\shortmid }\widehat{%
\mathcal{T}}^{\star }=\{\ ^{\shortmid }\widehat{\mathbf{T}}_{\ \star \beta
_{s}\gamma _{s}}^{\alpha _{s}}\},%
\mbox{ nonassociative  canonical
s-torsion };\   \label{mafgeomobn} \\
\ _{s}^{\shortmid }\mathcal{R} &\rightarrow &\ _{s}^{\shortmid }\widehat{%
\mathcal{R}}^{\star }=\{\ ^{\shortmid }\widehat{\mathbf{R}}_{\ \beta
_{s}\gamma _{s}\delta _{s}}^{\star \alpha _{s}}\},%
\mbox{nonassociative
canonical Riemannian s-curvature };  \notag \\
\ _{s}^{\shortmid }\mathcal{R}ic &\rightarrow &\ _{s}^{\shortmid }\widehat{%
\mathcal{R}}ic^{\star }=\{\ ^{\shortmid }\widehat{\mathbf{R}}_{\ \beta
_{s}\gamma _{s}}^{\star }:=\ ^{\shortmid }\widehat{\mathbf{R}}_{\ \beta
_{s}\gamma _{s}\alpha _{s}}^{\star \alpha _{s}}\neq \ ^{\shortmid }\widehat{%
\mathbf{R}}_{\ \gamma _{s}\beta _{s}}^{\star }\},%
\mbox{ nonassociative
canonical Ricci s-tensor};  \notag \\
\ _{s}^{\shortmid }\mathcal{R}sc &\rightarrow &\ _{s}^{\shortmid }\widehat{%
\mathcal{R}}sc^{\star }=\{\ ^{\shortmid }\mathbf{g}^{\beta _{s}\gamma _{s}}\
^{\shortmid }\widehat{\mathbf{R}}_{\ \beta _{s}\gamma _{s}}^{\star }\},%
\mbox{  nonassociative  canonical  Riemannian scalar };  \notag \\
\ _{s}^{\shortmid }\mathcal{Q} &\rightarrow &\ _{s}^{\shortmid }\mathcal{Q}%
^{\star }=\{\ ^{\shortmid }\widehat{\mathbf{Q}}_{\gamma _{s}\alpha _{s}\beta
_{s}\ }^{\star }=\ ^{\shortmid }\widehat{\mathbf{D}}_{\gamma _{s}}^{\star }\
_{\star }^{\shortmid }\mathbf{g}_{\alpha _{s}\beta _{s}}\},%
\mbox{ zero
nonassociative  canonical nonmetricity s-tensor }.  \notag
\end{eqnarray}

The nonassociative canonical Riemann s-tensor$\mathbf{\mathbf{\mathbf{%
\mathbf{\ ^{\shortmid }}}}}\widehat{\mathcal{\Re }}_{\quad }^{\star }=\{\
^{\shortmid }\widehat{\mathcal{\Re }}_{\quad \alpha _{s}\beta _{s}\gamma
_{s}}^{\star \mu _{s}}\}$ from (\ref{mafgeomobn}) can be defined and
computed \ for the data $(\ _{\star s}^{\shortparallel }\mathfrak{g=\{\
_{\star }^{\shortparallel }\mathfrak{\check{g}}_{\alpha _{s}\beta _{s}}=\
_{\star }^{\shortmid }\mathbf{g}_{\alpha _{s}\beta _{s}}\}},\
_{s}^{\shortmid }\widehat{\mathbf{D}}^{\star }=\{\ ^{\shortmid }\widehat{%
\mathbf{\Gamma }}_{\star \alpha _{s}\beta _{s}}^{\gamma _{s}}\}),$ see
details in \cite{partner02,partner04}, 
\begin{eqnarray}
\mathbf{\mathbf{\mathbf{\mathbf{\ ^{\shortmid }}}}}\widehat{\mathbf{R}}%
_{\quad \alpha _{s}\beta _{s}\gamma _{s}}^{\star \mu _{s}} &=&\mathbf{%
\mathbf{\mathbf{\mathbf{\ _{1}^{\shortmid }}}}}\widehat{\mathbf{R}}_{\quad
\alpha _{s}\beta _{s}\gamma _{s}}^{\star \mu _{s}}+\mathbf{\mathbf{\mathbf{%
\mathbf{\ _{2}^{\shortmid }}}}}\widehat{\mathbf{R}}_{\quad \alpha _{s}\beta
_{s}\gamma _{s}}^{\star \mu _{s}},\mbox{ where }  \label{nadriemhopfcan} \\
\mathbf{\mathbf{\mathbf{\mathbf{\ _{1}^{\ \shortmid }}}}}\widehat{\mathbf{R}}%
_{\quad \alpha _{s}\beta _{s}\gamma _{s}}^{\star \mu _{s}} &=&\ ^{\shortmid }%
\mathbf{e}_{\gamma _{s}}\ ^{\shortmid }\widehat{\Gamma }_{\star \alpha
_{s}\beta _{s}}^{\mu _{s}}-\ ^{\shortmid }\mathbf{e}_{\beta _{s}}\
^{\shortmid }\widehat{\Gamma }_{\star \alpha _{s}\gamma _{s}}^{\mu }+\
^{\shortmid }\widehat{\Gamma }_{\star \nu _{s}\tau _{s}}^{\mu _{s}}\star
_{s}(\delta _{\ \gamma _{s}}^{\tau _{s}}\ ^{\shortmid }\widehat{\Gamma }%
_{\star \alpha _{s}\beta _{s}}^{\nu _{s}}-\delta _{\ \beta _{s}}^{\tau
_{s}}\ ^{\shortmid }\widehat{\Gamma }_{\star \alpha _{s}\gamma _{s}}^{\nu
_{s}})+\ ^{\shortmid }w_{\beta _{s}\gamma _{s}}^{\tau _{s}}\star _{s}\
^{\shortmid }\widehat{\Gamma }_{\star \alpha _{s}\tau _{s}}^{\mu _{s}}, 
\notag \\
\ _{2}^{\shortmid }\widehat{\mathbf{R}}_{\quad \alpha _{s}\beta _{s}\gamma
_{s}}^{\star \mu _{s}} &=&i\kappa \ ^{\shortmid }\widehat{\Gamma }_{\star
\nu _{s}\tau _{s}}^{\mu _{s}}\star _{s}(\mathcal{R}_{\quad \gamma
_{s}}^{\tau _{s}\xi _{s}}\ ^{\shortmid }\mathbf{e}_{\xi _{s}}\ ^{\shortmid }%
\widehat{\Gamma }_{\star \alpha _{s}\beta _{s}}^{\nu _{s}}-\mathcal{R}%
_{\quad \beta _{s}}^{\tau _{s}\xi _{s}}\ ^{\shortmid }\mathbf{e}_{\xi _{s}}\
^{\shortmid }\widehat{\Gamma }_{\star \alpha _{s}\gamma _{s}}^{\nu _{s}}). 
\notag
\end{eqnarray}%
Using parametric decompositions of the star canonical s-connection in (\ref%
{nadriemhopfcan}), 
\begin{equation}
\ ^{\shortmid }\widehat{\mathbf{\Gamma }}_{\star \alpha _{s}\beta
_{s}}^{\gamma _{s}}=\ _{[0]}^{\shortmid }\widehat{\mathbf{\Gamma }}_{\star
\alpha _{s}\beta _{s}}^{\nu _{s}}+i\kappa \ _{[1]}^{\shortmid }\widehat{%
\mathbf{\Gamma }}_{\star \alpha _{s}\beta _{s}}^{\nu _{s}}=\
_{[00]}^{\shortmid }\widehat{\Gamma }_{\ast \alpha _{s}\beta _{s}}^{\nu
_{s}}+\ _{[01]}^{\shortmid }\widehat{\Gamma }_{\ast \alpha _{s}\beta
_{s}}^{\nu _{s}}(\hbar )+\ _{[10]}^{\shortmid }\widehat{\Gamma }_{\ast
\alpha _{s}\beta _{s}}^{\nu _{s}}(\kappa )+\ _{[11]}^{\shortmid }\widehat{%
\Gamma }_{\ast \alpha _{s}\beta _{s}}^{\nu _{s}}(\hbar \kappa )+O(\hbar
^{2},\kappa ^{2}...),  \label{paramscon}
\end{equation}%
we can compute such parametric decompositions of the nonassociative
canonical curvature tensor,%
\begin{equation*}
\ ^{\shortmid }\widehat{\mathbf{R}}_{\quad \alpha _{s}\beta _{s}\gamma
_{s}}^{\star \mu _{s}}=\mathbf{\mathbf{\mathbf{\mathbf{\ }}}}\
_{[00]}^{\shortmid }\widehat{\mathbf{R}}_{\quad \alpha _{s}\beta _{s}\gamma
_{s}}^{\star \mu _{s}}+\mathbf{\mathbf{\mathbf{\mathbf{\ }}}}\
_{[01]}^{\shortmid }\widehat{\mathbf{R}}_{\quad \alpha _{s}\beta _{s}\gamma
_{s}}^{\star \mu _{s}}(\hbar )+\mathbf{\mathbf{\mathbf{\mathbf{\ }}}}\
_{[10]}^{\shortmid }\widehat{\mathbf{R}}_{\quad \alpha _{s}\beta _{s}\gamma
_{s}}^{\star \mu _{s}}(\kappa )+\ _{[11]}^{\shortmid }\widehat{\mathbf{R}}%
_{\quad \alpha _{s}\beta _{s}\gamma _{s}}^{\star \mu _{s}}(\hbar \kappa
)+O(\hbar ^{2},\kappa ^{2},...).
\end{equation*}

Contracting the first and forth indices of (\ref{nadriemhopfcan}), we define
the nonassociative canonical Ricci s-tensor, 
\begin{eqnarray*}
\mathbf{\mathbf{\mathbf{\mathbf{\ _{s}^{\shortmid }}}}}\widehat{\mathcal{\Re 
}}ic^{\star } &=&\mathbf{\mathbf{\mathbf{\mathbf{\ ^{\shortmid }}}}}\widehat{%
\mathbf{\mathbf{\mathbf{\mathbf{R}}}}}ic_{\alpha _{s}\beta _{s}}^{\star
}\star _{s}(\ \mathbf{^{\shortmid }e}^{\alpha _{s}}\otimes _{\star s}\ 
\mathbf{^{\shortmid }e}^{\beta _{s}}),\mbox{ where } \\
&&\mathbf{\mathbf{\mathbf{\mathbf{\ ^{\shortmid }}}}}\widehat{\mathbf{%
\mathbf{\mathbf{\mathbf{R}}}}}ic_{\alpha _{s}\beta _{s}}^{\star }:=\
_{s}^{\shortmid }\widehat{\mathcal{\Re }}ic^{\star }(\ ^{\shortmid }\mathbf{e%
}_{\alpha _{s}},\ ^{\shortmid }\mathbf{e}_{\beta _{s}})=\mathbf{\langle }\ 
\mathbf{\mathbf{\mathbf{\mathbf{\ ^{\shortmid }}}}}\widehat{\mathbf{\mathbf{%
\mathbf{\mathbf{R}}}}}ic_{\mu _{s}\nu _{s}}^{\star }\star _{s}(\ \mathbf{%
^{\shortmid }e}^{\mu _{s}}\otimes _{\star _{s}}\ ^{\shortmid }\mathbf{e}%
^{\nu _{s}}),\mathbf{\mathbf{\ }\ ^{\shortmid }\mathbf{e}}_{\alpha _{s}}%
\mathbf{\otimes _{\star s}\ ^{\shortmid }\mathbf{e}}_{\beta _{s}}\mathbf{%
\rangle }_{\star _{s}},
\end{eqnarray*}%
\begin{eqnarray}
\mbox{and }\ ^{\shortmid }\widehat{\mathbf{R}}ic_{\alpha _{s}\beta
_{s}}^{\star }:= &&\mathbf{\mathbf{\mathbf{\mathbf{\ ^{\shortmid }}}}}%
\widehat{\mathcal{\Re }}_{\quad \alpha _{s}\beta _{s}\mu _{s}}^{\star \mu
_{s}}=\ _{[00]}^{\shortmid }\widehat{\mathbf{\mathbf{\mathbf{\mathbf{R}}}}}%
ic_{\alpha _{s}\beta _{s}}^{\star }+\mathbf{\mathbf{\mathbf{\mathbf{\ \ }}}}%
_{[01]}^{\shortmid }\widehat{\mathbf{\mathbf{\mathbf{\mathbf{R}}}}}%
ic_{\alpha _{s}\beta _{s}}^{\star }(\hbar )+\mathbf{\mathbf{\mathbf{\mathbf{%
\ }}}}_{[10]}^{\shortmid }\widehat{\mathbf{\mathbf{\mathbf{\mathbf{R}}}}}%
ic_{\alpha _{s}\beta _{s}}^{\star }(\kappa )  \notag \\
&&+\mathbf{\mathbf{\mathbf{\mathbf{\ }}}}_{[11]}^{\shortmid }\widehat{%
\mathbf{\mathbf{\mathbf{\mathbf{R}}}}}ic_{\alpha _{s}\beta _{s}}^{\star
}(\hbar \kappa )+O(\hbar ^{2},\kappa ^{2},...),\mbox{where}  \notag \\
&&\ _{[00]}^{\shortmid }\widehat{\mathbf{R}}ic_{\alpha _{s}\beta
_{s}}^{\star }=\ _{[00]}^{\shortmid }\widehat{\mathcal{\Re }}_{\quad \alpha
_{s}\beta _{s}\mu _{s}}^{\star \mu _{s}}\mathbf{\mathbf{\mathbf{\mathbf{\ ,}}%
}}\ _{[01]}^{\shortmid }\mathbf{\mathbf{\mathbf{\mathbf{\widehat{\mathbf{%
\mathbf{\mathbf{\mathbf{R}}}}}}}}}ic_{\alpha _{s}\beta _{s}}^{\star }=\
_{[01]}^{\shortmid }\mathbf{\mathbf{\mathbf{\mathbf{\widehat{\mathcal{\Re }}}%
}}}_{\quad \alpha _{s}\beta _{s}\mu _{s}}^{\star \mu _{s}},
\label{driccicanonstar1} \\
&&\ _{[10]}^{\shortmid }\mathbf{\mathbf{\mathbf{\mathbf{\widehat{\mathbf{%
\mathbf{\mathbf{\mathbf{R}}}}}}}}}ic_{\alpha _{s}\beta _{s}}^{\star }=\
_{[10]}^{\shortmid }\mathbf{\mathbf{\mathbf{\mathbf{\widehat{\mathcal{\Re }}}%
}}}_{\quad \alpha _{s}\beta _{s}\mu _{s}}^{\star \mu _{s}},\
_{[11]}^{\shortmid }\widehat{\mathbf{\mathbf{\mathbf{\mathbf{R}}}}}%
ic_{\alpha _{s}\beta _{s}}^{\star }=\ _{[11]}^{\shortmid }\mathbf{\mathbf{%
\mathbf{\mathbf{\widehat{\mathcal{\Re }}}}}}_{\quad \alpha _{s}\beta _{s}\mu
_{s}}^{\star \mu _{s}}.  \notag
\end{eqnarray}%
Because of nonholonomic structure, canonical Ricci s-tensors are not
symmetric for general (non) commutative and nonassociative cases.

Further h1-v2-c3-c4 decompositions in abstract and coefficient s-adapted
forms are also possible for formulas (\ref{nadriemhopfcan}) and (\ref%
{driccicanonstar1}) (we omit such details in this paper).

Contracting the indices of (\ref{driccicanonstar1}) with the inverse
nonassociative and nonsymmetric s-metric $\ _{\star }^{\shortmid }\mathfrak{g%
}^{\mu _{s}\nu _{s}},$ we define and compute the nonassociative nonholonomic
canonical Ricci scalar curvature:%
\begin{eqnarray}
\ _{s}^{\shortmid }\widehat{\mathbf{R}}sc^{\star }:= &&\ _{\star
}^{\shortmid }\mathfrak{g}^{\mu _{s}\nu _{s}}\mathbf{\mathbf{\mathbf{\mathbf{%
\ ^{\shortmid }}}}}\widehat{\mathbf{\mathbf{\mathbf{\mathbf{R}}}}}ic_{\mu
_{s}\nu _{s}}^{\star }=\left( \ _{\star }^{\shortmid }\mathfrak{\check{g}}%
^{\mu _{s}\nu _{s}}+\ _{\star }^{\shortmid }\mathfrak{a}^{\mu _{s}\nu
_{s}}\right) \left( \mathbf{\mathbf{\mathbf{\mathbf{\ ^{\shortmid }}}}}%
\widehat{\mathbf{\mathbf{\mathbf{\mathbf{R}}}}}ic_{(\mu _{s}\nu
_{s})}^{\star }+\mathbf{\mathbf{\mathbf{\mathbf{\ ^{\shortmid }}}}}\widehat{%
\mathbf{\mathbf{\mathbf{\mathbf{R}}}}}ic_{[\mu _{s}\nu _{s}]}^{\star
}\right) =\ _{s}^{\shortmid }\widehat{\mathbf{\mathbf{\mathbf{\mathbf{R}}}}}%
ss^{\star }+\ _{s}^{\shortmid }\widehat{\mathbf{\mathbf{\mathbf{\mathbf{R}}}}%
}sa^{\star },  \notag \\
&&\mbox{ where }\ _{s}^{\shortmid }\widehat{\mathbf{\mathbf{\mathbf{\mathbf{R%
}}}}}ss^{\star }=:\ _{\star }^{\shortmid }\mathfrak{\check{g}}^{\mu _{s}\nu
_{s}}\mathbf{\mathbf{\mathbf{\mathbf{\ ^{\shortmid }}}}}\widehat{\mathbf{%
\mathbf{\mathbf{\mathbf{R}}}}}ic_{(\mu _{s}\nu _{s})}^{\star }\mbox{ and }\
_{s}^{\shortmid }\widehat{\mathbf{\mathbf{\mathbf{\mathbf{R}}}}}sa^{\star
}:=\ _{\star }^{\shortmid }\mathfrak{a}^{\mu _{s}\nu _{s}}\mathbf{\mathbf{%
\mathbf{\mathbf{\ ^{\shortmid }}}}}\widehat{\mathbf{\mathbf{\mathbf{\mathbf{R%
}}}}}ic_{[\mu _{s}\nu _{s}]}^{\star }.  \label{ricciscsymnonsym}
\end{eqnarray}%
In (\ref{ricciscsymnonsym}), the respective symmetric $\left( ...\right) $
and anti-symmetric $\left[ ...\right] $ operators are defined using the
multiple $1/2$ when, for instance, $\mathbf{\mathbf{\mathbf{\mathbf{\
^{\shortmid }}}}}\widehat{\mathbf{\mathbf{\mathbf{\mathbf{R}}}}}ic_{\mu
_{s}\nu _{s}}^{\star }=\mathbf{\mathbf{\mathbf{\mathbf{\ ^{\shortmid }}}}}%
\widehat{\mathbf{\mathbf{\mathbf{\mathbf{R}}}}}ic_{(\mu _{s}\nu
_{s})}^{\star }+\mathbf{\mathbf{\mathbf{\mathbf{\ ^{\shortmid }}}}}\widehat{%
\mathbf{\mathbf{\mathbf{\mathbf{R}}}}}ic_{[\mu _{s}\nu _{s}]}^{\star }.$

The nonassociative phase space vacuum Einstein equations with a nontrivial
at least one shell cosmological constant ($\ _{s}^{\shortmid }\lambda \neq 0$
for any $s,$ or some shells) can be defined and computed for the canonical
s-connection $\ _{s}^{\shortmid }\widehat{\mathbf{D}}^{\star },$ 
\begin{equation}
\ ^{\shortmid }\widehat{\mathbf{R}}ic_{\alpha _{s}\beta _{s}}^{\star }-\frac{%
1}{2}\ _{\star }^{\shortmid }\mathfrak{g}_{\alpha _{s}\beta _{s}}\
_{s}^{\shortmid }\widehat{\mathbf{R}}sc^{\star }=\ _{s}^{\shortmid }\lambda
\ _{\star }^{\shortmid }\mathfrak{g}_{\alpha _{s}\beta _{s}},
\label{nonassocdeinst1}
\end{equation}%
where the nonassociative Ricci s-tensor and scalar curvature are defined
respectively by formulas (\ref{driccicanonstar1}) and (\ref{ricciscsymnonsym}%
).

\subsubsection{Parametric decomposition of nonassociative and vacuum
gravitational equations}

\label{appendixac}

The procedure of parametric decompositions of geometric s-objects $\mathbf{%
\mathbf{\mathbf{\mathbf{\ ^{\shortmid }}}}}\widehat{\mathbf{R}}_{\quad
\alpha _{s}\beta _{s}\gamma _{s}}^{\star \mu _{s}}$ (\ref{nadriemhopfcan}), $%
\ ^{\shortmid }\widehat{\mathbf{R}}ic_{\alpha _{s}\beta _{s}}^{\star }$ (\ref%
{driccicanonstar1}) \ and $\ _{s}^{\shortmid }\widehat{\mathbf{R}}sc^{\star
} $ (\ref{ricciscsymnonsym}) with $[01,10,11]:=\left\lceil \hbar ,\kappa
\right\rceil $ components, in parametric form of the canonical s-connections
(\ref{paramscon}) is elaborated in \cite{partner01,partner02}. Such
constructions extend the formalism for the LC-connections provided
originally in \cite{aschieri17}. In both cases, the nonassociative and
noncommutative of the Riemann and Ricci tensors contains contributions from
star product deformations which can be real or complex ones. In our
approach, we can consider nonholonomic distributions on phase space when the
(almost) complex structures are separated and for the real parts $\
_{[00]}^{\shortmid }\widehat{\mathbf{R}}ic_{\alpha _{s}\beta _{s}}^{\star
}=\ ^{\shortmid }\widehat{\mathbf{R}}_{\ \alpha _{s}\beta _{s}}$ and such
coefficients \ are determined by an associative and commutative s-adapted
canonical s-connection $\ _{s}^{\shortmid }\widehat{\mathbf{D}}$ (\ref%
{twocon}). As a result, the star s-deformed Ricci s-tensor (\ref%
{driccicanonstar1}) can be expressed in parametric form, 
\begin{equation}
\ _{s}^{\shortmid }\widehat{\mathcal{R}}ic^{\star }=\{\ ^{\shortmid }%
\widehat{\mathbf{R}}_{\ \beta _{s}\gamma _{s}}^{\star }\}=\ _{s}^{\shortmid }%
\widehat{\mathcal{R}}ic+\ _{s}^{\shortmid }\widehat{\mathcal{K}}%
ic\left\lceil \hbar ,\kappa \right\rceil =\{\ ^{\shortmid }\widehat{\mathbf{R%
}}_{\ \beta _{s}\gamma _{s}}+\ ^{\shortmid }\widehat{\mathbf{K}}_{\ \beta
_{s}\gamma _{s}}\left\lceil \hbar ,\kappa \right\rceil \},
\label{paramsricci}
\end{equation}%
there the distortion tensor 
\begin{equation*}
\ _{s}^{\shortmid }\widehat{\mathcal{K}}ic=\{\ ^{\shortmid }\widehat{\mathbf{%
K}}_{\ \beta _{s}\gamma _{s}}\left\lceil \hbar ,\kappa \right\rceil =\
_{[01]}^{\shortmid }\widehat{\mathbf{\mathbf{\mathbf{\mathbf{R}}}}}ic_{\beta
_{s}\gamma _{s}}^{\star }+\ _{[10]}^{\shortmid }\widehat{\mathbf{R}}%
ic_{\beta _{s}\gamma _{s}}^{\star }+\ _{[11]}^{\shortmid }\widehat{\mathbf{R}%
}ic_{\beta _{s}\gamma _{s}}^{\star }\}
\end{equation*}%
encodes nonassociative parametric deformations of the canonical Ricci
s-tenor.

We can adapt the nonholonomic s-structure that the nonassociative canonical
Ricci scalar is conventionally with a sum of some effective shell polarized
cosmological constants $\ ^{s}\Lambda (\ _{s}^{\shortmid }u)$ depending
respectively on shell coordinates, 
\begin{equation*}
\ ^{\shortmid }\widehat{\mathbf{R}}sc^{\star }=\ ^{1}\Lambda (\
_{1}^{\shortmid }u)+\ ^{2}\Lambda (\ _{2}^{\shortmid }u)+\ ^{3}\Lambda (\
_{3}^{\shortmid }u)+\ ^{4}\Lambda (\ _{4}^{\shortmid }u).
\end{equation*}%
Choosing effective $\ ^{s}\Lambda (\ _{s}^{\shortmid }u)$ and fixing, for
simplicity, $\ _{s}^{\shortmid }\lambda =\ ^{\shortmid }\lambda $ in a form
that $\ ^{\shortmid }\lambda +\frac{1}{2}\ _{\star }^{\shortmid }\mathfrak{g}%
_{\alpha _{s}\beta _{s}}\ _{s}^{\shortmid }\widehat{\mathbf{R}}sc^{\star
}=0, $ when the nonsymmetric metric (\ref{aux40a}) decouple into two
independent symmetric and antisymmetric computed respectively by formulas $\
_{\star }^{\shortmid }\mathfrak{\check{g}}_{\alpha _{s}\beta _{s}}$ (\ref%
{aux40b}) and $\ _{\star }^{\shortmid }\mathfrak{a}_{\alpha _{s}\beta _{s}}$
(\ref{aux40aa}), 
\begin{equation*}
\ _{\star }^{\shortmid }\mathfrak{g}_{\alpha _{s}\beta _{s}}=\ _{\star
}^{\shortmid }\mathfrak{\check{g}}_{\alpha _{s}\beta _{s}}+\ _{\star
}^{\shortmid }\mathfrak{a}_{\alpha _{s}\beta _{s}}=\ _{\star }^{\shortmid }%
\mathfrak{\check{g}}_{\alpha _{s}\beta _{s}}^{[0]}+\ _{\star }^{\shortmid }%
\mathfrak{a}_{\alpha _{s}\beta _{s}}^{[1]},
\end{equation*}
and determined in explicit form respectively by $\ _{\star }^{\shortmid }%
\mathfrak{\check{g}}_{\alpha _{s}\beta _{s}}^{[0]}=\ _{\star }^{\shortmid }%
\mathbf{g}_{\alpha _{s}\beta _{s}}$ and $\ _{\star }^{\shortmid }\mathfrak{a}%
_{\alpha _{s}\beta _{s}}^{[1]}=i\kappa \mathcal{R}_{\quad \lbrack \alpha
_{s}}^{\tau _{s}\xi _{s}}\ \mathbf{^{\shortmid }e}_{|\xi _{s}}\ ^{\shortmid }%
\mathbf{g}_{\tau _{s}|\beta _{s}]}$, where $|\xi _{s}\tau _{s}|$ means that
such indices are not involved in anti-symmetrization.

Using formulas (\ref{paramsricci}), we express the nonassociative vacuum
gravitational field equations (\ref{nonassocdeinst1}) in the form $\
^{\shortmid }\widehat{\mathbf{R}}ic_{\alpha _{s}\beta _{s}}=\ ^{\shortmid
}\Upsilon _{\alpha _{s}\beta _{s}}$ (\ref{seinsta}), where $\ ^{\shortmid
}\Upsilon _{\alpha _{s}\beta _{s}}=-\ _{s}^{\shortmid }\widehat{\mathcal{K}}%
ic_{\alpha _{s}\beta _{s}}.$ Such an effective source of type (\ref{sourca})
encodes nonassociative star R-flux deformations, 
\begin{eqnarray}
\ ^{\shortmid }\mathbf{K}_{_{\beta _{s}\gamma _{s}}} &=&\ _{[0]}^{\shortmid
}\Upsilon _{_{\beta _{s}\gamma _{s}}}+\ _{[1]}^{\shortmid }\mathbf{K}%
_{_{\beta _{s}\gamma _{s}}}\left\lceil \hbar ,\kappa \right\rceil ,%
\mbox{
where }  \label{nassocsr2} \\
&&\ _{[0]}^{\shortmid }\Upsilon _{_{\beta _{s}\gamma _{s}}}=\ ^{s}\Lambda (\
^{\shortmid }u^{\gamma _{s}})\ _{\star }^{\shortmid }\mathbf{g}_{\beta
_{s}\gamma _{s}}\ \mbox{and }\ _{[1]}^{\shortmid }\mathbf{K}_{_{\beta
_{s}\gamma _{s}}}\left\lceil \hbar ,\kappa \right\rceil =\ ^{s}\Lambda (\
^{\shortmid }u^{\gamma _{s}})\ _{\star }^{\shortmid }\mathfrak{\check{g}}%
_{\beta _{s}\gamma _{s}}^{[1]}(\kappa )-\ ^{\shortmid }\widehat{\mathbf{K}}%
_{\ \beta _{s}\gamma _{s}}\left\lceil \hbar ,\kappa \right\rceil ,  \notag
\end{eqnarray}%
is an effective parametric source with coefficients proportional to $\hbar
,\kappa $ and $\hbar \kappa .$ For computing R-flux $\kappa $-linear
effects, it is enough to consider nonholonomic distributions and effective
sources generated by data 
\begin{eqnarray*}
\ _{\star }^{\shortmid }\mathfrak{\check{g}}_{\alpha _{s}\beta _{s}}^{[0]}
&=&\ _{\star }^{\shortmid }\mathbf{g}_{\alpha _{s}\beta _{s}}=\ ^{\shortmid }%
\mathbf{g}_{\alpha _{s}\beta _{s}};\ _{\star }^{\shortmid }\mathfrak{\check{g%
}}_{\beta _{s}\gamma _{s}}^{[1]}(\kappa )=0,\ \ _{\star }^{\shortmid }%
\mathfrak{a}_{\alpha _{s}\beta _{s}}^{[0]}=0,\ _{\star }^{\shortmid }%
\mathfrak{a}_{\alpha _{s}\beta _{s}}^{[1]}=i\kappa \mathcal{R}_{\quad
\lbrack \alpha _{s}}^{\tau _{s}\xi _{s}}\ \mathbf{^{\shortmid }e}_{|\xi
_{s}}\ ^{\shortmid }\mathbf{g}_{\tau _{s}|\beta _{s}]}, \\
\mbox{ and }\ ^{\shortmid }\Upsilon _{\alpha _{s}\beta _{s}} &=&-\
_{s}^{\shortmid }\widehat{\mathcal{K}}ic_{\alpha _{s}\beta _{s}}=-\
^{\shortmid }\widehat{\mathbf{K}}_{\alpha _{s}\beta _{s}}.
\end{eqnarray*}%
The effective sources can be parameterized for nontrivial real
quasi-stationary 8-d configurations\footnote{\label{fnqs} A s-metric is 
\textsf{quasi-stationary} if the corresponding (non) associative phase
spacetime geometric s-objects possess a Killing symmetry on $\partial
_{4}=\partial _{t}$ on shell $s=2$ and on $\ ^{\shortparallel }\partial
_{7}, $ or $\ ^{\shortparallel }\partial _{8},$ for all shells up to $s=4.$}
of s-metrics using coordinates $(x^{k_{3}}, \ ^{\shortmid }p_{8}=E),$ with $%
\ _{\star }^{\shortmid }\mathbf{g}_{\beta _{s}\gamma _{s}\mid \hbar ,\kappa
=0}=\ ^{\shortmid }\mathbf{g}_{\beta _{s}\gamma _{s}},$ in various forms
depending on prescribed shell Killing symmetries. Nonassociative effects are
determined additionally as some induced nonsymmetric components $\ _{\star
}^{\shortmid }\mathfrak{a}_{\alpha _{s}\beta _{s}}^{[1]}.$

In this work, we consider such quasi-stationary shell by shell adapted
distributions on $\ _{s}^{\star }\mathcal{M}$ when 
\begin{equation}
\ ^{\shortmid }\mathbf{K}_{\ \beta _{s}}^{\alpha _{s}}~=[~_{1}^{\shortmid }%
\mathcal{K}(\kappa ,x^{k_{1}})\delta _{i_{1}}^{j_{1}},~_{2}^{\shortmid }%
\mathcal{K}(\kappa ,x^{k_{1}},x^{3})\delta _{b_{2}}^{a_{2}},~_{3}^{\shortmid
}\mathcal{K}(\kappa ,x^{k_{2}},\ ^{\shortmid }p_{6})\delta
_{a_{3}}^{b_{3}},~_{4}^{\shortmid }\mathcal{K}(\kappa ,x^{k_{3}},\
^{\shortmid }p_{8})\delta _{a_{4}}^{b_{4}}]  \label{cannonsymparamc2a}
\end{equation}%
contain as functionals certain $\kappa $-linear terms with $\mathcal{R}%
_{\quad \alpha _{s}}^{\tau _{s}\xi _{s}}.$ Prescribing certain values for
effective sources $\ ~_{s}^{\shortmid }\mathcal{K}$ (\ref{cannonsymparamc2a}%
) as \textbf{generating sources}, we constrain nonholonomically the
gravitational dynamics and effective and possible matter sources. Such
generating sources can be related to conventional cosmological constants via
nonlinear symmetries, when the nonsymmetric parts of the s-metrics and the
canonical Ricci s-tensors can be computed as R-flux deformations of some
off-diagonal symmetric metric configurations. Finally, we note that using
necessary types of frame s-adapted transform, $\ \ ^{\shortmid }\widehat{%
\Upsilon }_{\alpha _{s}^{\prime }\beta _{s}^{\prime }}=e_{\ \alpha
_{s}^{\prime }}^{\alpha _{s}}e_{\ \beta _{s}^{\prime }}^{\beta _{s}}\ \ \
^{\shortmid }\mathcal{K}_{\alpha _{s}\beta _{s}}$ we can transform certain
general sources into a subset of four generating soucres$\ ^{\shortmid }%
\mathcal{K}_{_{\beta _{s}\gamma _{s}}}=\{~_{s}^{\shortmid }\mathcal{K}\}.$
We use the label "$\widehat{}$" \ for such sources in order to emphasize
that they are determined by generating sources encoding in certain general
nonholonomic forms certain noncommutative data for star-product and R-flux
deformations.

\subsection{Nonassociative generalizations of Perelman's F- and W-functionals%
}

We consider families of nonassociative R-flux deformed phase spaces, $_{s}%
\mathcal{M}\rightarrow \ _{s}^{\star }\mathcal{M}(\tau )$ determined by star
product $\star _{s}(\tau )$ structure (\ref{starpn}) adapted to a
nonholonomic (2+2)+(2+2) decomposition (i.e. s-structure) of a cotangent
Lorentz bundle $\mathcal{M}=T^{\ast }V,\dim V=4,$ as in \cite%
{partner01,partner02,partner04}. Star product R-flux deformations of
fundamental geometry s-objects (\ref{mafgeomobn}), determined by
nonassociative geometric data $\left[ \ _{s}^{\shortmid }\mathfrak{g}^{\star
}(\tau ),\ _{s}^{\shortmid }\widehat{\mathbf{D}}^{\star }(\tau )\right] ,$
are performed following Convention 2 (\ref{conv2s}) with $\kappa $-linear
parametric decompositions when $\ _{\star }^{\shortmid }\mathfrak{\check{g}}%
_{\alpha _{s}\beta _{s}}^{[0]}(\tau )=\ _{\star }^{\shortmid }\mathbf{g}%
_{\alpha _{s}\beta _{s}}(\tau )=\ ^{\shortmid }\mathbf{g}_{\alpha _{s}\beta
_{s}}(\tau ).$ Geometric flows on a parameter $\tau $ are described in $[0]$%
-approximation (zero power on $\kappa $) by flows of some canonical data $(\
_{s}^{\shortmid }\mathbf{g}(\tau ),\ _{s}^{\shortmid }\widehat{\mathbf{D}}%
(\tau )),$ star product flows $\star _{s}(\tau ),$ determined by s-adapted
frames $\ ^{\shortmid }\mathbf{e}_{i_{s}}(\tau )$ in (\ref{starpn}), and
flows of volume elements 
\begin{equation}
d\ \ ^{\shortmid }\mathcal{V}ol(\tau )=\sqrt{|\ \ ^{\shortmid }\mathbf{g}%
_{\alpha _{s}\beta _{s}}\ (\tau )|}\ \delta ^{8}\ ^{\shortmid }u^{\gamma
_{s}}(\tau )  \label{volform}
\end{equation}%
are computed for N-elongated s-differentials $\delta ^{8}\ ^{\shortmid
}u^{\gamma _{s}}(\tau )$ using $\ ^{\shortmid }N_{\ i_{s}a_{s}}\ (\tau )$ as
in (\ref{nadapbdss}).

We can elaborate on relativistic thermodynamics models \ and their
nonassociative generalizations if $\ _{s}^{\shortmid }\mathbf{g=\{\
^{\shortmid }\mathbf{g}}_{\alpha _{s}\beta _{s}}\}$ is adapted to a causal
(3+1)+(3+1) splitting (in GR, such a formalism for Einstein manifolds is
considered, for instance, in \cite{misner}). Geometric flows of a s-metric
can be parameterized in the form 
\begin{eqnarray*}
\ _{s}^{\shortmid }\mathbf{g}(\tau ) &=&\ ^{\shortmid }\mathbf{g}_{\alpha
^{\prime }\beta ^{\prime }}(\tau ,\ _{s}^{\shortmid }u)d\ ^{\shortmid }%
\mathbf{e}^{\alpha ^{\prime }}(\tau )\otimes d\ ^{\shortmid }\mathbf{e}%
^{\beta ^{\prime }}(\tau ) \\
&=&q_{i}(\tau ,x^{k})dx^{i}\otimes dx^{i}+q_{3}(\tau ,x^{k},y^{3})\mathbf{e}%
^{3}(\tau )\otimes \mathbf{e}^{3}(\tau )-\breve{N}^{2}(\tau ,x^{k},y^{3})%
\mathbf{e}^{4}(\tau )\otimes \mathbf{e}^{4}(\tau ) \\
&&+\ ^{\shortmid }q^{a_{2}}(\tau ,x^{k},y^{3},p_{b_{2}})\ ^{\shortmid }%
\mathbf{e}_{a_{2}}(\tau )\otimes \ ^{\shortmid }\mathbf{e}_{a_{2}}(\tau ) \\
&&+\ ^{\shortmid }q^{7}(\tau ,x^{k},y^{3},p_{b_{2}},p_{b_{3}})\ ^{\shortmid }%
\mathbf{e}_{7}(\tau )\otimes \ ^{\shortmid }\mathbf{e}_{7}(\tau )-\
^{\shortmid }\breve{N}^{2}(\tau ,x^{k},y^{3},p_{b_{2}},p_{b_{3}})\
^{\shortmid }\mathbf{e}_{8}(\tau )\otimes \ ^{\shortmid }\mathbf{e}%
_{8}(\tau).
\end{eqnarray*}%
For computations in a phase space point, such an ansatz is written as an
extension of a couple of 3--d metrics, $q_{ij}=diag(q_{\grave{\imath}%
})=(q_{i},q_{3})$ on a hyper-surface $\widehat{\Xi }_{t}$ and $\ ^{\shortmid
}q^{\grave{a}\grave{b}}=diag(\ ^{\shortmid }q^{\grave{a}})=(\ ^{\shortmid
}q^{a_{2}},\ ^{\shortmid }q^{7})$ on a hyper-surface $\ ^{\shortmid }%
\widehat{\Xi }_{E},$ \ i.e. on $\ _{s}^{\shortmid }\widehat{\Xi }=\left( 
\widehat{\Xi }_{t},\ ^{\shortmid }\widehat{\Xi }_{E}\right) ,$ when 
\begin{equation}
q_{1}=g_{1},q_{2}=g_{2},q_{3}=g_{3},\breve{N}^{2}=-g_{4}\mbox{ and }\
^{\shortmid }q^{5}=\ ^{\shortmid }g^{5},\ ^{\shortmid }q^{6}=\ ^{\shortmid
}g^{6},\ ^{\shortmid }q^{7}=\ ^{\shortmid }g^{7},\ ^{\shortmid }\breve{N}%
^{2}=-\ ^{\shortmid }g^{8},  \label{shift1}
\end{equation}%
where $\breve{N}$ is a lapse function on the base manifold and $\
^{\shortmid }\breve{N}^{2}$ is a lapse function in the cofiber.

The Perelman type functionals (\ref{perelmfst}) can be generalized for
nonassociative canonical data $\left[ \ _{s}^{\shortmid }\mathbf{g}^{\star
}(\tau ),\ _{s}^{\shortmid }\widehat{\mathbf{D}}^{\star }(\tau )\right] $
following the Convention 2 (\ref{conv2s} and using formulas (\ref{mafgeomob}%
), 
\begin{eqnarray}
\ _{s}^{\shortmid }\widehat{\mathcal{F}}^{\star }(\tau ) &=&\int_{\
_{s}^{\shortmid }\widehat{\Xi }}(\ _{s}^{\shortmid }\widehat{\mathbf{R}}%
sc^{\star }+|\ _{s}^{\shortmid }\widehat{\mathbf{D}}^{\star }\
_{s}^{\shortmid }\widehat{f}|^{2})\star e^{-\ \ _{s}^{\shortmid }\widehat{f}%
}\ d\ ^{\shortmid }\mathcal{V}ol(\tau ),\mbox{ and }  \label{naffunct} \\
\ _{s}^{\shortmid }\widehat{\mathcal{W}}^{\star }(\tau ) &=&\int_{\
_{s}^{\shortmid }\widehat{\Xi }}\left( 4\pi \tau \right) ^{-4}\ [\tau (\
_{s}^{\shortmid }\widehat{\mathbf{R}}sc^{\star }+\sum\nolimits_{s}|\
_{s}^{\shortmid }\widehat{\mathbf{D}}^{\star }\star \ _{s}^{\shortmid }%
\widehat{f}|)^{2}+\ _{s}^{\shortmid }\widehat{f}-8]\star e^{-\ \
_{s}^{\shortmid }\widehat{f}}\ d\ ^{\shortmid }\mathcal{V}ol(\tau ),
\label{nawfunct}
\end{eqnarray}%
where the integrals and normalizing functions $\ _{s}^{\shortmid }\widehat{f}%
(\tau ,\ _{s}^{\shortmid }u)$ are stated to satisfy the condition 
\begin{equation}
\int_{\ _{s}^{\shortmid }\widehat{\Xi }}\ _{s}^{\shortmid }\widehat{\nu }\ \
d\ ^{\shortmid }\mathcal{V}ol(\tau ):=\int_{t_{1}}^{t_{2}}\int_{\widehat{\Xi 
}_{t}}\ \int_{\ ^{\shortmid }\widehat{\Xi }_{E}}\ _{s}^{\shortmid }\widehat{%
\nu }\ \ d\ ^{\shortmid }\mathcal{V}ol(\tau )=1,  \label{normcond}
\end{equation}%
for integration measures $\ _{s}^{\shortmid }\widehat{\nu }=\left( 4\pi \tau
\right) ^{-4}e^{-\ _{s}^{\shortmid }\widehat{f}}$ parameterized for shells
on 8-d phase spaces. The nonassociative canonical s-connection $\
_{s}^{\shortmid }\widehat{\mathbf{D}}^{\star }$ (\ref{twoconsstar}) and
respective canonical Ricci scalar $\ _{s}^{\shortmid }\widehat{\mathbf{R}}%
sc^{\star }$ (\ref{ricciscsymnonsym}) are computed for a parametrization (%
\ref{shift1}).\footnote{%
It should be noted that the F- and W-functionals were postulated \ by G.
Perelman \cite{perelman1} in such forms that they allowed him to perform a
variational calculus and prove certain forms of the R. Hamilton equations 
\cite{hamilton82}, to define an associated statistical/geometric
thermodynamics models for Ricci flows of Riemannian metrics, and to prove
the Poincar\'{e}-Thurston conjecture. Those constructions can be generalized
on nonholonomic Lorentz manifolds and (co) tangent bundles which allow to
prove relativistic variants of geometric flows and elaborated on respective
thermodynamics models even (as we emphasized in the Introduction and \cite%
{vacaru20,ib19,bubuianu19}) some relativistic variants of the Poincar\'{e}%
-Thurston conjecture were not formulated/proven in modern mathematics. The
Perelman functionals and respective thermodynamic models can be generalized
for various non-Riemannian geometries including nonassociative models if the 
$\tau $-parametric star product (\ref{starpn}) and Convention 2 are
considered. This allows us to elaborate on theories of nonassociative
geometric flows defined by R-flux deformations in string theory.} We can
consider star-deformations of the volume form (\ref{volform}), 
\begin{equation*}
e^{-\ \ _{s}^{\shortmid }\widehat{f}}d\ ^{\shortmid }\mathcal{V}ol(\tau
)\rightarrow e^{-\ \ _{s}^{\shortmid }\widehat{f}}d\ ^{\shortmid }\mathcal{V}%
ol^{\star }(\tau )=e^{-\ \ _{s}^{\shortmid }\widehat{f}}\sqrt{|\ _{\star
}^{\shortmid }\mathbf{g}_{\alpha _{s}\beta _{s}}\ (\tau )|}\delta \
^{\shortmid }u^{\gamma _{s}}(\tau ),
\end{equation*}%
for other types of adapted integration measures and nonholonomic s-shells
with $\ _{s}^{\shortmid }\mathfrak{g}^{\star }.$ Such transforms can be
encoded into a normalizing function $\ _{s}^{\shortmid }\widehat{f}$ and
respective separation of nonsymmetric components of s-metrics for $\kappa $%
--linear parameterizations. We simplify further computations if the star
products and integration measures, and the orders for performing covariant
derivations and integration, are stated as in functionals (\ref{naffunct})
and (\ref{nawfunct}).

\subsection{Nonassociative geometric flow and Ricci soliton equations}

\label{ssngfeq}In this section, we consider two methods for deriving
nonassociative geometric flow equations (using abstract geometric methods
and/or elaborating a nonassociative onholonomic s-adapted variational
procedure).

\subsubsection{Abstract nonassociative geometric star-deformations of R.
Hamilton equations}

\label{ss231}Following abstract geometric principles as in \cite{misner}, we
can derive necessary type geometric/physical important questions considering
in symbolic coordinate/frame forms corresponding fundamental associative
geometric objects. We can consider any variant of Ricci tensor and scalar
curvature defined by respective metric and covariant derivative structures
(and, if it is important for certain constructions, nonlinear Laplace,
d'Alambert operators). This way, for instance, we can derive the Einstein
equations in pure geometric form as in pseudo-Riemannian geometry. Such
gravitational filed equations can be written in terms of the Ricci tensor
(the left side) and postulating (for the right side) certain types of
(effective) sources determined by corresponding physically important
energy-momentum tensors. This geometric approach, can be generalized for the
canonical s-connection structure $\ _{s}^{\shortmid }\widehat{\mathbf{D}}$ (%
\ref{twocon}) which results in modified phase space Einstein equations (\ref%
{seinsta}). Written in "hat" variables, such nonholonomically distorted
gravitational field equations can be decoupled and integrated in certain
general forms applying the AFCDM. In terms of the LC-connection $\
^{\shortmid }\mathbf{\nabla }$ such systems of nonlinear PDEs do not possess
any general decoupling properties for generic off-diagonal metrics
depending, in principle, on all phase space coordinates.

The abstract geometric approach allows us to derive in symbolic form certain
(associative and commutative) nonholonomic geometric flow equations \cite%
{v07,lb18,ib19,bubuianu19,vacaru20} for $\tau $-families of geometric
s-objects $\left( \ _{s}^{\shortmid }\mathbf{g}(\tau ),\ _{s}^{\shortmid }%
\widehat{\mathbf{D}}(\tau )\right).$ For s-metrics $\ _{s}^{\shortmid }%
\mathbf{g}(\tau )=\{\ ^{\shortmid }\mathbf{g}_{\alpha _{s}\beta _{s}}(\tau
)\}$ (\ref{sdm}), we can construct nonholonomic canonical s-deformations of
the the R. Hamilton equations \cite{hamilton82} postulated for various
research in modern geometric analysis. Such equations were originally
considered in connection to string theory and condensed matter physics in 
\cite{friedan80}. Then, applying the Convention 2 (\ref{conv2s}) we can
analyze and solve the issue on deriving nonassociative geometric flow
equations for the star deformed data $\left( \ _{\star }^{\shortmid }%
\mathfrak{g}_{\alpha _{s}\beta _{s}}(\tau ),\ _{s}^{\shortmid }\widehat{%
\mathbf{D}}^{\star }(\tau)\right)$. Such equations can be postulated (using
appropriate diffeomorphysms and s-adapted frame structures) in the form 
\begin{eqnarray}
\partial _{\tau }\ _{\star }^{\shortmid }\mathfrak{g}_{\alpha _{s}\beta
_{s}}(\tau ) &=&-2\ ^{\shortmid }\widehat{\mathbf{R}}_{\ \alpha _{s}\beta
_{s}}^{\star }(\tau ),  \label{nonassocgeomfl} \\
\partial _{\tau }\ _{s}^{\shortmid }\widehat{f}(\tau ) &=&\ _{s}^{\shortmid }%
\widehat{\mathbf{R}}sc^{\star }(\tau )-\ ^{\star }\widehat{\bigtriangleup }%
(\tau )\star \ \ _{s}^{\shortmid }\widehat{f}(\tau )+(\ _{s}^{\shortmid }%
\widehat{\mathbf{D}}^{\star }(\tau )\star \ \ _{s}^{\shortmid }\widehat{f}%
(\tau ))^{2}(\tau ),  \notag
\end{eqnarray}%
where $\ ^{\star }\widehat{\bigtriangleup }$ is the Laplace operator
constructed for $\ _{s}^{\shortmid }\widehat{\mathbf{D}}^{\star }$ and the
nonsymmetric components of $\ _{\star }^{\shortmid }\mathfrak{g}_{\alpha
_{s}\beta _{s}}$ are computed using $\kappa $-linear parameterizations (\ref%
{aux40a})--(\ref{aux40aa}).

The nonassociative geometric flow equations (\ref{nonassocgeomfl}) include
as associative and commutative parts (for LC-configurations and Riemannian
signatures) certain phase space variants of the evolution equations (1.3)
studied in \cite{perelman1}. The noncommutative part of such equations is
different from that considered in \cite{v08} because that work was devoted
to a different type of noncommutative Ricci flow theory based on spectral
triples following the A. Connes approach. We postulated above nonassociative
system of nonlinear PDEs \ in such a form that it can be decoupled and
solved in certain general forms at least in $\kappa $-linear form (see
details below: in subsection \ref{ssparamflow}, section \ref{sec4} and
appendix \ref{appendixb}).

\subsubsection{A s-adapted variational procedure for deriving nonassociative
geometric flow equations}

\label{ss232}In paragraphs 1.1 and 1.2 of Section 1 in \cite{perelman1}, it
is considered in brief a variational procedure to prove the associative and
commutative variants of geometric flow equations (\ref{nonassocgeomfl})
using Riemannian data $(g_{\alpha \beta }(\tau ),\nabla _{\gamma }(\tau ))$
for a normalizing function $f(\tau )$ on a closed manifold $M$ of dimension, 
$\dim M=n$ and $\tau \in \lbrack 0,\tau _{1}].$ In this work, we use our
system of notations for nonholonomic manifolds and/or phase spaces, $\
_{s}^{\shortmid }\mathcal{M}\rightarrow \ _{s}^{\star }\mathcal{M},$ and
perform canonical s-adapted geometric constructions on certain closed
spacelike regions with respective double $(3+1)+(3+1)$ and $(2+2)+(2+2)$
fibrations extended under spacetime/ phase space paths covering such
regions. We note that formal definitions of geometric s-objects and
respective covariant and integral calculus do not depend on the signature of
s-metrics. Such signatures are important for providing proofs of the Poincar%
\'{e}-Thurston conjecture and generalizations (which is not a goal for this
work), see details in \cite{perelman1,perelman2,perelman3}.

In nonholonomic form, certain N-adapted variational geometric flow methods
were considered in \cite{v07} for various developments and applications in 
\cite{v07,lb18,ib19,bubuianu19,vacaru20} when $\nabla \rightarrow \widehat{%
\mathbf{D}}$ and d-metrics of arbitrary signatures subjected to modified/
generalized R. Hamilton equations can be found by applying the AFCDM. \ To
develop such a N-adapted variational calculus for nonholonomic Ricci flow
theories in a rigorous mathematical form is possible (such proofs are on
hundreds of pages, with respective distortions of d-connections and
geometric d-objects, like in monographs \cite{kleiner06,morgan06,cao06}). We
omit such technical details in this work and sketch the proof in a form
similar to \cite{perelman1} but with respective geometric symbolic
re-definitions of (non) associative nonholonomic s-objects following the
Convention 2 (\ref{conv2s}), stating formal nonholonomic measures and
performing integration on a closed region of $\ _{s}^{\star }\mathcal{M}$.

It should be noted that arbitrary deformations induced by a twist operator
are not general compatible with the variational principle to be used
directly, for instance, in nonassociative field theory. Nevertheless, this
is not an unsolved conceptual/ technical problem if we work with star
product R-flux deformations defined in s-adapted form as in (\ref{starpn}).
We can elaborate a well defined nonholonomic geometric flow theory (with a
self-consistent s-adapted variational calculus) on $\ _{s}^{\shortmid }%
\mathcal{M}$. Then, we can $\star $-deform the constructions on $\
_{s}^{\star }\mathcal{M}$; and, for at least for $\kappa $-linear parametric
solutions, compute respective deformations of $F$- and $W$-functionals and
their nonassociative geometric flow equations (\ref{nonassocgeomfl}). In
such a case, a star-functional $\ _{s}^{\shortmid }\widehat{\mathcal{F}}%
^{\star }$ (\ref{naffunct}) is constructed for an explicit class of $\kappa $%
-linear solutions of nonassociative R. Hamilton equations in canonical
s-variables, when the measure and volume forms can be assumed to satisfy the
condition (\ref{normcond}). A similar assumption is used for the proof of
1.2 Proposition and formulas (1.1) - (1.4) in \cite{perelman1}.

So, considering $\ _{s}^{\shortmid }\widehat{\mathcal{F}}^{\star }$, we can
define and compute a s-adapted variation $\delta \ _{s}^{\shortmid }\widehat{%
\mathcal{F}}^{\star }$ for some variations \newline
$\delta (\ _{\star }^{\shortmid }\mathbf{g}_{\alpha _{s}\beta _{s}})=\
\delta (\ ^{\shortmid }\mathbf{g}_{\alpha _{s}\beta _{s}})$ and $\delta \ \
_{s}^{\shortmid }\widehat{f}$ as follows:%
\begin{eqnarray*}
\delta \ _{s}^{\shortmid }\widehat{\mathcal{F}}^{\star }(\tau ) &=&\int_{\
_{s}^{\shortmid }\widehat{\Xi }}\{(-\ ^{\star }\widehat{\bigtriangleup }[\
^{\shortmid }\mathbf{g}^{\alpha _{s}\beta _{s}}\delta (\ ^{\shortmid }%
\mathbf{g}_{\alpha _{s}\beta _{s}})]+\ ^{\shortmid }\widehat{\mathbf{D}}%
_{\star }^{\alpha _{s}}\ ^{\shortmid }\widehat{\mathbf{D}}_{\star }^{\beta
_{s}}[\delta (\ ^{\shortmid }\mathbf{g}_{\alpha _{s}\beta _{s}})]-\
^{\shortmid }\widehat{\mathbf{R}}_{\ \star }^{\alpha _{s}\beta _{s}}\delta
(\ ^{\shortmid }\mathbf{g}_{\alpha _{s}\beta _{s}}) \\
&&-\delta (\ ^{\shortmid }\mathbf{g}_{\alpha _{s}\beta _{s}})\ ^{\shortmid }%
\widehat{\mathbf{D}}_{\star }^{\alpha _{s}}(\ _{s}^{\shortmid }\widehat{f})\
^{\shortmid }\widehat{\mathbf{D}}_{\star }^{\beta _{s}}(\ _{s}^{\shortmid }%
\widehat{f})+2\ ^{\shortmid }\widehat{\mathbf{D}}_{\alpha _{s}}^{\star }(\
_{s}^{\shortmid }\widehat{f})\ ^{\shortmid }\widehat{\mathbf{D}}_{\star
}^{\alpha _{s}}(\delta \ \ _{s}^{\shortmid }\widehat{f}) \\
&&+(\ _{s}^{\shortmid }\widehat{\mathbf{R}}sc^{\star }+|\ _{s}^{\shortmid }%
\widehat{\mathbf{D}}^{\star }\ _{s}^{\shortmid }\widehat{f}|^{2})(\frac{1}{2}%
\ ^{\shortmid }\mathbf{g}^{\alpha _{s}\beta _{s}}\delta (\ ^{\shortmid }%
\mathbf{g}_{\alpha _{s}\beta _{s}})-\delta \ \ _{s}^{\shortmid }\widehat{f}%
)\}\star e^{-\ \ _{s}^{\shortmid }\widehat{f}}\ 
\end{eqnarray*}%
\begin{eqnarray}
&=&\int_{\ _{s}^{\shortmid }\widehat{\Xi }}\{-\delta (\ ^{\shortmid }\mathbf{%
g}_{\alpha _{s}\beta _{s}})[\ ^{\shortmid }\widehat{\mathbf{R}}_{\ \star
}^{\alpha _{s}\beta _{s}}+\ ^{\shortmid }\widehat{\mathbf{D}}_{\star
}^{\alpha _{s}}\ ^{\shortmid }\widehat{\mathbf{D}}_{\star }^{\beta _{s}}(\
_{s}^{\shortmid }\widehat{f})]  \label{variationf} \\
&&+(\frac{1}{2}\ ^{\shortmid }\mathbf{g}^{\alpha _{s}\beta _{s}}\delta (\
^{\shortmid }\mathbf{g}_{\alpha _{s}\beta _{s}})-\delta \ \ _{s}^{\shortmid }%
\widehat{f})[2\ ^{\star }\widehat{\bigtriangleup }(\ _{s}^{\shortmid }%
\widehat{f})+\ _{s}^{\shortmid }\widehat{\mathbf{R}}sc^{\star }-|\
_{s}^{\shortmid }\widehat{\mathbf{D}}^{\star }\ _{s}^{\shortmid }\widehat{f}%
|^{2}]\}\star e^{-\ \ _{s}^{\shortmid }\widehat{f}}\   \notag
\end{eqnarray}%
Here we note that $\frac{1}{2}\ ^{\shortmid }\mathbf{g}^{\alpha _{s}\beta
_{s}}\delta (\ ^{\shortmid }\mathbf{g}_{\alpha _{s}\beta _{s}})-\delta \ \
_{s}^{\shortmid }\widehat{f})\equiv 0$ if the measure $d\ ^{\shortmid }%
\widehat{\mathcal{V}}:\mathcal{=}\ _{s}^{\shortmid }\widehat{\nu }\ d\
^{\shortmid }\mathcal{V}ol$ $\ e^{-\ \ _{s}^{\shortmid }\widehat{f}}=const.$
For such s-adapted configurations, we can prescribe the nonholonomic
structure and compute (using symmetric coordinate configurations and then
re-defining for N-connections) when the symmetric s-tensor 
\begin{equation*}
-[\ ^{\shortmid }\widehat{\mathbf{R}}_{\ \star }^{\alpha _{s}\beta _{s}}+\
^{\shortmid }\widehat{\mathbf{D}}_{\star }^{\alpha _{s}}\ ^{\shortmid }%
\widehat{\mathbf{D}}_{\star }^{\beta _{s}}(\ _{s}^{\shortmid }\widehat{f})]%
\mbox{ is the }L^{2}\mbox{ gradient of the functional }\ _{s}^{\shortmid }%
\widehat{\mathcal{F}}^{\star }[d\ ^{\shortmid }\widehat{\mathcal{V}}%
]=\int_{\ _{s}^{\shortmid }\widehat{\Xi }}(\ _{s}^{\shortmid }\widehat{%
\mathbf{R}}sc^{\star }+|\ _{s}^{\shortmid }\widehat{\mathbf{D}}^{\star }\
_{s}^{\shortmid }\widehat{f}|^{2}d\ ^{\shortmid }\widehat{\mathcal{V}}.
\end{equation*}%
In these formulas, $\ _{s}^{\shortmid }\widehat{f}$ can be considered as $%
\log (d\ ^{\shortmid }\mathcal{V}ol/d\ ^{\shortmid }\widehat{\mathcal{V}}).$
So, prescribing a measure $d\ ^{\shortmid }\widehat{\mathcal{V}},$ we can
model a nonassociative gradient flow subjected to equations 
\begin{equation*}
\partial _{\tau }\ \ _{\star }^{\shortmid }\mathbf{g}_{\alpha _{s}\beta
_{s}}=-2[\ ^{\shortmid }\widehat{\mathbf{R}}_{\ \alpha _{s}\beta
_{s}}^{\star }+\ ^{\shortmid }\widehat{\mathbf{D}}_{\alpha _{s}}^{\star }\
^{\shortmid }\widehat{\mathbf{D}}_{\beta _{s}}^{\star }(\ _{s}^{\shortmid }%
\widehat{f})],
\end{equation*}%
derived from a $\ _{s}^{\shortmid }\widehat{\mathcal{F}}^{\star }[d\
^{\shortmid }\widehat{\mathcal{V}}].$ Because the right part can be
constructed as a nonholonomic nonsymmetric star deformation, we can chose a
corresponding $\ _{s}^{\shortmid }\widehat{f}$ when $\ \ _{\star
}^{\shortmid }\mathbf{g}_{\alpha _{s}\beta _{s}}\rightarrow \ _{\star
}^{\shortmid }\mathfrak{g}_{\alpha _{s}\beta _{s}}$. Modifying by an
appropriate diffeomorphism and nonholonomic s-adapted structure, we obtain
this type of nonassociative geometric evolution equation: 
\begin{eqnarray*}
\partial _{\tau }\ _{\star }^{\shortmid }\mathfrak{g}_{\alpha _{s}\beta
_{s}} &=&-2\ ^{\shortmid }\widehat{\mathbf{R}}_{\ \alpha _{s}\beta
_{s}}^{\star }, \\
\partial _{\tau }\ _{s}^{\shortmid }\widehat{f} &=&\ _{s}^{\shortmid }%
\widehat{\mathbf{R}}sc^{\star }-\ ^{\star }\widehat{\bigtriangleup }\ \
_{s}^{\shortmid }\widehat{f}+|\ _{s}^{\shortmid }\widehat{\mathbf{D}}^{\star
}\ \ _{s}^{\shortmid }\widehat{f}|^{2}.
\end{eqnarray*}%
Such formulas are equivalent (up to certain nonholonomic transforms and
re-definitions of the normalizing functions) to the nonassociative geometric
flow equations (\ref{nonassocgeomfl}), which we postulated/ constructed
following abstract/ symbolic principles.

It is important to note that above formulas obtained from a s-adapted
variational principle with (\ref{variationf}) still have a geometric
symbolic character if we consider general star product R-flux deformations.
Even in the associative and commutative Riemannian case, gradient flows may
not exist for general measures (see related explanations before paragram 1.2
Proposition and formulas (1.1) - (1.4) in \cite{perelman1}). So, we can not
prove that (non) associative geometric flow evolution equations of type (\ref%
{nonassocgeomfl}) can be proven in a general, or s-adapted form from a
functional $\ _{s}^{\shortmid }\widehat{\mathcal{F}}^{\star }$ (\ref%
{naffunct}). This results in various un-determined variants of
nonassociative functionals and symbolic geometric flow equations which to
not allow even to formulate certain nonassociative variants of the Poincar%
\'{e}-Thurston conjecture. Nevertheless, well-defined star-deformed
functionals $\ _{s}^{\shortmid }\widehat{\mathcal{F}}^{\star }$ and $\
_{s}^{\shortmid }\widehat{\mathcal{W}}^{\star }$ and related variants of
nonassociative geometric evolution equations can be introduced in a
self-consistent geometric form (and with various important implications in
modern gravity and quantum information theories) if we consider $\kappa $%
-linear parametric deformations. In such cases, all geometric and physical
objects (and related N-adapted variational procedures) allow to define and
compute such values and finding of exact/parametric solutions in explicit
forms. This is possible when we apply the AFCDM and construct generic
off-diagonal solutions for certain (associative and commutative)
nonholonomic s-adapted configurations and then subject such geometric/
physical data to star product deformations (\ref{starpn}). If such R-flux
deformations are computed following the procedure described in section \ref%
{sec4} and appendix \ref{appendixb}), the s-adapted variational procedure (%
\ref{variationf}) becomes well-defined mathematically even the constructions
involve a twist operator. All such tedious and technical computations are
performed using the third line of (\ref{starpn}) and $\kappa $-parametric
decompositions of geometric s-objects as in \cite%
{aschieri17,partner01,partner02,partner04}. We may have certain undetermined
values for general classes of nonholonomic bases and s-connections. But if
we chose certain canonical variables and deformations of physical important
and well-defined solutions (for instance, for BH in any nonassociative or
associative variant), then the corresponding nonassociative geometric flows
can be modelled in a nonassociative gradient form with a well-defined
s-adapted variational procedure.

In our series of works \cite{partner01,partner02,partner04}, we consider
that for the nonassociative geometric flow theories the abstract geometric
symbolic principles are more fundamental and efficient that standard
variational procedures (similar to classical and quantum field theories)
which became un-determined, for instance, for general star product R-flux
deformations. Such an assumption is based on ideas from \cite{misner} that
having certain data for a necessary set of fundamental geometric objects
(metric-affine structures, nonlinear and linear connections, respective
Riemannian, Ricci tensors etc.) we can elaborate always on respective
gravity/ geometric flow models following standard geometric principles. The
constructions are symbolic, but for explicit variants of nonassociative/
noncommutative/ supersymmetric etc. generalizations we can state an
equivalent variational procedure if certain well-defined $\kappa $%
-parametric decompositions are stated for such models following a typical
"deformation philosophy".

Nonassociative geometric flow equations can be derived in similar forms if,
for instance, a s-adapted variational procedure is performed for $\
_{s}^{\shortmid }\widehat{\mathcal{W}}^{\star }(\tau )$ (\ref{nawfunct}).
Such details for the LC-connection are provided in \cite{perelman1}, see
analogous constructions for Riemannian geometric flows; all described by
respective formulas 3.1 - 3.4 in section 3 of that work. In s-adapted form,
the approach was generalized in \cite{v07,lb18,ib19,bubuianu19,vacaru20}. We
omit such technical details in this work because they can be derived in
abstract form following geometric principles and the Convention 2 (\ref%
{conv2s}). For recent applications in high energy physics, we cite \cite%
{kehagias19,biasio20,biasio21,lueben21,biasio22} where the normalizing
function is postulated as a dilaton field and associative and commutative
versions of metric-dilaton Ricci flows are investigated. Certain geometric
flow equations can be also motivated as star product R-flux deformations of
a two-dimensional sigma model with beta functions and dilaton field (see
equations (79) and (80) in \cite{kehagias19}). Here we note that in section
1.4$^{\ast }$ of \cite{perelman1} it is mentioned that the F-functional and
"its variation formula can be found in the literature on the string theory
...", when $f$ can be treated as a dilaton filed. In our works on
nonassociative geometric flows, we show that G. Perelman constructions can
be generalized for star products with R-flux deformations from string
theory. The AFCDM allows to elaborate on such nonassociative geometric flow
models in explicit forms considering various classes of physically important
solutions.

\subsubsection{Nonassociative Ricci soliton equations in canonical
s-variables}

Ricci solitons are defined as self-similar configurations of gradient
geometric flows for a fixed parameter $\tau _{0}.$ For Riemannian and
Kaehler Ricci flows, such geometries are studied in details in \cite%
{kleiner06,morgan06,cao06} (where different types of Ricci soliton equations
are formulated). In \cite{vacaru20,ib19,bubuianu19} and references therein,
the approach was extended to nonholonomic s-adapted constructions. In
canonical s-variables on $\ _{s}\mathcal{M},$ the Ricci soliton s-equations
derived from a W-functional are of type 
\begin{equation*}
\ ^{\shortmid }\widehat{\mathbf{R}}_{\ \alpha _{s}\beta _{s}}+\ ^{\shortmid }%
\widehat{\mathbf{D}}_{\alpha _{s}}\ ^{\shortmid }\widehat{\mathbf{D}}_{\beta
_{s}}\ ^{\shortmid }\varpi (\ _{s}^{\shortmid }u)=\lambda \ ^{\shortmid }%
\mathbf{g}_{\alpha _{s}\beta _{s}},
\end{equation*}%
where $^{\shortmid }\varpi $ is a smooth potential function on every shell $%
s=1,2,3,4$ and $\lambda =const.$ Following the Convention 2 (\ref{conv2s}),
such systems of nonlinear PDEs can be deformed by star products and R-fluxes
to nonassociative Ricci solitons defined by equations 
\begin{equation}
\ ^{\shortmid }\widehat{\mathbf{R}}_{\ \alpha _{s}\beta _{s}}^{\star }+\
^{\shortmid }\widehat{\mathbf{D}}_{\alpha _{s}}^{\star }\ ^{\shortmid }%
\widehat{\mathbf{D}}_{\beta _{s}}^{\star }\ ^{\shortmid }\varpi (\
_{s}^{\shortmid }u)=\lambda \ \ \ _{\star }^{\shortmid }\mathfrak{g}_{\alpha
_{s}\beta _{s}}.  \label{naricsol}
\end{equation}%
Similar equations (certain differences can be related to different
nonholonomic structures and/or different normalizing functions) can derived
from a respective s-adapted variational calculus with $\tau = \tau _{0}$ for 
$\ _{s}^{\shortmid }\widehat{\mathcal{W}}^{\star }(\tau )$ (\ref{nawfunct})
and/or from (\ref{nonassocgeomfl}). We omit such technical details. Here we
note that the nonassociative phase space vacuum gravitational equations (\ref%
{nonassocdeinst1}) defined for the canonical s-connection $\ ^{\shortmid }%
\widehat{\mathbf{D}}_{\alpha _{s}}^{\star }$ consist an example of
nonassociative Ricci soliton ones (\ref{naricsol}).

The nonassociative geometric flow constructions provided in this section can
be re-defined in terms of respective LC-connections, $\ _{\star}^{\shortmid
}\nabla $ and $\ ^{\shortmid }\nabla ,$ if we impose additional nonholonomic
constraints of type (\ref{lccondnonass}), when $\ _{s}^{\shortmid }\widehat{%
\mathbf{D}}_{\mid \ _{s}^{\shortmid }\widehat{\mathbf{T}}=0}^{\star }= \
_{\star }^{\shortmid }\nabla .$ As a result, the nonassociative equations (%
\ref{nonassocgeomfl}) and (\ref{naricsol}) transform respectively into (\ref%
{nonassocheq}) and (\ref{nonassocveq}) postulated in the Introduction
section. For $\ _{\star }^{\shortmid}\nabla ,$ such nonassociative geometric
flow and Ricci soliton equations could be postulated just having the results
of papers \cite{blumenhagen16,aschieri17}, where the nonassociative Ricci
tensors were defined and computed for $\ _{\star }^{\shortmid }\nabla $ (in
our notations). The main motivation for elaborating such theories in terms
of $\ ^{\shortmid }\widehat{\mathbf{D}}_{\alpha _{s}}^{\star }$ (and, with
nonholonomic constraints, of $\ _{\star }^{\shortmid }\nabla $) is that
using nonholonomic s-adapted variables we can decouple and solve in very
general forms such systems of nonlinear PDEs \cite%
{partner02,partner03,partner04}. This is possible if we apply the AFCDM (see
main ideas and important formulas in Appendix \ref{appendixb}). Constructing
exact and parametric solutions of nonassociative Ricci flow/ soliton
equations, we analyze how the results and methods of nonassociative geometry
can be applied in modern particle physics, gravity and information theory.

\subsection{Parametric decomposition of nonassociative functionals and
geometric flow equations}

\label{ssparamflow}To elaborate on possible applications in modern gravity
and cosmology, the nonassociative F- and W-functionals and related geometric
flow equations are considered for a $\kappa $-linear parametric
decomposition. Using formulas (\ref{paramscon}), (\ref{driccicanonstar1}), (%
\ref{ricciscsymnonsym}) and (\ref{paramsricci}), for respective parametric
formulas for the canonical s-connection, nonsymmetric Ricci s-tensor and
scalar curvature, we write (\ref{naffunct}) and (\ref{nawfunct}) in the
forms: 
\begin{eqnarray}
\ _{s}^{\shortmid }\widehat{\mathcal{F}}_{\kappa }^{\star }(\tau )
&=&\int_{\ _{s}^{\shortmid }\widehat{\Xi }}(\ \ _{s}^{\shortmid }\widehat{%
\mathbf{R}}sc+\ _{s}^{\shortmid }\widehat{\mathbf{K}}sc+|\ _{s}^{\shortmid }%
\widehat{\mathbf{D}}\ _{s}^{\shortmid }\widehat{f}|^{2})e^{-\ \
_{s}^{\shortmid }\widehat{f}}\ d\ ^{\shortmid }\mathcal{V}ol(\tau ),%
\mbox{
and }  \label{naffunctp} \\
\ _{s}^{\shortmid }\widehat{\mathcal{W}}_{\kappa }^{\star }(\tau )
&=&\int_{\ _{s}^{\shortmid }\widehat{\Xi }}\left( 4\pi \tau \right) ^{-4}\
[\tau (\ \ _{s}^{\shortmid }\widehat{\mathbf{R}}sc+\ _{s}^{\shortmid }%
\widehat{\mathbf{K}}sc+\sum\nolimits_{s}|\ _{s}^{\shortmid }\widehat{\mathbf{%
D}}\ _{s}^{\shortmid }\widehat{f}|)^{2}+\ _{s}^{\shortmid }\widehat{f}%
-8]e^{-\ \ _{s}^{\shortmid }\widehat{f}}\ d\ ^{\shortmid }\mathcal{V}ol(\tau
),  \label{nawfunctp}
\end{eqnarray}%
where $_{s}^{\shortmid }\widehat{\mathbf{R}}sc^{\star }=\ _{s}^{\shortmid }%
\widehat{\mathbf{R}}sc+\ _{s}^{\shortmid }\widehat{\mathbf{K}}sc,$ for $\
_{s}^{\shortmid }\widehat{\mathbf{K}}sc:=\ _{\star }^{\shortmid }\mathfrak{g}%
^{\mu _{s}\nu _{s}}\ ^{\shortmid }\widehat{\mathbf{K}}_{\ \beta _{s}\gamma
_{s}}\left\lceil \hbar ,\kappa \right\rceil $ and the normalizing function $%
\ \ _{s}^{\shortmid }\widehat{f}$ is re-defined to include $\left\lceil
\hbar ,\kappa \right\rceil $-terms from $\ ^{\shortmid }\widehat{\mathbf{D}}%
^{\star }\rightarrow \ _{s}^{\shortmid }\widehat{\mathbf{D}}$ and remaining
terms from $\kappa $-parametric decompositions.

There are two ways for deriving nonassociative $\kappa $-linear
generalizations of the R. Hamilton equations. In the first case, we consider 
$\kappa $-parametric decompositions of (\ref{nonassocgeomfl}) and, in the
second case, we apply a s-adapted nonholonomic variational procedure to $\
_{s}^{\shortmid }\widehat{\mathcal{F}}_{\kappa }^{\star }(\tau )$ (\ref%
{naffunctp}), or $\ _{s}^{\shortmid }\widehat{\mathcal{W}}_{\kappa }^{\star
}(\tau )$ (\ref{nawfunctp}). In all cases, adapting corresponding the
nonholonomic structure, we obtain such phase geometric flow equations
encoding $\kappa $-terms, 
\begin{eqnarray}
\partial _{\tau }\ ^{\shortmid }\mathbf{g}_{\alpha _{s}\beta _{s}}(\tau )
&=&-2(\ ^{\shortmid }\widehat{\mathbf{R}}_{\ \alpha _{s}\beta _{s}}(\tau )+\
^{\shortmid }\widehat{\mathbf{K}}_{\ \alpha _{s}\beta _{s}}(\tau
,\left\lceil \hbar ,\kappa \right\rceil )),  \label{nonassocgeomflp} \\
\partial _{\tau }\ \ _{s}^{\shortmid }\widehat{f}(\tau ) &=&\
_{s}^{\shortmid }\widehat{\mathbf{R}}sc(\tau )+\ _{s}^{\shortmid }\widehat{%
\mathbf{K}}sc(\tau )-\ \widehat{\bigtriangleup }(\tau )\ \ _{s}^{\shortmid }%
\widehat{f}(\tau )+(\ _{s}^{\shortmid }\widehat{\mathbf{D}}(\tau )\ \
_{s}^{\shortmid }\widehat{f}(\tau ))^{2}(\tau ),  \notag
\end{eqnarray}%
where $\widehat{\bigtriangleup }$ is the Laplace operator constructed for $\
_{s}^{\shortmid }\widehat{\mathbf{D}}.$ Positively, the s-adapted
variational procedure with\ $\kappa $-linear decompositions in (\ref%
{nonassocgeomfl}) can be performed in a well-defined mathematical form by
involving the AFCDM for constructing respective classes of exact/parametric
solutions.

For self-similar configurations with $\tau =\tau _{0},$ the equations (\ref%
{nonassocgeomflp}) transform into a system of nonlinear PDEs for $\kappa $%
--parametric canonical shell Ricci solitons, 
\begin{equation}
\ ^{\shortmid }\widehat{\mathbf{R}}_{\ \alpha _{s}\beta _{s}}+\ ^{\shortmid }%
\widehat{\mathbf{K}}_{\ \alpha _{s}\beta _{s}}(\tau ,\left\lceil \hbar
,\kappa \right\rceil )+\ ^{\shortmid }\widehat{\mathbf{D}}_{\alpha _{s}}\
^{\shortmid }\widehat{\mathbf{D}}_{\beta _{s}}\ ^{\shortmid }\varpi (\
_{s}^{\shortmid }u)=\lambda \ \ ^{\shortmid }\mathbf{g}_{\alpha _{s}\beta
_{s}}.  \label{naricsolp}
\end{equation}%
Similar equations can be also obtained from a corresponding $\kappa $-linear
decompositions of the nonassociative Ricci soliton equations (\ref{naricsol}%
). Re-defining $\ ^{\shortmid }\varpi (\ _{s}^{\shortmid }u)$ for some
particular choices and for corresponding nonholonomic structures, we obtain
from (\ref{naricsolp}) phase space modified gravitational equations $\
^{\shortmid }\widehat{\mathbf{R}}ic_{\alpha _{s}\beta _{s}}=\
^{\shortmid}\Upsilon _{\alpha _{s}\beta _{s}}$ (\ref{seinsta}), where $\
^{\shortmid}\Upsilon _{\alpha _{s}\beta _{s}}=-\ _{s}^{\shortmid }\widehat{%
\mathcal{K}}ic_{\alpha _{s}\beta _{s}}.$ Such systems of nonlinear PDEs can
be decoupled and integrated in general off-diagonal forms using the AFCDM if
the effective source $\ ^{\shortmid }\Upsilon _{\alpha _{s}\beta _{s}}$ is
parameterized following the conventions (\ref{nassocsr2}) and $\
_{s}^{\shortmid }\mathcal{K}$ (\ref{cannonsymparamc2a}).

\section{Nonassociative geometric thermodynamics}

\label{sec3}For the Ricci flows of Riemannian metrics, the W-functional (\ref%
{perelmfst}) can be treated as a "minus entropy" which allows to formulate a
statistical thermodynamic model with thermodynamic variables determined by $%
\tau $-running fundamental geometric objects on Riemann manifolds. In \cite%
{vacaru20,ib19,bubuianu19} (see also references therein), we investigated
possibilities to extend such constructions to relativistic geometric
thermodynamic models and (modified) gravity and quantum information
theories. A very important result was that modified G. Perelman
thermodynamic models can be associated to various classes generic
off-diagonal solutions (in general, with non-Riemannian connections and
without conventional horizons) when the concept of Bekenstein--Hawking
thermodynamics is not applicable.

The goal of this section is to show how nonholonomic geometric thermodynamic
models can be elaborated for nonassociative geometric flows and Ricci
solitons determined by certain data $[\ _{\star }^{\shortmid }\mathfrak{g}%
_{\alpha _{s}\beta _{s}}(\tau ), \ _{s}^{\shortmid}\widehat{\mathbf{D}}%
^{\star }(\tau ),\ _{s}^{\shortmid }\widehat{f}(\tau )]$ and $\kappa $%
-linear parametric decompositions.

\subsection{Star product and R-flux deformed statistical thermodynamic
variables}

On $_{s}\mathcal{M}$ with an additional nonholonomic (3+1)+(3+1) splitting,
we introduce the statistical partition function 
\begin{equation}
\ _{s}^{\shortmid }\widehat{Z}(\tau )=\exp [\int_{\ _{s}^{\shortmid }%
\widehat{\Xi }}[-\ _{s}^{\shortmid }\widehat{f}+4]\ \left( 4\pi \tau \right)
^{-4}e^{-\ _{s}^{\shortmid }\widehat{f}}\ ^{\shortmid }\delta \ ^{\shortmid }%
\mathcal{V}(\tau ),  \label{spf}
\end{equation}%
where the volume element (\ref{volform}) are computed for a s-metric (\ref%
{shift1}) with "shift and lapse" functions, {\small 
\begin{equation}
\ ^{\shortmid }\delta \ ^{\shortmid }\mathcal{V}(\tau )=\sqrt{|q_{1}(\tau
)q_{2}(\tau )q_{3}(\tau )\breve{N}^{2}(\tau )\ ^{\shortmid }q^{5}(\tau )\
^{\shortmid }q^{6}(\tau )\ ^{\shortmid }q^{7}(\tau )\ ^{\shortmid }\breve{N}%
^{2}(\tau )|}dx^{1}dx^{2}\delta y^{3}\delta y^{4}\ ^{\shortmid }\delta \
^{\shortmid }u_{5}(\tau )\ ^{\shortmid }\delta \ ^{\shortmid }u_{6}(\tau )\
^{\shortmid }\delta \ ^{\shortmid }u_{7}(\tau )\ ^{\shortmid }\delta \
^{\shortmid }u_{8}(\tau ).  \label{volume}
\end{equation}%
} The Convention 2 (\ref{conv2s}) on star product R-flux deformations of
s-adapted geometric objects into respective nonassociative ones has to be
adapted for parameterizations of type $\ _{s}^{\shortmid t}\widehat{\mathbf{D%
}}=\ _{s}^{\shortmid }\widehat{\mathbf{D}}_{\mid \widehat{\Xi }%
_{t}}\rightarrow \ _{s}^{\shortmid }\widehat{\mathbf{D}}_{\mid \widehat{\Xi }%
_{t}}^{\star }$ and $\ _{s}^{\shortmid E}\widehat{\mathbf{D}}=\
_{s}^{\shortmid }\widehat{\mathbf{D}}_{\mid \widehat{\Xi }_{E}}\rightarrow \
_{s}^{\shortmid }\widehat{\mathbf{D}}_{\mid \widehat{\Xi }_{E}}^{\star }.$
Such transforms have to be considered for hyper-surface restrictions of the
canonical s-connections $\ _{s}^{\shortmid }\widehat{\mathbf{D}}$ $%
\rightarrow \ _{s}^{\shortmid }\widehat{\mathbf{D}}^{\star }$ and computing
integrals with volume forms (\ref{volume}). We can define and compute Ricci
s-tensors and scalar curvatures determined by $\ _{s}^{\shortmid t}\widehat{%
\mathbf{D}}^{\star }$ and $\ _{s}^{\shortmid E}\widehat{\mathbf{D}}^{\star }$
are denoted $\ _{s}^{\shortmid t}\widehat{\mathbf{R}}_{\grave{\imath}\grave{j%
}}^{\star },\ _{s}^{\shortmid E}\widehat{\mathbf{R}}_{\star }^{\grave{a}%
\grave{b}}$ and $\ ^{\shortmid t}\widehat{R}^{\star },\ ^{\shortmid E}%
\widehat{R}^{\star }$ which are useful for computing hyper-surface geometric
s-objects for quasi-stationary configurations with redefined normalizing
functions (see examples with $\kappa $-parametric nonassociative Ricci
solitons in \cite{partner04} \ and, for nonholonomic associative and
commutative configurations in \cite{vacaru20,ib19,bubuianu19}) and next
sections).

Using $\ _{s}^{\shortmid }\widehat{Z}$ (\ref{spf}) and $\ _{s}^{\shortmid }%
\widehat{\mathcal{W}}^{\star }(\tau )$ (\ref{nawfunct}) as the respective
partition function and W-entropy functional on $\ _{s}^{\star }\mathcal{M},$%
\footnote{%
To formulate a statistical thermodynamic model, we can consider a partition
function $Z=\int \exp (-\beta E)d\omega (E)$ for the canonical ensemble at
temperature $\beta ^{-1}=\tau $ being defined by the measure taken to be the
density of states $\omega (E).$ The thermodynamical variables are computed
as the average energy, $\ \left\langle E\right\rangle :=-\partial \log
Z/\partial \beta ,$ the entropy $S:=\beta \left\langle E\right\rangle +\log
Z $ and the fluctuation parameter $\sigma :=\left\langle \left(
E-\left\langle E\right\rangle \right) ^{2}\right\rangle =\partial ^{2}\log
Z/\partial \beta ^{2}.$} we can define and compute respective thermodynamic
variables (average energy, $\ _{s}^{\shortmid }\widehat{\mathcal{E}}^{\star
} $, entropy, $\ \ _{s}^{\shortmid }\widehat{S}^{\star },$ and fluctuation, $%
\ _{s}^{\shortmid }\widehat{\sigma }^{\star }$): 
\begin{eqnarray}
\ _{s}^{\shortmid }\widehat{\mathcal{E}}^{\star }\ &=&-\tau ^{2}\int_{\
_{s}^{\shortmid }\widehat{\Xi }}\ \left( 4\pi \tau \right) ^{-4}\left( \
_{s}^{\shortmid }\widehat{\mathbf{R}}sc^{\star }+|\ _{s}^{\shortmid }%
\widehat{\mathbf{D}}^{\star }\ _{s}^{\shortmid }\widehat{f}|^{2}-\frac{4}{%
\tau }\right) \star e^{-\ _{s}^{\shortmid }\widehat{f}}\ ^{\shortmid }\delta
\ ^{\shortmid }\mathcal{V}(\tau ),  \label{nagthermodvalues} \\
\ \ _{s}^{\shortmid }\widehat{S}^{\star } &=&-\int_{\ _{s}^{\shortmid }%
\widehat{\Xi }}\left( 4\pi \tau \right) ^{-4}\left( \tau (\ _{s}^{\shortmid }%
\widehat{\mathbf{R}}sc^{\star }+|\ _{s}^{\shortmid }\widehat{\mathbf{D}}%
^{\star }\ _{s}^{\shortmid }\widehat{f}|^{2})+\ _{s}^{\shortmid }\tilde{f}%
-8\right) \star e^{-\ _{s}^{\shortmid }\widehat{f}}\ ^{\shortmid }\delta \
^{\shortmid }\mathcal{V}(\tau ),  \notag \\
\ _{s}^{\shortmid }\widehat{\sigma }^{\star } &=&2\ \tau ^{4}\int_{\
_{s}^{\shortmid }\widehat{\Xi }}\left( 4\pi \tau \right) ^{-4}|\ \
^{\shortmid }\widehat{\mathbf{R}}_{\alpha _{s}\beta _{s}}^{\star }+\
^{\shortmid }\widehat{\mathbf{D}}_{\alpha _{s}}^{\star }\ ^{\shortmid }%
\widehat{\mathbf{D}}_{\beta _{s}}^{\star }\ \ _{s}^{\shortmid }\widehat{f}-%
\frac{1}{2\tau }\mathbf{g}_{\alpha _{s}\beta _{s}}^{\star }|^{2}\star e^{-\
_{s}^{\shortmid }\widehat{f}}\ ^{\shortmid }\delta \ ^{\shortmid }\mathcal{V}%
(\tau ).  \notag
\end{eqnarray}%
To prove these formulas we can apply a tedious variational s-adapted
calculus on nonassocitative phase space. Following the abstract geometric
formalism, such formulas can be derived in a simplified symbolic form when
geometric s-objects are generalized into similar ones with star labels.

We can restrict such formulas to 4--d and 6-d shell configurations, for
respective redefinitions of $\ _{s}^{\shortmid }\widehat{f}$ into a
convenient $\ _{s}^{\shortmid }\tilde{f}$, in order to adapt the geometric
thermodynamic constructions to a prescribed both shell and (3+1)+(3+1)
splitting. For corresponding fixed value $\tau =\tau _{0},$ the formulas (%
\ref{nagthermodvalues}) characterize nonassociative Ricci soliton (\ref%
{naricsol}) (and, in particular, nonassociative vacuum gravitational (\ref%
{nonassocdeinst1})) equations. Such thermodynamic values can be computed for
any example of exact/parametric solution of nonassociative geometric flow
equations (\ref{nonassocgeomfl}).

\subsection{Parametric decompositions in nonassociative geometric
thermodynamics}

For $\kappa $-linear parametric decompositions as in section \ref%
{ssparamflow} (following again the Convention 2 (\ref{conv2s}) and $\tau $%
-parametric formulas (\ref{nadriemhopfcan})-(\ref{ricciscsymnonsym}) and (%
\ref{paramsricci})-(\ref{nassocsr2})), the formulas for thermodynamic
variables (\ref{nagthermodvalues}) encoding data for nonassociative
geometric flows transform respectively into 
\begin{eqnarray}
\ _{s}^{\shortmid }\widehat{\mathcal{E}}_{\kappa }^{\star } &=&-\tau
^{2}\int_{\ _{s}^{\shortmid }\widehat{\Xi }}\ \left( 4\pi \tau \right)
^{-4}\left( \ _{s}^{\shortmid }\widehat{\mathbf{R}}sc(\tau )+\
_{s}^{\shortmid }\widehat{\mathbf{K}}sc(\tau )+|\ _{s}^{\shortmid }\widehat{%
\mathbf{D}}\ _{s}^{\shortmid }\widehat{f}|^{2}(\tau )-\frac{4}{\tau }\right)
e^{-\ _{s}^{\shortmid }\widehat{f}}\ ^{\shortmid }\delta \ ^{\shortmid }%
\mathcal{V}(\tau ),  \label{nagthermodvaluesp} \\
\ \ _{s}^{\shortmid }\widehat{S}_{\kappa }^{\star } &=&-\int_{\
_{s}^{\shortmid }\widehat{\Xi }}\left( 4\pi \tau \right) ^{-4}\left( \tau (\
_{s}^{\shortmid }\widehat{\mathbf{R}}sc(\tau )+\ _{s}^{\shortmid }\widehat{%
\mathbf{K}}sc(\tau )+|\ _{s}^{\shortmid }\widehat{\mathbf{D}}\
_{s}^{\shortmid }\widehat{f}|^{2}(\tau ))+\ _{s}^{\shortmid }\widehat{f}%
(\tau )-8\right) e^{-\ _{s}^{\shortmid }\widehat{f}}\ ^{\shortmid }\delta \
^{\shortmid }\mathcal{V}(\tau ),  \notag \\
\ _{s}^{\shortmid }\widehat{\sigma }_{\kappa }^{\star } &=&2\ \tau
^{4}\int_{\ _{s}^{\shortmid }\widehat{\Xi }}\left( 4\pi \tau \right) ^{-4}|\
\ ^{\shortmid }\widehat{\mathbf{R}}_{\alpha _{s}\beta _{s}}(\tau )+\
^{\shortmid }\widehat{\mathbf{K}}_{\alpha _{s}\beta _{s}}(\tau )+\
^{\shortmid }\widehat{\mathbf{D}}_{\alpha _{s}}\ ^{\shortmid }\widehat{%
\mathbf{D}}_{\beta _{s}}\ \ _{s}^{\shortmid }\widehat{f}(\tau )-\frac{1}{%
2\tau }\mathbf{g}_{\alpha _{s}\beta _{s}}(\tau )|^{2}e^{-\ _{s}^{\shortmid }%
\widehat{f}}\ ^{\shortmid }\delta \ ^{\shortmid }\mathcal{V}(\tau ).  \notag
\end{eqnarray}%
We emphasize that such variables encode certain nonassociative data in $\
_{s}^{\shortmid }\widehat{\mathbf{K}}sc$ and $\ ^{\shortmid }\widehat{%
\mathbf{K}}_{\alpha _{s}\beta _{s}}$ and certain dependencies in nontrivial $%
\kappa $-linear terms in $\ _{s}^{\shortmid }\widehat{\mathbf{D}}(\tau ),%
\mathbf{g}_{\alpha _{s}\beta _{s}}(\tau ),^{\shortmid }\delta \ ^{\shortmid }%
\mathcal{V}(\tau )$ and $\ _{s}^{\shortmid }\widehat{f}(\tau )$ defined for
solutions of canonical nonholonomic Ricci flow (\ref{nonassocgeomflp}), or
Ricci soliton (\ref{naricsolp}), equations.

The formulas (\ref{nagthermodvaluesp}) can be derived alternatively using on 
$\ _{s}^{\star }\mathcal{M}$ a) a s-adapted variational calculus for the
statistical generating function $\ _{s}^{\shortmid }\widehat{Z}(\tau )$ (\ref%
{volform}) and W-entropy $\ _{s}^{\shortmid }\widehat{\mathcal{W}}_{\kappa
}^{\star }(\tau )$ (\ref{nawfunctp}), with a $\kappa $-linear parametric
decomposition of nonholonomic geometric and thermodynamic variables and/or
b) a corresponding abstract nonassociative geometric calculus.

We note that, in general, the nonassociative geometric flow thermodynamic
variables may be not well-defined as physical values, for instance, one
could be obtained negative entropies etc. This depends on the classes of
solutions we consider and compute such values. In $\kappa $-linear
parametric form we can investigate such issues and select certain
self-consistent and relativistic causal nonassociative cosmological
scenarios or for some BH like configurations.

Finally, for the last two subsections, it should be noted that similar
thermodynamic variables for nonassociative geometric flows or Ricci solitons
can be formulated/ in terms of respective LC-connections, $\
_{\star}^{\shortmid }\nabla $ and $\ ^{\shortmid }\nabla ,$ if we impose
additional nonholonomic constraints of type (\ref{lccondnonass}), when $\
_{s}^{\shortmid }\widehat{\mathbf{D}}_{\mid \ _{s}^{\shortmid }\widehat{%
\mathbf{T}}=0}^{\star }=\ _{\star }^{\shortmid }\nabla .$ Such constraints
result into standard G. Perelman's functionals (\ref{perelmfst}) and related
thermodynamic variables but on phase space $\ _{s}^{\star }\mathcal{M}.$ In
many cases, various distortion of s-connection terms can be encoded into a
new type of normalization functions $\ _{s}^{\shortmid }\widehat{f}(\tau ),$
or in respective classes of generating functions and generating effective
sources.

\section{Parametric geometric flows and off-diagonal quasi-stationary
solutions}

\label{sec4}The goal of this section is to prove that $\kappa $-linear
parametric geometric flow equations (\ref{nonassocgeomflp}) (and, in
particular, the nonassociative Ricci soliton equations (\ref{naricsolp}))
can be decoupled and integrated in general off-diagonal forms for effective
sources encoding nonassociative star product and R-flux data and additional $%
\tau $-induced coefficients. We follow the AFCDM \cite%
{vacaru20,ib19,bubuianu19,partner02,partner03,partner04} outlined in
Appendix \ref{appendixb}. There are provided four possible parameterizations
of such quasi-stationary solutions and analyzed their nonlinear symmetries.

An effective $\tau $-depending source $\ ^{\shortmid }\Im _{\alpha _{s}\beta
_{s}}(\tau )=-(\ ^{\shortmid }\widehat{\mathbf{K}}_{\alpha _{s}\beta
_{s}}(\tau )+\frac{1}{2}\partial _{\tau }\ ^{\shortmid }\mathbf{g}_{\alpha
_{s}\beta _{s}}(\tau ))$, parameterized on $\ _{s}^{\star }\mathcal{M}$ in
s-shell adapted form 
\begin{equation}
\ ^{\shortmid }\Im _{\ \beta _{s}}^{\alpha _{s}}~(\tau ,\ ^{\shortmid
}u^{\gamma _{s}})=[~_{1}^{\shortmid }\Im (\kappa ,\tau ,x^{k_{1}})\delta
_{i_{1}}^{j_{1}},~_{2}^{\shortmid }\Im (\kappa ,\tau ,x^{k_{1}},y^{3})\delta
_{b_{2}}^{a_{2}},~_{3}^{\shortmid }\Im (\kappa ,\tau ,x^{k_{2}},\
^{\shortmid }p_{5})\delta _{a_{3}}^{b_{3}},~_{4}^{\shortmid }\Im (\kappa
,\tau ,x^{k_{3}},\ ^{\shortmid }p_{7})\delta _{a_{4}}^{b_{4}}],
\label{cannonsymparamc2b}
\end{equation}%
can be used for generating quasi-stationary solutions with Killing symmetry
on $\partial _{4}=\partial _{t}.$ For other types of Killing symmetries, we
need corresponding type parameterizations. Such families of effective
sources contain as functionals certain $\kappa $-linear terms with $\mathcal{%
R}_{\quad \alpha _{s}}^{\tau _{s}\xi _{s}}(\tau )$ which for any fixed $\tau 
$ are similar to $\ _{s}^{\shortmid }\mathcal{K}$ (\ref{cannonsymparamc2a}).
We suppose that parameterizations of type (\ref{cannonsymparamc2b}) can be
obtained for certain frame s-adapted transform, $\ ^{\shortmid }\widehat{\Im 
}_{\alpha _{s}^{\prime }\beta _{s}^{\prime }}=e_{\ \alpha _{s}^{\prime
}}^{\alpha _{s}}e_{\ \beta _{s}^{\prime }}^{\beta _{s}}\ ^{\shortmid }\Im
_{\alpha _{s}\beta _{s}},$ when some general sources are transformed into a
subset of four generating sources $\ ^{\shortmid }\Im _{_{\beta _{s}\gamma
_{s}}}=diag\{\ _{s}^{\shortmid }\Im \}.$ Any prescribed $\ _{s}^{\shortmid
}\Im ~(\tau ,\ ^{\shortmid }u^{\gamma _{s}})$ imposes a s-shell nonholonomic
constraint for $\tau $-derivatives of the metrics s-coefficients $\partial
_{\tau }\ ^{\shortmid }\mathbf{g}_{\alpha _{s}\beta _{s}}(\tau ).$ For small
parametric deformations, such constraints can be solved in explicit general
forms. In other cases, we have to search for some special classes of
generating and integration functions which allow to find some examples of
exact/ parametric solutions.

Using effective sources (\ref{cannonsymparamc2b}), we can write the $\kappa $%
-linear parametric geometric flow equations (\ref{nonassocgeomflp}) in the
form 
\begin{equation}
\ ^{\shortmid }\widehat{\mathbf{R}}ic_{\alpha _{s}\beta _{s}}(\tau )=\
^{\shortmid }\Im _{\alpha _{s}\beta _{s}}(\tau ),  \label{nonassocgeomflef}
\end{equation}%
which are very similar to the modified Einstein equations (\ref{seinsta})
(see (\ref{modeinst}) for a running source, $\ ^{\shortmid }\widehat{\mathbf{%
R}}_{\ \ \gamma _{s}}^{\beta _{s}}(\tau )={\delta }_{\ \ \gamma _{s}}^{\beta
_{s}}\ _{s}^{\shortmid }\Im (\tau )$). In these equations, there is an
additional dependence geometric objects on the geometric flow parameter $%
\tau $ and when the sources of type (\ref{sourca}) are corrected with terms
of type $\frac{1}{2}\partial _{\tau }\ _{s}^{\shortmid }\mathbf{g.}$ Such
systems of nonlinear PDEs can be solved in very general off-diagonal forms
using the same formulas as in Appendix \ref{appendixb} but with additional
assumptions when all coefficients of s-metrics depend additionally on $\tau $
and (considering nonlinear symmetries) on running effective cosmological
constants $\ _{s}^{\shortmid }\Lambda ~(\tau ).$ We consider that
corresponding classes of generic off-diagonal solutions are physically
important if they satisfy well-defined causality conditions and
self-consistent G. Perelman like thermodynamic variables (\ref%
{nagthermodvaluesp}) in certain phase space finite regions. In general, it
is not possible to express in explicit functional form all coefficients $%
^{\shortmid }\mathbf{g}_{\alpha _{s}\beta _{s}}(\tau)=$ $^{\shortmid }%
\mathbf{g}_{\alpha _{s}\beta _{s}}[\ _{s}^{\shortmid }\Im (\tau ,\
^{\shortmid }u^{\gamma _{s}})].$ Nevertheless, using decompositions on small
parameters (like on $\kappa $ and other physical constants) and
corresponding s-adapting of geometric constructions, we can construct in
explicit forms exact solutions in nonassociative gravity at least to certain
levels of approximation including $\kappa $-linear terms.

The ansatz for generating quasi-stationary solutions of nonassociative
geometric flow equations (\ref{nonassocgeomflef}) can be chosen as (\ref%
{ans1dm}) and (\ref{ans1n}) but with additional dependencies on $\tau ,$ 
{\small 
\begin{eqnarray}
d\widehat{s}^{2}(\tau ) &=&g_{i_{1}}(\tau
,x^{k_{1}})(dx^{i_{1}})^{2}+g_{a_{2}}(\tau ,x^{i_{1}},y^{3})(\mathbf{e}%
^{a_{2}}(\tau ))^{2}+\ ^{\shortmid }g^{a_{3}}(\tau ,x^{i_{2}},p_{6})(\
^{\shortmid }\mathbf{e}_{a_{3}}(\tau ))^{2}+\ ^{\shortmid }g^{a_{4}}(\tau ,\
^{\shortmid }x^{i_{3}},p_{7})(\ ^{\shortmid }\mathbf{e}_{a_{4}}(\tau ))^{2},
\notag \\
&&\mbox{where }\mathbf{e}^{a_{2}}(\tau )=dy^{a_{2}}+N_{k_{1}}^{a_{2}}(\tau
,x^{i_{1}},y^{3})dx^{k_{1}},\ ^{\shortmid }\mathbf{e}_{a_{3}}(\tau
)=dp_{a_{3}}+\ ^{\shortmid }N_{a_{3}k_{2}}(\tau ,x^{i_{2}},p_{5})dx^{k_{2}},
\notag \\
&&\ ^{\shortmid }\mathbf{e}_{a_{4}}(\tau )=dp_{a_{4}}+\ ^{\shortmid
}N_{a_{4}k_{3}}(\tau ,\ ^{\shortmid }x^{i_{3}},p_{7})d\ ^{\shortmid
}x^{k_{3}}\   \label{ans1rf}
\end{eqnarray}%
} are determined by geometric flows of N-connection coefficients.

\subsection{Geometric evolution of quasi-stationary solutions with effective
sources}

\label{sssol1rf} Applying the AFCDM for temperature running of sources in (%
\ref{sol1}), $\ _{s}^{\shortmid }\mathcal{K}\rightarrow \ _{s}^{\shortmid
}\Im (\tau )$ (\ref{cannonsymparamc2b}), and introducing $\tau $%
-dependencies for the coefficients of s-metric and N-connection in (\ref%
{sol1}), we construct a class of quasi-stationary solutions for
nonassociative $\kappa $-parametric geometric flows: 
\begin{eqnarray}
d\widehat{s}^{2}(\tau ) &=&e^{\psi (\hbar ,\kappa ;\tau
,x^{k_{1}})}[(dx^{1})^{2}+(dx^{2})^{2}]+  \label{sol1rf} \\
&&\frac{[\partial _{3}(\ _{2}\Psi (\tau ))]^{2}}{4(~_{2}^{\shortmid }\Im
(\tau ))^{2}\{g_{4}^{[0]}(\tau )-\int dy^{3}\frac{\partial _{3}[(\ _{2}\Psi
(\tau ))^{2}]}{4(\ ~_{2}^{\shortmid }\Im (\tau ))}\}}(\mathbf{e}^{3}(\tau
))^{2}+(g_{4}^{[0]}(\tau )-\int dy^{3}\frac{\partial _{3}[(\ _{2}\Psi (\tau
))^{2}]}{4(~_{2}^{\shortmid }\Im (\tau ))})(\mathbf{e}^{4}(\tau ))^{2} 
\notag
\end{eqnarray}%
\begin{equation*}
+\frac{[\ ^{\shortmid }\partial ^{5}(\ _{3}^{\shortmid }\Psi (\tau ))]^{2}}{%
4(~_{3}^{\shortmid }\Im )^{2}\{\ ^{\shortmid }g_{[0]}^{6}(\tau )-\int dp_{5}%
\frac{\ ^{\shortmid }\partial ^{6}[(\ _{3}^{\shortmid }\Psi (\tau ))^{2}]}{%
4(~_{3}^{\shortmid }\Im (\tau ))}\}}(\ ^{\shortmid }\mathbf{e}_{5}(\tau
))^{2}+(\ ^{\shortmid }g_{[0]}^{6}(\tau )-\int dp_{5}\frac{\ ^{\shortmid
}\partial ^{5}[(\ _{3}^{\shortmid }\Psi (\tau ))^{2}]}{4(\ _{3}^{\shortmid
}\Im )})(\ ^{\shortmid }\mathbf{e}_{6}(\tau ))^{2}
\end{equation*}%
\begin{equation*}
+\frac{[\ ^{\shortmid }\partial ^{7}(\ _{4}^{\shortmid }\Psi (\tau ))]^{2}}{%
4(~_{4}^{\shortmid }\Im (\tau ))^{2}\{\ ^{\shortmid }g_{[0]}^{8}(\tau )-\int
dp_{7}\frac{\ ^{\shortmid }\partial ^{7}[(\ _{4}^{\shortmid }\Psi (\tau
))^{2}]}{4(~_{4}^{\shortmid }\Im (\tau ))}\}}(\ ^{\shortmid }\mathbf{e}%
_{7}(\tau ))^{2}+(\ ^{\shortmid }g_{[0]}^{8}(\tau )-\int dp_{7}\frac{\
^{\shortmid }\partial ^{7}[(\ _{4}^{\shortmid }\Psi (\tau ))^{2}]}{4(\
_{4}^{\shortmid }\Im (\tau ))})(\ ^{\shortmid }\mathbf{e}_{8}(\tau ))^{2}.
\end{equation*}%
The nonholonomic s-frames in this formula are computed:%
\begin{eqnarray}
\mathbf{e}^{3}(\tau ) &=&dy^{3}+w_{k_{1}}(\hbar ,\kappa ,\tau
,x^{i_{1}},y^{3})dx^{k_{1}}=dy^{3}+\frac{\partial _{k_{1}}(\ _{2}\Psi (\tau
))}{\partial _{3}(\ _{2}\Psi (\tau ))}dx^{k_{1}},  \notag \\
\mathbf{e}^{4}(\tau ) &=&dt+n_{k_{1}}(\hbar ,\kappa ,\tau
,x^{i_{1}},y^{3})dx^{k_{1}}  \label{sol1nrf} \\
&=&dy^{4}+(\ _{1}n_{k_{1}}(\tau )+\ _{2}n_{k_{1}}(\tau )\int dy^{3}\frac{%
\partial _{3}[(\ _{2}\Psi (\tau ))^{2}]}{4(\ ~_{2}^{\shortmid }\Im (\tau
))^{2}|g_{4}^{[0]}(\tau )-\int dy^{3}\frac{\partial _{3}[(\ _{2}\Psi (\tau
))^{2}]}{4(~_{2}^{\shortmid }\Im (\tau ))}|^{5/2}})dx^{k_{1}},  \notag
\end{eqnarray}%
\begin{eqnarray*}
\ ^{\shortmid }\mathbf{e}_{5}(\tau ) &=&dp_{5}+\ ^{\shortmid
}w_{k_{2}}(\hbar ,\kappa ,\tau ,x^{i_{2}},p_{5})dx^{k_{2}}=dp_{5}+\frac{%
\partial _{k_{2}}(\ _{3}^{\shortmid }\Psi (\tau ))}{\ ^{\shortmid }\partial
^{5}(\ _{3}^{\shortmid }\Psi (\tau ))}dx^{k_{2}}, \\
\ ^{\shortmid }\mathbf{e}_{6}(\tau ) &=&dp_{6}+\ ^{\shortmid
}n_{k_{2}}(\hbar ,\kappa ,\tau ,x^{i_{2}},p_{5})dx^{k_{2}} \\
&=&dp_{6}+(\ _{1}^{\shortmid }n_{k_{2}}(\tau )+\ _{2}^{\shortmid
}n_{k_{2}}(\tau )\int dp_{5}\frac{\ ^{\shortmid }\partial ^{5}[(\
_{3}^{\shortmid }\Psi (\tau ))^{2}]}{4(\ _{3}^{\shortmid }\Im (\tau ))^{2}|\
^{\shortmid }g_{[0]}^{6}(\tau )-\int dp_{5}\frac{\ ^{\shortmid }\partial
^{5}[(\ _{3}^{\shortmid }\Psi (\tau ))^{2}]}{4(\ _{3}^{\shortmid }\Im (\tau
))}|^{5/2}})dx^{k_{2}},
\end{eqnarray*}%
\begin{eqnarray*}
\ ^{\shortmid }\mathbf{e}_{7}(\tau ) &=&dp_{7}+\ ^{\shortmid
}w_{k_{3}}(\hbar ,\kappa ,\tau ,x^{i_{2}},p_{5},p_{7})d\ ^{\shortmid
}x^{k_{3}}=dp_{7}+\frac{\ ^{\shortmid }\partial _{k_{3}}(\ _{4}^{\shortmid
}\Psi (\tau ))}{\ ^{\shortmid }\partial ^{7}(\ _{4}^{\shortmid }\Psi (\tau ))%
}d\ ^{\shortmid }x^{k_{3}}, \\
\ ^{\shortmid }\mathbf{e}_{8}(\tau ) &=&dp_{8}+\ ^{\shortmid
}n_{k_{3}}(\hbar ,\kappa ,\tau ,x^{i_{2}},p_{5},p_{7})d\ ^{\shortmid
}x^{k_{3}} \\
&=&dp_{8}+(\ _{1}^{\shortmid }n_{k_{3}}(\tau )+\ _{2}^{\shortmid
}n_{k_{3}}(\tau )\int dp_{7}\frac{\ ^{\shortmid }\partial ^{7}[(\
_{4}^{\shortmid }\Psi (\tau ))^{2}]}{4(\ _{4}^{\shortmid }\Im (\tau ))^{2}|\
^{\shortmid }g_{[0]}^{8}(\tau )-\int dp_{7}\frac{\ ^{\shortmid }\partial
^{7}[(\ _{4}^{\shortmid }\Psi (\tau ))^{2}]}{4(\ _{4}^{\shortmid }\Im (\tau
))}|^{5/2}})\ d\ ^{\shortmid }x^{k_{3}}.
\end{eqnarray*}%
The generating and integration functions for the class of solutions (\ref%
{sol1rf}) with N-coefficients (\ref{sol1nrf}) are similar to (\ref%
{integrfunct}) but with extended $\tau $-parametric dependence: 
\begin{eqnarray}
\mbox{generating functions: } &&\psi (\tau )\simeq \psi (\hbar ,\kappa ;\tau
,x^{k_{1}});\ _{2}\Psi (\tau )\simeq \ _{2}\Psi (\hbar ,\kappa ;\tau
,x^{k_{1}},y^{3});  \label{integrfunctrf} \\
&&\ _{3}^{\shortmid }\Psi (\tau )\simeq \ _{3}^{\shortmid }\Psi (\hbar
,\kappa ;\tau ,x^{k_{2}},p_{5});\ _{4}^{\shortmid }\Psi (\tau )\simeq \
_{4}^{\shortmid }\Psi (\hbar ,\kappa ;\tau ,\ ^{\shortmid }x^{k_{3}},p_{7});
\notag \\
\mbox{generating sources: \quad} &&\ _{1}^{\shortmid }\Im \mathcal{(\tau )}%
\simeq \ ~_{1}^{\shortmid }\Im (\hbar ,\kappa ;\tau ,x^{k_{1}});\
~_{2}^{\shortmid }\Im \mathcal{(\tau )}\simeq \ ~_{2}^{\shortmid }\Im (\hbar
,\kappa ;\tau ,x^{k_{1}},y^{3});  \notag \\
&&\ _{3}^{\shortmid }\Im \mathcal{(\tau )}\simeq \ ~_{3}^{\shortmid }\Im
(\hbar ,\kappa ;\tau ,x^{k_{2}},p_{5});~_{4}^{\shortmid }\Im \mathcal{(\tau )%
}\simeq \ ~~_{4}^{\shortmid }\Im (\hbar ,\kappa ;\tau ,\ ^{\shortmid
}x^{k_{3}},p_{7});  \notag \\
\mbox{integrating functions: } &&  \notag \\
g_{4}^{[0]}(\tau ) &\simeq &g_{4}^{[0]}(\hbar ,\kappa ;\tau ,x^{k_{1}}),\
_{1}n_{k_{1}}\mathcal{(\tau )}\simeq \ _{1}n_{k_{1}}(\hbar ,\kappa ;\tau
,x^{j_{1}}),\ _{2}n_{k_{1}}\mathcal{(\tau )}\simeq \ _{2}n_{k_{1}}(\hbar
,\kappa ;\tau ,x^{j_{1}});  \notag \\
\ ^{\shortmid }g_{[0]}^{6}\tau ) &\simeq &\ ^{\shortmid }g_{[0]}^{6}(\hbar
,\kappa ;\tau ,x^{k_{2}}),\ _{1}n_{k_{2}}\mathcal{(\tau )}\simeq \
_{1}n_{k_{2}}(\hbar ,\kappa ;\tau ,x^{j_{2}}),\ _{2}n_{k_{2}}\mathcal{(\tau )%
}\simeq \ _{2}n_{k_{2}}(\hbar ,\kappa ;\tau ,x^{j_{2}});  \notag \\
\ ^{\shortmid }g_{[0]}^{8}(\tau ) &\simeq &\ ^{\shortmid }g_{[0]}^{8}(\hbar
,\kappa ;\tau ,\ ^{\shortmid }x^{j_{3}}),\ _{1}n_{k_{3}}\mathcal{(\tau )}%
\simeq \ _{1}^{\shortmid }n_{k_{3}}(\hbar ,\kappa ;\tau ,\ ^{\shortmid
}x^{j_{3}}),\ _{2}n_{k_{3}}\mathcal{(\tau )}\simeq \ _{2}^{\shortmid
}n_{k_{3}}(\hbar ,\kappa ;\tau ,\ ^{\shortmid }x^{j_{3}}).  \notag
\end{eqnarray}%
The family of generating functions $\psi (\tau )$ are solutions of a
respective family of 2-d Poisson equations, 
\begin{equation}
\partial _{11}^{2}\psi (\hbar ,\kappa ;\tau ,x^{k_{1}})+\partial
_{22}^{2}\psi (\hbar ,\kappa ;\tau ,x^{k_{1}})=2\ _{1}\Im (\hbar ,\kappa
;\tau ,x^{k_{1}}),  \label{poas2d}
\end{equation}%
encoding geometric flows of nonassociative data if, in general, $\ _{1}\Im 
\mathcal{(\tau )}$ contains such nonholonomic dependencies.

Geometric evolution scenarios of quasi-stationary configurations defined
above are characterized by four types of additional geometric and
thermodynamic flow variables:

\begin{enumerate}
\item The geometric evolution of nonsymmetric metrics $\ _{\star
}^{\shortmid }\mathfrak{a}_{\alpha _{s}\beta _{s}}(\tau )= \ _{\star
}^{\shortmid }\mathfrak{a}_{\alpha _{s}\beta _{s}}(\hbar ,\kappa ;\tau , \
^{\shortmid}u^{\gamma _{s}})$ is computed in explicit form by introducing in
(\ref{aux40aa}) the s-metric and N-connection coefficients, respectively, (%
\ref{sol1rf}) and (\ref{sol1nrf}) (we omit such formulas in this work). For
flow evolution of quasi-stationary configurations, it is possible to
decouple the symmetric and nonsymmetric components of s-metrics (proven in 
\cite{partner02,partner03,partner04}). This allows us to study independently
their nonassociative geometric evolution models.

\item In general, such solutions are with nontrivial geometric flows of
nonholonomic torsion. Nevertheless, we can always constrain such geometric
flows of s-metrics to subclasses of $\tau $-parametric families of
generating data solving the equations (\ref{lccondnonass}) and/or (\ref%
{lccond}), which allow to extract configurations with zero torsion. The
remarks at the end of appendix \ref{sssol1} state how to restrict the
generating data (\ref{integrfunctrf}) in order to restrict the nonolonomic
flows to families of LC-connections $\ _{\star}^{\shortmid}\nabla (\tau )$
and$\ ^{\shortmid }\nabla (\tau ).$

\item We can compute respective thermodynamic variables (\ref%
{nagthermodvaluesp}) associated to such quasi-stationary solutions. In next
section, we shall provide such examples for BH configurations and their
nonassociative star product R-flux deformatins.

\item The solutions for nonassociative Ricci soliton equations (\ref%
{naricsolp}) consist self-similar configurations if the geometric flow
constant is fixed, $\tau =\tau _{0},$ after a class of generic off-diagonal
solutions is constructed in a form (\ref{sol1rf}) and (\ref{sol1nrf}). Such
s-metrics are characterized by fixed $\tau _{0}$ geometric s-adapted values
and thermodynamic variables stated in 1-3. Such Ricci soliton configurations
can be generated equivalently by solutions constructed using the AFCDM as it
is outlined in appendix \ref{appendixb}.
\end{enumerate}

The class of off-diagonal solutions defined by (\ref{sol1rf}) and (\ref%
{sol1nrf}) involves non-explicit nonholonomic constraints on temperature
derivatives of certain s-metric coefficients, $\partial _{\tau }\
^{\shortmid }\mathbf{g}_{\alpha _{s}\beta _{s}}(\tau ),$ encoded in
effective sources $\ ^{\shortmid }\Im _{\alpha _{s}\beta _{s}}(\tau )$ (\ref%
{cannonsymparamc2b}). To decouple completely such formulas is possible only
for more special classes of nonholonomic distributions or additional
decompositions on small parameters, for instance, for certain types of BH
nonassociative deformations to BE ones, ore other type configurations.

Finally, we note that the families of quasi-stationary s-metrics (\ref%
{sol1rf}) are of type (\ref{ans1rf}) with s-shell Killing symmetries on $%
\partial _{t},\ ^{\shortmid }\partial ^{6},\ ^{\shortmid }\partial ^{8}.$
They are defined by s-adapted coefficients with respective "symbolic" phase
coordinates and $\tau $--dependencies of coefficients and
generating/integration functions and effective sources. In a similar form
(we have to change the "symbolic" phase coordinates and parametric
dependencies of s-coefficients and respective functions), we can construct
solutions with Killing symmetries on $\partial _{t},\ ^{\shortmid }\partial
^{6},\ ^{\shortmid }\partial ^{7},$ or $\partial _{t},\ ^{\shortmid
}\partial ^{5},\ ^{\shortmid }\partial ^{8},$ or $\partial _{t},\
^{\shortmid }\partial ^{5},\ ^{\shortmid }\partial ^{7}.$ Such formulas have
very similar physical interpretations if they are for metrics of the same
signature. If we consider solutions, for instance, with symmetries on $%
\partial _{3},\ ^{\shortmid }\partial ^{6},\ ^{\shortmid }\partial ^{8},$
the s-metric and s-connection coefficients depend generically on a time like
variable $y^{4}=t.$ This allows to study the nonassociative geometric
evolutions of locally anisotropic cosmological solutions. We plan to
elaborate on nonassociative cosmological scenarios in our further partner
works (see \cite{lb18} and references therein for associative and
commutative generic off-diagonal cosmological models).

\subsection{Nonlinear symmetries and temperature running cosmological
constants}

Quasi-stationary s-metric (\ref{sol1rf}) with N-connection coefficients (\ref%
{ans1rf}) posses very important nonlinear symmetries which generalize for
nonassociative geometric flow equations the nonlinear symmetries for Ricci
solitons and vacuum gravitational equations stated in appendix \ref{bassnsym}%
. Corresponding nonlinear transforms allow:

\begin{itemize}
\item to construct nonassociative nonholonomic geometric flow deformations
of families of \textbf{prime} s-metrics $\ _{s}^{\shortmid }\mathbf{%
\mathring{g}}(\tau )$ (they can be arbitrary ones, i.e. not solutions of
some (modified) Einstein equations) into a corresponding family of \textbf{%
target} s-metrics $\ _{s}^{\shortmid }\mathbf{g}(\tau )$ defining a
nonassociative geometric flow evolution scenarios of quasi-stationary
metrics on $\ _{s}^{\star }\mathcal{M},$ 
\begin{equation}
\ _{s}^{\shortmid }\mathbf{\mathring{g}}(\tau )\rightarrow \ _{s}^{\shortmid
}\mathbf{g}(\tau )=[\ ^{\shortmid }g_{\alpha _{s}}(\tau )=\ ^{\shortmid
}\eta _{\alpha _{s}}(\tau )\ ^{\shortmid }\mathring{g}_{\alpha _{s}}(\tau
),\ ^{\shortmid }N_{i_{s-1}}^{a_{s}}(\tau )=\ ^{\shortmid }\eta
_{i_{s-1}}^{a_{s}}(\tau )\ ^{\shortmid }\mathring{N}_{i_{s-1}}^{a_{s}}(\tau
)],  \label{offdiagdefr}
\end{equation}%
where the deformations with gravitational running on $\tau $ polarization
functions are defined be formulas (\ref{offdiagdef}), (\ref{dmpolariz}) and (%
\ref{espolariz}) generalized for $\tau $-dependencies of respective
s-coefficients;

\item to re-define the geometric flows of generating functions and relate
the effective sources to certain effective shell $\tau $-running
cosmological constants, 
\begin{eqnarray}
&&(\ _{s}\Psi (\tau ),\ _{s}^{\shortmid }\Im (\tau )) \leftrightarrow (\
_{s}^{\shortmid }\mathbf{g}(\tau ),\ _{s}^{\shortmid }\Im (\tau
))\leftrightarrow (\ _{s}^{\shortmid }\eta (\tau ) \ ^{\shortmid }\mathring{g%
}_{\alpha _{s}}(\tau )\sim (\ ^{\shortmid }\zeta _{\alpha _{s}}(\tau)
(1+\kappa \ ^{\shortmid }\chi _{\alpha _{s}}(\tau )) \ ^{\shortmid }%
\mathring{g}_{\alpha _{s}}(\tau ),\ _{s}^{\shortmid }\Im
(\tau))\leftrightarrow  \notag \\
&& (\ _{s}\Phi (\tau ),\ _{s}^{\shortmid }\Lambda (\tau )) \leftrightarrow
(\ _{s}^{\shortmid }\mathbf{g}, \ _{s}^{\shortmid }\Lambda
(\tau))\leftrightarrow (\ _{s}^{\shortmid }\eta (\tau )\ ^{\shortmid }%
\mathring{g}_{\alpha _{s}}(\tau )\sim (\ ^{\shortmid }\zeta _{\alpha
_{s}}(\tau)(1+\kappa \ ^{\shortmid }\chi _{\alpha _{s}}(\tau )) \
^{\shortmid } \mathring{g}_{\alpha _{s}}(\tau ), \ _{s}^{\shortmid }\Lambda
(\tau )),  \label{nonlinsymr}
\end{eqnarray}%
where $\ _{s}^{\shortmid }\Lambda _{0}=\ _{s}^{\shortmid }\Lambda (\tau_{0}) 
$ for nonassociative Ricci soliton symmetries of type (\ref{nonlinsym}).
\end{itemize}

For simplicity, we study in this work only geometric flows with running of
effective cosmological constants even the nonlinear symmetries of
quasi-stationary solutions can be formulated for generalized phase space
polarizations of cosmological constants $\ _{s}^{\shortmid }\Lambda (\tau ,\
^{\shortmid }u^{\gamma _{s}}).$ Such models involve re-definitions of
effective generating sources $\ _{s}^{\shortmid }\Im (\tau ,\
^{\shortmid}u^{\gamma _{s}})$ into another classes of effective sources/
cosmological constants $\ _{s}^{\shortmid }\Lambda (\tau ,\
^{\shortmid}u^{\gamma _{s}})$.

Nonlinear transforms of flows of quasi-stationary s-metric (\ref{sol1rf})
into equivalent ones with different classes of generating functions are
described by introducing additional $\tau $-dependencies in (\ref%
{smetrfunctionals}), when 
\begin{eqnarray}
\partial _{3}[(\ _{2}\Psi (\tau ))^{2}] &=&-\int dy^{3}(~_{2}^{\shortmid
}\Im (\tau ))\partial _{3}g_{4}(\tau )\simeq -\int dy^{3}(~_{2}^{\shortmid
}\Im (\tau ))\partial _{3}(\ ^{\shortmid }\eta _{4}(\tau )\ \mathring{g}%
_{4}(\tau ))  \notag \\
&\simeq &-\int dy^{3}(~_{2}^{\shortmid }\Im (\tau ))\partial _{3}[\
^{\shortmid }\zeta _{4}(\tau )(1+\kappa \ ^{\shortmid }\chi _{4}(\tau ))\ 
\mathring{g}_{4}(\tau )],  \label{nonlinsymrex} \\
(\ _{2}\Phi (\tau ))^{2} &=&-4\ _{2}\Lambda (\tau )g_{4}(\tau )\simeq -4\
_{2}\Lambda (\tau )\ ^{\shortmid }\eta _{4}(\tau )\ \mathring{g}_{4}(\tau ) 
\notag \\
&\simeq &-4\ _{2}\Lambda (\tau )\ ^{\shortmid }\zeta _{4}(\tau )(1+\kappa \
^{\shortmid }\chi _{4}(\tau ))\ \mathring{g}_{4}(\tau );  \notag
\end{eqnarray}%
\begin{eqnarray*}
~\ ^{\shortmid }\partial ^{5}[(\ _{3}^{\shortmid }\Psi (\tau ))^{2}]
&=&-\int dp_{5}(~_{3}^{\shortmid }\Im (\tau ))\ ^{\shortmid }\partial ^{5}\
^{\shortmid }g^{6}(\tau )\simeq -\int dp_{5}(~_{3}^{\shortmid }\Im (\tau ))\
^{\shortmid }\partial ^{5}(\ ^{\shortmid }\eta ^{6}(\tau )\ ^{\shortmid }%
\mathring{g}^{6}(\tau )) \\
&\simeq &-\int dp_{5}(~_{3}^{\shortmid }\Im (\tau ))\ ^{\shortmid }\partial
^{5}[\ ^{\shortmid }\zeta ^{6}(\tau )(1+\kappa \ ^{\shortmid }\chi ^{6}(\tau
))\ \mathring{g}^{6}(\tau )], \\
(\ _{3}^{\shortmid }\Phi (\tau ))^{2} &=&-4\ _{3}^{\shortmid }\Lambda (\tau
)\ ^{\shortmid }g^{6}(\tau )\simeq \ -4\ _{3}^{\shortmid }\Lambda (\tau )\
^{\shortmid }\eta ^{6}(\tau )\ ^{\shortmid }\mathring{g}^{6}(\tau ) \\
&\simeq &-4\ _{3}^{\shortmid }\Lambda (\tau )\ ^{\shortmid }\zeta ^{6}(\tau
)(1+\kappa \ ^{\shortmid }\chi ^{6}(\tau ))\ ^{\shortmid }\mathring{g}%
^{6}(\tau );
\end{eqnarray*}%
\begin{eqnarray*}
~\ ^{\shortmid }\partial ^{7}[(\ _{4}^{\shortmid }\Psi (\tau ))^{2}]
&=&-\int dp_{7}(~_{4}^{\shortmid }\Im (\tau ))\ ^{\shortmid }\partial ^{7}\
^{\shortmid }g^{8}(\tau )\simeq -\int dp_{7}(~_{4}^{\shortmid }\Im (\tau ))\
^{\shortmid }\partial ^{7}(\ ^{\shortmid }\eta ^{8}(\tau )\ ^{\shortmid }%
\mathring{g}^{8}(\tau )) \\
&\simeq &-\int dp_{7}(~_{4}^{\shortmid }\Im (\tau ))\ ^{\shortmid }\partial
^{7}[\ ^{\shortmid }\zeta ^{8}(\tau )(1+\kappa \ ^{\shortmid }\chi ^{8}(\tau
))\ \mathring{g}^{8}(\tau )], \\
(\ _{4}^{\shortmid }\Phi (\tau ))^{2} &=&-4\ _{4}^{\shortmid }\Lambda (\tau
)\ ^{\shortmid }g^{8}(\tau )\simeq \ -4\ _{4}^{\shortmid }\Lambda (\tau )\
^{\shortmid }\eta ^{8}(\tau )\ ^{\shortmid }\mathring{g}^{8}(\tau ) \\
&\simeq &-4\ _{4}^{\shortmid }\Lambda (\tau )\ ^{\shortmid }\zeta ^{8}(\tau
)(1+\kappa \ ^{\shortmid }\chi ^{8}(\tau ))\ ^{\shortmid }\mathring{g}%
^{8}(\tau ).
\end{eqnarray*}%
We present the corresponding quadratic line elements for quasi-stationary
geometric flow solutions defined by such transforms in next subsections.

\subsection{Parametric solutions for nonassociative geometric flows with
running cosmological constants}

Nonlinear symmetries (\ref{nonlinsymr}) allow us to change the generating
functions and generating sources into certain new types of generating
functions and effective cosmological data, $\ [\ _{s}\Psi (\tau ),\
_{s}^{\shortmid }\Im (\tau )]\rightarrow \lbrack \ _{s}\Phi (\tau ),\
_{s}^{\shortmid }\Lambda (\tau )].$ In result, the $\kappa $-linear
parametric nonassociative geometric flow equations (\ref{nonassocgeomflp})
(written as $\ ^{\shortmid }\widehat{\mathbf{R}}_{\ \ \gamma _{s}}^{\beta
_{s}}(\tau ,\ _{s}\Psi (\tau ))= {\delta }_{\ \ \gamma _{s}}^{\beta _{s}}\
_{s}^{\shortmid }\Im (\tau )\ $ (\ref{nonassocgeomflef}) and integrated in
quasi-stationary form using $\ _{s}\Psi (\tau )$) can be re-defined
equivalently as a system of functional equations with $\ _{s}\Phi (\tau )$, 
\begin{equation}
\ ^{\shortmid }\widehat{\mathbf{R}}_{\ \ \gamma _{s}}^{\beta _{s}}(\tau ,\
_{s}\Phi (\tau ),\ _{s}^{\shortmid }\Im (\tau ))= {\delta }_{\ \ \gamma
_{s}}^{\beta _{s}}\ _{s}^{\shortmid }\Lambda (\tau ).
\label{nonassocgeomflefcc}
\end{equation}%
We suppose that such nonlinear systems of PDEs are derived for certain shell
effective $\tau $-running constants $\ _{s}^{\shortmid }\Lambda (\tau)$
introduced for modelling geometric flow evolution processes. The solutions
of (\ref{nonassocgeomflefcc}) are $\tau $-parametric generalizations of
s-metrics (\ref{sol1a}) and generating data (\ref{gensol1a}), \newline
$\ _{s}^{\shortmid }\mathbf{g}[\hbar ,\kappa ,\tau ,\psi (\tau),\ _{s}\Psi
(\tau ),\ _{s}^{\shortmid }\Im (\tau )]\rightarrow \ _{s}^{\shortmid }%
\mathbf{g}[\hbar ,\kappa ,\tau ,\psi (\tau ),\ _{s}\Phi (\tau ),\
_{s}^{\shortmid }\Lambda (\tau )].$ They are defined equivalently by the
quasi-stationary quadratic element (\ref{sol1rf}) with nonholonomic frames (%
\ref{sol1nrf}) transformed respectively into such values:

Using nonlinear formulas (\ref{nonlinsymrex}), the quasi-stationary
solutions of $\tau $-parametric running $\kappa $-linear modified Einstein
equations (\ref{nonassocgeomflefcc}) are defined by such quadratic linear
elements: 
\begin{eqnarray}
d\widehat{s}^{2}(\tau ) &=&e^{\psi (\hbar ,\kappa ;\tau
,x^{k_{1}})}[(dx^{1})^{2}+(dx^{2})^{2}]+  \label{sol2rf} \\
&&-\frac{1}{g_{4}^{[0]}(\tau )-\frac{(\ _{2}\Phi (\tau ))^{2}}{4\
_{2}\Lambda (\tau )}}\frac{(\ _{2}\Phi (\tau ))^{2}[\partial _{3}(\ _{2}\Phi
(\tau ))]^{2}}{|\ _{2}\Lambda (\tau )\int dy^{3}(~_{2}^{\shortmid }\Im (\tau
))[\partial _{3}(\ _{2}\Phi (\tau ))^{2}]|}(\mathbf{e}^{3}(\tau
))^{2}+\left( g_{4}^{[0]}(\tau )-\frac{(\ _{2}\Phi (\tau ))^{2}}{4\
_{2}\Lambda (\tau )}\right) (\mathbf{e}^{4}(\tau ))^{2}  \notag
\end{eqnarray}%
\begin{equation*}
-\frac{1}{\ ^{\shortmid }g_{[0]}^{6}(\tau )-\frac{(\ _{3}^{\shortmid }\Phi
(\tau ))^{2}}{4\ _{3}^{\shortmid }\Lambda (\tau )}}\frac{(\ _{3}^{\shortmid
}\Phi (\tau ))^{2}[\ ^{\shortmid }\partial ^{5}(\ _{3}^{\shortmid }\Phi
(\tau ))]^{2}}{|\ _{3}^{\shortmid }\Lambda (\tau )\int
dp_{5}(~_{3}^{\shortmid }\Im (\tau ))\ \ ^{\shortmid }\partial ^{5}[(\
_{3}^{\shortmid }\Phi (\tau ))^{2}]|}(\ ^{\shortmid }\mathbf{e}_{5}(\tau
))^{2}+\left( \ ^{\shortmid }g_{[0]}^{6}(\tau )-\frac{(\ _{3}^{\shortmid
}\Phi (\tau ))^{2}}{4\ _{3}^{\shortmid }\Lambda (\tau )}\right) (\
^{\shortmid }\mathbf{e}_{6}(\tau ))^{2}
\end{equation*}%
\begin{equation*}
-\frac{1}{\ ^{\shortmid }g_{[0]}^{8}(\tau )-\frac{(\ _{4}^{\shortmid }\Phi
(\tau ))^{2}}{4\ _{4}^{\shortmid }\Lambda (\tau )}}\frac{(\ _{4}^{\shortmid
}\Phi (\tau ))^{2}[\ ^{\shortmid }\partial ^{7}(\ _{4}^{\shortmid }\Phi
(\tau ))]^{2}}{|\ _{4}^{\shortmid }\Lambda (\tau )\int
dp_{7}(~_{4}^{\shortmid }\Im )\ \ ^{\shortmid }\partial ^{7}[(\
_{4}^{\shortmid }\Phi (\tau ))^{2}]|}(\ ^{\shortmid }\mathbf{e}_{7}(\tau
))^{2}+\left( \ ^{\shortmid }g_{[0]}^{8}(\tau )-\frac{(\ _{4}^{\shortmid
}\Phi (\tau ))^{2}}{4\ _{4}^{\shortmid }\Lambda (\tau )}\right) (\
^{\shortmid }\mathbf{e}_{8}(\tau ))^{2},
\end{equation*}%
where the nonholonomic s-frames and respective off-diagonal terms are
computed:%
\begin{eqnarray}
\mathbf{e}^{3}(\tau ) &=&dy^{3}+\frac{\partial _{k_{1}}\ \int
dy^{3}(~_{2}^{\shortmid }\Im (\tau ))\ \partial _{3}[(\ _{2}\Phi (\tau
))^{2}]}{(~_{2}^{\shortmid }\Im (\tau ))\ \partial _{3}[(\ _{2}\Phi (\tau
))^{2}]}dx^{k_{1}},  \notag \\
\mathbf{e}^{4}(\tau ) &=&dt+(\ _{1}n_{k_{1}}(\tau )+\ _{2}n_{k_{1}}(\tau )%
\frac{\int dy^{3}\frac{(\ _{2}\Phi (\tau ))^{2}[\ \partial _{3}(\ _{2}\Phi
(\tau ))]^{2}}{|\ _{2}\Lambda (\tau )\int dy^{3}(~_{2}^{\shortmid }\Im (\tau
))[\ \partial _{3}(\ _{2}\Phi (\tau ))^{2}]|}}{\left\vert g_{4}^{[0]}(\tau )-%
\frac{(\ _{2}\Phi (\tau ))^{2}}{4\ _{2}\Lambda (\tau )}\right\vert ^{5/2}}%
)dx^{k_{1}},  \label{sol2nrf}
\end{eqnarray}%
\begin{eqnarray*}
\ ^{\shortmid }\mathbf{e}_{5}(\tau ) &=&dp_{5}+\frac{\partial _{k_{2}}\ \int
dp_{5}(~_{3}^{\shortmid }\Im (\tau ))\ \ ^{\shortmid }\partial ^{5}[(\
_{3}^{\shortmid }\Phi (\tau ))^{2}]}{(~_{3}^{\shortmid }\Im (\tau ))\ \ \
^{\shortmid }\partial ^{5}[(\ _{3}^{\shortmid }\Phi (\tau ))^{2}]}dx^{k_{2}},
\\
\ ^{\shortmid }\mathbf{e}_{6}(\tau ) &=&dp_{6}+(\ _{1}^{\shortmid
}n_{k_{2}}(\tau )+\ _{2}^{\shortmid }n_{k_{2}}(\tau )\frac{\int dp_{5}\frac{%
(\ _{3}^{\shortmid }\Phi (\tau ))^{2}[\ ^{\shortmid }\partial ^{5}(\
_{3}^{\shortmid }\Phi (\tau ))]^{2}}{|\ _{3}^{\shortmid }\Lambda (\tau )\int
dp_{5}(~_{3}^{\shortmid }\Im (\tau ))[\ ^{\shortmid }\partial ^{5}(\
_{3}^{\shortmid }\Phi (\tau ))^{2}]|}}{\left\vert \ ^{\shortmid
}g_{[0]}^{6}(\tau )-\frac{(\ _{3}^{\shortmid }\Phi (\tau ))^{2}}{4\
_{3}^{\shortmid }\Lambda (\tau )}\right\vert ^{5/2}})dx^{k_{2}},
\end{eqnarray*}%
\begin{eqnarray*}
\ ^{\shortmid }\mathbf{e}_{7}(\tau ) &=&dp_{7}+\frac{\partial _{k_{3}}\ \int
dp_{7}(~_{4}^{\shortmid }\Im (\tau ))\ \ ^{\shortmid }\partial ^{7}[(\
_{4}^{\shortmid }\Phi (\tau ))^{2}]}{(~_{4}^{\shortmid }\Im (\tau ))\ \ \
^{\shortmid }\partial ^{7}[(\ _{4}^{\shortmid }\Phi (\tau ))^{2}]}d\
^{\shortmid }x^{k_{3}}, \\
\ ^{\shortmid }\mathbf{e}_{8}(\tau ) &=&dE+(\ _{1}^{\shortmid
}n_{k_{3}}(\tau )+\ _{2}^{\shortmid }n_{k_{3}}(\tau )\frac{\int dp_{7}\frac{%
(\ _{4}^{\shortmid }\Phi (\tau ))^{2}[\ ^{\shortmid }\partial ^{7}(\
_{4}^{\shortmid }\Phi (\tau ))]^{2}}{|\ _{4}^{\shortmid }\Lambda (\tau )\int
dp_{7}(~_{4}^{\shortmid }\Im (\tau ))[\ ^{\shortmid }\partial ^{7}(\
_{4}^{\shortmid }\Phi (\tau ))^{2}]|}}{\left\vert \ ^{\shortmid
}g_{[0]}^{8}(\tau )-\frac{(\ _{4}^{\shortmid }\Phi (\tau ))^{2}}{4\
_{4}^{\shortmid }\Lambda (\tau )}\right\vert ^{5/2}})d\ ^{\shortmid
}x^{k_{3}}.
\end{eqnarray*}%
The conventions for the generating and integration functions and effective
sources in s-metric coefficients (\ref{sol2rf}) with nonholonomic frames (%
\ref{sol2nrf}) are those from (\ref{gensol1a}) but with $\ _{s}^{\shortmid }%
\mathcal{K}\rightarrow \ _{s}^{\shortmid }\Im (\tau )$ and for the data $(\
_{s}^{\shortmid }\Phi (\tau ),\ _{s}^{\shortmid }\Lambda (\tau )).$
Respective functionals can be constrained to define LC-configurations and
model their nonassociative geometric evolution as we explain in point 2 of
subsection \ref{sssol1rf}. For all cases of nonlinear transforms (\ref%
{nonlinsymr}), the functional representations of off-diagonal solutions
allow to encode possible contributions from effective cosmological constants
when certain evolution of effective sources is re-distributed into
off-diagonal terms of s-metrics and with modifications of the diagonal
s-adapted terms. The effective $\ _{s}^{\shortmid }\Im (\tau )$ are not
completely substituted by effective $\tau $-running constants $\
_{s}^{\shortmid }\Lambda (\tau )$ and both types of values are present in
integrals for certain s-connection coefficients $g_{3}(\tau ),\ ^{\shortmid
}g^{5}(\tau ),\ ^{\shortmid }g^{8}(\tau )$ and all N-connection coefficients
in (\ref{sol2nrf}). Nevertheless, we can simplify substantially certain
classes of solutions using $\ _{s}^{\shortmid}\Lambda (\tau )$ and then to
speculate on their physical properties etc.

\subsection{Flows with some s-metric coefficients as generating functions}

We can generate quasi-stationary $\tau $-running ansatz of type (\ref{ans1dm}%
) prescribing $g_{4}(\tau ),\ ^{\shortmid }g^{6}(\tau )$ and $\
^{\shortmid}g^{8}(\tau )$ as generating functions. For constructing $\kappa $%
-parametric Ricci soliton configurations, we can apply directly the
procedure described in appendix \ref{assmetrgf} and s-metrics (\ref{sol1b}).
To study nonassociative geometric flow evolution we can consider such
functionals for the generating functions and their nonlinear transforms (\ref%
{nonlinsymrex}): 
\begin{eqnarray*}
g_{4}(\tau ) &=&g_{4}(\tau ,x^{k_{1}},y^{3})=g_{4}[\ _{2}\Psi (\tau), \
_{2}^{\shortmid }\Im (\tau )]=g_{4}[\ _{2}\Phi (\tau ),\ _{2}\Lambda (\tau
)]; \\
\ ^{\shortmid }g^{6}(\tau ) &=&\ ^{\shortmid }g^{6}(\tau ,x^{i_{2}},p_{5})=\
^{\shortmid }g^{6}[\ _{3}\Psi (\tau ),~_{3}^{\shortmid }\Im (\tau )]=\
^{\shortmid }g^{6}[\ _{3}^{\shortmid }\Phi (\tau ),\ _{3}\Lambda (\tau )]; \\
\ ^{\shortmid }g^{8}(\tau ) &=&\ ^{\shortmid }g^{8}(\tau
,x^{i_{2}},p_{6},E)=\ ^{\shortmid }g^{8}[\ _{4}\Psi (\tau ),~_{4}^{\shortmid
}\Im (\tau )]=\ ^{\shortmid }g^{8}[\ _{4}^{\shortmid }\Phi (\tau ),\
_{4}\Lambda (\tau )].
\end{eqnarray*}

For instance, expressing $\ _{s}\Psi (\tau )=\ _{s}\Psi \lbrack \
_{s}^{\shortmid }\Im (\tau ),g_{4}(\tau ),\ ^{\shortmid }g^{6}(\tau ), \
^{\shortmid }g^{8}(\tau )],$ we can compute the un-known coefficients of a
s-metric (\ref{sol1rf}) and N-coefficients (\ref{sol1nrf}). The resulting
quadratic linear element is 
\begin{eqnarray}
d\widehat{s}^{2}(\tau ) &=&e^{\psi (\tau )}[(dx^{1})^{2}+(dx^{2})^{2}]-\frac{%
(\partial _{3}g_{4}(\tau ))^{2}}{|\int dy^{3}\partial _{3}[(~_{2}^{\shortmid
}\Im (\tau ))g_{4}(\tau )]|\ g_{4}(\tau )}(\mathbf{e}^{3}(\tau
))^{2}+g_{4}(\tau )(\mathbf{e}^{4}(\tau ))^{2}  \label{sol3rf} \\
&&-\frac{[\ ^{\shortmid }\partial ^{5}(\ ^{\shortmid }g^{6}(\tau ))]^{2}}{%
|\int dp_{5}\ ^{\shortmid }\partial ^{5}[(~_{3}^{\shortmid }\Im (\tau ))\
^{\shortmid }g^{6}(\tau )]\ |\ ^{\shortmid }g^{6}(\tau )}(\ ^{\shortmid }%
\mathbf{e}_{5}(\tau ))^{2}+\ ^{\shortmid }g^{6}(\tau )(\ ^{\shortmid }%
\mathbf{e}_{6}(\tau ))^{2}-  \notag \\
&&-\frac{[~^{\shortmid }\partial ^{7}(~^{\shortmid }g^{7}(\tau ))]^{2}}{%
|\int dp_{7}\ ~^{\shortmid }\partial ^{7}[(~_{4}^{\shortmid }\Im (\tau
))~^{\shortmid }g^{8}(\tau )]\ |\ ~^{\shortmid }g^{8}(\tau )}(\ ^{\shortmid }%
\mathbf{e}_{7}(\tau ))^{2}+\ ^{\shortmid }g^{8}(\tau )(\ ^{\shortmid }%
\mathbf{e}_{8}(\tau ))^{2},  \notag
\end{eqnarray}%
where the nonholonomic s-frames and respective off-diagonal terms defined by
N-connection coefficients are computed:%
\begin{eqnarray}
\mathbf{e}^{3}(\tau ) &=&dy^{3}+\frac{\partial _{k_{1}}[\int
dy^{3}(~_{2}^{\shortmid }\Im (\tau ))\ \partial _{3}g_{4}(\tau )]}{%
(~_{2}^{\shortmid }\Im (\tau ))\ \partial _{3}g_{4}(\tau )}dx^{k_{1}},
\label{sol3ncf} \\
\mathbf{e}^{4}(\tau ) &=&dt+(\ _{1}n_{k_{1}}(\tau )+\ _{2}n_{k_{1}}(\tau
)\int dy^{3}\frac{(\partial _{3}g_{4}(\tau ))^{2}}{|\int dy^{3}\partial
_{3}[(~_{2}^{\shortmid }\Im (\tau ))g_{4}(\tau )]|\ [g_{4}(\tau )]^{5/2}}%
)dx^{k_{1}},  \notag
\end{eqnarray}%
\begin{eqnarray*}
\ ^{\shortmid }\mathbf{e}_{5}(\tau ) &=&dp_{5}+\frac{\partial _{k_{2}}[\int
dp_{5}(~_{3}^{\shortmid }\Im (\tau ))\ ~^{\shortmid }\partial
^{5}(~^{\shortmid }g^{6}(\tau ))]}{(~_{3}^{\shortmid }\Im (\tau
))~^{\shortmid }\partial ^{5}(~^{\shortmid }g^{6}(\tau ))}dx^{k_{2}}, \\
\ ^{\shortmid }\mathbf{e}_{6}(\tau ) &=&dp_{6}+(\ _{1}^{\shortmid
}n_{k_{2}}(\tau )+\ _{2}^{\shortmid }n_{k_{2}}(\tau )\int dp_{5}\frac{%
[~^{\shortmid }\partial ^{5}(~^{\shortmid }g^{6}(\tau ))]^{2}}{|\int dp_{5}\
~^{\shortmid }\partial ^{5}[(~_{3}^{\shortmid }\Im (\tau ))~^{\shortmid
}g^{6}(\tau )]|\ [~^{\shortmid }g^{6}(\tau )]^{5/2}})dx^{k_{2}},
\end{eqnarray*}%
\begin{eqnarray*}
\ ^{\shortmid }\mathbf{e}_{7}(\tau ) &=&dp_{7}+\frac{~^{\shortmid }\partial
_{k_{3}}[\int dp_{7}(\ ~_{4}^{\shortmid }\Im (\tau ))\ ~^{\shortmid
}\partial ^{7}(~^{\shortmid }g^{8}(\tau ))]}{(~_{4}^{\shortmid }\Im (\tau
))~^{\shortmid }\partial ^{7}(~^{\shortmid }g^{8}(\tau ))}d\ ^{\shortmid
}x^{k_{3}}, \\
\ ^{\shortmid }\mathbf{e}_{8}(\tau ) &=&dE+(\ _{1}^{\shortmid
}n_{k_{3}}(\tau )+\ _{2}^{\shortmid }n_{k_{3}}(\tau )\int dp_{7}\frac{%
[~^{\shortmid }\partial ^{7}(~^{\shortmid }g^{8}(\tau ))]^{2}}{|\int dp_{7}\
~^{\shortmid }\partial ^{7}[(~_{4}^{\shortmid }\Im (\tau ))~^{\shortmid
}g^{8}(\tau )]|\ [~^{\shortmid }g^{8}(\tau )]^{5/2}})d\ ^{\shortmid
}x^{k_{3}},
\end{eqnarray*}%
The s-adapted coefficients (\ref{sol3rf}) and (\ref{sol3ncf}) define $\tau $%
-flows of quasi-stationary solutions of type \newline
$\ _{s}^{\shortmid }\mathbf{g}[\hbar ,\kappa ,\psi ,~_{s}^{\shortmid }%
\mathcal{K},g_{4},\ ^{\shortmid }g^{6},\ ^{\shortmid }g^{8}]$ in (\ref%
{smetrfunctionals}) evolving in s-adapted $\kappa $-linear parametric form
to s-metrics \newline
$\ _{s}^{\shortmid }\mathbf{g}(\tau )\simeq $ $\ _{s}^{\shortmid }\mathbf{g}%
[\hbar ,\kappa ,\psi (\tau ),~_{s}^{\shortmid }\Im (\tau ),g_{4}(\tau ),\
^{\shortmid }g^{6}(\tau ),\ ^{\shortmid }g^{8}(\tau )]$.

Above linear quadratic elements with respective s- and N-coefficients can be
re-defined to include functional dependencies on running cosmological
constants $\ _{s}^{\shortmid }\Lambda (\tau )$ if we begin with (\ref{sol2rf}%
) and (\ref{sol2nrf}) and express, using respective formulas from (\ref%
{nonlinsymrex}), $\ _{s}\Phi (\tau )=\ _{s}\Phi \lbrack \
_{s}^{\shortmid}\Lambda (\tau ), \ _{s}^{\shortmid }\Im ,g_{4}(\tau ),\
^{\shortmid}g^{6}(\tau ),\ ^{\shortmid }g^{8}(\tau )].$ This allows us to
model the $\tau $-evolution of nonassociative Ricci solitons of type $\
_{s}^{\shortmid}\mathbf{g}[\hbar ,\kappa ,\psi ,\ _{s}^{\shortmid }\Lambda
_{0}, \ _{s}^{\shortmid }\mathcal{K},g_{4},\ ^{\shortmid }g^{6},\
^{\shortmid}g^{8}],$ see (\ref{smetrfunctionals}), into generic off-diagonal
solutions of nonassociative $\kappa $-linear parametric geometric flow
equations (\ref{nonassocgeomflef}) determined by families of
quasi-stationary s-metrics $\ _{s}^{\shortmid }\mathbf{g}(\tau )\simeq \
_{s}^{\shortmid }\mathbf{g}[\hbar ,\kappa ,\psi (\tau ),\ _{s}^{\shortmid
}\Lambda (\tau ),~_{s}^{\shortmid}\Im (\tau ),g_{4}(\tau ),\ ^{\shortmid
}g^{6}(\tau ),\ ^{\shortmid}g^{8}(\tau )]$.

\subsection{Quasi-stationary nonassociative evolution via gravitational
polarizations}

\label{ssqsgrpol}How to construct quasi-stationary solutions for
nonassociative Ricci soliton and vacuum gravitational equations using the
AFCDM with gravitational polarization functions is discussed in appendix \ref%
{assgrpolfl}. Here, we extend the approach for generating $\tau$-parametric
quasi-stationary solutions for nonassociative geometric flow equations (\ref%
{nonassocgeomflef}).

Off-diagonal deformations of a family of prescribed prime metric into other
families of target ones, \newline
$\ _{s}^{\shortmid }\mathbf{\mathring{g}}(\tau )=[\ ^{\shortmid }\mathring{g}%
_{\alpha _{s}}(\tau ), \ ^{\shortmid }\mathring{N}_{i_{s-1}}^{a_{s}}(\tau)]%
\rightarrow \ _{s}^{\shortmid }\mathbf{g}(\tau )$ (\ref{offdiagdefr})
described by $\eta $-polarizations (\ref{dmpolariz}), can be defined by such 
$\tau $-parametric generating functions%
\begin{equation}
\psi (\tau )\simeq \psi (\hbar ,\kappa ;\tau ,x^{k_{1}}),\eta _{4}(\tau )\
\simeq \eta _{4}(\tau ,x^{k_{1}},y^{3}),\ ^{\shortmid }\eta ^{6}(\tau
)\simeq \ ^{\shortmid }\eta ^{6}(\tau ,x^{i_{2}},p_{5}),\ ^{\shortmid }\eta
^{8}(\tau )\simeq \ ^{\shortmid }\eta ^{8}(\tau ,x^{i_{2}},p_{5},p_{7}).
\label{etapolgen}
\end{equation}%
As a result, we generalize in real variables and for nonassociative
geometric flows the quadratic line element constructed in Appendix B2,
formula (B2), to \cite{partner02}: 
\begin{eqnarray}
d\ \ ^{\shortmid }\widehat{s}^{2}(\tau ) &=&\ \ ^{\shortmid }g_{\alpha
_{s}\beta _{s}}(\hbar ,\kappa ,\tau ,x^{k},y^{3},p_{a_{3}},p_{a_{4}};\
^{\shortmid }\mathring{g}_{\alpha _{s}}(\tau );\eta _{4}(\tau ),\
^{\shortmid }\eta ^{6}(\tau ),\ ^{\shortmid }\eta ^{8}(\tau ),\
_{s}^{\shortmid }\Lambda (\tau );\ _{s}^{\shortmid }\Im (\tau ))d~\
^{\shortmid }u^{\alpha _{s}}d~\ ^{\shortmid }u^{\beta _{s}}
\label{offdiagpolfr} \\
&=&e^{\psi (\tau )}[(dx^{1})^{2}+(dx^{2})^{2}]-  \notag \\
&&\frac{[\partial _{3}(\eta _{4}(\tau )\ \mathring{g}_{4}(\tau ))]^{2}}{%
|\int dy^{3}~_{2}\Im (\tau )\partial _{3}(\eta _{4}(\tau )\ \mathring{g}%
_{4}(\tau ))|\ (\eta _{4}(\tau )\mathring{g}_{4}(\tau ))}\{dy^{3}+\frac{%
\partial _{i_{1}}[\int dy^{3}\ _{2}\Im (\tau )\ \partial _{3}(\eta _{4}(\tau
)\mathring{g}_{4}(\tau ))]}{~_{2}\Im (\tau )\partial _{3}(\eta _{4}(\tau )%
\mathring{g}_{4}(\tau ))}dx^{i_{1}}\}^{2}+  \notag \\
&&\eta _{4}(\tau )\mathring{g}_{4}(\tau ))\{dt+[\ _{1}n_{k_{1}}(\tau )+\
_{2}n_{k_{1}}(\tau )\int dy^{3}\frac{[\partial _{3}(\eta _{4}(\tau )%
\mathring{g}_{4}(\tau ))]^{2}}{|\int dy^{3}\ _{2}\Im (\tau )\partial
_{3}(\eta _{4}(\tau )\mathring{g}_{4}(\tau ))|\ (\eta _{4}(\tau )\mathring{g}%
_{4}(\tau ))^{5/2}}]dx^{k_{1}}\}  \notag
\end{eqnarray}%
\begin{eqnarray*}
&&-\frac{[\ ^{\shortmid }\partial ^{5}(\ ^{\shortmid }\eta ^{6}(\tau )\
^{\shortmid }\mathring{g}^{6}(\tau ))]^{2}}{|\int dp_{5}~_{3}^{\shortmid
}\Im (\tau )\ \ ^{\shortmid }\partial ^{5}(\ ^{\shortmid }\eta ^{6}(\tau )\
\ ^{\shortmid }\mathring{g}^{6}(\tau ))\ |\ (\ ^{\shortmid }\eta ^{6}(\tau
)\ ^{\shortmid }\mathring{g}^{6}(\tau ))}\{dp_{5}+\frac{\ \ ^{\shortmid
}\partial _{i_{2}}[\int dp_{5}\ _{3}^{\shortmid }\Im (\tau )\ ^{\shortmid
}\partial ^{5}(\ ^{\shortmid }\eta ^{6}(\tau )\ ^{\shortmid }\mathring{g}%
^{6}(\tau ))]}{~_{3}^{\shortmid }\Im (\tau )\ ^{\shortmid }\partial ^{5}(\
^{\shortmid }\eta ^{6}(\tau )\ ^{\shortmid }\mathring{g}^{6}(\tau ))}%
dx^{i_{2}}\}^{2} \\
&&+(\ ^{\shortmid }\eta ^{6}(\tau )\ ^{\shortmid }\mathring{g}^{6}(\tau
))\{dp_{6}+[\ _{1}^{\shortmid }n_{k_{2}}(\tau )+ \\
&& \ _{2}^{\shortmid}n_{k_{2}}(\tau )\int dp_{5}\frac{[\ ^{\shortmid
}\partial ^{5}(\ ^{\shortmid }\eta ^{6}(\tau )\ ^{\shortmid }\mathring{g}%
^{6}(\tau ))]^{2}}{|\int dp_{5}~_{3}^{\shortmid }\Im (\tau )\ \partial
^{5}(\ ^{\shortmid }\eta ^{6}(\tau )\ ^{\shortmid }\mathring{g}^{6}(\tau
))|\ (\ ^{\shortmid }\eta ^{6}(\tau )\ ^{\shortmid }\mathring{g}^{6}(\tau
))^{5/2}}]dx^{k_{2}}\}
\end{eqnarray*}%
\begin{eqnarray*}
&&-\frac{[\ ^{\shortmid }\partial ^{7}(\ ^{\shortmid }\eta ^{8}(\tau )\
^{\shortmid }\mathring{g}^{8}(\tau ))]^{2}}{|\int dp_{7}~_{4}^{\shortmid
}\Im (\tau )\ ^{\shortmid }\partial ^{8}(\ ^{\shortmid }\eta ^{7}(\tau )\
^{\shortmid }\mathring{g}^{7}(\tau ))\ |\ (\ ^{\shortmid }\eta ^{7}(\tau )\
^{\shortmid }\mathring{g}^{7}(\tau ))}\{dp_{7}+\frac{~\ ^{\shortmid
}\partial _{i_{3}}[\int dp_{7}~_{4}^{\shortmid }\Im (\tau )\ \ ^{\shortmid
}\partial ^{7}(\ ^{\shortmid }\eta ^{8}(\tau )\ \ ^{\shortmid }\mathring{g}%
^{8}(\tau ))]}{~_{4}^{\shortmid }\Im (\tau )\ ^{\shortmid }\partial ^{7}(\
^{\shortmid }\eta ^{8}(\tau )\ ^{\shortmid }\mathring{g}^{8}(\tau ))}d\
^{\shortmid }x^{i_{3}}\}^{2} \\
&&+(\ ^{\shortmid }\eta ^{8}(\tau )\ ^{\shortmid }\mathring{g}^{8}(\tau
))\{dE+[\ _{1}n_{k_{3}}(\tau )+ \\
&& \ _{2}n_{k_{3}}(\tau )\int dp_{7}\frac{[\ ^{\shortmid }\partial ^{7}(\
^{\shortmid }\eta ^{8}(\tau )\ ^{\shortmid }\mathring{g}^{8}(\tau ))]^{2}}{%
|\int dp_{7}~_{4}^{\shortmid }\Im (\tau )[\ ^{\shortmid }\partial ^{7}(\
^{\shortmid }\eta ^{8}(\tau )\ ^{\shortmid }\mathring{g}^{8}(\tau ))]|\ [(\
^{\shortmid }\eta ^{8}(\tau )\ ^{\shortmid }\mathring{g}^{8}(\tau ))]^{5/2}}%
]d\ ^{\shortmid }x^{k_{3}}\}.
\end{eqnarray*}

The gravitational polarization $\eta$--functions describe transforms of
certain classes of prime s-metrics into other types of target s-metrics. We
can prescribe respective geometric/ physical properties and investigate how
geometric flows may relate the evolution of such configurations. Respective
formulas for small parametric nonassociative geometric flow deformations of
type (\ref{offdiagpolfr}) when $\ _{s}^{\shortmid }\eta (\tau )\ ^{\shortmid}%
\mathring{g}_{\alpha _{s}}(\tau )\sim \ ^{\shortmid }\zeta _{\alpha
_{s}}(\tau ) (1+\kappa \ ^{\shortmid }\chi _{\alpha _{s}}(\tau ))\ \
^{\shortmid }\mathring{g}_{\alpha _{s}}(\tau )$ (\ref{nonlinsymr}) are
provided in appendix \ref{assgrpolfl}.

\section{Modified Bekenstein--Hawking and G. Perelman thermodynamics of BH
solutions \newline deformed by nonassociative geometric flows}

\label{sec5}Applying nonholonomic geometric flow methods \cite%
{vacaru20,ib19,bubuianu19}, we concluded \cite{partner04} that solutions for
nonassociative star product R-flux deformations of the Tangerlini higher
dimension BHs and double Schwarzschild BHs can be described in the framework
of a (modified) $\kappa $-linear parametric Perelman's statistical
thermodynamic model encoding nonassociative data. In section \ref{sec3}, we
generalized the approach to a theory of nonassociative Ricci flows and
formulated a statistical thermodynamic model using decoupling properties of
respective systems of nonlinear PDEs involving for the canonical
s-connection structure.

\vskip5pt The goal of this section is to elaborate on explicit models of
nonassociative flow evolution of defined by quasi-stationary parametric
solutions. We construct and analyze important geometric and physical
properties of two new classes of nonassociative modified BH solutions in the
framework of nonassociative geometric evolution theory. For a fixed
evolution parameter $\tau _{0}$, corresponding s-metrics describe $\kappa $%
-linear deformations of the double Reisner-Nordstr\"{o}m de Sitter, RN-dS,
metrics and their 'dissipation' into off-diagonal terms; or couples of
Schwarzschild - AdS BHs deformed to black ellipsoid, BE, configurations; or
higher dimension RN anti de Sitter, RN-AdS, configurations. Such
quasi-stationary generic off-diagonal solutions present explicit examples of
nonassociative Ricci solitons, which can be extended to describe geometric
flow evolution scenarios on temperature like $\tau $-parameter.

\vskip5pt For very special cases of nonholonomic deformations (for instance,
defining BE configurations and/or other variants with geometric evolving
hyper-surface horizons), we can apply the concept of Bekenstein-Hawking
entropy \cite{bek1,bek2,haw1,haw2}. In another turn, the generalized G.
Perelman W-entropy and thermodynamic variables \cite%
{perelman1,vacaru20,ib19,bubuianu19,partner04} can be defined and computed
for all classes of solutions in nonassociative geometric flow and (modified)
gravity theories. We prove this by providing explicit examples how to
compute thermodynamic variables for general quasi-stationary $\tau $%
--deformations of nonassociative modified RN-(A)dS, BE deformations and
Schwarzschild-(A)dS metrics.

\subsection{Geometric flow thermodynamics of nonassociative quasi-stationary
solutions with running \newline cosmological constants}

Statistical G. Perelman thermodynamic models can be defined for any
nonassociative geometric flow data $[\ _{s}^{\shortmid }\mathfrak{g}^{\star
}(\tau ),\ _{s}^{\shortmid }\widehat{\mathbf{D}}^{\star }(\tau ),\
_{s}^{\shortmid }\widehat{f}(\tau )] $ as we explain in section \ref{sec2}.
For $\kappa $-linear parametric decompositions, the corresponding
thermodynamic variables are computed using the formulas for $\
_{s}^{\shortmid }\widehat{\mathcal{E}}_{\kappa }^{\star }$ and $\
_{s}^{\shortmid }\widehat{S}_{\kappa }^{\star }$ from (\ref%
{nagthermodvaluesp}). In this work, we do not provide cumbersome formulas
for computing quadratic fluctuation parameter $\ _{s}^{\shortmid }\widehat{%
\sigma }_{\kappa }^{\star }).$ The nonassociative thermodynamic variables
are derived for a W-entropy $\ _{s}^{\shortmid }\widehat{\mathcal{W}}%
_{\kappa}^{\star }(\tau )$ (\ref{nawfunctp}) using a respective statistical
generating function $\ _{s}^{\shortmid }\widehat{Z}(\tau )$ (\ref{spf}). We
can prescribe the nonholonomic structure on phase space $\ _{s}^{\star }%
\mathcal{M}$ and the normalizing functions $\ _{s}^{\shortmid }\widehat{f}%
(\tau )$ in such forms that all basic formulas are determined by a volume
form $^{\shortmid}\delta \ ^{\shortmid }\mathcal{V}(\tau )$ (\ref{volume}).

For exact/parametric solutions of nonassociative geometric flow equations,
we can compute the thermodynamic variables corresponding to $\tau$--modified
Einstein equations (\ref{nonassocgeomflef}), with effective sources $\
_{s}^{\shortmid }\Im (\tau ),$ and/or (\ref{nonassocgeomflefcc}) with
running effective cosmological constants $\ _{s}^{\shortmid }\Lambda (\tau).$
In the first approach, we can formulate a G. Perelman thermodynamic model
for quasi-stationary solutions taken in the form (\ref{sol1rf}) with
N-coefficients (\ref{sol1nrf}). This results in cumbersome formulas for
thermodynamic variables which are not appropriate for investigating, for
instance, physical problems related to the swampland conjecture \cite%
{oog06,kehagias19,biasio20,biasio21,lueben21,biasio22}. Using nonlinear
symmetries (\ref{nonlinsymr}), with $\ _{s}^{\shortmid }\mathbf{g}%
[\hbar,\kappa ,\tau ,\psi (\tau ),\ _{s}\Psi (\tau ), \ _{s}^{\shortmid }\Im
(\tau)]\rightarrow \ _{s}^{\shortmid }\mathbf{g}[\hbar ,\kappa ,\tau
,\psi(\tau),\ _{s}\Phi (\tau ), \ _{s}^{\shortmid }\Lambda (\tau )],$ we may
develop a second approach involving quasi-stationary solutions of type (\ref%
{sol1a}). This allows us to simplify the formulas for explicit computation
of geometric thermodynamic variables and elaborate on physical models with
running/fixed effective cosmological constants $\ _{s}^{\shortmid }\Lambda
(\tau ).$

The nonassociative geometric flow thermodynamic variables defined for a
temperature parameter $\tau ,0<\tau ^{\prime }\leq \tau ,$ with prescribed
constant normalizing functions and for a volume form $\ ^{\shortmid }\delta\
^{\shortmid }\mathcal{V}(\tau )$ computed for quasi-stationary data $%
[_{s}\Phi (\tau ),\ _{s}^{\shortmid }\Lambda (\tau )]$, are expressed in the
form 
\begin{eqnarray}
\ _{s}^{\shortmid }\widehat{\mathcal{W}}_{\kappa }^{\star }(\tau )
&=&\int\nolimits_{\tau ^{\prime }}^{\tau }\frac{d\tau }{(4\pi \tau )^{4}}%
\int_{\ _{s}^{\shortmid }\widehat{\Xi }}\left( \tau \lbrack
\sum\nolimits_{s}\ _{s}^{\shortmid }\Lambda (\tau )]^{2}-8\right) \
^{\shortmid }\delta \ ^{\shortmid }\mathcal{V}(\tau ),  \label{thvcann} \\
\ _{s}^{\shortmid }\widehat{\mathcal{Z}}_{\kappa }^{\star }(\tau ) &=&\exp %
\left[ \int\nolimits_{\tau ^{\prime }}^{\tau }\frac{d\tau }{(2\pi \tau )^{4}}%
\int_{\ _{s}^{\shortmid }\widehat{\Xi }}\ ^{\shortmid }\delta \ ^{\shortmid }%
\mathcal{V}(\tau )\right] ,  \notag \\
\ _{s}^{\shortmid }\widehat{\mathcal{E}}_{\kappa }^{\star }(\tau )
&=&-\int\nolimits_{\tau ^{\prime }}^{\tau }\frac{d\tau }{(4\pi )^{4}\tau ^{2}%
}\int_{\ _{s}^{\shortmid }\widehat{\Xi }}\left( [\sum\nolimits_{s}\
_{s}^{\shortmid }\Lambda (\tau )]-\frac{4}{\tau }\right) \ ^{\shortmid
}\delta \ ^{\shortmid }\mathcal{V}(\tau ),  \notag \\
\ _{s}^{\shortmid }\widehat{\mathcal{S}}_{\kappa }^{\star }(\tau )
&=&-\int\nolimits_{\tau ^{\prime }}^{\tau }\frac{d\tau }{(4\pi \tau )^{4}}%
\int_{\ _{s}^{\shortmid }\widehat{\Xi }}\left( \tau \lbrack
\sum\nolimits_{s}\ _{s}^{\shortmid }\Lambda (\tau )]-8\right) \ ^{\shortmid
}\delta \ ^{\shortmid }\mathcal{V}(\tau ).  \notag
\end{eqnarray}

The integration on $\tau $ parameter and respective 8-d hyper-surface
integrals in (\ref{thvcann}) should be defined for nonholonomic
s-distributions which result in well-defined relativistic thermodynamic
values.\footnote{%
For instance, we must exclude un-physical configurations with negative
entropy; additionally, we should analyze and select on more optimal
energetic regimes and stability conditions; and construct thermodynamic
models with causal evolution on cotangent Lorentz bundles etc.} Such
constructions were considered for a fixed temperature parameter $\tau _{0}$
and nonassociative Ricci solitons in \cite{partner04} (see formulas (60) and
(61) in that work).

For any prescribed data $(\tau ^{\prime },\tau ),\ _{s}^{\shortmid }\widehat{%
\Xi }$ and $\ _{s}^{\shortmid }\Lambda (\tau ),$ the thermodynamic variables
(\ref{thvcann}) are determined by a volume form $\ ^{\shortmid }\delta \
^{\shortmid }\mathcal{V}(\tau )$ which must be computed for a chosen class
of exact/parametric solutions of nonassociative geometric flow equations. We
can elaborate an explicit geometric integration formalism adapted to
nonassociative and nonholonomic distributions, and parametric deformations,
if we choose, for instance some primary data $\ _{s}^{\shortmid }\mathfrak{%
\mathring{g}}^{\star }$ and study possible flow evolution scenarios to
certain target $\ _{s}^{\shortmid }\mathfrak{g}^{\star }(\tau )$. For a $%
\tau $-family of s-metrics (\ref{sol2rf}) with respect to nonholonomic
frames (\ref{sol2nrf}), we can define the volume functional 
\begin{eqnarray}
\ ^{\shortmid }\delta \ ^{\shortmid }\mathcal{V}(\tau ) &=&\ ^{\shortmid
}\delta \ ^{\shortmid }\mathcal{V[}\tau ,\ _{s}^{\shortmid }\Lambda (\tau
),\ \ _{s}^{\shortmid }\Im (\tau );\psi (\tau ),\ _{s}^{\shortmid }\Phi
(\tau )]  \label{volumf} \\
&=&e^{\psi (\tau )}\frac{|\ _{2}\Phi (\tau )\partial _{3}[\ _{2}\Phi (\tau
)]^{2}|}{|\ _{2}\Lambda (\tau )\int dy^{3}\ _{2}\Im (\tau )\{\partial _{3}[\
_{2}\Phi (\tau )]^{2}\}^{2}|^{1/2}\ }[dy^{3}+\frac{\partial _{i_{1}}\left(
\int dy^{3}\ _{2}\Im (\tau )\partial _{3}[\ _{2}\Phi (\tau )]^{2}\right) }{\
_{2}\Im (\tau )\partial _{3}[\ _{2}\Phi (\tau )]^{2}}%
dx^{i_{1}}]dx^{1}dx^{2}dt  \notag \\
&&\frac{|\ _{3}^{\shortmid }\Phi (\tau )\ ^{\shortmid }\partial ^{5}[\
_{3}^{\shortmid }\Phi (\tau )]^{2}|}{|\ _{3}^{\shortmid }\Lambda (\tau )\int
dp_{5}\ _{3}^{\shortmid }\Im (\tau )\ ^{\shortmid }\partial ^{5}[\
_{3}^{\shortmid }\Phi (\tau )]^{2}|^{1/2}\ }[dp_{5}+\frac{\ ^{\shortmid
}\partial _{i_{2}}\left( \int dp_{5}\ _{3}^{\shortmid }\Im (\tau )\
^{\shortmid }\partial ^{5}[\ _{3}^{\shortmid }\Phi (\tau )]^{2}\right) }{\
_{3}^{\shortmid }\Im (\tau )\ ^{\shortmid }\partial ^{5}[\ _{3}^{\shortmid
}\Phi (\tau )]^{2}}dx^{i_{2}}]dp_{6}  \notag \\
&&\frac{|\ _{4}^{\shortmid }\Phi (\tau )\ ^{\shortmid }\partial ^{7}[\
_{4}^{\shortmid }\Phi (\tau )]^{2}|}{|\ _{4}^{\shortmid }\Lambda (\tau )\int
dp_{7}\ \ _{4}^{\shortmid }\Im (\tau )\ ^{\shortmid }\partial ^{7}[\ \
_{4}^{\shortmid }\Phi (\tau )]^{2}|^{1/2}\ }[dp_{7}+\frac{\ ^{\shortmid
}\partial _{i_{3}}\left( \int dp_{7}\ _{4}^{\shortmid }\Im (\tau )\
^{\shortmid }\partial ^{7}[\ _{4}^{\shortmid }\Phi (\tau )]^{2}\right) }{\ \
_{4}^{\shortmid }\Im (\tau )\ ^{\shortmid }\partial ^{7}[\ _{4}^{\shortmid
}\Phi (\tau )]^{2}}dx^{i_{3}}]dE.  \notag
\end{eqnarray}%
To compute in explicit forms and study properties of such volume functionals
we can consider, for simplicity, nonholonomic evolution models with trivial
integration functions $\ _{1}^{\shortmid }n_{k_{s}}=0$ and $\
_{2}^{\shortmid }n_{k_{s}}=0.$ The formulas for $\ ^{\shortmid }\delta \
^{\shortmid }\mathcal{V}(\tau )$ (\ref{volumf}) can be computed for other
classes of solutions determined by $g$-generating functions and/or $\eta $- /%
$\chi $-polarization functions using (\ref{nonlinsymrex}), when 
\begin{eqnarray}
\ _{2}\Phi (\tau ) &=&2\sqrt{|\ _{2}\Lambda (\tau )\ g_{4}(\tau )|}=\ 2\sqrt{%
|\ _{2}\Lambda (\tau )\ \eta _{4}(\tau )\mathring{g}_{4}(\tau )|}\simeq 2%
\sqrt{|\ _{2}\Lambda (\tau )\ \zeta _{4}(\tau )\mathring{g}_{4}(\tau )|}[1-%
\frac{\kappa }{2}\chi _{4}(\tau )],  \label{genf1} \\
\ \ _{3}^{\shortmid }\Phi (\tau ) &=&2\sqrt{|\ _{3}^{\shortmid }\Lambda
(\tau )~^{\shortmid }g^{6}(\tau )|}=\ 2\sqrt{|\ _{3}^{\shortmid }\Lambda
(\tau )~^{\shortmid }\eta ^{6}(\tau )~^{\shortmid }\mathring{g}^{6}(\tau )|}%
\simeq 2\sqrt{|\ _{3}^{\shortmid }\Lambda (\tau )~^{\shortmid }\zeta
^{6}(\tau )~^{\shortmid }\mathring{g}^{6}(\tau )|}[1-\frac{\kappa }{2}%
~^{\shortmid }\chi ^{6}(\tau )],  \notag \\
\ \ _{4}^{\shortmid }\Phi (\tau ) &=&2\sqrt{|\ _{4}^{\shortmid }\Lambda
(\tau )~^{\shortmid }g^{8}(\tau )|}=\ 2\sqrt{|\ _{4}^{\shortmid }\Lambda
(\tau )~^{\shortmid }\eta ^{8}(\tau )~^{\shortmid }\mathring{g}^{8}(\tau )|}%
\simeq 2\sqrt{|\ _{4}^{\shortmid }\Lambda (\tau )~^{\shortmid }\zeta
^{8}(\tau )~^{\shortmid }\mathring{g}^{8}(\tau )|}[1-\frac{\kappa }{2}%
~^{\shortmid }\chi ^{8}(\tau )],  \notag
\end{eqnarray}%
for a prime s-metric $\ _{\shortmid }\mathring{g}_{\alpha }(\tau )=(%
\mathring{g}_{i}(\tau ),\ _{\shortmid }\mathring{g}^{a}(\tau )).$
Introducing formulas (\ref{genf1}) involving $\eta $-polarizations in (\ref%
{volumf}), then separating terms with shell $\tau $-running cosmological
constants, we express: 
\begin{eqnarray*}
\ ^{\shortmid }\delta \ ^{\shortmid }\mathcal{V} &=&\ ^{\shortmid }\delta \
^{\shortmid }\mathcal{V}[\tau ,\ _{s}^{\shortmid }\Lambda (\tau ),\
_{s}^{\shortmid }\Im (\tau );\psi (\tau ),\ g_{4}(\tau ),\ ^{\shortmid
}g^{6}(\tau ),\ ^{\shortmid }g^{8}(\tau )]=\ ^{\shortmid }\delta \
^{\shortmid }\mathcal{V}(\ _{s}\Im (\tau ),\ _{s}^{\shortmid }\Lambda (\tau
),\ ^{\shortmid }\eta _{\alpha _{s}}(\tau )\ ^{\shortmid }\mathring{g}%
_{\alpha _{s}}) \\
&=&\frac{1}{\sqrt{|\ _{1}\Lambda (\tau )\ _{2}\Lambda (\tau )\
_{3}^{\shortmid }\Lambda (\tau )\ _{4}^{\shortmid }\Lambda (\tau )|}}\ \
^{\shortmid }\delta \ _{\eta }^{\shortmid }\mathcal{V},\mbox{ where }\
^{\shortmid }\delta \ _{\eta }^{\shortmid }\mathcal{V}=\ ^{\shortmid }\delta
\ _{\eta }^{1}\mathcal{V}\times \ ^{\shortmid }\delta \ _{\eta }^{2}\mathcal{%
V}\times \ ^{\shortmid }\delta \ _{\eta }^{3}\mathcal{V}\times \ ^{\shortmid
}\delta \ _{\eta }^{4}\mathcal{V}.
\end{eqnarray*}%
In these formulas, there are used the functionals: 
\begin{eqnarray}
\ ^{\shortmid }\delta \ _{\eta }^{1}\mathcal{V} &=&\ ^{\shortmid }\delta \
_{\eta }^{1}\mathcal{V}[\ _{1}\Im (\tau ),\eta _{2}(\tau )\ \mathring{g}_{2}]
\label{volumfuncts} \\
&=&\frac{16}{3}e^{\widetilde{\psi }(\tau )}dx^{1}dx^{2}=\frac{3}{16}\sqrt{|\
_{1}\Lambda (\tau )|}e^{\psi (\tau )}dx^{1}dx^{2},\mbox{ for }\psi (\tau )%
\mbox{ being a solution of  }(\ref{poas2d}),  \notag \\
\ ^{\shortmid }\delta \ _{\eta }^{2}\mathcal{V} &=&\ ^{\shortmid }\delta \
_{\eta }^{2}\mathcal{V}[\ _{2}\Im (\tau ),\eta _{4}(\tau )\ \mathring{g}_{4}]
\notag \\
&=&\frac{16}{3}\frac{\partial _{3}|\ \eta _{4}(\tau )\ \mathring{g}%
_{4}|^{3/2}}{\ \sqrt{|\int dy^{3}\ _{2}\Im (\tau )\{\partial _{3}|\ \eta
_{4}(\tau )\ \mathring{g}_{4}|\}^{2}|}}[dy^{3}+\frac{\partial _{i_{1}}\left(
\int dy^{3}\ _{2}\Im (\tau )\partial _{3}|\ \eta _{4}(\tau )\ \mathring{g}%
_{4}|\right) dx^{i_{1}}}{\ _{2}\Im (\tau )\partial _{3}|\ \eta _{4}(\tau )\ 
\mathring{g}_{4}|}]dt,  \notag \\
\ ^{\shortmid }\delta \ _{\eta }^{3}\mathcal{V} &=&\ ^{\shortmid }\delta \
_{\eta }^{3}\mathcal{V}[\ _{3}\Im (\tau ),~^{\shortmid }\eta ^{6}(\tau
)~^{\shortmid }\mathring{g}^{6}]  \notag \\
&=&\frac{16}{3}\frac{\ ^{\shortmid }\partial ^{5}|\ ~^{\shortmid }\eta
^{6}(\tau )~^{\shortmid }\mathring{g}^{6}|^{3/2}}{\ \sqrt{|\int dp_{5}\
_{3}\Im (\tau )\{\ ^{\shortmid }\partial ^{5}|~^{\shortmid }\eta ^{6}(\tau
)~^{\shortmid }\mathring{g}^{6}|\}^{2}|}}[dp_{5}+\frac{\partial
_{i_{2}}\left( \int dp_{5}\ _{2}\Im (\tau )\ ^{\shortmid }\partial ^{5}|\
~^{\shortmid }\eta ^{6}(\tau )~^{\shortmid }\mathring{g}^{6}|\right)
dx^{i_{2}}}{\ _{2}\Im (\tau )\ ^{\shortmid }\partial ^{5}|~^{\shortmid }\eta
^{6}(\tau )~^{\shortmid }\mathring{g}^{6}|}]dp_{6},  \notag
\end{eqnarray}
\begin{eqnarray*}
\ ^{\shortmid }\delta \ _{\eta }^{4}\mathcal{V} &=&\ ^{\shortmid }\delta \
_{\eta }^{4}\mathcal{V}[\ _{4}\Im (\tau ),~^{\shortmid }\eta ^{8}(\tau
)~^{\shortmid }\mathring{g}^{8}]  \notag \\
&=&\frac{16}{3}\frac{\ ^{\shortmid }\partial ^{7}|\ ~^{\shortmid }\eta
^{8}(\tau )~^{\shortmid }\mathring{g}^{8}|^{3/2}}{\sqrt{\ |\int dp_{7}\
_{4}\Im (\tau )\{\ ^{\shortmid }\partial ^{7}|~^{\shortmid }\eta ^{8}(\tau
)~^{\shortmid }\mathring{g}^{8}|\}^{2}|}}[dp_{7}+\frac{\partial
_{i_{3}}\left( \int dp_{7}\ _{3}\Im (\tau )\ ^{\shortmid }\partial ^{7}|\
~^{\shortmid }\eta ^{8}(\tau )~^{\shortmid }\mathring{g}^{8}|\right) }{\
_{2}\Im (\tau )\ ^{\shortmid }\partial ^{7}|~^{\shortmid }\eta ^{8}(\tau
)~^{\shortmid }\mathring{g}^{8}|}dx^{i_{3}}]dE.  \notag
\end{eqnarray*}

The G. Perelman thermodynamic variables (\ref{thvcann})\ computed for the
volume functionals can be expressed as thermodynamic functionals: 
\begin{eqnarray}
\ _{s}^{\shortmid }\widehat{\mathcal{W}}_{\kappa }^{\star }(\tau )
&=&\int\nolimits_{\tau ^{\prime }}^{\tau }\frac{d\tau }{(4\pi \tau )^{4}}%
\frac{\tau \lbrack \ _{1}\Lambda (\tau )+\ _{2}\Lambda (\tau )+\
_{3}^{\shortmid }\Lambda (\tau )+\ _{4}^{\shortmid }\Lambda (\tau )]^{2}-8}{%
\sqrt{|\ _{1}\Lambda (\tau )\ _{2}\Lambda (\tau )\ _{3}^{\shortmid }\Lambda
(\tau )\ _{4}^{\shortmid }\Lambda (\tau )|}}\ _{\eta }^{\shortmid }\mathcal{%
\mathring{V}}(\tau ),  \label{thvcannpd} \\
\ _{s}^{\shortmid }\widehat{\mathcal{Z}}_{\kappa }^{\star }(\tau ) &=&\exp %
\left[ \int\nolimits_{\tau ^{\prime }}^{\tau }\frac{d\tau }{(2\pi \tau )^{4}}%
\frac{1}{\sqrt{|\ _{1}\Lambda (\tau )\ _{2}\Lambda (\tau )\ _{3}^{\shortmid
}\Lambda (\tau )\ _{4}^{\shortmid }\Lambda (\tau )|}}\ _{\eta }^{\shortmid }%
\mathcal{\mathring{V}}(\tau )\right] ,  \notag \\
\ _{s}^{\shortmid }\widehat{\mathcal{E}}_{\kappa }^{\star }(\tau )
&=&-\int\nolimits_{\tau ^{\prime }}^{\tau }\frac{d\tau }{(4\pi )^{4}\tau ^{3}%
}\frac{\tau \lbrack \ _{1}\Lambda (\tau )+\ _{2}\Lambda (\tau )+\
_{3}^{\shortmid }\Lambda (\tau )+\ _{4}^{\shortmid }\Lambda (\tau )]-4}{%
\sqrt{|\ _{1}\Lambda (\tau )\ _{2}\Lambda (\tau )\ _{3}^{\shortmid }\Lambda
(\tau )\ _{4}^{\shortmid }\Lambda (\tau )|}}\ _{\eta }^{\shortmid }\mathcal{%
\mathring{V}}(\tau ),  \notag \\
\ _{s}^{\shortmid }\widehat{\mathcal{S}}_{\kappa }^{\star }(\tau )
&=&-\int\nolimits_{\tau ^{\prime }}^{\tau }\frac{d\tau }{(4\pi \tau )^{4}}%
\frac{\tau \lbrack \ _{1}\Lambda (\tau )+\ _{2}\Lambda (\tau )+\
_{3}^{\shortmid }\Lambda (\tau )+\ _{4}^{\shortmid }\Lambda (\tau )]-8}{%
\sqrt{|\ _{1}\Lambda (\tau )\ _{2}\Lambda (\tau )\ _{3}^{\shortmid }\Lambda
(\tau )\ _{4}^{\shortmid }\Lambda (\tau )|}}\ _{\eta }^{\shortmid }\mathcal{%
\mathring{V}}(\tau ).  \notag
\end{eqnarray}%
In these formulas, we use the running phase space volume functional 
\begin{equation}
\ _{\eta }^{\shortmid }\mathcal{\mathring{V}}(\tau )=\int_{\ _{s}^{\shortmid
}\widehat{\Xi }}\ ^{\shortmid }\delta \ _{\eta }^{\shortmid }\mathcal{V}(\
_{s}^{\shortmid }\Im (\tau ),\ ~^{\shortmid }\mathring{g}_{\alpha _{s}})
\label{volumfpsp}
\end{equation}%
determined by prescribed classes of generating $\eta $-functions, effective
generating sources $\ _{s}^{\shortmid }\Im (\tau ),$ coefficients of a prime
s-metric $\ ^{\shortmid }\mathring{g}_{\alpha _{s}}$ and nonholonomic
distributions defining the hyper-surface $\ _{s}^{\shortmid }\widehat{\Xi }.$

We can define the effective volume functionals (\ref{volumfuncts}) and
geometric thermodynamic variables (\ref{thvcannpd}) for further parametric
decompositions with $\kappa $-linear approximations (\ref{genf1}) and $\chi $%
-polarizations and find parametric formulas for $\tau $-flows and
nonassociative R-flux deformations of prime metrics, 
\begin{eqnarray}
\ ^{\shortmid }\delta \ ^{\shortmid }\mathcal{V} &=&\ ^{\shortmid }\delta \
^{\shortmid }\mathcal{V}_{0}\mathcal{[}\tau ,\ _{s}^{\shortmid }\Lambda
(\tau ),\ \ _{s}^{\shortmid }\Im (\tau );\psi (\tau ),\ \mathring{g}%
_{i_{1}},\ \mathring{g}_{a_{2}},~^{\shortmid }\mathring{g}%
^{a_{3}},~^{\shortmid }\mathring{g}^{a_{4}};\zeta _{4}(\tau ),~^{\shortmid
}\zeta ^{6}(\tau ),~^{\shortmid }\zeta ^{8}(\tau )]+  \label{volumfd} \\
&&\kappa \ ^{\shortmid }\delta \ ^{\shortmid }\mathcal{V}_{1}[\tau ,\
_{s}^{\shortmid }\Lambda (\tau ),\ \ _{s}^{\shortmid }\Im (\tau );\psi (\tau
),\ \mathring{g}_{i_{1}},\ \mathring{g}_{a_{2}},~^{\shortmid }\mathring{g}%
^{a_{3}},~^{\shortmid }\mathring{g}^{a_{4}};\zeta _{4}(\tau ),~^{\shortmid
}\zeta ^{6}(\tau ),~^{\shortmid }\zeta ^{8}(\tau ),\chi _{4}(\tau
),~^{\shortmid }\chi ^{6}(\tau ),~^{\shortmid }\chi ^{8}(\tau )].  \notag
\end{eqnarray}%
Introducing a quasi-stationary parametric solution in (\ref{volumfd}) for
nonassociative R. Hamilton equations, we can compute corresponding $\kappa $%
-decompositions of the thermodynamic variables (\ref{thvcann}), 
\begin{equation}
\ _{s}^{\shortmid }\widehat{\mathcal{W}}_{\kappa }^{\star }(\tau )=\
_{s}^{\shortmid }\widehat{\mathcal{W}}_{0}+\kappa \ _{s}^{\shortmid }%
\widehat{\mathcal{W}}_{1}^{\star }(\tau ),\ _{s}^{\shortmid }\widehat{%
\mathcal{Z}}_{\kappa }^{\star }(\tau )=\ _{s}^{\shortmid }\widehat{\mathcal{Z%
}}_{0}\ _{s}^{\shortmid }\widehat{\mathcal{Z}}_{1}^{\star }(\tau ),\
_{s}^{\shortmid }\widehat{\mathcal{E}}_{\kappa }^{\star }(\tau )=\
_{s}^{\shortmid }\widehat{\mathcal{E}}_{0}+\kappa \ _{s}^{\shortmid }%
\widehat{\mathcal{E}}_{1}^{\star }(\tau ),\ _{s}^{\shortmid }\widehat{%
\mathcal{S}}_{\kappa }^{\star }(\tau )=\ _{s}^{\shortmid }\widehat{\mathcal{S%
}}_{0}+\kappa \ _{s}^{\shortmid }\widehat{\mathcal{S}}_{1}^{\star }(\tau ).
\label{klinthvar}
\end{equation}%
In this work, there are not presented cumbersome computations and
incremental formulas with $\kappa $--linear decomposition for $\
^{\shortmid}\delta \ ^{\shortmid }\mathcal{V}=\ ^{\shortmid }\delta \
^{\shortmid }\mathcal{V}_{0}+ \kappa \ ^{\shortmid }\delta \ ^{\shortmid }%
\mathcal{V}_{1}$ (\ref{volumfd}) and (\ref{klinthvar}) (considered for
solutions of type (\ref{2rotsol}), with $\chi $-polarization functions). We
consider a more general and compact approach when the thermodynamic
variables (\ref{thvcann}) are computed for volume forms (\ref{volumf}) using
quasi-stationary solutions of type (\ref{sol2rf}), (\ref{sol3rf}), or (\ref%
{offdiagpolfr}).

\subsection{Geometric evolution of nonassociative double Reisner-Nordstr\"{o}%
m-(A)dS BHs in phase spaces}

\label{ssgfevdbh}In a series of recent works \cite%
{kehagias19,biasio20,biasio21,lueben21,biasio22}, certain models of
geometric flows of the Schwarzschild-AdS, RN and other type metrics were
studied in connection to the swampland program \cite%
{vaf05,oog06,pal19,gomez19}. Those papers are devoted to associative and
commutative geometric and physical theories with solutions which can be
characterized by the Bekenstein-Hawking entropy. In this subsection, we
consider nonassociative generalizations and nonholonomic geometric flow
deformations of the 4-d RN-dS metrics dubbed both on the base spacetime and
typical cofiber and star-deformed by a 8-d phase spaces evolution. We cite 
\cite{reis16,nord18}, for fundamental results on RN BHs (see also monographs 
\cite{misner,hawking73,wald82,kramer03}), and \cite{cao08,konoplya14} and
references therein, on higher dimension extensions for RN-(A)dS.

\vskip5pt The geometric thermodynamic variables (\ref{nabhthermod}) and (\ref%
{hbbhtherman}) can be used for constructing an effective thermodynamic model
for quasi-stationary evolution in a conventional phase space media of double
BHs into respective BE configurations. Such constructions for nonassociative
geometric flow and/or off-diagonal deformation scenarios can be performed
for very special classes of nonholonomic constraints, when the existence of
corresponding horizon hyper-surfaces allows us to describe the $\tau$%
-evolution of such physical objects in the framework of generalized
Bekenstein-Hawking thermodynamics. For explicit models non involving
prescribed hyper-horizons and/or duality conditions for Ricci solitons
and/or geometric flows, even we work with quasi-stationary s-metrics, the
concept of Hawking entropy is not applicable and we have to elaborate on
other types of statistical and geometric thermodynamic theories. In section
4 of \cite{partner04}, we concluded that in general form we have to change
the paradigm and characterize nonassociative geometric flow and
gravitational theories (and various classes of related physically important
solutions) in the framework of modified G. Perelman thermodynamics \cite%
{perelman1}.

\subsubsection{Prime metrics for phase space double RN-dS BHs}

Let us consider a 4-d base Lorentz spacetime manifold $V$ on which the
Einstein-Maxwell theory is defined by the action for a metric $g_{\alpha
\beta}$ and electromagnetic field $A_{\mu },$%
\begin{equation}
S=\int\nolimits_{V}d^{4}x\sqrt{|g|}[\frac{1}{16\pi G_{4}}(R-2\check{\Lambda}%
)]-\frac{1}{4e_{0}^{2}}F_{\mu \nu }F^{\mu \nu }],  \label{emt4}
\end{equation}%
where $e_{0}$ is the electromagnetic constant, $F_{\mu \nu }$ is the
anti-symmetric strength tensor of $A_{\mu };$ $G_{4}$ is the 4-d
gravitational constant, and $\check{\Lambda}>0$ is the de Sitter, dS,
cosmological constant. In this theory, a Reisner-Nordstr\"{o}m, RN, BH is
constructed as a spherically symmetric and static solution of corresponding
gravitational and electromagnetic field equations in GR with zero
cosmological constant. The corresponding quadratic line element for such a
RN-dS solution with positive cosmological constant, describing an
electrically charged BH in an asymptotic dS spacetime, can be parameterized
as a prime spacetime metric 
\begin{eqnarray}
d\ ^{\flat }s_{[4d]}^{2} &=&\ ^{\flat }g_{1}(r)dr^{2}+\ ^{\flat
}g_{2}(r)d\theta ^{2}+\ ^{\flat }g_{3}(r,\theta )d\varphi ^{2}+\ ^{\flat
}g_{4}(r)dt^{2}\mbox{ and }A_{\mu }=(Q/r,0,0,0),\mbox{ for }  \notag \\
\ \ ^{\flat }g_{4}(r) &=&-(1-\frac{r_{s}}{r}+\frac{r_{Q}^{2}}{r^{2}}-\frac{%
r^{2}}{r_{\check{\Lambda}}^{2}})=-[\ ^{\flat }g_{1}(r)]^{-1},\ \ ^{\flat
}g_{2}(r)=r^{2},\ \ ^{\flat }g_{3}(r,\theta )=r^{2}\sin ^{2}\theta ,
\label{pmrnds}
\end{eqnarray}%
where the constant velocity of light is stated $c=1;$ the Schwarzschild
radius $r_{s}=2Gm$ is determined by the BH mass; the characteristic electric
length $r_{Q}^{2}=Q^{2}G/4\pi e_{0}^{2};$ and $r_{\check{\Lambda}}^{2}=3/%
\check{\Lambda}.$ The local spherical coordinates are parameterized $%
x^{1}=r,x^{2}=\theta ,y^{3}=\varphi $ and $y^{4}=t,$ with $d\Omega
_{2}^{2}=d\theta ^{2}+ \sin ^{2}\theta d\varphi ^{2}$ being the metric on
the unity 2-d sphere, when the conditions for the causal horizon $r=r_{h}$
are stated for quadric polynomial $\ ^{\flat }g_{4}(r_{h})=0.$

In a 8-d phase space $\ _{s}\mathcal{M},$ we use double 4-d local spherical
coordinates, both on base spacetime manifold $V$ (as in $d\
^{\flat}s_{[4d]}^{2}$ (\ref{pmrnds})) and 4-d spherical momentum type
coordinates $p_{1}=p_{r}, p_{2}=p_{\theta }, p_{3}=p_{\varphi }$ and $%
p_{4}=E $; and consider the quadratic linear element 
\begin{eqnarray}
d\ _{\shortmid }^{\flat }s_{[8d]}^{2} &=&d\ ^{\flat }s_{[4d]}^{2}+d\ \
_{\shortmid }^{\flat }s^{2},  \label{pmrndsf} \\
&& \mbox{ with } d\ _{\shortmid }^{\flat }s^{2}=\ _{\shortmid }^{\flat
}g^{5}(p)dp^{2}+\ _{\shortmid }^{\flat }g^{6}(p)dp_{\theta }{}^{2}+\
_{\shortmid }^{\flat }g^{7}(p,p_{\theta })dp_{\varphi }^{2}+\ _{\shortmid
}^{\flat }g^{8}(p)dE^{2}, \mbox{ for }  \notag \\
&&\ _{\shortmid }^{\flat }g^{8}(p)=-(1-\frac{p_{s}}{p}+\frac{p_{Q}^{2}}{p^{2}%
}-\frac{p^{2}}{p_{\ ^{\shortmid }\check{\Lambda}}^{2}})=-[\
_{\shortmid}^{\flat }g^{5}(p)]^{-1}, \ _{\shortmid }^{\flat }\mathring{g}%
^{6}(p)=p^{2},\ _{\shortmid }^{\flat }\mathring{g}^{7}(p,p_{\theta })=
p^{2}\sin ^{2}p_{\theta }.  \notag
\end{eqnarray}%
In such formulas, the dimension for $\ p=\sqrt{%
(p_{1})^{2}+(p_{2})^{2}+(p_{3})}$ is stated (via multiplication on a
constant parameter, or working in natural units with $G=c=\hbar =1$) to be
the same as for $r=\sqrt{(x^{1})^{2}+(x^{2})^{2}+(x^{3})}$. There are also
considered some conventional "horizontal", $\check{\Lambda},$ and
"covertical", $\ ^{\shortmid }\check{\Lambda},$ cosmological constants. To
define a double 4-d BH configuration on $\ _{s}\mathcal{M}$ we can introduce
in the typical co-fiber space a conventional Schwarzschild radius $p_{s}=2\
^{\shortmid }G\ ^{\shortmid }m$ is determined by the co-fiber BH mass $\
^{\shortmid }m$ with a conventional (it can be different from $G$)
gravitational constant $\ ^{\shortmid }G$; the characteristic electric
length $p_{Q}^{2}=\ ^{\shortmid }Q^{2}\ ^{\shortmid }G/4\pi \
^{\shortmid}e_{0}^{2}$ with conventional electric charge in co-fiber, $\
^{\shortmid}e_{0};$ and $p_{\check{\Lambda}}^{2}=3/\ ^{\shortmid }\check{%
\Lambda}.$

For arbitrary frame/coordinate transforms, a diagonal phase space s-metric $%
\ _{\shortmid }^{\flat }g_{\alpha }=(\ ^{\flat }g_{i},\
_{\shortmid}^{\flat}g^{a})$ (\ref{pmrndsf}) can be written in off-diagonal
form $\ _{\shortmid}^{\flat }g_{\alpha \beta }(\ ^{\shortmid }u)$ with a
prime shell structure $\ _{\shortmid }^{\flat }\mathbf{g}$ adapted to a
prime and trivial N-connection splitting $\ _{\shortmid }^{\flat }\mathbf{N}$%
. In general, such data $(\ _{\shortmid }^{\flat }\mathbf{g,\ _{\shortmid
}^{\flat }N})$ do not define a solution of vacuum (non)
associative/commutative gravitational equations even their horizontal/vacuum
components, for instance, (\ref{pmrnds}) can determine certain 4-d
electro-vacuum or RNdS BH configurations.

\subsubsection{Nonassociative geometric $\protect\kappa $-linear evolution
of double phase space BH configurations}

In this subsection, we construct $\kappa $-linear parametric solutions of
nonassociative geometric flow equations (\ref{nonassocgeomflefcc})
describing the evolution of a double BH phase space metric $\
_{\shortmid}^{\flat }g_{\alpha }=(\ ^{\flat }g_{i},\ _{\shortmid
}^{\flat}g^{a})$ (\ref{pmrndsf}). In a trivial s-adapted form (with
N-connection coefficients such way defined by some coordinate transforms
when the solutions do not contain coordinate singularities), such a primary
s-metric can be parameterized by corresponding s-coefficients $\ _{\shortmid
s}^{\flat }\mathbf{g}=(\ _{s}^{\flat }g,\ _{\shortmid s}^{\flat}g)= \{\
_{\shortmid }^{\flat }g_{\alpha _{s}}=(\ ^{\flat }g_{i_{1}}, \
^{\flat}g_{a_{2}},\ _{\shortmid }^{\flat }g^{a_{3}},)\}.$ Using nonlinear
symmetries (\ref{nonlinsymr}) and respective nonlinear transforms (\ref%
{nonlinsymrex}), we can re-define the generating functions to define
nonassociative R-flux deformations of such prime BH s-metrics into $\tau $%
-families of target quasi-stationary ones, $\ _{\shortmid s}^{\flat }\mathbf{%
g} \rightarrow \ _{s}^{\shortmid }\mathbf{g}(\tau )$ (\ref{sol2rf}) with
N-connection coefficients $\ ^{\shortmid }N_{i_{s-1}}^{a_{s}}(\tau )$ in
nonholonomic s-frames (\ref{sol2nrf}).

We can express the parametric solutions for nonassociative $\kappa $-linear
geometric flow deformations of type $\ _{\shortmid s}^{\flat }\mathbf{g}%
\rightarrow \ _{s}^{\shortmid }\mathbf{g}(\tau )$ (\ref{sol2rf}) in terms of
gravitational $\eta $-polarizations and generating functions (\ref{etapolgen}%
) using phase space local coordinates as for the prime s-metric (\ref%
{pmrndsf}). The corresponding class of quasi-stationary $\tau $-running
s-metrics are defined by quadratic linear elements of type (\ref%
{offdiagpolfr}), with $\mathring{g}_{4}(\tau)\rightarrow \ ^{\flat }g_{4},\
^{\shortmid }\mathring{g}^{6}(\tau)\rightarrow \ _{\shortmid }^{\flat
}g^{6}, \ ^{\shortmid }\mathring{g}^{8}(\tau )\rightarrow \ _{\shortmid
}^{\flat }g^{8},$ when the primary s-metrics do not depend on $\tau .$ To
avoid singular coordinate evolution scenarios for a necessary $\tau $%
-interval we can prescribe a primary s-metric written in respective
coordinates and further frame transforms to define certain well-defined data%
\begin{equation*}
\left( \ _{\shortmid s}^{\flat }\mathbf{g;}\ _{\shortmid s}^{\flat }\mathbf{N%
}\right) = (\ _{s}^{\flat }g,\ _{\shortmid s}^{\flat }g; \
_{\shortmid}^{\flat }N_{i_{s-1}}^{a_{s}})=\{\ _{\shortmid }^{\flat
}g_{\alpha _{s}}= (\ ^{\flat }g_{i_{1}},\ ^{\flat }g_{a_{2}},\ _{\shortmid
}^{\flat }g^{a_{3}};\ \ ^{\flat }N_{i_{1}}^{a_{2}},\ \ _{\shortmid }^{\flat
}N_{i_{2}a_{3}},\ _{\shortmid }^{\flat }N_{i_{3}a_{4}})\}.
\end{equation*}%
In $\eta $--polarized nonsymmetric $\kappa $-linear $\tau $-evolving
quasi-stationary phase space backgrounds, double RNdS BH configurations are
described by quadratic elements: 
\begin{eqnarray}
d\ \ ^{\shortmid }\widehat{s}^{2}(\tau ) &=& \ ^{\shortmid }g_{\alpha
_{s}\beta _{s}}(\hbar ,\kappa ,\tau ,x^{k},y^{3},p_{a_{3}},p_{a_{4}};\
_{\shortmid }^{\flat }g_{\alpha _{s}}(\tau );\eta _{4}(\tau ),\ ^{\shortmid
}\eta ^{6}(\tau ),\ ^{\shortmid }\eta ^{8}(\tau ),\ _{s}^{\shortmid }\Lambda
(\tau );\ _{s}^{\shortmid }\Im (\tau ))d~\ ^{\shortmid }u^{\alpha _{s}}d~\
^{\shortmid }u^{\beta _{s}}  \notag \\
&=&e^{\psi (\tau )}[(dx^{1})^{2}+(dx^{2})^{2}]-  \label{dbhpspol}
\end{eqnarray}%
\begin{eqnarray*}
&&\frac{[\partial _{3}(\eta _{4}(\tau )\ ^{\flat }g_{4})]^{2}}{|\int
dy^{3}~_{2}\Im (\tau )\partial _{3}(\eta _{4}(\tau )\ \ ^{\flat }g_{4})|\
(\eta _{4}(\tau )\ ^{\flat }g_{4})}\{dy^{3}+\frac{\partial _{i_{1}}[\int
dy^{3}\ _{2}\Im (\tau )\ \partial _{3}(\eta _{4}(\tau )\ ^{\flat }g_{4})]}{%
~_{2}\Im (\tau )\partial _{3}(\eta _{4}(\tau )\ ^{\flat }g_{4})}%
dx^{i_{1}}\}^{2}+ \\
&&\eta _{4}(\tau )\ ^{\flat }g_{4})\{dt+[\ _{1}n_{k_{1}}(\tau )+\
_{2}n_{k_{1}}(\tau )\int dy^{3}\frac{[\partial _{3}(\eta _{4}(\tau )\
^{\flat }g_{4})]^{2}}{|\int dy^{3}\ _{2}\Im (\tau )\partial _{3}(\eta
_{4}(\tau )\ ^{\flat }g_{4})|\ (\eta _{4}(\tau )\ ^{\flat }g_{4})^{5/2}}%
]dx^{k_{1}}\}
\end{eqnarray*}%
\begin{eqnarray*}
&&-\frac{[\ ^{\shortmid }\partial ^{5}(\ ^{\shortmid }\eta ^{6}(\tau )\
_{\shortmid }^{\flat }g^{6})]^{2}}{|\int dp_{5}~_{3}^{\shortmid }\Im (\tau
)\ \ ^{\shortmid }\partial ^{5}(\ ^{\shortmid }\eta ^{6}(\tau )\ \
^{\shortmid }\mathring{g}^{6}(\tau ))\ |\ (\ ^{\shortmid }\eta ^{6}(\tau )\
\ _{\shortmid }^{\flat }g^{6})}\{dp_{5}+\frac{\ \ ^{\shortmid }\partial
_{i_{2}}[\int dp_{5}\ _{3}^{\shortmid }\Im (\tau )\ ^{\shortmid }\partial
^{5}(\ ^{\shortmid }\eta ^{6}(\tau )\ \ _{\shortmid }^{\flat }g^{6})]}{%
~_{3}^{\shortmid }\Im (\tau )\ ^{\shortmid }\partial ^{5}(\ ^{\shortmid
}\eta ^{6}(\tau )\ \ _{\shortmid }^{\flat }g^{6})}dx^{i_{2}}\}^{2} \\
&&+(\ ^{\shortmid }\eta ^{6}(\tau )\ \ _{\shortmid
}^{\flat}g^{6})\{dp_{6}+[\ _{1}^{\shortmid }n_{k_{2}}(\tau )+ \
_{2}^{\shortmid}n_{k_{2}}(\tau )\int dp_{5}\frac{[\ ^{\shortmid }\partial
^{5}(\ ^{\shortmid }\eta ^{6}(\tau ) \ _{\shortmid }^{\flat }g^{6})]^{2}}{%
|\int dp_{5} \ _{3}^{\shortmid }\Im (\tau) \partial ^{5}(\ ^{\shortmid }\eta
^{6}(\tau ) \ _{\shortmid }^{\flat }g^{6})|\ (\ ^{\shortmid }\eta
^{6}(\tau)\ _{\shortmid }^{\flat }g^{6})^{5/2}}]dx^{k_{2}}\}-
\end{eqnarray*}%
\begin{eqnarray*}
&&\frac{[\ ^{\shortmid }\partial ^{7}(\ ^{\shortmid }\eta ^{8}(\tau )\
_{\shortmid }^{\flat }g^{8})]^{2}} {|\int dp_{7}\ _{4}^{\shortmid }\Im
(\tau)\ ^{\shortmid }\partial ^{7}(\ ^{\shortmid }\eta ^{8}(\tau ) \
_{\shortmid}^{\flat }g^{8})\ |\ (\ ^{\shortmid }\eta ^{8}(\tau )\
_{\shortmid}^{\flat }g^{8})}\{dp_{7}+ \frac{\ ^{\shortmid }\partial
_{i_{3}}[\int dp_{7} \ _{4}^{\shortmid }\Im (\tau )\ ^{\shortmid }\partial
^{7}(\ ^{\shortmid }\eta ^{8}(\tau ) \ _{\shortmid }^{\flat }g^{8})]}{\
_{4}^{\shortmid }\Im (\tau )\ ^{\shortmid }\partial ^{7}(\ ^{\shortmid}\eta
^{8}(\tau ) \ _{\shortmid }^{\flat }g^{8})}d\ ^{\shortmid}x^{i_{3}}\}^{2} +
\\
&&(\ ^{\shortmid }\eta ^{8}(\tau )\ _{\shortmid }^{\flat }g^{8})\{dE+[\
_{1}n_{k_{3}}(\tau )+ \ _{2}n_{k_{3}}(\tau )\int dp_{7}\frac{[\
^{\shortmid}\partial ^{7}(\ ^{\shortmid }\eta ^{8}(\tau )\ _{\shortmid
}^{\flat }g^{8})]^{2}}{|\int dp_{7}\ _{4}^{\shortmid }\Im (\tau )[\
^{\shortmid}\partial ^{7}(\ ^{\shortmid }\eta ^{8}(\tau )\ _{\shortmid
}^{\flat }g^{8})]|\ [(\ ^{\shortmid }\eta ^{8}(\tau )\ _{\shortmid
}^{\flat}g^{8})]^{5/2}}]d\ ^{\shortmid }x^{k_{3}}\}.
\end{eqnarray*}%
The integration functions in (\ref{dbhpspol}) are of type (\ref{gensol1a})
but extended to $\tau $-dependencies and written for coordinates used in (%
\ref{pmrndsf}), 
\begin{eqnarray*}
&&g_{4}^{[0]}(\hbar ,\kappa ,\tau ,r,\theta ),\ _{1}n_{k_{1}}(\hbar ,\kappa
,\tau ,r,\theta ),\ _{2}n_{k_{1}}(\hbar ,\kappa ,\tau ,r,\theta ); \\
&&\ ^{\shortmid }g_{[0]}^{5}(\hbar ,\kappa ,\tau ,r,\theta ,\varphi ,p),\
_{1}n_{k_{2}}(\hbar ,\kappa ,\tau ,r,\theta ,\varphi ,p),\
_{2}n_{k_{2}}(\hbar ,\kappa ,\tau ,r,\theta ,\varphi ,p); \\
&&\ ^{\shortmid }g_{[0]}^{7}(\hbar ,\kappa ,\tau ,r,\theta ,\varphi
,p,p_{\varphi }),\ _{1}^{\shortmid }n_{k_{3}}(\hbar ,\kappa ,\tau
,p,p_{\varphi }),\ _{2}^{\shortmid }n_{k_{3}}(\hbar ,\kappa ,\tau
,p,p_{\varphi }).
\end{eqnarray*}%
We can consider additional conditions when such generic off-diagonal
gravitational interactions and nonassociative geometric evolution flows
transform a prime s-metric for double conventional 4-d electro-vacuum or
RNdS BHs into vacuum quasi-stationary configurations with $\tau $-evolution,
when the electromagnetic interactions are "dissipated" into a nonholonomic
vacuum gravitational structure encoding star product R-flux data.

\subsubsection{Computing the Bekenstein-Hawking entropy for double phase
space BE configurations}

\label{ss523}We study two classes of nonassociative couples BE/BH s-metrics
which are characterized by phase space Bekenstein-Hawking type thermodynamic
models.

\paragraph{Example 1:\ Nonassociative $\protect\tau $-deformed double RNdS
BHs with dissipation into BEs and Schwar\-zschild BHs \newline
}

Any solution (\ref{dbhpspol}) can be decomposed in terms of $\chi $%
--generating functions as for the quadratic line element (\ref%
{epsilongenfdecomp}) if the prime s-metric coefficients $\ _{\shortmid
s}^{\flat }\mathbf{g}$ are used for a prime metric $\ _{s}^{\shortmid }%
\mathbf{\mathring{g}}$. In such cases, the generating and integration
functions are written in $\kappa $--linearized form (\ref{offdncelepsilon}), 
\begin{eqnarray*}
\psi (\tau ) &\simeq &\psi (\hbar ,\kappa ;\tau ,r,\theta )\simeq \psi
_{0}(\hbar ,\tau ,r,\theta ) (1+\kappa \ _{\psi }\chi (\hbar ,\tau
,r,\theta)),\mbox{ for } \\
\ \eta _{2}(\tau ) &\simeq &\eta _{2}(\hbar ,\kappa ;\tau ,r,\theta )\simeq
\zeta _{2}(\hbar ,\tau ,r,\theta ) (1+\kappa \chi _{2}(\hbar ,\tau
,r,\theta)),\mbox{ we can consider }\ \eta _{2}(\tau )=\ \eta _{1}(\tau ); \\
\eta _{4}(\tau ) &\simeq &\ \eta _{4}(\hbar ,\kappa ;\tau ,r, \theta,\varphi
)\simeq \zeta _{4}(\hbar ,\tau ,r,\theta ,\varphi ) (1+\kappa \ \chi
_{4}(\hbar ,\tau ,r,\theta ,\varphi )), \\
\ ^{\shortmid }\eta ^{6}(\tau ) &\simeq & \ ^{\shortmid }\eta ^{6}(\hbar,
\kappa ; \tau ,r,\theta ,\varphi ,p) \simeq \ ^{\shortmid }\zeta ^{6}(\hbar
,\kappa ;\tau ,r,\theta ,\varphi ,p) (1+\kappa \ ^{\shortmid }\chi
^{6}(\hbar ,\kappa ;\tau ,r,\theta ,\varphi ,p)), \\
\ ^{\shortmid }\eta ^{8}(\tau ) &\simeq & \ ^{\shortmid }\eta ^{8}(\hbar
,\kappa ;\tau ,r,\theta ,\varphi ,p,p_{\varphi }) \simeq \ ^{\shortmid
}\zeta ^{8}(\hbar ,\kappa ;\tau ,r,\theta ,\varphi ,p,p_{\varphi })
(1+\kappa \ ^{\shortmid }\chi ^{8}(\hbar ,\kappa ;\tau ,r,\theta ,\varphi
,p,p_{\varphi})).
\end{eqnarray*}%
We may construct $\tau $-families of quasi-stationary solutions (\ref%
{epsilongenfdecomp} with conventional horizons when the $\chi $%
-polarizations satisfy the conditions 
\begin{eqnarray}
\zeta _{4}(1+\kappa \ \chi _{4})\ ^{\flat }g_{4}(r) &=& \zeta _{4}(1- \frac{%
r_{s}}{r} + \frac{r_{Q}^{2}}{r^{2}}-\frac{r^{2}}{r_{\check{\Lambda}}^{2}}+
\kappa \ \chi _{4}) = \tilde{\zeta}_{4}(1-\frac{r_{s}}{r}+\kappa \ \tilde{%
\chi}_{4})=0,  \label{auxalgeq} \\
\ ^{\shortmid }\zeta ^{8}(1+\ ^{p}\kappa \ ^{\shortmid }\chi ^{8})\
_{\shortmid }^{\flat }g^{8}(p) &=&\ ^{\shortmid }\zeta ^{8}(1-\frac{p_{s}}{p}%
+\frac{p_{Q}^{2}}{p^{2}}-\frac{p^{2}}{p_{\ ^{\shortmid }\check{\Lambda}}^{2}}%
+\kappa \ ^{\shortmid }\chi ^{8})=\ ^{\shortmid }\tilde{\zeta}^{8}(1-\frac{%
p_{s}}{p}+\kappa \ ^{\shortmid }\tilde{\chi}^{8})=0,  \notag
\end{eqnarray}%
for non-zero $\chi _{4}(\tau )$ and $\ ^{\shortmid }\chi ^{8}(\tau); \
^{\shortmid }\zeta ^{6}(\tau )=1$ and $\ \zeta _{4}(\tau )\simeq 1,$ $\
^{\shortmid }\zeta ^{8}(\tau )\simeq 1$ and re-definition of the generating
data, 
\begin{eqnarray*}
\left( \chi _{4}(\tau ),\ \zeta _{4}(\tau );\ ^{\shortmid }\chi ^{8}(\tau),\
^{\shortmid }\zeta ^{8}(\tau )\right) &\rightarrow &(\tilde{\chi}_{4}(\tau
)=\ \zeta _{4}(\tau )\chi _{4}(\tau ),\tilde{\zeta}_{4}(\tau )= (1+\frac{%
r_{Q}^{2}r_{\check{\Lambda}}^{2}-r^{4}}{r(r-r_{s})r_{\check{\Lambda}}^{2}})\
\zeta _{4}(\tau ); \\
&& \ ^{\shortmid }\tilde{\chi}^{8}(\tau ) = \ ^{\shortmid }\tilde{\zeta}%
^{8}(\tau )\ ^{\shortmid }\chi ^{8}(\tau ), \ ^{\shortmid }\tilde{\zeta}%
^{8}(\tau )=(1+\frac{r_{Q}^{2}r_{\check{\Lambda}}^{2}-r^{4}}{r(r-r_{s})r_{%
\check{\Lambda}}^{2}})\ ^{\shortmid }\zeta ^{8}(\tau )).
\end{eqnarray*}%
Viable physical models with off-diagonal solutions are generated if the
integration functions for N-coefficients are chosen to tend, for instance,
to zero for $r\rightarrow \infty $ and $p\rightarrow \infty .$

Geometric evolution models of two black ellipsoid, BE, phase space
configurations are defined if we prescribe such generating functions: 
\begin{eqnarray}
\ \tilde{\chi}_{4}(\tau ) &=&\ ^{e}\chi _{4}(\tau ,r,\theta ,\varphi )=2 
\underline{\chi }(\tau ,r,\theta )\sin (\omega _{0}\varphi +\varphi _{0});
\label{delipgenf} \\
\ ^{\shortmid }\tilde{\chi}^{6}(\tau ) &=&\ ^{\shortmid }\chi ^{6}(\tau )=0, %
\mbox{ for } \ ^{\shortmid }\tilde{\zeta}^{6}(\tau )=\ ^{\shortmid }\zeta
^{6}(\tau )=1;  \notag \\
\ \ ^{\shortmid }\tilde{\chi}^{8}(\tau ) &=&\ \ ^{e}\chi (\tau ,p,p_{\theta
},p_{\varphi })=2\overline{\chi }(\tau ,p,p_{\theta })\sin (\ _{\shortmid
}^{p}\omega _{0}\ p_{\varphi }+\ p_{\varphi }^{0}),  \notag
\end{eqnarray}%
where $\underline{\chi }(\tau ,r,\theta )$ and $\overline{\chi }%
(\tau,p,p_{\theta })$ are smooth functions (or $\tau $-running constants);
the smooth $\zeta $-functions can be approximated to unity, and $(\omega
_{0},\varphi _{0})$ and $(\ _{\shortmid }^{p}\omega _{0},\ p_{\varphi }^{0})$
are couples of constants. To define all possible horizons we have to solve
the system of two independent forth order algebraic equations for $r$ and $p$
and stated by gravitational polarizations $\chi _{4}(\tau )$ and $\
^{\shortmid }\chi ^{8}(\tau )$ (we omit such technical details). For this
subclass of quasi-stationary phase space solutions, we can consider small
parametric deformations and regions when 
\begin{equation}
r(\tau )\approx \ r_{s}/(1-\kappa \tilde{\chi}_{4}(\tau ,\varphi )) 
\mbox{
and }\ ^{p}r(\tau )\approx p_{s}/ (1-\kappa \ ^{\shortmid }\tilde{\chi}%
^{8}(\tau ,p_{\varphi })).  \label{param2be}
\end{equation}%
This describes a scenarios when phase space two RNdS BHs evolve under
nonassociative geometric flows into BE deformations of certain base
spacetime and co-fiber Schwarzschild solutions.

For prescribed gravitational $\chi $-polarizations (\ref{delipgenf}), the
parametric formulas (\ref{param2be}) define for rotoid configurations
running on $\tau .$ Corresponding nonholonomic structures can be chosen to
define certain nonassociative geometric evolution of more general black
ellipsoid, BE, configurations. The corresponding parametric formulas for
respective BE horizons are defined by small gravitational polarizations
determined by nonassociative star product R-flux deformations. In the limits
of zero eccentricity $\kappa ,$ such double BE configurations transform into
prime double BH ones.

Putting together above formulas, we construct two BE target phase space
quadratic linear element with $\tau $--evolution, 
\begin{eqnarray}
d\ _{\shortmid }^{\flat }s_{[8d]}^{2}(\tau ) &=& e^{\psi _{0}}(1+\kappa \
^{\psi (\tau )}\ ^{\shortmid }\chi (\tau )) [\ ^{\flat }g_{1}dr^{2}+\
^{\flat}g_{2}d\theta ^{2}]  \label{2rotsol} \\
&&-\{\frac{4[\partial _{3}(|\zeta _{4}(\tau )\ ^{\flat }g_{4}|^{1/2})]^{2}}{%
\ ^{\flat }g_{4}|\int dy^{3}\{\ _{2}\Im (\tau )\partial _{3}(\zeta _{4}(\tau
)\ ^{\flat }g_{4})\}|}-\kappa \lbrack \frac{\partial _{3}(\chi _{4}(\tau
)|\zeta _{4}(\tau )\ ^{\flat }g_{4}|^{1/2})}{4\partial _{3}(|\zeta _{4}(\tau
)\ ^{\flat }g_{4}|^{1/2})}  \notag \\
&&-\frac{\int dy^{3}\{\ _{2}\Im (\tau )\partial _{3}[(\zeta _{4}(\tau )\
^{\flat }g_{4})\chi _{4}(\tau )]\}}{\int dy^{3}\{\ _{2}\Im (\tau )\partial
_{3}(\zeta _{4}(\tau )\ ^{\flat }g_{4})\}}]\}\ ^{\flat }g_{3}(\mathbf{e}%
^{3}(\tau ))^{2}  \notag
\end{eqnarray}
\begin{eqnarray*}
&&+\ \tilde{\zeta}_{4}(\tau )(1-\frac{r_{s}}{r}+\kappa \ \tilde{\chi}%
_{4}(\tau ))(\mathbf{e}^{4}(\tau ))^{2}+\ _{\shortmid }^{\flat
}g^{5}dp^{2}+\ _{\shortmid }^{\flat }g^{6}dp_{\theta }{}^{2}  \notag \\
&&-\{\frac{4[\ ^{\shortmid }\partial ^{7}(|\ ^{\shortmid }\zeta ^{8}(\tau )\
\ _{\shortmid }^{\flat }g^{8}|^{1/2})]^{2}}{~\ _{\shortmid }^{\flat
}g^{7}|\int dp_{7}\{\ _{4}^{\shortmid }\Im (\tau )\ ^{\shortmid }\partial
^{7}(\ ^{\shortmid }\zeta ^{8}(\tau )~\ _{\shortmid }^{\flat }g^{8})\}|}%
-\kappa \lbrack \frac{\ ^{\shortmid }\partial ^{7}(\ ^{\shortmid }\chi
^{8}(\tau )|\ ^{\shortmid }\zeta ^{8}(\tau )\ _{\shortmid }^{\flat
}g^{8}|^{1/2})}{4\ ^{\shortmid }\partial ^{7}(|\ ^{\shortmid }\zeta
^{8}(\tau )\ _{\shortmid }^{\flat }g^{8}|^{1/2})}-  \notag \\
&&\frac{\int dp_{7}\{\ _{4}^{\shortmid }\Im (\tau )\ ^{\shortmid }\partial
^{7}[(\ ^{\shortmid }\zeta ^{8}(\tau )\ _{\shortmid }^{\flat }g^{8})\
^{\shortmid }\chi ^{8}(\tau )]\}}{\int dp_{7}\{\ _{4}^{\shortmid }\Im (\tau
)\ ^{\shortmid }\partial ^{7}[(\ ^{\shortmid }\zeta ^{8}(\tau )\ _{\shortmid
}^{\flat }g^{8})]\}}]\}\ _{\shortmid }^{\flat }g^{7}(\ ^{\shortmid }\mathbf{e%
}_{7}(\tau ))^{2}+\ ^{\shortmid }\tilde{\zeta}^{8}(\tau )(1-\frac{p_{s}}{p}%
+\kappa \ ^{\shortmid }\tilde{\chi}^{8}(\tau ))(\ ^{\shortmid }\mathbf{e}%
_{8}(\tau ))^{2},  \notag
\end{eqnarray*}%
where%
\begin{equation*}
\mathbf{e}^{3}(\tau )=d\varphi +[\frac{\partial _{i_{1}}\int dy^{3}\ _{2}\Im
(\tau )\ \partial _{3}\zeta _{4}(\tau )}{\ ^{\flat }N_{i_{1}}^{3}\ _{2}\Im
(\tau )\partial _{3}\zeta _{4}(\tau )}+\kappa (\frac{\partial _{i_{1}}[\int
dy^{3}\ _{2}\Im (\tau )\partial _{3}(\zeta _{4}(\tau )\chi _{4}(\tau ))]}{%
\partial _{i_{1}}\ [\int dy^{3}\ _{2}\Im (\tau )\partial _{3}\zeta _{4}(\tau
)]}-\frac{\partial _{3}(\zeta _{4}(\tau )\chi _{4}(\tau ))}{\partial
_{3}\zeta _{4}(\tau )})]\ ^{\flat }N_{i_{1}}^{3}dx^{i_{1}},
\end{equation*}%
\begin{eqnarray*}
\mathbf{e}^{4}(\tau ) &=&dt+[\ _{1}n_{k_{1}}+16\ _{2}n_{k_{1}}\int dy^{3}%
\frac{\left( \partial _{3}[(\zeta _{4}(\tau )\ ^{\flat
}g_{4})^{-1/4}]\right) ^{2}}{|\int dy^{3}\partial _{3}[\ _{2}\Im (\tau
)(\zeta _{4}(\tau )\ ^{\flat }g_{4})]|}+ \\
&&\kappa \frac{16\ _{2}n_{k_{1}}\int dy^{3}\frac{\left( \partial _{3}[(\zeta
_{4}(\tau )\ ^{\flat }g_{4})^{-1/4}]\right) ^{2}}{|\int dy^{3}\partial
_{3}[\ _{2}\Im (\tau )(\zeta _{4}(\tau )\ ^{\flat }g_{4})]|}(\frac{\partial
_{3}[(\zeta _{4}(\tau )\ ^{\flat }g_{4})^{-1/4}\chi _{4})]}{2\partial
_{3}[(\zeta _{4}(\tau )\ ^{\flat }g_{4})^{-1/4}]}+\frac{\int dy^{3}\partial
_{3}[\ _{2}\Im (\tau )(\zeta _{4}(\tau )\chi _{4}(\tau )\ ^{\flat }g_{4})]}{%
\int dy^{3}\partial _{3}[\ _{2}\Im (\tau )(\zeta _{4}(\tau )\ ^{\flat
}g_{4})]})}{\ _{1}n_{k_{1}}+16\ _{2}n_{k_{1}}[\int dy^{3}\frac{\left(
\partial _{3}[(\zeta _{4}(\tau )\ ^{\flat }g_{4})^{-1/4}]\right) ^{2}}{|\int
dy^{3}\partial _{3}[\ _{2}\Im (\tau )(\zeta _{4}(\tau )\ ^{\flat }g_{4})]|}]}%
]dx^{\acute{k}_{1}},
\end{eqnarray*}%
\begin{eqnarray*}
\ ^{\shortmid }\mathbf{e}_{7}(\tau ) &=&dp_{\varphi }+[\frac{\ ^{\shortmid
}\partial _{i_{3}}\ \int dp_{7}\ _{4}^{\shortmid }\Im (\tau )\ \ ^{\shortmid
}\partial ^{7}(\ ^{\shortmid }\zeta ^{8}(\tau ))}{\ ^{\shortmid }\mathring{N}%
_{i_{3}7}\ _{4}^{\shortmid }\Im (\tau )\ ^{\shortmid }\partial ^{7}(\
^{\shortmid }\zeta ^{8}(\tau ))}+ \\
&&\kappa (\frac{\ ^{\shortmid }\partial _{i_{3}}[\int dp_{7}\
_{4}^{\shortmid }\Im (\tau )\ ^{\shortmid }\partial ^{7}(\ ^{\shortmid
}\zeta ^{8}(\tau )\ _{\shortmid }^{\flat }g^{8})]}{\ ^{\shortmid }\partial
_{i_{3}}\ [\int dp_{7}\ _{4}^{\shortmid }\Im (\tau )\ ^{\shortmid }\partial
^{7}(\ ^{\shortmid }\zeta ^{8}(\tau ))]}-\frac{\ ^{\shortmid }\partial
^{7}(\ ^{\shortmid }\zeta ^{8}(\tau )\ _{\shortmid }^{\flat }g^{8})}{\
^{\shortmid }\partial ^{7}(\ ^{\shortmid }\zeta ^{8}(\tau ))})]\ _{\shortmid
}^{\flat }N_{i_{3}7}d\ ^{\shortmid }x^{i_{3}}, \\
\ ^{\shortmid }\mathbf{e}_{8}(\tau ) &=&dE+[\ _{1}^{\shortmid }n_{i_{3}}+16\
_{2}^{\shortmid }n_{i_{3}}\int dp_{7}\frac{\left( \ ^{\shortmid }\partial
^{7}[(\ ^{\shortmid }\zeta ^{8}(\tau )\ _{\shortmid }^{\flat
}g^{8})^{-1/4}]\right) ^{2}}{|\int dp_{7}\ _{4}^{\shortmid }\Im (\tau )~\
^{\shortmid }\partial ^{7}(\ ^{\shortmid }\zeta ^{8}(\tau )\ _{\shortmid
}^{\flat }g^{8})|}+ \\
&&\kappa \times 16\ _{2}^{\shortmid }n_{i_{3}}\int dp_{7}\frac{\left( \
^{\shortmid }\partial ^{7}[(\ ^{\shortmid }\zeta ^{8}(\tau )\ _{\shortmid
}^{\flat }g^{8})^{-1/4}]\right) ^{2}}{|\int dp_{7}\ \ _{4}^{\shortmid }\Im
(\tau )\ ^{\shortmid }\partial ^{7}[(\ ^{\shortmid }\zeta ^{8}(\tau )\ \
_{\shortmid }^{\flat }g^{8})]|}(\frac{\ ^{\shortmid }\partial ^{7}[(\
^{\shortmid }\zeta ^{8}(\tau )\ _{\shortmid }^{\flat }g^{8})^{-1/4}\
^{\shortmid }\chi ^{8}(\tau ))]}{2\ \ ^{\shortmid }\partial ^{7}[(\
^{\shortmid }\zeta ^{8}(\tau )\ _{\shortmid }^{\flat }g^{8})^{-1/4}]}+ \\
&&\frac{\int dp_{7}\ \ _{4}^{\shortmid }\Im (\tau )\ ^{\shortmid }\partial
^{7}[(\ ^{\shortmid }\zeta ^{8}(\tau )\ _{\shortmid }^{\flat }g^{8})\
^{\shortmid }\chi ^{8}(\tau )]}{\int dp_{7}\ \ _{4}^{\shortmid }\Im (\tau )\
^{\shortmid }\partial ^{7}(\ ^{\shortmid }\zeta ^{8}(\tau )\ _{\shortmid
}^{\flat }g^{8})}) \\
&&\left( \ _{1}^{\shortmid }n_{i_{3}}+16\ _{2}^{\shortmid }n_{i_{3}}[\int
dp_{7}\frac{\left( \ ^{\shortmid }\partial ^{7}\ [(\ ^{\shortmid }\zeta
^{8}(\tau )\ _{\shortmid }^{\flat }g^{8})^{-1/4}]\right) ^{2}}{|\int dp_{7}\
\ _{4}^{\shortmid }\Im (\tau )\ ^{\shortmid }\partial ^{7}(\ ^{\shortmid
}\zeta ^{8}(\tau )\ _{\shortmid }^{\flat }g^{8})|}]\right) ^{-1}]dx^{i_{3}}.
\end{eqnarray*}

For any double rotoid configuration (\ref{2rotsol}) with $\tau =\tau _{0},$
we can define and compute phase space generalizations of the Hawking
temperature, $T,$ and the Bekenstein-Hawking entropy, $S,$ as in section
4.1.2 of \cite{partner04}. Corresponding basic formulas for the 'standard'
BH thermodynamics can be generalized for $\tau $--running nonassociative
Ricci solitons for effective locally anisotropic thermodynamic variables
depending on angular spacetime and co-fiber coordinates: 
\begin{eqnarray}
T(\tau ,r,\theta ,\varphi ;p,p_{\theta },p_{\varphi }) &=&\frac{1}{4\pi }%
\left( \frac{\Omega _{2}}{4}\right) ^{1/2}\{S_{0}^{-1/2}(\tau ,r,\theta
,\varphi )+\ _{\shortmid }S_{0}^{-1/2}(\tau ,p,p_{\theta },p_{\varphi })\},%
\mbox{ with }  \notag \\
S_{0}(\tau ,r,\theta ,\varphi ) &=&\frac{\Omega _{2}\times \ (r_{s})^{4}}{4}%
[1+\frac{4\kappa }{3}\underline{\chi }(\tau ,r,\theta )\sin (\omega
_{0}\varphi +\varphi _{0})]\mbox{ and }  \notag \\
\ _{\shortmid }S_{0}(\tau ,p,p_{\theta },p_{\varphi }) &=&\frac{\Omega
_{2}\times (p_{s})^{4}}{4}[1+\frac{4\kappa }{3}\overline{\chi }(\tau
,p,p_{\theta })\sin (\ _{\shortmid }^{p}\omega _{0}\ p_{\varphi }+\
p_{\varphi }^{0})].  \label{nabhthermod}
\end{eqnarray}%
In such formulas, we used the volume of unit -sphere, $\Omega _{2}=\frac{\pi
^{2}}{(2)!},$ when the respective horizons in the 4-d spacetime and 4-d
co-fiber are $r_{s}$ and $p_{s}$ as in formulas (\ref{pmrndsf}), and (\ref%
{delipgenf}) and (\ref{param2be}). If $\underline{\chi }\kappa \rightarrow 0$
and/or $\overline{\chi }\kappa \rightarrow 0,$ the phase space double BEs
transform into respective BH configurations with conventional different
isotropic temperatures in the spacetime base and the co-fiber space.

\paragraph{Example 2: Nonassociative $\protect\tau $-running couples of
Schwarzschild-AdS BHs and BEs deformations \newline
}

Another example of phase space double BE configurations for which the
concept of Bekenstein-Hawking thermodynamics is applicable consists from
nonassociative star product R-flux deformations of corresponding couples of
Schwarzschild-AdS BHs. Considering prime 8-d metrics (\ref{pmrndsf}) with 
\begin{eqnarray}
\ \ ^{\flat }g_{4}(r) &\rightarrow &\ \ ^{\epsilon }g_{4}(r)=-(\epsilon -%
\frac{2m_{+}}{r}+\check{\Lambda}r^{2})=-[\ ^{\epsilon }g_{1}(r)]^{-1}%
\mbox{
and }  \notag \\
\ _{\shortmid }^{\flat }g^{8}(p) &\rightarrow &\ _{\shortmid }^{\epsilon
}g^{8}(p)=-(\epsilon -\frac{2\ _{\shortmid }m_{+}}{p}+\ ^{\shortmid }\check{%
\Lambda}p^{2})=-[\ _{\shortmid }^{\epsilon }g^{5}(p)]^{-1},  \label{metrf2}
\end{eqnarray}%
for mass parameters $m_{+}$ and $\ _{\shortmid }m_{+},$ where $\epsilon
=(+1,0,-1)$ which corresponding, respectively, to spherical/ planar/
hyperbolic horizon geometries, we define prime s-metrics as solutions of the
phase modified nonholonomic Einstein equations (\ref{modeinst}), see also (%
\ref{nonassocrsol}), with nonlinear symmetries (\ref{nonlinsym}) relating
certain prescribed effective sources $\ _{s}^{\shortmid }\mathcal{K}$ to
effective cosmological constants $\ _{1}^{\shortmid }\Lambda _{0}=\
_{2}^{\shortmid }\Lambda _{0}=\check{\Lambda}\geq 0$ and $\
_{3}^{\shortmid}\Lambda _{0}= \ _{4}^{\shortmid }\Lambda _{0}= \ ^{\shortmid
}\check{\Lambda}\geq 0.$ We can extend such s-metrics for running
cosmological constants $\ _{s}^{\shortmid }\Lambda (\tau )=[\check{\Lambda}%
(\tau ),\ ^{\shortmid }\check{\Lambda}(\tau )] \geq 0$ as solutions of (\ref%
{nonassocgeomflefcc}), parameterized in the form%
\begin{eqnarray}
d\ \ _{\shortmid }^{\epsilon }s_{[8d]}^{2}(\tau ) &=& d\
^{\epsilon}s^{2}(\tau )+d\ _{\shortmid }^{\epsilon }s^{2}(\tau ), 
\mbox{
with }  \label{pmrndsf2} \\
&&d\ ^{\epsilon }s^{2}(\tau ) =\ ^{\epsilon }g_{1}(\tau ,r)dr^{2}+\
^{\epsilon }g_{2}(r)d\theta ^{2}+ \ ^{\epsilon }g_{3}(r,\theta ) d\varphi
^{2}+\ ^{\epsilon }g_{4}(\tau ,r)dt^{2},  \notag \\
&&d\ _{\shortmid }^{\epsilon }s^{2} = \
_{\shortmid}^{\epsilon}g^{5}(\tau,p)dp^{2}+ \ _{\shortmid }^{\epsilon
}g^{6}(p)dp_{\theta }^{2} + \ _{\shortmid }^{\epsilon
}g^{7}(p,p_{\theta})dp_{\varphi }^{2}+ \ _{\shortmid }^{\epsilon }g^{8}(\tau
,p)dE^{2},  \notag
\end{eqnarray}%
when the coefficients are determined as respective metric functions (\ref%
{metrf2}) modified for running constants, 
\begin{eqnarray*}
\ \ ^{\epsilon }g_{4}(\tau ,r) &=&-(\epsilon -\frac{2m_{+}(\tau )}{r}+\check{%
\Lambda}(\tau )r^{2})= -[\ ^{\epsilon }g_{1}(\tau ,r)]^{-1}\mbox{ and } \\
\ _{\shortmid }^{\epsilon }g^{8}(\tau ,p) &=&-(\epsilon -\frac{2\
_{\shortmid }m_{+}(\tau )}{p}+ \ ^{\shortmid }\check{\Lambda}%
(\tau)p^{2})=-[\ _{\shortmid }^{\epsilon }g^{5}(\tau ,p)]^{-1};
\end{eqnarray*}
and (for instance, for spherical horizon geometries) $\
^{\epsilon}g_{2}(r)=r^{2},\ ^{\epsilon }g_{3}(r,\theta )=r^{2}\sin
^{2}\theta $ and \newline
$\ _{\shortmid }^{\epsilon }\mathring{g}^{6}(p)=p^{2},\
_{\shortmid}^{\epsilon }\mathring{g}^{7}(p,p_{\theta })= p^{2}\sin
^{2}p_{\theta }.$

We considered above a prime s-metric (\ref{pmrndsf}) which is not a
nonassociative vacuum solution but $\tau$-evolves into nonassociative
geometric flows of quasi-stationary solutions. In this subsection, the prime
s-metric (\ref{pmrndsf2}) defines already $\tau $-families of double BH
solutions of nonassociative vacuum Einstein equations. For a fixed $\tau
_{0},$ such a nonassociative Ricci soliton is defined by a stationary
configuration is defined by metrics with a conventional horizon radius $%
r_{+},$ in the base manifold, and $p_{+},$ in the co-fiber space. The
thermodynamic quantities (entropy and temperature) for such couples of phase
space Schwarzschild-AdS BHs are defined and computed in standard forms:%
\begin{eqnarray}
\ \ ^{\epsilon }S_{0}(\tau ) &=&\frac{\pi \ ^{\epsilon }\Theta }{4}%
r_{+}^{2}(\tau ), \ ^{\epsilon }T_{0}(\tau )= \frac{\epsilon +3\check{\Lambda%
}(\tau )r_{+}^{2}(\tau )}{4\pi r_{+}(\tau )}\mbox{ and }  \notag \\
\ _{\shortmid }^{\epsilon }S_{0}(\tau ) &=&\frac{\pi \ \
_{\shortmid}^{\epsilon }\Theta }{4}p_{+}^{2}(\tau ), \ \ _{\shortmid
}^{\epsilon}T_{0}(\tau )=\frac{\ ^{\shortmid }\epsilon + 3\ ^{\shortmid }%
\check{\Lambda}(\tau )p_{+}^{2}(\tau )}{4\pi p_{+}(\tau )},
\label{hbbhtherm}
\end{eqnarray}%
where running of conventional mass parameters are computed following
formulas 
\begin{equation*}
m_{+}(\tau )=\frac{r_{+}(\tau )\ ^{\epsilon }\Theta }{8}[\epsilon + \check{%
\Lambda}(\tau )r_{+}^{2}(\tau )]\mbox{ and }\ _{\shortmid }m_{+}(\tau )= 
\frac{p_{+}(\tau )\ _{\shortmid }^{\epsilon}\Theta }{8}[\ ^{\shortmid
}\epsilon + \ ^{\shortmid }\check{\Lambda}(\tau)p_{+}^{2}(\tau )].
\end{equation*}%
In above equations, we use respective areas of constant-curvatures spaces, $%
\pi \ ^{\epsilon }\Theta $ and $\pi \ _{\shortmid }^{\epsilon }\Theta .$ For
instance: $\ ^{1}\Theta =4,$ for a sphere; $\ ^{0}\Theta =XY,$ using $X$ and 
$Y$ as sides of the torus; there is not a simple example to compute $\
^{-1}\Theta .$ Here we note that following the standard BH thermodynamics
for solutions with hyper-surfaces we can define and compute a conventional
pressure, $P,$ and volume, $Vol$ (for instance, $P=\frac{3}{8\pi }\check{%
\Lambda}$ and $Vol=\frac{\pi \ ^{\epsilon }\Theta }{3}$ $r_{+}^{3}).$

Rotoid deformations with $\chi $-polarizations (\ref{delipgenf}) under
nonassociative geometric information flows of families of s-metrics (\ref%
{pmrndsf2}), $\ _{\shortmid }^{\epsilon }g_{\alpha }\rightarrow \
_{\shortmid }^{\epsilon \chi }g_{\alpha },$ can be generated if we change
the data for the prime s-metrics, $\ _{\shortmid }^{\flat }g_{\alpha }=(\
^{\flat }g_{i},\ _{\shortmid }^{\flat }g^{a})$ (\ref{pmrndsf}) $\rightarrow
\ _{\shortmid }^{\epsilon }g_{\alpha }=(\ ^{\epsilon }g_{i},\
_{\shortmid}^{\epsilon }g^{a}),$ into (\ref{2rotsol}). We do not write in
explicit form such quadratic linear elements, but provide the formulas for
the effective locally anisotropic thermodynamic variables depending on
angular spacetime and co-fiber variables:%
\begin{eqnarray}
\ ^{\epsilon \chi }S(\tau ,r,\theta ,\varphi ) &=&\ ^{\epsilon }S_{0}(\tau) 
\left[ 1+4\kappa \underline{\chi }(\tau ,r,\theta )\sin (\omega _{0}\varphi
+\varphi _{0})\right] \ ,  \label{hbbhtherman} \\
\ ^{\epsilon \chi }T(\tau ,r,\theta ,\varphi ) &=&\ ^{\epsilon }T_{0}(\tau) 
\left[ 1+2\kappa \underline{\chi }(\tau ,r,\theta )\sin (\omega _{0}\varphi
+\varphi _{0})\right] \mbox{ and }  \notag \\
\ _{\shortmid }^{\epsilon \chi }S(\tau ,p,p_{\theta },p_{\varphi }) &=&\
_{\shortmid }^{\epsilon }S_{0}(\tau ) \left[ 1+ 4\kappa \overline{\chi }%
(\tau,p,p_{\theta })\sin (\ _{\shortmid }^{p}\omega _{0}\ p_{\varphi }+\
p_{\varphi }^{0})\right] \ ,  \notag \\
\ _{\shortmid }^{\epsilon \chi }T(\tau ,p,p_{\theta },p_{\varphi }) &=& \
_{\shortmid }^{\epsilon }T_{0}(\tau ) \left[ 1+ 2\kappa \overline{\chi }%
(\tau, p, p_{\theta })\sin (\ _{\shortmid }^{p}\omega _{0}\ p_{\varphi }+ \
p_{\varphi }^{0})\right] .  \notag
\end{eqnarray}%
For $\kappa \rightarrow 0$, these formulas transform into locally isotropic
ones (\ref{hbbhtherm}). As in the thermodynamics of moving media, we have
two temperature like values, $\tau $ and $\ ^{\epsilon \chi }T.$ In the case
of ellipsoidal deformations of couples of Schwarzschild-AdS metrics, we have
to consider two different temperature like variables $\ ^{\epsilon \chi }T$
and $\ _{\shortmid }^{\epsilon \chi }T.$ This is different from the case
described by the formulas (\ref{nabhthermod}) defined and computed for prime
s-metrics which are not solutions of certain nonassociative vacuum Einstein
equations, but they became such ones for the target s-metrics after $\tau $%
-parametric dissipation of the electromagnetic components into the
N-connection coefficients.

Above phase space BHs and BEs are characterized by nonassociative locally
anisotropic degrees of freedom determined by R-flux deformations with
spherical/rotoid symmetries. We omit the technical details and cumbersome
formulas for computing $\ _{\star }^{\shortmid }\mathfrak{g}_{\alpha
_{s}\beta _{s}}(\tau )= \ _{\star }^{\shortmid }\mathfrak{\check{g}}_{\alpha
_{s}\beta _{s}}(\tau )+ \ _{\star }^{\shortmid }\mathfrak{a}_{\alpha
_{s}\beta _{s}}(\tau )$ (\ref{aux40b}) using corresponding prime data (\ref%
{pmrndsf}), or (\ref{pmrndsf2}), and respective off-diagonal parametric
solutions for double BEs and/or BHs.

\subsubsection{Geometric thermodynamic variables for nonassociative flows of
RN-dS BHs}

\label{ss524}The G. Perelman thermodynamic variables (\ref{thvcann}) can be
computed in explicit form for any class of $\tau $-running quasi-stationary
solutions (\ref{dbhpspol}) and/ or (\ref{2rotsol}) defining nonassociative
geometric evolution and star product R-flux deformations of double RN-dS BH
configurations with $_{1}\Lambda (\tau )=\ _{2}\Lambda (\tau )=$ $\check{%
\Lambda}$ and $\ _{3}^{\shortmid }\Lambda (\tau )=\ _{4}^{\shortmid }\Lambda
(\tau )=$ $\ ^{\shortmid }\check{\Lambda}.$ For simplicity, we consider a
nonholonomic geometric model with fixed values of cosmological h- and
c-cosmologival constants. Changing the prime data, $\ _{s}^{\shortmid }%
\mathbf{\mathring{g}}$ $\rightarrow $ $\ _{\shortmid s}^{\flat }\mathbf{g}$ (%
\ref{pmrndsf}), in the volume forms (\ref{volumfuncts}), we express 
\begin{eqnarray*}
\ ^{\shortmid }\delta \ ^{\shortmid }\mathcal{V}(\tau ) &=&\ ^{\shortmid
}\delta \ ^{\shortmid }\mathcal{V}(\ \ _{s}^{\shortmid }\Im (\tau ),\ \check{%
\Lambda},\ ^{\shortmid }\check{\Lambda},\ ^{\shortmid }\eta _{\alpha
_{s}}(\tau )\ \ _{\shortmid }^{\flat }g_{\alpha _{s}})=\frac{1}{|\check{%
\Lambda}\ ^{\shortmid }\check{\Lambda}|}\ \ ^{\shortmid }\delta \ _{\eta
}^{\shortmid }\mathcal{V}(\ _{s}^{\shortmid }\Im (\tau ),\ _{\shortmid
}^{\flat }g_{\alpha _{s}}),\mbox{ where } \\
\ ^{\shortmid }\delta \ _{\eta }^{\shortmid }\mathcal{V}(\ _{s}^{\shortmid
}\Im (\tau ),\ _{\shortmid }^{\flat }g_{\alpha _{s}}) &=&\ ^{\shortmid
}\delta \ _{\eta }^{1}\mathcal{V}[\ _{1}\Im (\tau ),\eta _{2}(\tau )\ \
^{\flat }g_{2}]\times \ ^{\shortmid }\delta \ _{\eta }^{2}\mathcal{V}[\
_{2}\Im (\tau ),\eta _{4}(\tau )\ \ ^{\flat }g_{4}]\times \\
&&\ ^{\shortmid }\delta \ _{\eta }^{3}\mathcal{V}[\ _{3}\Im (\tau
),~^{\shortmid }\eta ^{6}(\tau )~\ _{\shortmid }^{\flat }g^{6}]\times \
^{\shortmid }\delta \ _{\eta }^{4}\mathcal{V}[\ _{4}\Im (\tau ),\
^{\shortmid }\eta ^{8}(\tau )~\ _{\shortmid }^{\flat }g^{8}].
\end{eqnarray*}%
This allows to compute, using formulas (\ref{thvcannpd})\ and (\ref%
{volumfpsp}) , such thermodynamic variables and, respective, volume
functionals: 
\begin{eqnarray}
\ _{s}^{\shortmid }\widehat{\mathcal{W}}_{\kappa }^{\star }(\tau )
&=&\int\nolimits_{\tau ^{\prime }}^{\tau }\frac{d\tau }{64(\pi \tau )^{4}}%
\frac{\tau (\check{\Lambda}+\ ^{\shortmid }\check{\Lambda})^{2}-2}{|\check{%
\Lambda}\ ^{\shortmid }\check{\Lambda}|}\ _{\eta }^{\shortmid \flat }%
\mathcal{V}(\tau ),\ _{s}^{\shortmid }\widehat{\mathcal{Z}}_{\kappa }^{\star
}(\tau )=\exp \left[ \int\nolimits_{\tau ^{\prime }}^{\tau }\frac{d\tau }{%
(2\pi \tau )^{4}}\frac{1}{|\check{\Lambda}\ ^{\shortmid }\check{\Lambda}|}\
_{\eta }^{\shortmid \flat }\mathcal{V}(\tau )\right] ,  \label{thvcannpd1} \\
\ _{s}^{\shortmid }\widehat{\mathcal{E}}_{\kappa }^{\star }(\tau )
&=&-\int\nolimits_{\tau ^{\prime }}^{\tau }\frac{d\tau }{128\pi ^{4}\tau ^{3}%
}\frac{\tau (\check{\Lambda}+\ ^{\shortmid }\check{\Lambda})-2}{|\check{%
\Lambda}\ ^{\shortmid }\check{\Lambda}|}\ _{\eta }^{\shortmid \flat }%
\mathcal{V}(\tau ),\ _{s}^{\shortmid }\widehat{\mathcal{S}}_{\kappa }^{\star
}(\tau )=-\int\nolimits_{\tau ^{\prime }}^{\tau }\frac{d\tau }{128(\pi \tau
)^{4}}\frac{\tau (\check{\Lambda}+\ ^{\shortmid }\check{\Lambda})-4}{|\check{%
\Lambda}\ ^{\shortmid }\check{\Lambda}|}\ _{\eta }^{\shortmid \flat }%
\mathcal{V}(\tau ),  \notag
\end{eqnarray}%
\begin{equation*}
\mbox{ for }\ _{\eta }^{\shortmid \flat }\mathcal{V}(\tau )=\int_{\
_{s}^{\shortmid }\widehat{\Xi }}\ ^{\shortmid }\delta \ _{\eta }^{\shortmid }%
\mathcal{V}(\ _{s}^{\shortmid }\Im (\tau ),\ _{\shortmid }^{\flat }g_{\alpha
_{s}}).
\end{equation*}%
Such values are well-defined for $\tau $-running nonholonomic configurations
with respective $\eta $- or $\chi $-generating functions when $\
_{s}^{\shortmid }\widehat{\mathcal{E}}_{\kappa }^{\star }(\tau )$ and $\
_{s}^{\shortmid }\widehat{\mathcal{S}}_{\kappa }^{\star }(\tau )$ can be
treated as respective effective energy and entropy flow transports in a
phase space media evolving in the interval $\tau ^{\prime }<\tau .$

\subsection{Nonassociative flows of phase space Reisner-Nordstr\"{o}m-AdS BHs%
}

In this subsection, we construct a different class of nonassociative
geometric flow deformations of RN BHs, when the effective cosmological
constants are determined by negative cosmological constants and respective
prime metric configurations, which correspond to different models than those
stated by (\ref{emt4}) and (\ref{pmrndsf}).

\subsubsection{Prime metrics for higher dimension phase space RN-AdS BHs}

As shown in \cite{chamb99}, a $d=5$ dimensional Einstein-Maxwell action 
\begin{equation}
S=\frac{1}{16\pi G_{[5]}}\int\nolimits_{V_{[5]}}d^{5}x\sqrt{|g_{[5]}|}
[R_{[5]}-l_{[5]}^{2}F_{[4]}^{2}+\frac{12}{l_{[5]}^{2}}],  \label{emt5}
\end{equation}%
with a negative constant $\Lambda _{\lbrack 5]}=-6/l_{[5]}^{2}$ determined
by the AdS radius $l_{[5]},$ can be naturally viewed (with an additional
Chern-Simons term) as an effective truncation of the IIB supergravity on a
5-d sphere, $\mathbb{S}^{5}.$ In \cite{zhang15}, the thermodynamic geometry
of 5-d Reisner-Nordstr\"{o}m-AdS BHs existing in the theory (\ref{emt5}) was
studied for extensions to models of phase spaces determined by the scalar
curvatures of the thermodynamic Weinhold/ Ruppeinder / Quevedo metrics.

In our approach to nonassociative gravity and geometric flow theory, we can
consider a $d=5$ dimensional analog of the Reisner-Nordstr\"{o}m AdS metric
trivially embedded into a 8-d phase space $\ _{s}\mathcal{M},$%
\begin{equation}
d\ \breve{s}_{[5+3]}^{2}=\ ^{\shortmid }\breve{g}_{\alpha _{s}}(\
^{\shortmid }u^{\gamma _{s}})(\mathbf{\breve{e}}^{\alpha _{s}})^{2}=\frac{d%
\breve{r}^{2}}{\breve{f}(\breve{r})}-\breve{f}(\breve{r})dt^{2}+\breve{r}%
^{2}[(d^{2}\hat{x}^{2})^{2}+(d\hat{x}%
^{3})^{2}+(dp_{5})^{5}]+(dp_{6})^{2}+(dp_{7})^{2}-dE^{2},  \label{pm5d8d}
\end{equation}%
where in natural units $\hat{x}^{1}=\breve{r}=\sqrt{%
(x^{1})^{2}+(x^{2})^{2}+(x^{3})^{2}+(p_{5})^{2}};$ for simplicity, we choose 
$\hat{x}^{2}=\hat{x}^{2}(x^{2},x^{3},p_{5}),$ $\hat{x}^{3}=\hat{x}%
^{3}(x^{2},x^{3},p_{5})$ and $\hat{x}^{5}=\hat{x}^{5}(x^{2},x^{3},p_{5})$ as
coordinates for a diagonal metric on an effective 3-d Einstein phase space $%
V_{[3]}$ of constant scalar curvature (let say, $6\hat{k},$ for $\hat{k}=1$%
). The metric function in (\ref{pm5d8d}) is given by 
\begin{equation*}
\breve{f}(\breve{r})=1-\frac{\hat{m}}{\breve{r}^{2}}+\frac{\breve{r}^{2}}{%
l_{[5]}^{2}}+\frac{\hat{q}^{2}}{\breve{r}^{4}},
\end{equation*}%
where the integration constant $\hat{m}$ is related to the mass of BH, $\hat{%
M}=3\omega _{\lbrack 3]}\hat{m}/16\pi G_{[5]},$ for $\omega _{\lbrack 3]}$
denoting the volume of $V_{[3]};$ and the parameter $\hat{q}$ is related to
the physical charge $\hat{Q}$ of the RN-AdS BH via formula $\hat{q}=4\pi
G_{[5]}\hat{Q}/\sqrt{3}\omega _{\lbrack 3]}.$

The prime metric coefficients $\ \breve{g}_{1}=\breve{f}(\breve{r})^{-1},\
^{\shortmid }\breve{g}_{2}=\ ^{\shortmid }\breve{g}_{3}= \ ^{\shortmid }%
\breve{g}^{5}=\breve{r}^{2},\ \breve{g}_{4}=-\breve{f}(\breve{r}),\
^{\shortmid }\breve{g}^{6}= \ ^{\shortmid }\breve{g}^{7}=-\ ^{\shortmid }%
\breve{g}^{8}=1$ and $\ ^{\shortmid }\breve{g}_{i_{s-1}}^{a_{s}}(\breve{r},t,%
\hat{x}^{2},\hat{x}^{3},\hat{x}^{5},p_{6,}p_{7},E)=0$ from (\ref{pm5d8d})
can be subjected to s-adapted frame/coordinate transforms into certain data $%
(\ ^{\shortmid }\mathbf{\breve{g}}_{\alpha _{s}}(\ ^{\shortmid }u^{\gamma
_{s}});\ ^{\shortmid }\mathbf{\breve{N}}_{i_{s-1}}^{a_{s}}(\
^{\shortmid}u^{\gamma _{s}}))$ which allow to apply the AFCDM for
constructing exact and parametric solutions. The 5-d part of the 8-d metric (%
\ref{pm5d8d}) can be uplifted to ten dimensions and viewed as the near
horizon geometry of $\check{N}$ rotating black D3-branes in type IIB
supergravity \cite{chamb99,zhang15}, when $l_{[10]}^{4}=2\check{N}\ell
_{p}^{4}/\pi ^{2}\equiv \alpha ^{2}\check{N}$ , where $\ell _{p}$ is the
10-d Planck length. The nonassociative geometric constructions with star
product and R-flux deformations involve different types of constants on 8-d
phase spaces. It should be noted here that we can elaborate similar
nonholonomic geometric constructions with $_{s}\mathcal{M}\rightarrow \
_{s}^{\star }\mathcal{M}$, when for a corresponding cotangent Lorentz bundle 
$\mathcal{M}=T^{\ast }V,\dim V=5$ and $\dim $ $\mathcal{M}=10.$

\subsubsection{Nonassociative geometric $\protect\kappa $-linear evolution
of phase space RN-AdS BHs}

We consider nonassociative generic off-diagonal generalizations of s-metric $%
\ ^{\shortmid }\mathbf{\breve{g}}_{\alpha _{s}}$ (\ref{pm5d8d}) under $%
\kappa $-linear geometric flow evolution with fixed 8-d phases space
cosmological constant $_{s}\Lambda (\tau )=\Lambda _{\lbrack 5]}<0,$ for $%
s=1,2,3,4.$ Considering $\ ^{\shortmid }\breve{g}_{\alpha _{s}}$ as the
coefficients of the prime s-metric in (\ref{offdiagpolfr}) (instead of $\
_{s}^{\shortmid }\mathbf{\mathring{g}}(\tau )$) for respective effective
sources $\ _{s}^{\shortmid }\Im (\tau )$ related via nonlinear symmetries (%
\ref{nonlinsymr}) to $\ \Lambda _{\lbrack 5]},$ we generate such a $\tau $%
-family of quasi-stationary solutions of nonassociative geometric flow
equations (\ref{nonassocgeomflefcc}), $\ ^{\shortmid }\breve{g}_{\alpha
_{s}}\rightarrow \ _{s}^{\shortmid }\mathbf{\mathring{g}}(\tau ),$
parameterized, for instance, using $\Phi $-generating functions as in
quadratic linear elements (\ref{sol2rf}) with nonholonomic frames (\ref%
{sol2nrf}), 
\begin{eqnarray}
d\widehat{s}^{2}(\tau ) &=&e^{\psi (\hbar ,\kappa ;\tau ,\breve{r},\hat{x}%
^{2},\Lambda _{\lbrack 5]})}[(d\breve{r})^{2}+(d\hat{x}^{2})^{2}]-
\label{narnbhs} \\
&&\frac{1}{g_{4}^{[0]}(\tau )-\frac{(\ _{2}\Phi (\tau ))^{2}}{4\ \Lambda
_{\lbrack 5]}}} \frac{(\ _{2}\Phi (\tau ))^{2}[\partial _{3}(\ _{2}\Phi
(\tau))]^{2}}{|\ \Lambda _{\lbrack 5]}\int dy^{3}(~_{2}^{\shortmid }\Im
(\tau ))[\partial _{3}(\ _{2}\Phi (\tau ))^{2}]|}(\mathbf{e}%
^{3}(\tau))^{2}+\left( g_{4}^{[0]}(\tau )-\frac{(\ _{2}\Phi (\tau ))^{2}}{4\
\Lambda _{\lbrack 5]}}\right) (\mathbf{e}^{4}(\tau ))^{2}  \notag
\end{eqnarray}%
\begin{equation*}
-\frac{1}{\ ^{\shortmid }g_{[0]}^{6}(\tau )-\frac{(\ _{3}^{\shortmid }\Phi
(\tau ))^{2}}{4\ \Lambda _{\lbrack 5]}}}\frac{(\ _{3}^{\shortmid }\Phi (\tau
))^{2}[\ ^{\shortmid }\partial ^{5}(\ _{3}^{\shortmid }\Phi (\tau ))]^{2}}{%
|\ \Lambda _{\lbrack 5]}\int dp_{5}(~_{3}^{\shortmid }\Im (\tau ))\ \
^{\shortmid }\partial ^{5}[(\ _{3}^{\shortmid }\Phi (\tau ))^{2}]|}(\
^{\shortmid }\mathbf{e}_{5}(\tau ))^{2}+\left( \ ^{\shortmid
}g_{[0]}^{6}(\tau )-\frac{(\ _{3}^{\shortmid }\Phi (\tau ))^{2}}{4\ \Lambda
_{\lbrack 5]}}\right) (\ ^{\shortmid }\mathbf{e}_{6}(\tau ))^{2}
\end{equation*}%
\begin{equation*}
-\frac{1}{\ ^{\shortmid }g_{[0]}^{8}(\tau )-\frac{(\ _{4}^{\shortmid }\Phi
(\tau ))^{2}}{4\ \Lambda _{\lbrack 5]}}}\frac{(\ _{4}^{\shortmid }\Phi (\tau
))^{2}[\ ^{\shortmid }\partial ^{7}(\ _{4}^{\shortmid }\Phi (\tau ))]^{2}}{%
|\ \Lambda _{\lbrack 5]}\int dp_{7}(~_{4}^{\shortmid }\Im )\ \ ^{\shortmid
}\partial ^{7}[(\ _{4}^{\shortmid }\Phi (\tau ))^{2}]|}(\ ^{\shortmid }%
\mathbf{e}_{7}(\tau ))^{2}+\left( \ ^{\shortmid }g_{[0]}^{8}(\tau )-\frac{(\
_{4}^{\shortmid }\Phi (\tau ))^{2}}{4\ \Lambda _{\lbrack 5]}}\right) (\
^{\shortmid }\mathbf{e}_{8}(\tau ))^{2}.
\end{equation*}%
In these formulas, there are used local coordinates $\ ^{\shortmid}u^{\gamma
_{s}}=(\breve{r},t,\hat{x}^{2},\hat{x}^{3},\hat{x}^{5},p_{6,}p_{7},E)$ and
s-adapted frames: 
\begin{eqnarray*}
\mathbf{e}^{3}(\tau ) &=&d\hat{x}^{3}+\frac{\partial _{k_{1}}\ \int d\hat{x}%
^{3}(~_{2}^{\shortmid }\Im (\tau )) \hat{\partial}_{3}[(\ _{2}\Phi (\tau
))^{2}]}{(~_{2}^{\shortmid }\Im (\tau ))\ \hat{\partial}_{3}[(\ _{2}\Phi
(\tau ))^{2}]}dx^{k_{1}}, \\
\mathbf{e}^{4}(\tau ) &=&dt+(\ _{1}n_{k_{1}}(\tau )+\ _{2}n_{k_{1}}(\tau )%
\frac{\int d\hat{x}^{3}\frac{(\ _{2}\Phi (\tau ))^{2}[\ \hat{\partial}_{3}(\
_{2}\Phi (\tau ))]^{2}}{|\ _{2}\Lambda (\tau )\int d\hat{x}%
^{3}(~_{2}^{\shortmid }\Im (\tau ))[\ \hat{\partial}_{3}(\ _{2}\Phi (\tau
))^{2}]|}}{\left\vert g_{4}^{[0]}(\tau )-\frac{(\ _{2}\Phi (\tau ))^{2}}{4\
_{2}\Lambda (\tau )}\right\vert ^{5/2}})dx^{k_{1}},
\end{eqnarray*}%
\begin{eqnarray*}
\ ^{\shortmid }\mathbf{e}_{5}(\tau ) &=&d\hat{x}^{5}+\frac{\partial
_{k_{2}}\ \int d\hat{x}^{5}(~_{3}^{\shortmid }\Im (\tau ))\ \hat{\partial}%
[(\ _{3}^{\shortmid }\Phi (\tau ))^{2}]}{(~_{3}^{\shortmid }\Im (\tau ))\ \
\ \hat{\partial}_{5}[(\ _{3}^{\shortmid }\Phi (\tau ))^{2}]}dx^{k_{2}}, \\
\ ^{\shortmid }\mathbf{e}_{6}(\tau ) &=&dp_{6}+(\ _{1}^{\shortmid
}n_{k_{2}}(\tau )+\ _{2}^{\shortmid }n_{k_{2}}(\tau )\frac{\int dp_{5}\frac{%
(\ _{3}^{\shortmid }\Phi (\tau ))^{2}[\ ^{\shortmid }\partial ^{5}(\
_{3}^{\shortmid }\Phi (\tau ))]^{2}}{|\ \Lambda _{\lbrack 5]}\int
dp_{5}(~_{3}^{\shortmid }\Im (\tau ))[\ ^{\shortmid }\partial ^{5}(\
_{3}^{\shortmid }\Phi (\tau ))^{2}]|}}{\left\vert \ ^{\shortmid
}g_{[0]}^{6}(\tau )-\frac{(\ _{3}^{\shortmid }\Phi (\tau ))^{2}}{4\ \Lambda
_{\lbrack 5]}}\right\vert ^{5/2}})dx^{k_{2}},
\end{eqnarray*}%
\begin{eqnarray*}
\ ^{\shortmid }\mathbf{e}_{7}(\tau ) &=&dp_{7}+\frac{\partial _{k_{3}}\ \int
dp_{7}(~_{4}^{\shortmid }\Im (\tau ))\ \ ^{\shortmid }\partial ^{7}[(\
_{4}^{\shortmid }\Phi (\tau ))^{2}]}{(~_{4}^{\shortmid }\Im (\tau ))\ \ \
^{\shortmid }\partial ^{7}[(\ _{4}^{\shortmid }\Phi (\tau ))^{2}]}d\
^{\shortmid }x^{k_{3}}, \\
\ ^{\shortmid }\mathbf{e}_{8}(\tau ) &=&dE+(\ _{1}^{\shortmid
}n_{k_{3}}(\tau )+\ _{2}^{\shortmid }n_{k_{3}}(\tau )\frac{\int dp_{7}\frac{%
(\ _{4}^{\shortmid }\Phi (\tau ))^{2}[\ ^{\shortmid }\partial ^{7}(\
_{4}^{\shortmid }\Phi (\tau ))]^{2}}{|\ \Lambda _{\lbrack 5]}\int
dp_{7}(~_{4}^{\shortmid }\Im (\tau ))[\ ^{\shortmid }\partial ^{7}(\
_{4}^{\shortmid }\Phi (\tau ))^{2}]|}}{\left\vert \ ^{\shortmid
}g_{[0]}^{8}(\tau )-\frac{(\ _{4}^{\shortmid }\Phi (\tau ))^{2}}{4\ \Lambda
_{\lbrack 5]}}\right\vert ^{5/2}})d\ ^{\shortmid }x^{k_{3}}.
\end{eqnarray*}%
The integration functions considered above are defined in the form: 
\begin{eqnarray*}
&&g_{4}^{[0]}(\hbar ,\kappa ,\tau ,\breve{r},\hat{x}^{2}),\
_{1}n_{k_{1}}(\hbar ,\kappa ,\tau ,\breve{r},\hat{x}^{2}),\
_{2}n_{k_{1}}(\hbar ,\kappa ,\tau ,\breve{r},\hat{x}^{2}); \\
&&\ ^{\shortmid }g_{[0]}^{5}(\hbar ,\kappa ,\tau ,\breve{r},\hat{x}^{2},\hat{%
x}^{3},\hat{x}^{5}),\ _{1}n_{k_{2}}(\hbar ,\kappa ,\tau ,\breve{r},\hat{x}%
^{2},\hat{x}^{3},\hat{x}^{5}),\ _{2}n_{k_{2}}(\hbar ,\kappa ,\tau ,\breve{r},%
\hat{x}^{2},\hat{x}^{3},\hat{x}^{5}); \\
&&\ ^{\shortmid }g_{[0]}^{7}(\hbar ,\kappa ,\tau ,\breve{r},\hat{x}^{2},\hat{%
x}^{3},\hat{x}^{5},p_{7}),\ _{1}^{\shortmid }n_{k_{3}}(\hbar ,\kappa ,\tau ,%
\breve{r},\hat{x}^{2},\hat{x}^{3},\hat{x}^{5},p_{7}),\ _{2}^{\shortmid
}n_{k_{3}}(\hbar ,\kappa ,\tau ,\breve{r},\hat{x}^{2},\hat{x}^{3},\hat{x}%
^{5},p_{7}).
\end{eqnarray*}

The solutions (\ref{narnbhs}) can be written in terms of $g$-generating
functions and/or $\eta $- /$\chi $-polarization functions using nonlinear
transforms (\ref{nonlinsymrex}) for 
\begin{eqnarray}
\ _{2}\Phi (\tau ) &=&2\sqrt{|\ \Lambda _{\lbrack 5]}\ g_{4}(\tau )|}=\ 2%
\sqrt{|\ \Lambda _{\lbrack 5]}\ \eta _{4}(\tau )\breve{g}_{4})|}\simeq 2%
\sqrt{|\ \Lambda _{\lbrack 5]}\ \zeta _{4}(\tau )\breve{g}_{4}|}[1-\frac{%
\kappa }{2}\chi _{4}(\tau )],  \label{genf2} \\
\ \ _{3}^{\shortmid }\Phi (\tau ) &=&2\sqrt{|\ \Lambda _{\lbrack
5]}~^{\shortmid }g^{6}(\tau )|}=\ 2\sqrt{|\ \Lambda _{\lbrack
5]}~^{\shortmid }\eta ^{6}(\tau )~^{\shortmid }\breve{g}^{6}|}\simeq 2\sqrt{%
|\ \Lambda _{\lbrack 5]}~^{\shortmid }\zeta ^{6}(\tau )~^{\shortmid }\breve{g%
}^{6}|}[1-\frac{\kappa }{2}~^{\shortmid }\chi ^{6}(\tau )],  \notag \\
\ \ _{4}^{\shortmid }\Phi (\tau ) &=&2\sqrt{|\ \Lambda _{\lbrack
5]}~^{\shortmid }g^{8}(\tau )|}=\ 2\sqrt{|\ \Lambda _{\lbrack
5]}~^{\shortmid }\eta ^{8}(\tau )~^{\shortmid }\breve{g}^{8}|}\simeq 2\sqrt{%
|\ \Lambda _{\lbrack 5]}~^{\shortmid }\zeta ^{8}(\tau )~^{\shortmid }\breve{g%
}^{8}|}[1-\frac{\kappa }{2}~^{\shortmid }\chi ^{8}(\tau )],  \notag
\end{eqnarray}%
for a prime s-metric $\ ^{\shortmid }\mathbf{\breve{g}}_{\alpha _{s}}$ (\ref%
{pm5d8d}).$\ $For such transforms, the generating and integration functions
are written in $\kappa $--linearized form (\ref{offdncelepsilon}), 
\begin{eqnarray*}
\psi (\tau ) &\simeq &\psi (\hbar ,\kappa ;\tau ,\breve{r},\hat{x}%
^{2})\simeq \psi _{0}(\hbar ,\tau ,\breve{r},\hat{x}^{2})(1+\kappa \ _{\psi
}\chi (\hbar ,\tau ,\breve{r},\hat{x}^{2})),\mbox{ for } \\
\ \eta _{2}(\tau ) &\simeq &\eta _{2}(\hbar ,\kappa ;\tau ,\breve{r},\hat{x}%
^{2})\simeq \zeta _{2}(\hbar ,\tau ,\breve{r},\hat{x}^{2})(1+\kappa \chi
_{2}(\hbar ,\tau ,\breve{r},\hat{x}^{2})),\mbox{ we can consider }\ \eta
_{2}(\tau )=\ \eta _{1}(\tau ); \\
\eta _{4}(\tau ) &\simeq &\ \eta _{4}(\hbar ,\kappa ;\tau ,\breve{r},\hat{x}%
^{2},\hat{x}^{3})\simeq \zeta _{4}(\hbar ,\tau ,\breve{r},\hat{x}^{2},\hat{x}%
^{3})(1+\kappa \ \chi _{4}(\hbar ,\tau ,\breve{r},\hat{x}^{2},\hat{x}^{3})),
\\
\ ^{\shortmid }\eta ^{6}(\tau ) &\simeq &\ ^{\shortmid }\eta ^{6}(\hbar
,\kappa ;\tau ,\breve{r},\hat{x}^{2},\hat{x}^{3},\hat{x}^{5})\simeq \
^{\shortmid }\zeta ^{6}(\hbar ,\kappa ;\tau ,\breve{r},\hat{x}^{2},\hat{x}%
^{3},\hat{x}^{5})(1+\kappa \ ^{\shortmid }\chi ^{6}(\hbar ,\kappa ;\tau ,%
\breve{r},\hat{x}^{2},\hat{x}^{3},\hat{x}^{5})), \\
\ ^{\shortmid }\eta ^{8}(\tau ) &\simeq &\ ^{\shortmid }\eta ^{8}(\hbar
,\kappa ;\tau ,\breve{r},\hat{x}^{2},\hat{x}^{3},\hat{x}^{5},p_{7})\simeq \
^{\shortmid }\zeta ^{8}(\hbar ,\kappa ;\tau ,\breve{r},\hat{x}^{2},\hat{x}%
^{3},\hat{x}^{5},p_{7})(1+\kappa \ ^{\shortmid }\chi ^{8}(\hbar ,\kappa
;\tau ,\breve{r},\hat{x}^{2},\hat{x}^{3},\hat{x}^{5},p_{7})).
\end{eqnarray*}

Using formulas (\ref{genf2}), we can extract solutions with rotoid spacetime
configurations determined by nonassociative star product R-flux deformation
(considering $\chi $-polarizations), or to compute volume forms (\ref{volumf}%
) for $\eta $-polarizations.

\subsubsection{Bekenstein-Hawking entropy of $\protect\tau $-running phase
space RN-AdS BEs configurations}

\label{ss533}For a subclass of nonholonomic configurations, the s-metrics (%
\ref{narnbhs}) define higher dimension BH and/or BE configurations with
conventional horizons which can be characterized variables in the framework
of generalized Bekenstein-Hawking thermodynamics \cite{bek1,bek2,haw1,haw2}.
For simplicity, we shall consider solutions for 6-d $\tau $-running
quasi-stationary configurations evolving in a 8-d phase space, for
simplicity, with trivial integration functions of type $\ _{1}^{\shortmid
}n_{k_{s}}=0$ and $\ _{2}^{\shortmid }n_{k_{s}}=0.$ The corresponding
nonlinear quadratic elements are parameterized in the form: 
\begin{eqnarray}
d\ _{\shortmid }^{\chi }s_{[6\subset 8d]}^{2}(\tau ) &=&e^{\psi
_{0}}(1+\kappa \ ^{\psi (\tau )}\ ^{\shortmid }\chi (\tau ))[\ \breve{g}_{1}(%
\breve{r})d\breve{r}^{2}+\breve{g}_{2}(\breve{r})(d\hat{x}^{2})]
\label{sol4of} \\
&&-\{\frac{4[\hat{\partial}_{3}(|\zeta _{4}(\tau )\breve{g}_{4}(\breve{r}%
)|^{1/2})]^{2}}{\ \breve{g}_{4}(\breve{r})|\int d\hat{x}^{3}\{\ _{2}\Im
(\tau )\hat{\partial}_{3}(\zeta _{4}(\tau )\ \breve{g}_{4}(\breve{r}))\}|}%
-\kappa \lbrack \frac{\hat{\partial}_{3}(\chi _{4}(\tau )|\zeta _{4}(\tau )\ 
\breve{g}_{4}(\breve{r})|^{1/2})}{4\hat{\partial}_{3}(|\zeta _{4}(\tau )\ 
\breve{g}_{4}(\breve{r})|^{1/2})}  \notag \\
&&-\frac{\int d\hat{x}^{3}\{\ _{2}\Im (\tau )\hat{\partial}_{3}[(\zeta
_{4}(\tau )\ \breve{g}_{4}(\breve{r}))\chi _{4}(\tau )]\}}{\int d\hat{x}%
^{3}\{\ _{2}\Im (\tau )\hat{\partial}_{3}(\zeta _{4}(\tau )\ \breve{g}_{4}(%
\breve{r}))\}}]\}\ \breve{g}_{3}(\mathbf{e}^{3}(\tau ))^{2}+\ \zeta
_{4}(\tau )(1+\kappa \ \chi _{4}(\tau ))\breve{g}_{4}(\breve{r})dt^{2} 
\notag
\end{eqnarray}%
\begin{eqnarray*}
&&-\{\frac{4[\hat{\partial}_{5}(|\ ^{\shortmid }\zeta ^{6}(\tau )\ \ \breve{g%
}^{6}|^{1/2})]^{2}}{~\breve{g}_{5}(\breve{r})|\int d\hat{x}^{5}\{\
_{3}^{\shortmid }\Im (\tau )\ ^{\shortmid }\partial ^{7}(\ ^{\shortmid
}\zeta ^{6}(\tau )~\breve{g}^{6})\}|}-\kappa \lbrack \frac{\ \hat{\partial}%
_{5}(\ ^{\shortmid }\chi ^{6}(\tau )|\ ^{\shortmid }\zeta ^{6}(\tau )\ 
\breve{g}^{6}|^{1/2})}{4\hat{\partial}_{5}(|\ ^{\shortmid }\zeta ^{6}(\tau
)\ \breve{g}^{6}|^{1/2})} \\
- &&\frac{\int d\hat{x}^{5}\{\ _{3}^{\shortmid }\Im (\tau )\ \hat{\partial}%
_{5}[(\ ^{\shortmid }\zeta ^{6}(\tau )\breve{g}^{6})\ ^{\shortmid }\chi
^{8}(\tau )]\}}{\int d\hat{x}^{5}\{\ _{3}^{\shortmid }\Im (\tau )\ \hat{%
\partial}_{5}[(\ ^{\shortmid }\zeta ^{6}(\tau )\breve{g}^{6})]\}}]\}\ \breve{%
g}_{5}(\breve{r})(\mathbf{e}^{5}(\tau ))^{2}+\ ^{\shortmid }\zeta ^{6}(\tau
)\ (1+\kappa \ ^{\shortmid }\chi ^{6}(\tau
))(dp_{6})^{2}+(dp_{7})^{2}-dE^{2},
\end{eqnarray*}%
where%
\begin{equation*}
\mathbf{e}^{3}(\tau )=d\hat{x}^{3}+[\frac{\hat{\partial}_{i_{1}}\int d\hat{x}%
^{3}\ _{2}\Im (\tau )\ \hat{\partial}_{3}\zeta _{4}(\tau )}{\breve{N}%
_{i_{1}}^{3}\ _{2}\Im (\tau )\hat{\partial}_{3}\zeta _{4}(\tau )}+\kappa (%
\frac{\hat{\partial}_{i_{1}}[\int d\hat{x}^{3}\ _{2}\Im (\tau )\hat{\partial}%
_{3}(\zeta _{4}(\tau )\chi _{4}(\tau ))]}{\hat{\partial}_{i_{1}}\ [\int d%
\hat{x}^{3}\ _{2}\Im (\tau )\hat{\partial}_{3}\zeta _{4}(\tau )]}-\frac{\hat{%
\partial}_{3}(\zeta _{4}(\tau )\chi _{4}(\tau ))}{\hat{\partial}_{3}\zeta
_{4}(\tau )})]\ \breve{N}_{i_{1}}^{3}dx^{i_{1}},
\end{equation*}%
\begin{eqnarray*}
\ \mathbf{e}^{5}(\tau ) &=&d\hat{x}^{5}+[\frac{\hat{\partial}_{i_{2}}\ \int d%
\hat{x}^{5}\ _{3}^{\shortmid }\Im (\tau )\ \ \hat{\partial}_{5}(\
^{\shortmid }\zeta ^{6}(\tau ))}{\ ^{\shortmid }\breve{N}_{i_{2}}^{5}\
_{3}^{\shortmid }\Im (\tau )\ \ \hat{\partial}_{5}(\ ^{\shortmid }\zeta
^{6}(\tau ))}+ \\
&&\kappa (\frac{\hat{\partial}_{i_{2}}[\int d\hat{x}^{5}\ _{3}^{\shortmid
}\Im (\tau )\ \ \hat{\partial}_{5}(\ ^{\shortmid }\zeta ^{6}(\tau )\ \ 
\breve{g}^{6})]}{\hat{\partial}_{i_{2}}\ [\int d\hat{x}^{5}\ _{3}^{\shortmid
}\Im (\tau )\ \ \hat{\partial}_{5}(\ ^{\shortmid }\zeta ^{6}(\tau ))]}-\frac{%
\ \hat{\partial}_{5}(\ ^{\shortmid }\zeta ^{6}(\tau )\ \ \breve{g}^{6})}{\ 
\hat{\partial}_{5}(\ ^{\shortmid }\zeta ^{6}(\tau ))})]\ \ ^{\shortmid }%
\breve{N}_{i_{2}}^{5}d\ ^{\shortmid }x^{i_{2}}.
\end{eqnarray*}

A subclass of solutions (\ref{sol4of}) generates $\tau $-families of rotoid
configurations in coordinates $(\breve{r},\hat{x}^{2},\hat{x}^{3})$ (as
nonholonomic deformations of the phase BH solution (\ref{pm5d8d})) if we
chose such generating functions: 
\begin{equation}
\ \ \chi _{4}(\tau )=\hat{\chi}_{4}(\tau ,\breve{r},\hat{x}^{2},\hat{x}%
^{3})=2\underline{\chi }(\tau ,\breve{r},\hat{x}^{2})\sin (\omega _{0}\hat{x}%
^{3}+\hat{x}_{0}^{3}),  \label{5dbe}
\end{equation}%
where $\underline{\chi }(\tau ,\breve{r},\hat{x}^{2})$ are smooth functions
(or constants), and $(\omega _{0},\hat{x}_{0}^{3})$ is a couple of
constants. In a conventional 5-d phase space on shells $s=1,2,3,$ trivially
imbedded into a 8-d phase space posses a distinct ellipsoidal type horizon
with respective eccentricity $\kappa $ a stated by the equations 
\begin{equation*}
\zeta _{4}(\tau )(1+\kappa \ \chi _{4}(\tau ))\breve{g}_{4}(\breve{r})=0%
\mbox{ i.e. }(1+\kappa \ \chi _{4})\breve{f}(\breve{r})=(1-\frac{\hat{m}}{%
\breve{r}^{2}}-\frac{\Lambda _{\lbrack 5]}}{6}\breve{r}^{2}+\frac{\hat{q}^{2}%
}{\breve{r}^{4}}+\kappa \ \chi _{4})=0,
\end{equation*}%
for $\zeta _{4}\neq 0.$ For small parametric deformations and configurations
with $-\frac{\Lambda _{\lbrack 5]}}{6}\breve{r}^{2}+ \frac{\hat{q}^{2}}{%
\breve{r}^{4}}\approx 0,$ we can approximate for a fixed $\tau _{0},$ $%
\breve{r}\simeq \hat{m}^{1/2}/(1- \frac{\kappa \ }{2}\hat{\chi}_{4}).$ These
are parametric formulas for a rotoid horizon defined by small gravitational
R-flux polarizations. In the limits of zero eccentricity, such e BE
configurations transform into a 5-d BH embedded into nonassociative 8-d
phase space.

Extending the concept of Bekenstein-Hawking entropy for phase spaces
determined by quadratic linear elements (\ref{pm5d8d}), we can define such
thermodynamic values (computations are similar to those for formulas
(8)-(15) \ in \cite{zhang15} but with different constants and following our
notations):%
\begin{eqnarray}
\ ^{0}\breve{S} &=&\frac{\ ^{0}\breve{A}}{4G_{[5]}}=\frac{\omega _{\lbrack
3]}\breve{r}_{h}}{4G_{[5]}}\mbox{ and }\ ^{0}\breve{T}=\frac{1}{2\pi \breve{r%
}_{h}}(\epsilon +2\frac{\breve{r}_{h}^{2}}{l_{[5]}^{2}})-\frac{2G_{[10]}^{2}%
\hat{Q}^{2}}{3\pi ^{9}l_{[5]}^{8}\breve{r}_{h}^{5}},\mbox{ for }  \notag \\
\hat{M} &=&\frac{3\omega _{\lbrack 3]}\hat{m}}{16\pi G_{[5]}}(\epsilon 
\breve{r}_{h}^{2}+\frac{\breve{r}_{h}^{4}}{l_{[5]}^{2}}+\frac{4G_{[5]}\hat{Q}%
^{2}l_{[5]}^{2}}{3\pi ^{2}\breve{r}_{h}^{2}}),  \label{bhth58}
\end{eqnarray}%
where $\breve{r}_{h}$ and $^{0}\breve{A}$ are, respectively the horizon and
area of horizon of 5-d BH, $G_{[5]}=G_{[10]}/(\pi ^{3}l_{[5]}^{5})$ and $%
G_{[10]}=\ell _{p}^{8}.$ Using these formulas for rotoid deformations $%
\breve{r}_{h}\rightarrow \hat{m}^{1/2}/(1-\frac{\kappa \ }{2}\hat{\chi}_{4})$
and $^{0}\breve{A}$ $\rightarrow \ ^{rot}\breve{A},$ with $\hat{\chi}%
_{4}(\tau )$ (\ref{5dbe}), we compute for respective BE configurations:%
\begin{equation}
\breve{S}(\tau )=\ ^{0}\breve{S}(1+\frac{\kappa \ }{2}\hat{\chi}_{4}(\tau ))%
\mbox{ and }\breve{T}(\tau )=\ ^{0}\breve{T}+\kappa \left( -\frac{\epsilon }{%
4\pi \breve{r}_{h}}+\frac{\breve{r}_{h}}{2\pi l_{[5]}^{2}}-\frac{%
5G_{[10]}^{2}\hat{Q}^{2}}{3\pi ^{9}l_{[5]}^{8}\breve{r}_{h}^{5}}\right) \hat{%
\chi}_{4}(\tau ).  \label{rotbhbhthermv}
\end{equation}%
The modified Hawking temperatures $\breve{T}(\tau )$ and$\ ^{0}\breve{T}$
are stated by requiring the absence of the potential conical singularity of
the Euclidean BH at the horizon in the phase space.

\subsubsection{G. Perelman thermodynamics of nonassociative flows of phase
RN-AdS BHs}

\label{ss534}We can not apply the Bekenstein-Hawking thermodynamic paradigm
in order to characterize physical properties of general classes of
quasi-stationary solutions of type (\ref{narnbhs}) and/or (\ref{sol4of})
excepting very special cases of nonassociative deformations, for instance,
to BE configurations of type (\ref{5dbe}). The G. Perelman approach is more
general and allows to define and compute statistical thermodynamic variables
of type (\ref{thvcann}). Let us sketch how to compute such values for any
data $\ ^{\shortmid }\mathbf{\breve{g}}_{\alpha _{s}}$ (\ref{pm5d8d}), $\
_{s}^{\shortmid }\Im (\tau )$ related via nonlinear symmetries (\ref%
{nonlinsymr}) to $\ \Lambda _{\lbrack 5]},$ and nontrivial (on shells $%
s=1,2,3,$ see also formulas (\ref{genf2})). For simplicity, we can consider
the same value of cosmological constant on such shells when 
\begin{equation}
|\ \Lambda _{\lbrack 5]}\ \eta _{4}(\tau )\breve{g}_{4})|=|\ \Lambda
_{\lbrack 5]}\ \zeta _{4}(\tau )\breve{g}_{4}|(1-\kappa \chi _{4}(\tau )),|\
\Lambda _{\lbrack 5]}~^{\shortmid }\eta ^{6}(\tau )~^{\shortmid }\breve{g}%
^{6}|=|\ \Lambda _{\lbrack 5]}~^{\shortmid }\zeta ^{6}(\tau )~^{\shortmid }%
\breve{g}^{6}|(1-\kappa ~^{\shortmid }\chi ^{6}(\tau ))  \label{aux6}
\end{equation}%
for a a subclass of s-metrics of type (\ref{sol4of}). We obtain such
thermodynamic functionals: 
\begin{eqnarray}
\ _{s}^{\shortmid }\widehat{\mathcal{W}}_{\kappa }^{\star }(\tau )
&=&\int\nolimits_{\tau ^{\prime }}^{\tau }\frac{d\tau }{32(\pi \tau )^{4}}%
\frac{2\tau \Lambda _{\lbrack 5]}^{2}-1}{\Lambda _{\lbrack 5]}^{2}}\ _{\eta
}^{\shortmid }\mathcal{\breve{V}}(\tau ),\ \ _{s}^{\shortmid }\widehat{%
\mathcal{Z}}_{\kappa }^{\star }(\tau )=\exp \left[ \int\nolimits_{\tau
^{\prime }}^{\tau }\frac{d\tau }{(2\pi \tau )^{4}}\frac{1}{\Lambda _{\lbrack
5]}^{2}}\ _{\eta }^{\shortmid }\mathcal{\breve{V}}(\tau )\right] ,
\label{thvcannpd2} \\
\ _{s}^{\shortmid }\widehat{\mathcal{E}}_{\kappa }^{\star }(\tau )
&=&-\int\nolimits_{\tau ^{\prime }}^{\tau }\frac{d\tau }{64\pi ^{4}\tau ^{3}}%
\frac{\tau \Lambda _{\lbrack 5]}-1}{\Lambda _{\lbrack 5]}^{2}}\ _{\eta
}^{\shortmid }\mathcal{\breve{V}}(\tau ),\ _{s}^{\shortmid }\widehat{%
\mathcal{S}}_{\kappa }^{\star }(\tau )=-\int\nolimits_{\tau ^{\prime
}}^{\tau }\frac{d\tau }{64(\pi \tau )^{4}}\frac{\tau \Lambda _{\lbrack 5]}-2%
}{\Lambda _{\lbrack 5]}^{2}}\ _{\eta }^{\shortmid }\mathcal{\breve{V}}(\tau
).  \notag
\end{eqnarray}%
In these formulas, we use the running phase space volume functional 
\begin{equation}
\ _{\eta }^{\shortmid }\mathcal{\breve{V}}(\tau )=\int_{\ _{s}^{\shortmid }%
\widehat{\Xi }}\ ^{\shortmid }\delta \ _{\eta }^{\shortmid }\mathcal{V}(\
_{s}^{\shortmid }\Im (\tau ),~^{\shortmid }\breve{g}_{\alpha _{s}}),%
\mbox{
for }s=1,2,3.  \label{aux7}
\end{equation}%
Above presented values are determined by prescribed classes of generating $%
\eta $-functions (\ref{aux6}), effective generating sources $\
_{s}^{\shortmid }\Im (\tau ),$ coefficients of a prime s-metric $\
^{\shortmid }\breve{g}_{\alpha _{s}}$ and nonholonomic distributions
defining the hyper-surface $\ _{s}^{\shortmid }\widehat{\Xi },$ when the
volume forms are computed 
\begin{eqnarray*}
&&\ ^{\shortmid }\delta \ ^{\shortmid }\mathcal{V}(\tau )=\ ^{\shortmid
}\delta \ ^{\shortmid }\mathcal{V}(\ \ _{s}^{\shortmid }\Im (\tau ),\
\Lambda _{\lbrack 5]},\ ^{\shortmid }\eta _{\alpha _{s}}(\tau )~^{\shortmid }%
\breve{g}_{\alpha _{s}})=\frac{1}{\Lambda _{\lbrack 5]}^{2}}\ ^{\shortmid
}\delta \ _{\eta }^{\shortmid }\mathcal{V}(\ _{s}^{\shortmid }\Im (\tau ),\
^{\shortmid }\breve{g}_{\alpha _{s}}),\mbox{ where } \\
&&\ ^{\shortmid }\delta \ _{\eta }^{\shortmid }\mathcal{V}(\ _{s}^{\shortmid
}\Im (\tau ),~^{\shortmid }\breve{g}_{\alpha _{s}})=\ ^{\shortmid }\delta \
_{\eta }^{1}\mathcal{V}[_{1}\Im (\tau ),\eta _{2}(\tau )\ \breve{g}%
_{2}]\times \ ^{\shortmid }\delta \ _{\eta }^{2}\mathcal{V}[_{2}\Im (\tau
),\eta _{4}(\tau )\ \breve{g}_{4}]\times \ ^{\shortmid }\delta \ _{\eta }^{3}%
\mathcal{V}[_{3}\Im (\tau ),~^{\shortmid }\eta ^{6}(\tau )~^{\shortmid }%
\breve{g}^{6}].
\end{eqnarray*}

G. Perelman thermodynamic variables (\ref{thvcannpd2}) can be computed in
explicit form if we prescribe certain nonholonomic distributions in $\
_{\eta }^{\shortmid }\mathcal{\breve{V}}(\tau )$ (\ref{aux7}) which results
in physically viable geometric thermodynamic models. For instance, we can
study solutions with more general $\kappa $-parametric deformations than
those determined by rotoid generating functions (\ref{5dbe}). Prescribing
generating sources $\ _{s}^{\shortmid }\Im (\tau )$ and $\eta $-generating
functions for $\ ^{\shortmid }\delta \ _{\eta }^{\shortmid }\mathcal{V},$ we
define explicit thermodynamic models of deformed phase space BH objects
propagating on a local temperature like parameter $\tau $ like, for
instance, in the hydrodynamic of moving media. This analogy is very rough
because instead of hydrodynamic flow equations we consider nonassociative
geometric flow equations and certain special classes of solutions. The
physical interpretation of corresponding entropy and temperature from (\ref%
{thvcannpd2}) is different from that for the hyper-surface variables (\ref%
{rotbhbhthermv}) in the Bekenstein-Hawking approach. Finally we note that
the effective running of thermodynamic variables on shell cosmological
constants in (\ref{thvcannpd1}) and (\ref{thvcannpd2}) is very different. We
have to analyze such different implications, for instance, in extending the
swampland conjecture for nonassociative BHs and $\tau $-running
quasi-stationary solutions.

\section{Extending swampland conjectures for nonassociative geometric flows
and \newline nonholonomic BHs deformations}

\label{sec6} Connections between the swampland conjectures in string theory,
QG, exact solutions in MGTs , effective QFTs and (non) commutative geometric
flow equations, with a number of examples with BH solutions, were studied in
a series of works \cite%
{luest19,kehagias19,biasio20,biasio21,lueben21,biasio22}. We also cite such
papers and references therein for recent developments and applications in
modern gravity, cosmology and astrophysics; and studies of various types of
Hawking--Page phase transitions under Ricci, Yamanabe, Ricci-Bourguignon
flows etc. It should be remembered that the general aim of the swampland
program \cite{vaf05,oog06,pal19,gomez19} is to elaborate on mathematical and
physical criteria for distinguishing (low-energy) effective classical and
quantum theories, QG and MGTs, which can be completed in the UV, from other
classes of theories which cannot.

\vskip5pt The approach to nonassociative geometry and gravity \cite%
{blumenhagen16,aschieri17} and nonholonomic generalizations with the AFCDM
for constructing exact/parametric solutions in nonassociative/noncommutative
Ricci flow theories and MGTs \cite%
{partner01,partner02,partner03,partner04,vacaru20,ib19,bubuianu19} was
developed for star product R-flux deformation models in string/M-theory.
Respective fundamental nonassociative geometric flow and (modified) vacuum
gravitational (Ricci soliton) equations consist very sophisticate systems of
coupled nonlinear PDEs encoding nonassociative data as effective sources and
generic off-diagonal terms of nonholonomic shell adapted metrics. Such
equations can be integrated in very general exact/parametric forms, for
instance, for quasi-stationary configurations running on $\tau $-flow
parameter as we proved in section \ref{sec4}, see also related results
examples in section 4 of \cite{partner04}. Positively, for certain classes
of solutions with respective nonlinear symmetries and $\tau $-running, or
prescribed/fixed, shell effective cosmological constants, such solutions and
respective nonassociative geometric thermodynamic models can be constructed
to be compatible with certain refined versions of the swampland conjectures
(both on higher dimension spacetimes and/or phase space). Nevertheless,
generalized classes of solutions do not obligatory involve certain effective
(running) cosmological constants, for instance, for nonassociative BH
deformations, see subsections \ref{ss524} and \ref{ss534}. Even for such
nonassociative geometric flow configurations, we can always compute the
modified G. Perelman F- and W-functionals and related thermodynamic
variables, the issues of compatibility or not compatibility with the
swampland ideas and conjectures have to be studied respectively for any
explicit class of solutions.

\vskip5pt In this section, we analyze and discuss two kinds of
nonassociative generalized swampland and Ricci flow conjectures. First,
there are outlined refined versions on the phase space of typical results
obtained for high dimensional (A)dS spaces, for instance, modifications of
the Black Hole Entropy Distance Conjecture (BHEDC) to the case of
nonassociative black ellipsoid, BE, configurations etc. Such constructions
are possible for nonassociative $\tau $-running quasi-stationary solutions
with conventional hyper-surface horizons and associated modified
Bekenstein-Hawking thermodynamic models studied above in subsections \ref%
{ss523} and \ref{ss533}. Second, we elaborate on some main points of this
paper when the swampland program is extended to nonassociative geometric
flow and MGTs using the concept of W-entropy and G. Perelman thermodynamics.
There are considered certain well-defined conditions when positively
swampland and nonassociative Ricci flow conjectures can formulated for such
generalizations. We also analyze explicit examples of $\tau $-running
quasi-stationary solutions when it is not clear and we can't conclude
without an additional analysis if such low-energy effective (encoding
nonassociative data) geometric flow and gravity models can be completed into
QG in the UV and distinguished from those that cannot.

\subsection{The generalized distance conjecture, nonassociative Ricci flows
and gravity}

\label{ss61}The goal of this subsection is to show how the swampland,
gradient flows and infinite distance conjectures \cite{luest19,kehagias19}
can be extended on nonassociative phase spaces for solutions of
nonassociative Ricci flow/soliton equations derived in sections \ref{ssngfeq}
and \ref{ssparamflow} and $\kappa $-parameterized in the form (\ref%
{nonassocgeomfl}).

\subsubsection{Phase space generalized and Weyl distances, and the AdS
distance conjecture}

For a phase space with star product R-flux nonassociative deformation, $_{s}%
\mathcal{M}\rightarrow \ _{s}^{\star }\mathcal{M}$, restricted to a finite
region $\ _{s}^{\star }\mathcal{U}\subset \ _{s}^{\star }\mathcal{M},$ we
consider a $\tau $-family of\ symmetric s-metrics $\ _{s}^{\shortmid }%
\mathbf{g}(\tau )$ following Convention 2 (\ref{conv2s}). There are used $%
\kappa $-linear parametric decomposition when $\ _{\star }^{\shortmid }%
\mathfrak{\check{g}}_{\alpha _{s}\beta _{s}}^{[0]}(\tau )=\
_{\star}^{\shortmid }\mathbf{g}_{\alpha _{s}\beta _{s}}(\tau )=\ ^{\shortmid
}\mathbf{g}_{\alpha _{s}\beta _{s}}(\tau )$, for star product flows $%
\star_{s}(\tau )$ determined by s-adapted frames $\ ^{\shortmid }\mathbf{e}%
_{i_{s}}(\tau )$ in (\ref{starpn}) and canonical data $(\ _{s}^{\shortmid }%
\mathbf{g}(\tau ),\ _{s}^{\shortmid }\widehat{\mathbf{D}}(\tau )).$ We can
consider arbitrary variations of such a s-metric, or to choose a class of
solutions of nonassociative geometric flow equations (\ref{nonassocgeomfl})
for the data $[\ _{\star }^{\shortmid}\mathfrak{g}_{\alpha _{s}\beta
_{s}}(\tau ),\ _{s}^{\shortmid }\widehat{\mathbf{D}}^{\star }(\tau )]$, when
the nonsymmetric $\ _{\star }^{\shortmid }\mathfrak{g}_{\alpha _{s}\beta
_{s}}$ are computed using $\kappa $-linear parameterizations (\ref{aux40a}%
)--(\ref{aux40aa}). On $\ _{s}^{\star }\mathcal{U}$, a path $\gamma (\tau
),\tau _{i}\leq $ $\tau \leq \tau _{f},$ is considered (for which a proper
metric distance $\aleph $ is defined by $\ ^{\shortmid }\mathbf{g}_{\alpha
_{s}\beta _{s}}(\tau ))$ (\ref{sdm}).\footnote{%
In this work, we write $\aleph $ instead of $\bigtriangleup $ considered,
for instance, in formula (1) from \cite{kehagias19} for formulating the
general distance conjecture. On nonassociative phase spaces, we elaborate a
different system of notations when the symbol $\ ^{\star }\widehat{%
\bigtriangleup }$ is the Laplace operator constructed for $\ _{s}^{\shortmid}%
\widehat{\mathbf{D}}^{\star }.$ Here we also note that in field theories,
the parameter $\tau $ can be an arbitrary one, parameterizing some
curves/geodesic etc. In this section, we are interested in models when $\tau$
can be always related to a geometric flow / temperature like parameter.} We
introduce 
\begin{equation}
\ _{g}^{s}\aleph =\aleph \lbrack \ ^{\shortmid }\mathbf{g}_{\alpha _{s}\beta
_{s}}(\tau ),\ _{s}^{\shortmid }\Lambda (\tau )]=\mathcal{C}%
\int\nolimits_{\tau _{i}}^{\tau _{f}}\left( \frac{1}{\mathcal{V}_{\mathcal{U}%
}}\int_{\mathcal{U}}\ ^{\shortmid }\mathbf{g}^{\alpha _{s}\mu _{s}}\
^{\shortmid }\mathbf{g}^{\beta _{s}\nu _{s}}\frac{\ \partial \ ^{\shortmid }%
\mathbf{g}_{\alpha _{s}\beta _{s}}}{\partial \tau }\ \ \frac{\partial \ \
^{\shortmid }\mathbf{g}_{\mu _{s}\nu _{s}}}{\partial \tau }d\ ^{\shortmid }%
\mathcal{V}ol(\tau )\right) ^{1/2}d\tau  \label{mfdist}
\end{equation}%
for the volume form $d\ ^{\shortmid }\mathcal{V}ol(\tau )$ (\ref{volform}),
where the constant $\mathcal{C}\sim \mathcal{O}(1)$ depending on dimension
of $\mathcal{U}$ and $\mathcal{V}_{\mathcal{U}}$ is the volume of $\mathcal{U%
}\rightarrow \ _{s}^{\star }\mathcal{U}.$ In this work, we consider that $\
_{g}^{s}\aleph $ is determined by a $\tau $-family of $\kappa $-parametric
solutions of geometric flow equations (\ref{nonassocgeomflefcc}) for some
prescribed running shell cosmological constants $\
_{s}^{\shortmid}\Lambda(\tau ).$

For phase spaces, the generalized distance conjecture states that there must
be an infinite tower of states with an effective mass scale $\
^{\shortmid}m(\tau ),$ when 
\begin{equation}
\ ^{\shortmid }m(\tau _{f})\sim \ ^{\shortmid }m(\tau _{i})e^{-\hat{\alpha}%
|\ _{g}^{s}\aleph |},\mbox{ where }\hat{\alpha}\sim \mathcal{O}(1).
\label{masstf}
\end{equation}%
This formula was introduced without labels "$\ ^{\shortmid }$" for
corresponding notations for higher dimension spacetimes in \cite%
{luest19,kehagias19}, when 
\begin{equation}
m\sim M_{p}e^{-\alpha |\ _{g}\aleph |}.  \label{masst}
\end{equation}%
In field theories, it is considered that generically the masses $m(\tau
_{i}) $ at some initial values $\tau _{i}$ are of order of $M_{p}$ (the
Planck mass constant, determined by $\hbar ).$ We have to introduce such
constants with additional labels also for phase spaces $\ _{s}^{\star }%
\mathcal{M}$, when we elaborate on (nonassociative and noncommutative)
quantum geometric flow models and/or QFTs. On 4-d spacetime base with shells 
$s=1,2,$ the formula $\ ^{\shortmid }m$ (\ref{masstf}) still contains
nonassociative star product and R-flux data encoded into $\ _{g}^{s}\aleph .$
Only for very special nonholonomic configurations, non involving the effects
of possible off-diagonal and generalized effective sources, and omitting $%
\kappa $-parametric terms, we may follow the assumptions for $m$ (\ref{masst}%
). We note that the mass scale $m$ in associative/commutative classical and
quantum theories states a natural cut-off above which the effective field
description is not valid. In such cases, for large distance variations in
the space of metrics, when $|\ _{g}\aleph |\rightarrow \infty ,$ we get for
QG models a respective massless tower of states. This breaks down/
invalidates the effective field theory description. In modern literature 
\cite{vaf05,oog06,pal19,gomez19,luest19,kehagias19}, it is argued that for
respective conditions such theories belong to the swampland. The conjectures
and related results have to be revised for nonassociative geometric flow and
gravity/matter field theories when exact/parametric generic off-diagonal
solutions of effective nonholonomic Ricci flow and/or Ricci soliton
equations are considered following the AFCDM. We formulate:

\vskip5pt \textbf{The generalized distance conjecture and claim for
nonassociative phase spaces, CCL1:}

\begin{description}
\item[a)] \textsf{\ Conjecture:} \emph{For nonassociative geometric flow,
gravity, and classical field theories, and respective QG and QGTs encoding
star product and R-flux data from string/M-theory, there are classes of
exact/parametric solutions with nonlinear phase space symmetries of type (%
\ref{nonlinsym}) connecting effective sources to effective shell
cosmological constants (for $\tau $-running or fixed $\tau _{0}$
quasi-stationary and/or locally anisotropic cosmological configurations with
a time like coordinate ), when the corresponding effective
geometric/physical models belong to the swampland.}

\item[b)] \textsf{\ Claim:} \emph{In a general context, nonassociative/
noncommutative/ nonholonomic / generic off-diagonal/ generalized (non)
linear connection modifications of gravity theories contain also large
classes of exact/parametric solutions (involving, or not, nonlinear
symmetries of type (\ref{nonlinsym})) defining effective $\tau $-running, or
fixed $\tau_{0})$, configurations which are physically well defined and
characterized by a respective modified G. Perelman thermodynamics with
variables of type (\ref{nagthermodvalues}). We have to analyze additionally
for which conditions such models belong or not to the swampland.}
\end{description}

It should be noted here that above a) \textsf{Conjecture} is a
nonassociative phase space version of the generalized distance conjecture
for (associative and commutative) d-dimensional manifolds studied in \cite%
{luest19,kehagias19}. In this work, we show that such a conjecture (and
related ones, see next subsections) can be formulated for respective
conditions on nonassociative geometric flows when $\ _{g}^{s}\aleph $(\ref%
{mfdist})$\rightarrow \infty $ in the space of $\tau $-running, or fixed $%
\tau _{0},$ pase space s-metrics. Nevertheless, using the AFCDM, we can
construct large classes of exact/parametric solutions when $\ _{g}^{s}\aleph 
$ is "freezed" someway, for respective nonholonomic configurations, and do
not result in an infinite tower of effective phase masses $\ ^{\shortmid }m$
(\ref{masstf}). This reflects a more rich propriety of respective classes of
solutions of corresponding systems of nonlinear PDEs (describing the
geometric flow evolution and/or corresponding field equations) when generic
off-diagonal interactions and evolution scenarios are modeled in a more
general way than in the case when the analysis is performed in the framework
of diagonalizable solutions and ODEs. Such results (we provided very general
classes of quasi-stationary solutions and explicit examples in the previous
section) motivate the b) \textsf{\ Claim}. So, a (nonassociative) modified
geometric flow, gravity, and/or classical/quantum field theory may involve
large classes of solutions, and respective effective models which belong, or
not to the swampland. It depends on the type of nonholonomic constraints,
nonlinear symmetries and nonlinear interactions, circumstances of the
considered coupling, prescribed generating functions and effective sources,
and integration functions/constants. In all cases, for various models with
solutions satisfying the conditions of above stated CCL1, we are able to
elaborate on modified G. Perelman thermodynamic models (\ref%
{nagthermodvalues}), for general assumptions on nonassociative geometric
flows; or (\ref{nagthermodvaluesp}), for (\ref{pm5d8d}), thermodynamic
variables; and (\ref{thvcannpd1}), or (\ref{thvcannpd2}), for explicit
examples of general nonassociative BH/BE deformations.

\vskip5pt At the next step, we analyze how keeping the conditions of CCL1,
with notions formulated for the (phase) spaces of s-metrics, it is possible
to formulate a respective conjecture and claim involving the so-called Weyl
distance \cite{luest19} but generalized for nonassociative phase spaces.
Specifically, we can consider an external conformal re-scaling of s-metrics
on $_{s}\mathcal{M}$, 
\begin{equation}
\ ^{\shortmid }\mathbf{\tilde{g}}_{\alpha _{s}\beta _{s}}(\bar{\tau})=e^{2%
\bar{\tau}}\ ^{\shortmid }\mathbf{g}_{\alpha _{s}\beta _{s}},  \label{wresc}
\end{equation}%
where the (re-scaling) parameter $\bar{\tau}$, in general, is different from
a geometric flow/temperature like parameter $\tau .$ Introducing (\ref{wresc}%
) in (\ref{mfdist}), we compute $\ _{g}^{s}\bar{\aleph}\simeq \bar{\tau}_{f}-%
\bar{\tau}_{i}\simeq \bar{\tau}.$ Following a) \textsf{Conjecture,} there is
a corresponding tower of phase space states $\tilde{m}$ (\ref{masstf}) with
a scale of effective masses as $\tilde{m}\sim e^{-\hat{\alpha}\bar{\tau}}.$
We obtain light masses in respective towers if $\ _{g}^{s}\bar{\aleph}\simeq 
\bar{\tau} \rightarrow \infty $ (we can consider also an opposite large
distance limit, $\ _{g}^{s}\bar{\aleph}\simeq \bar{\tau}\rightarrow -\infty$%
, when light states are with masses $\tilde{m}^{\prime }\sim e^{\hat{\alpha}%
\bar{\tau}}$). For a Weyl re-scaling, we can consider variations of the
cosmological constant, $\ ^{\shortmid }\bar{\Lambda}_{i}\leq \ ^{\shortmid }%
\bar{\Lambda}(\bar{\tau}) \leq \ ^{\shortmid}\bar{\Lambda}_{f}$ for
corresponding families of (A)dS phase space vacua. We can express the $\bar{%
\tau}$-transforms of the cosmological constant for an arbitrary dimension $%
\check{d}$ of $_{s}\mathcal{M}$ ( in this work, typical constructions are
performed for $\check{d}$ =8), 
\begin{equation}
\ ^{\shortmid }\bar{\Lambda}=-\frac{1}{2}(\check{d}-1)(\check{d}%
-2)M_{p}^{2}e^{-2\bar{\tau}}.  \label{weylcosmc}
\end{equation}%
The phase space metric distance between any initial, $\ ^{\shortmid }\Lambda
_{i}, $ and finial, $\ ^{\shortmid }\Lambda _{f},$ values of such running
cosmological constant can be approximated 
\begin{equation}
\ _{g}^{s}\bar{\aleph}\simeq \bar{\tau}_{f}-\bar{\tau}_{i}\simeq \log (\
^{\shortmid }\Lambda _{i}/\ ^{\shortmid }\Lambda _{f}),  \label{logsc}
\end{equation}%
which states that the limit $\ ^{\shortmid }\Lambda _{f}\rightarrow 0$ is at
infinite distance with respect to the Weyl re-scaling (\ref{wresc}) of an
(A)dS phase s-metric. This corresponds to the limit $\bar{\tau}\rightarrow
\infty .$

Combining the behaviour determined by the Weyl re-scaling formulas (\ref%
{wresc}) - (\ref{logsc}) and \textbf{CCL1}, we extend the AdS Distance
conjecture from \cite{luest19,kehagias19} in such a form:

\vskip5pt \textbf{The AdS distance conjecture and claim for nonassociative
phase spaces, CCL2:}

\begin{description}
\item[a)] \textsf{\ Conjecture:} \emph{For models of (non) associative phase
space geometric flow and gravity theories, and QG on a $\check{d}$%
-dimensional phase space $_{s}\mathcal{M}$ with cosmological constant $\
^{\shortmid }\bar{\Lambda}$ (\ref{weylcosmc}), there exists an infinite
tower of phase space states with mass scale $\ ^{\shortmid }m$ which, as $\
^{\shortmid }\Lambda \rightarrow 0,$ behave as $\ ^{\shortmid }m$ $\sim |\
^{\shortmid }\Lambda |^{\bar{\alpha}},$ where $\bar{\alpha}$ is a positive
order-one constant. Such a behavior can be described by phase space
s-metrics (\ref{pm5d8d}) and their nonassociative quasi-stationary
deformations encoding star product and R-flux data from string/M-theory with
nonlinear phase space symmetries of type (\ref{nonlinsym}) connecting
effective sources to effective shell cosmological constants $\
_{s}^{\shortmid }\Lambda (\tau )=\ ^{\shortmid }\bar{\Lambda}.$ }

\item[b)] \textsf{\ Claim:} \emph{In general, in QG models with
nonassociative/ noncommutative/ nonholonomic / generic off-diagonal/
generalized (non) linear connection modifications, there are large classes
of exact/parametric solutions (involving, or not, nonlinear symmetries of
type (\ref{nonlinsym}) when the approximation (\ref{logsc}) is not valid.
Nevertheless, corresponding classes of solutions defining effective $\tau $%
-running, or fixed $\tau _{0}$, configurations, are physically well defined
and characterized by a respective modified G. Perelman thermodynamics with
variables (\ref{nagthermodvalues})}.
\end{description}

We formulate the b) \textsf{\ Claims } in \textbf{CCL2} because for (non)
associative geometric flow theories and gravity, and their generic
off-diagonal solutions, the conformal symmetry of s-metric is not a general
property. In some models, one speculates on possible duality between AdS and
dS configurations (see footnote 3 in \cite{kehagias19} on necessary exchange
of the mass-towers and powers of cosmological constants). Such duality
properties and re-scaling behaviour are not important for the procedure of
generating exact/parametric solutions using the AFCDM.

Finally, it should be noted that we can also consider a "dual" infinite
distance limit when the associated curvatures of phase space/ spacetime and
(effective) cosmological constants became very large for $\bar{\tau}%
\rightarrow -\infty .$ Such conditions for the string/M-theory and various
coupling constants are analyzed in section III of \cite{kehagias19}. In
general, it is not clear if a tower of light states may appear for such
models large curvature/ cosmological constants. Similar conclusions can be
drawn for nonassociative phase models determined, for instance, by diagonal
s-metrics of type (\ref{pm5d8d}) and certain $\kappa $-parametric
deformations. For more general classes of solutions, limits of type $\bar{%
\tau}\rightarrow -\infty $ may be not obligatory correlated to certain large
curvature/cosmological constants values because of generic off-diagonal
interactions and imposed nonholonomic constraints on the geometric flow
evolution and/or gravitational and matter field dynamics. Such issues have
to be investigated for any class of well-defined physical solutions
determined by respective generating functions and effective sources,
integration functions and prescribed (non) linear symmetries.

\subsubsection{Nonassociative Ricci flows and infinite phase space distance
conjectures and claims}

For associative and commutative Riemannian metrics, the geometric flow
equations imply the properties that for positive Ricci curvature the
manifold is contracting and for regions of negative curvature the manifold
is extending. Such properties exist for space like configurations in
pseudo-Riemannian geometry and nonassociative generalizations when, in
general, the evolution scenarios depend on the type of solutions for
respective systems of nonlinear PDEs. Let us begin with some properties of
the associative and commutative R. Hamilton equations for the LC-connection $%
\nabla ,$ 
\begin{equation}
\frac{\partial \mathbf{g}(\tau )}{\partial \tau }=-2\ \mathcal{R}ic[\nabla
](\tau ).  \label{hameqst}
\end{equation}%
The Ricci flow ends in certain fixed points $\tau _{\odot },$ when 
\begin{equation*}
\frac{\partial \mathbf{g}(\tau )}{\partial \tau }\mid _{\tau =\tau _{\odot
}}=0\mbox{ and }\mathcal{R}ic(\tau _{\odot })=0,
\end{equation*}%
i.e. a fixed point defines a flat spacetime with vanishing curvature and
zero cosmological constant, $\Lambda =0$. Because a metric with vanishing
cosmological constant is at infinite Weyl distance in the space of all AdS
metrics, it was conjectured \cite{luest19,kehagias19} that in QG on a family
of background metrics $\mathbf{g}(\tau)$ under Ricci flows "there exists an
infinite tower of states which become massless when following the flow
towards a fixed point $\mathbf{g}_{\odot }=\mathbf{g}(\tau _{\odot })$ at
infinite distance".

In nonassociative geometric flow theory, the equations (\ref{hameqst}) can
be generalized in the form (\ref{nonassocheq}) for a family of
nonassociative metrics $\ ^{\shortmid }\mathfrak{g}^{\star }(\tau )$ and
respective family of nonassociative LC-connections $\ ^{\shortmid }\nabla
^{\star }(\tau )$ on phase space $\ ^{\star }\mathcal{M}.$ Such
nonassociative Ricci flow equations involve $\kappa $-parametric
decompositions of geometric objects. Corresponding decompositions of
nonsymmetric metric, canonical s-connection, and Ricci s-tensor structures
are given, for instance, in formulas (\ref{aux40a}), (\ref{paramscon}) and (%
\ref{driccicanonstar1}); for $\ ^{\shortmid }\nabla ,$ such formulas were
proven in abstract and coordinate forms for LC-configurations in \cite%
{aschieri17} and generalized in nonholonomic s-adapted forms in \cite%
{partner01,partner02}. The issue on definition of fixed points for
nonassociative Ricci flows is more sophisticate and requests a rigorous
analysis of ceratin families of exact/parametric solutions of nonlinear
systems of PDEs of type (\ref{nonassocgeomfl}) and their $\kappa $-linear
parametric decompositions (\ref{nonassocgeomflp}). For $\tau $-families of
configurations with nonlinear symmetries (\ref{nonlinsym}) and respectively
prescribed running/ fixed shell cosmological constants $\
_{s}^{\shortmid}\Lambda (\tau ),$ we can analyze fixed point properties of $%
\kappa $-parametric of geometric flow equations (\ref{nonassocgeomflefcc}),
when 
\begin{equation}
\frac{\partial \ _{s}^{\shortmid }\mathbf{g}(\tau )}{\partial \tau }\mid
_{\tau =\tau _{\odot }}=0 \mbox{ and } \ ^{\shortmid }\widehat{\mathbf{R}}%
_{\ \ \gamma _{s}}^{\beta _{s}}(\tau _{\odot },\ _{s}\Phi (\tau _{\odot }),\
_{s}^{\shortmid }\Im (\tau _{\odot }))={\delta }_{\ \ \gamma _{s}}^{\beta
_{s}}\ _{s}^{\shortmid }\Lambda (\tau _{\odot })=0.  \label{nafixpointc}
\end{equation}%
Here we note that fixed points determined by such equations for the
canonical s-connection do not define, in general, fixed points for
nonassociative geometric flows with $\ ^{\shortmid }\nabla ^{\star }(\tau ).$
Nevertheless, we can extract and study properties of LC-configurations
considering additional nonholonomic constraints (\ref{lccondnonass}),%
\begin{equation}
\ _{\star s}^{\shortmid }\widehat{\mathbf{Z}}(\tau _{\odot })=0,%
\mbox{ which
is equivalent to }\ _{s}^{\shortmid }\widehat{\mathbf{D}}_{\mid \
_{s}^{\shortmid }\widehat{\mathbf{T}}(\tau _{\odot })=0}^{\star }(\tau
_{\odot })=\ \ _{\star }^{\shortmid }\nabla (\tau _{\odot }).
\label{nafixpoinlc}
\end{equation}

Above formulas and observations lead to:

\vskip5pt \textbf{The fixed points of nonassociative geometric flows
conjecture and claim, CCL3:}

\begin{description}
\item[a)] \textsf{\ Conjecture:} \emph{Consider a QG model on a family of
background s-metric $\ _{s}^{\shortmid }\mathbf{g}(\tau )$ satisfying the $%
\kappa $-parametric nonassociative geometric flow equations (\ref%
{nonassocgeomflefcc}). There exists an infinite tower of phase space states
with zero effective masses when following the nonassociative geometric flow
evolution toward a fixed point $\ _{s}^{\shortmid }\mathbf{g}_{\odot }=\
_{s}^{\shortmid }\mathbf{g}(\tau _{\odot })$ at infinite distance. }

\item[b)] \textsf{\ Claim:} \emph{For QG models with nonassociative star
product and R-flux modifications, there are large classes of
exact/parametric solutions (involving, or not, nonlinear symmetries of type (%
\ref{nonlinsym})) when conditions of type (\ref{nafixpointc}) do not hold,
for instance, for LC-configurations (\ref{nafixpoinlc}). The conditions of
existence at fixed points, of infinite towers of phase spaces and behaviour
of corresponding effective masses must be analyzed correspondingly for any
class of solutions of $\kappa $-parametric of geometric flow equations (\ref%
{nonassocgeomflefcc}) or certain their equivalents. }
\end{description}

We can compute the metric distance (\ref{mfdist}) along the nonassociative
Ricci flows between a $\tau $ and a fixed point $\tau _{\odot }$ as function 
$_{g}^{s}\aleph (\tau ,\tau _{\odot })$ determined by a background solution $%
\ _{s}^{\shortmid }\mathbf{g}(\tau )$ (we can compute respective
nonsymmetric and symmetric components of the nonassociative s-metric) in
formula (\ref{mfdist}). For such configurations, possible associated towers
of phase space states as in a) \textsf{Conjecture} in \textbf{CCL3 } scale
as 
\begin{equation*}
\ ^{\shortmid }m(\tau _{\odot })\sim \ ^{\shortmid }m(\tau )e^{-\hat{\alpha}%
|\ _{g}^{s}\aleph (\tau ,\tau _{\odot })|}.
\end{equation*}%
For many examples, a fixed point will be of the form $\tau _{\odot }=\pm
\infty .$ The tower of nonholonomic states can only appear when the
nonassociative Ricci flow evolves to a fixed point. In general, a model/
off-diagonal solution is not necessary characterized by an infinite tower of
effective massless states in phase space. If $_{g}^{s}\aleph (\tau ,\tau
_{\odot })\rightarrow \pm \infty ,$ a fixed point/ solution can never be
reached. This means the transition between (non) associative geometric flow
theories/ Ricci solitons/ gravity and/or respective classes of solutions
along the geometric flows to the fixed point is discontinuous.

The a) \textsf{Conjecture} in \textbf{CCL3} was checked in section 2.1 of 
\cite{kehagias19} (see Conjecture A in that work) for simple cases of
associative and commutative Ricci flows, simple cases of Einstein spaces
(for instance, for (A) dS with non-zero cosmological constant) and string
theory and QG realizations. Similar behavior can be stated by phase space
s-metrics (\ref{pm5d8d}) and their nonassociative quasi-stationary
deformations encoding star product and R-flux data when nonlinear phase
space symmetries (\ref{nonlinsym}) are considered for connecting effective
sources to effective shell cosmological constants $\ _{s}^{\shortmid}\Lambda
(\tau ),$ in particular, we can approximate $\ _{s}^{\shortmid}\Lambda (\tau
)=\ ^{\shortmid }\bar{\Lambda}.$

\subsubsection{Gibbons-Hawking entropy, scalar curvatures, and generalized
distances for \newline nonassociative geometric flows}

\label{ss613} Let us consider an example of nonassociative Ricci flow
equations (\ref{nonassocgeomflefcc}) defined with nonlinear symmetries (\ref%
{nonlinsym}) resulting in a conventional phase space cosmological constant $%
\ _{s}^{\shortmid }\Lambda (\tau )=\ ^{\shortmid }\Lambda _{0}.$ Considering
classes of solutions involving conventional phase space horizons, we can
apply the concept of Gibbons-Hawking, GB, entropy \cite{gibbons77} and write 
$\ ^{\shortmid }\mathcal{S}_{GH}=(\ ^{\shortmid }\Lambda _{0})^{-1}$. For
quantum models, $\ ^{\shortmid }\mathcal{S}_{GH}$ can be related to the
dimension of the Hilbert space of $\ ^{\shortmid }\mathcal{H}$ for
observer's causal domain \cite{banks01,witten01}, when 
\begin{equation*}
\dim \ ^{\shortmid }\mathcal{H}=e^{1/\ ^{\shortmid }\Lambda _{0}}\rightarrow
\infty \mbox{ for } \ ^{\shortmid }\Lambda _{0}\rightarrow 0.
\end{equation*}%
This property was used in \cite{ooguri19} to relate the modified de Sitter
conjecture to the GB entropy of tower of massless states (for infinite
distance at $\ ^{\shortmid }\Lambda _{0}\rightarrow 0$). This property can
be extended to nonassociative phase spaces $\ ^{\star}\mathcal{M}$ for
respective conditions with $\ _{s}^{\shortmid }\Lambda (\tau )=\
^{\shortmid}\Lambda _{0}$ and, for instance, for s-metrics of type (\ref%
{pm5d8d}). As explained above, see formula (\ref{logsc}), we can write $\
_{g}^{s}\bar{\aleph}\simeq \log |\ ^{\shortmid }\Lambda _{0}|,$ which for de
Sitter configurations with positive cosmological constant results in $\
_{g}^{s}\bar{\aleph}\simeq \log \ ^{\shortmid }\mathcal{S}_{GH}.$ We
conclude that in the large distance limit the GB entropy becomes large and
this leads to a large tower of light states.

The concepts of Gibbons-Hawking and/or Bekenstein-Hawking entropies can be
applied only for very special classes of solutions (with conventional
horizons) in geometric flow and gravity theories. In next subsections we
shall consider modified G. Perelman functionals in order characterize more
general classes of generic off-diagonal solutions.

Here, we analyze another type of geometric properties of the nonassociative
phase space geometric distance. To work directly with formula $\
_{g}^{s}\aleph $ (\ref{mfdist}) is quite sophisticate and not completely
clear how to formulate certain consistent geometric criteria to decide which
families of background s-metrics can be used as backgrounds in a consistent
QG theory not involving an additional tower of light states, with or not
generic off-diagonal terms, in the infinite distance limit. That why we
define in this work alternative, more general and more abstract distance
measures, which can encode nonassociative star product R-flux contributions.

Let us consider nonassociative Ricci flows for a family of geometric data $%
[\ _{\star }^{\shortmid }\mathfrak{g}_{\alpha _{s}\beta _{s}}(\tau ),\
_{s}^{\shortmid }\widehat{\mathbf{D}}^{\star }(\tau )]$ subjected to the
condition that they are determined by a solution of nonassociative R.
Hamilton equations (\ref{nonassocgeomfl}). Such equations involve complex
like variables and canonical nonholonomic structures modeling generic
off-diagonal evolution processes and, for fixed Ricci soliton
configurations, gravitational interactions subjected to non-integrable
constraints. Important geometric and possible physical properties are
encoded in the corresponding families of canonical nonassociative Ricci
scalars $\ _{s}^{\shortmid }\widehat{\mathbf{R}}sc^{\star }(\tau )$ (\ref%
{ricciscsymnonsym}). In partner works \cite{partner01,partner02}, it is
elaborated a procedure of parametric decompositions of related fundamental
geometric s-objects $\ ^{\shortmid }\widehat{\mathbf{R}}_{\quad \alpha
_{s}\beta _{s}\gamma _{s}}^{\star \mu _{s}}$ (\ref{nadriemhopfcan}), $\
^{\shortmid }\widehat{\mathbf{R}}ic_{\alpha _{s}\beta _{s}}^{\star }$ (\ref%
{driccicanonstar1}) and $\ _{s}^{\shortmid }\widehat{\mathbf{R}}sc^{\star }$
(\ref{ricciscsymnonsym}) with $[01,10,11]:=\left\lceil \hbar ,\kappa
\right\rceil $ components, using the parametric form of the canonical
s-connections (\ref{paramscon}). This procedure is summarized in by formulas
(\ref{nadriemhopfcan}) - (\ref{ricciscsymnonsym}) in Appendix.

The Conjecture B1 from \cite{kehagias19} states that the distance $\ \aleph
_{R}$ in the field space of the background Riemannian metrics along the
Ricci flows on a $\check{d}$-dimensional Riemannian manifold is determined
by the scalar curvature of the LC-connection, $R(\tau ):=\mathcal{R}%
sc[\nabla ](\tau ),$ when at $R=0$, there is an infinite tower of additional
massless states in QG. That Conjecture do not have a straightforward
extension for nonassociative geometric flows and related gravity theories
because of such reasons:

\begin{enumerate}
\item Nonassociative phase spaces $\ _{s}^{\star }\mathcal{M}$ can be
endowed with different classes of (non) linear connections and respective
curvature and Ricci s-tensors, and respective scalars. There are necessary
additional assumptions and physical motivations on how, for instance, the
canonical geometric objects are related to similar ones for
LC-configurations.

\item The generalized distance functional $\ _{g}^{s}\aleph $ (\ref{mfdist})
and, for instance, the canonical scalar curvature $\ _{s}^{\shortmid }%
\widehat{\mathbf{R}}sc^{\star }(\tau )$ (\ref{ricciscsymnonsym}), and their
restrictions for LC-configurations considering additional nonholonomic
constraints (\ref{lccondnonass}), involve complex coordinates and geometric
variables. It is not clear, in general, how to find a physical
interpretation for such non-quantum models even the complex variables are
important in QG.
\end{enumerate}

We can solve in general forms the problems stated above in paragraphs 1 and
2 if we work with canonical geometric data when the terms with complex
variables are transformed into almost complex and almost symplectic ones. If
necessary, such nonholonomic geometric objects can be considered as certain
canonical distortions of certain LC-models. Respective $\kappa $-linear
decompositions to models of parametric (real and effective associative and
commutative, but nevertheless nonholonomic) geometric flows, with
nonholonomic R. Hamilton equations (\ref{nonassocgeomflp}), also allow to
elaborate on physical viable theories with corresponding Conjectures and
Claims:

\vskip5pt \textbf{The canonical distance - scalar curvature conjecture and
claim for nonassociative geometric flows, CCL4:}

\begin{description}
\item[a)] \textsf{\ Conjecture:} \emph{Consider a model of nonassociative
geometric flows determined by $\kappa $-parametric R. Hamilton equations (%
\ref{nonassocgeomflp}). The generalized distance $\ _{g}^{s}\aleph $ (\ref%
{mfdist}) can be defined in a form $\ _{R}^{s}\widehat{\aleph }$ determined
by the canonical scalar curvature $\ _{s}^{\shortmid }\widehat{\mathbf{R}}%
sc[\ _{s}^{\shortmid }\mathbf{g}(\tau),\ _{s}^{\shortmid }\widehat{\mathbf{D}%
}(\tau ), \ ^{\shortmid }\Im _{\alpha _{s}\beta _{s}}(\tau )]$ for certain
canonical data for the equivalent system of PDEs (\ref{nonassocgeomflef}).
For QG models at $\ _{s}^{\shortmid }\widehat{\mathbf{R}}sc=0,$ there exists
a canonical infinite tower of phase space states with zero effective masses.}

\item[b)] \textsf{\ Claim:} \emph{For configurations with nonlinear
symmetries (\ref{nonlinsymr}) and transforms of generating functions and
generating sources, $[\ _{s}\Psi (\tau ),\ _{s}^{\shortmid }\Im
(\tau)]\rightarrow [\ _{s}\Phi (\tau ),\ _{s}^{\shortmid }\Lambda (\tau)],$
we can construct exact/parametric solutions of $\tau $-running $\kappa $%
-linear modified Einstein equations (\ref{nonassocgeomflefcc}) satisfying
the conditions stated by above conjecture. LC-configurations can be
extracted if additional nonholonomic constraints (\ref{lccondnonass}) are
imposed. }
\end{description}

Generalizing for nonassociative canonical geometric flows on $\ ^{\star }%
\mathcal{M}$ the formula (51) from \cite{kehagias19}, we consider an
alternative definition of (\ref{mfdist}) when 
\begin{equation*}
\ _{R}^{s}\widehat{\aleph }\simeq \log (\ _{s}^{\shortmid }\widehat{\mathbf{R%
}}sc(\tau _{i})/\ _{s}^{\shortmid }\widehat{\mathbf{R}}sc(\tau _{f})).
\end{equation*}
For some classes of solutions, we can consider flow evolution models with $%
\tau _{f}<\tau _{i}.$ If in the vicinity of a fixed point $\
_{s}^{\shortmid} \widehat{\mathbf{R}}sc(\tau _f )=0$ but $\ _{s}^{\shortmid }%
\widehat{\mathbf{R}}sc(\tau _{i})\neq 0,$ we can approximate%
\begin{equation}
\ _{R}^{s}\widehat{\aleph }(\tau )\simeq \log (\ _{s}^{\shortmid }\widehat{%
\mathbf{R}}sc(\tau ))\simeq \log (\sum\nolimits_{s} \
_{s}^{\shortmid}\Lambda (\tau )),  \label{gdchccl4}
\end{equation}%
for configurations with $\tau $-running shell cosmological constants. The
nonassociative off-diagonal geometric flux evolution and gravitational
interactions define in the infinite canonical distance limit (when $\
_{R}^{s}\widehat{\aleph }\rightarrow \infty $ with $\ _{s}^{\shortmid }%
\widehat{\mathbf{R}}sc\rightarrow 0)$ a canonical tower of effective
massless states%
\begin{equation}
\ ^{\shortmid }m\sim M_{p}e^{-\hat{\alpha}|\ \ _{R}^{s}\widehat{\aleph }%
|}\simeq (\ _{s}^{\shortmid}\widehat{\mathbf{R}}sc)^{\hat{\alpha}} \simeq
(\sum\nolimits_{s}\ _{s}^{\shortmid }\Lambda )^{\hat{\alpha}}.
\label{masstf1}
\end{equation}

The statements of CCL4 can be reformulated for LC-configurations. Here we
note that, in general, the condition $\ _{s}^{\shortmid }Rsc[\
^{\shortmid}\nabla ]=0$ is characterized by $\ _{s}^{\shortmid }\widehat{%
\mathbf{R}}sc\neq 0$ which reflects the nonholonomic, locally anisotropic
and off-diagonal characters of effective masses $\ ^{\shortmid}m $ (\ref%
{masstf}) and/or (\ref{masstf1}). We prefer to work with canonical s-adapted
geometric variables because this allows us to apply the AFCDM and construct
general classes of solutions. The problem to consider $\ ^{\shortmid }\nabla$
as a more fundamental than the (auxiliary) canonical s-connection, $\
_{s}^{\shortmid }\widehat{\mathbf{D}},$ or other not/or s-adapted linear
connections, has to be analysed using certain additional theoretical
arguments and experimental/observational data. For pseudo-Riemannian
spacetimes, the associated entropy $\mathcal{S}\simeq \exp (\ _{R}^{s}%
\widehat{\aleph }(\tau ))$ become infinite in the flat spacetime limit. This
was also pointed out in \cite{dvali16} for the BH entropy considerations
using the Bekenstein-Hawking paradigm. Restricting above formulas for base
spacetime LC-configurations, we conclude that the flat space limit should be
accompanied by an infinite number of massless stated and that, for instance,
the Minikowki spacetime is infinitely far away from the curved manifolds
along the geometric flow. This means that the flat space never be reached,
and that the transition to flat space is discontinuous. Nevertheless, such
results have to be revised, for instance, if we consider contributions from
nonassociative star products and R-fluxes and work with more general classes
of generic off-diagonal solutions.

\subsubsection{Generalized distances and nonassociative geometric flow G.
Perelman functionals}

\label{ss614} We have to introduce and study more complicate distance
functionals and test respective claims and conjectures for nonassociative
geometric flows and related classical gravity and QG models. Let us begin
with original considerations for geometric flows on $\check{d}$-dimensional
Riemannian manifolds, see Conjecture B2 in \cite{kehagias19} for the
F-functional $F(\tau )=F(g(\tau),f(\tau ))$ (\ref{perelmfst}) and (in our
notations) respective distance functional $\aleph _{F}.$ It was conjectured
that in the background field space along the combined dilaton-metric flow $\
\aleph _{F}$ is determined by F-functional when in QG models related to such
Ricci flows there is an infinite tower of additional massless states at $%
F=0. $ Here we note that $f(\tau )$ is not obligatory a dilaton type field
and originally it was considered as a normalizing function \cite{perelman1},
see also the footnote \ref{fnnormf}.

In the case of nonassociative geometric flow, we can elaborate on physical
viable theories with:

\vskip5pt

\textbf{The canonical distance - F-functional conjecture and claim for
nonassociative geometric flows, CCL5:}

\begin{description}
\item[a)] \textsf{\ Conjecture:} \emph{Consider a model of nonassociative
geometric flows determined by a $\kappa $-parametric functional $\
_{s}^{\shortmid }\widehat{\mathcal{F}}_{\kappa }^{\star }(\tau )$ (\ref%
{naffunctp}). The generalized distance $\ _{g}^{s}\aleph $ (\ref{mfdist})
can be defined and computed in a form $\ _{F}^{s}\widehat{\aleph }$
determined by the canonical scalar curvature for certain canonical data $[\
_{s}^{\shortmid }\mathbf{g}(\tau ), \ _{s}^{\shortmid }\widehat{\mathbf{D}}%
(\tau ), \ ^{\shortmid }\Im _{\alpha _{s}\beta _{s}}(\tau )]$ as solutions
of nonassociative geometric flow equations (\ref{nonassocgeomflef}). For QG
models at $_{s}^{\shortmid }\widehat{\mathcal{F}}_{\kappa }^{\star }=0,$
there exists a canonical infinite tower of phase space states with zero
effective masses.}

\item[b)] \textsf{\ Claim:} \emph{For nonassociative geometric flow models
and gravitational configurations with nonlinear symmetries (\ref{nonlinsymr}%
), we can construct exact/ parametric solutions of $\tau $-parametric
running $\kappa $-linear modified Einstein equations (\ref%
{nonassocgeomflefcc}) satisfying, or not, the conditions stated by above
conjecture. We can extract LC-configurations for additional nonholonomic
constraints (\ref{lccondnonass}). }
\end{description}

We can define and compute a generalized distance functional if 
\begin{equation*}
\ _{F}^{s}\widehat{\aleph }\simeq [\ _{s}^{\shortmid}\widehat{\mathcal{F}}%
_{\kappa }^{\star }(\tau _{i})/ \ _{s}^{\shortmid }\widehat{\mathcal{F}}%
_{\kappa }^{\star }(\tau _{f})].
\end{equation*}
In the case when $\ _{s}^{\shortmid }\widehat{\mathcal{F}}_{\kappa
}^{\star}(\tau _{f})$ is a fixed point of the nonassociative flow equations (%
\ref{nonassocgeomflef}), the canonical tower of effective massless stated
scale for $\ _{F}^{s}\widehat{\aleph }\rightarrow \infty $ as 
\begin{equation*}
\ ^{\shortmid }m\sim M_{p}e^{-\hat{\alpha}| \ _{F}^{s}\widehat{\aleph }%
|}\simeq \left( \ _{s}^{\shortmid }\widehat{\mathcal{F}}_{\kappa
}^{\star}(\tau )\right) ^{\hat{\alpha}}\simeq \left(\ _{s}^{\shortmid }%
\widehat{\mathbf{R}}sc(\tau ) + \ _{s}^{\shortmid}\widehat{\mathbf{K}}%
sc(\tau )\right) ^{\hat{\alpha}}.
\end{equation*}%
A normalizing function $\widehat{f}(\tau )$ can be prescribed in a form when
the distortion $\ _{s}^{\shortmid }\widehat{\mathbf{K}}sc(\tau)$ is absorbed
in such a form that the nonassociative geometric flow is determined by $\
_{F}^{s}\widehat{\aleph }\simeq \ _{R}^{s}\widehat{\aleph }\simeq \log (\
_{s}^{\shortmid }\widehat{\mathbf{R}}sc)$ as in (\ref{gdchccl4}). For other
type models, for instance, with $f$ treated as a dilaton field, and
considering only a base spacetime flow of dilaton, we can approximate $\
_{F}^{s}\widehat{\aleph }\simeq \log g_{s},$ where $g_{s}$ is the string
coupling (see formula (65) and footnote 8 in \cite{kehagias19}).
Nonassociative geometric flow configurations with nonlinear symmetries (\ref%
{nonlinsymr}) can be characterized by $_{F}^{s}\widehat{\aleph }\simeq -\log
\left\vert \sum\nolimits_{s}\ _{s}^{\shortmid }\Lambda (\tau)\right\vert ,$
which is compatible with the conditions of previous \textbf{CCL4.}

\vskip5pt

G. Perelman \cite{perelman1} defined also another important functional, the
W-functional, which is a "minus entropy" and can be used for treating the
case of collapsing cycles of the Ricci flow theory of Riemannian metrics. In
connection to the generalized distance functional, the Conjecture B3 from 
\cite{kehagias19} states that for geometric flows of Riemannian metrics we
can introduce a generalized distance $\ _{W}^{s}\widehat{\aleph }$ in the
background field space along the Ricci flow evolution which is determined by
the W-entropy. For QG models in the points where this variable vanishes,
there is an infinite tower of massless states. Such constructions involve
more rich nonholonomic geometric structures for nonassociative geometric
flows, and results in physical models with local anisotropy.

For $\kappa $-parametric decompositions, we can introduce a re-scaling
factor $\chi (\tau )$ in the definition of $\ _{s}^{\shortmid }\widehat{%
\mathcal{W}}_{\kappa }^{\star }(\tau )$ (\ref{nawfunctp}), when 
\begin{eqnarray}
&& \ _{s}^{\shortmid }\widehat{\mathcal{W}}_{\kappa }^{\star }[\
_{s}^{\shortmid }\mathbf{g}(\tau ), \ _{s}^{\shortmid }\widehat{\mathbf{D}}%
(\tau ),\chi (\tau);\ ^{\shortmid }\Im _{\alpha _{s}\beta _{s}}(\tau )] =
\label{nawfunctp1} \\
&& \int_{\ _{s}^{\shortmid }\widehat{\Xi }}\delta \ ^{\shortmid }u^{\gamma
_{s}}(\chi )\sqrt{|\ ^{\shortmid }\mathbf{g}_{\alpha _{s}\beta _{s}}\ (\chi
)|} \left[ \chi \left( \ _{s}^{\shortmid }\widehat{\mathbf{R}}sc(\chi )+(\
_{s}^{\shortmid }\widehat{\mathbf{D}}(\chi )\ \ _{s}^{\shortmid }\widehat{f}%
(\chi ))^{2}\right) +\ \ _{s}^{\shortmid }\widehat{f}(\chi )-8\right] \frac{%
e^{- \ _{s}^{\shortmid }\widehat{f}(\chi )}}{(4\pi \chi )^{4}},  \notag
\end{eqnarray}%
under constraint $\int_{\ _{s}^{\shortmid }\widehat{\Xi }}\delta \
^{\shortmid }u^{\gamma _{s}}(\chi ) \sqrt{|\ ^{\shortmid }\mathbf{g}_{\alpha
_{s}\beta _{s}}\ (\chi )|}e^{- \ _{s}^{\shortmid }\widehat{f}(\chi )}/(4\pi
\chi )^{4}=1.$ The normalizing function $\ _{s}^{\shortmid }\widehat{f}%
(\chi) $ and re-scaling factor $\chi (\tau )$ can be chosen in such a form
that the resulting variational geometric flow equations involve only the
canonical Ricci s-tensor and respective scalar curvature. As a result, the
nonassociative geometric flow equations (\ref{nonassocgeomflp}) can be
written equivalently as 
\begin{eqnarray}
\partial _{\tau }\ ^{\shortmid }\mathbf{g}_{\alpha _{s}\beta _{s}}(\tau )
&=& -2\ ^{\shortmid }\widehat{\mathbf{R}}_{\ \alpha _{s}\beta _{s}}(\tau ),
\label{nonassocgeomflp1} \\
\partial _{\tau }\ \ _{s}^{\shortmid }\widehat{f}(\tau ) &=&\
_{s}^{\shortmid }\widehat{\mathbf{R}}sc(\tau )- \widehat{\bigtriangleup }%
(\tau )\ _{s}^{\shortmid }\widehat{f}(\tau )+(\ _{s}^{\shortmid }\widehat{%
\mathbf{D}}(\tau ) \ _{s}^{\shortmid }\widehat{f}(\tau ))^{2}(\tau )+4/\chi
(\tau ),  \notag \\
\partial _{\tau }\ \chi (\tau ) &=& -1.  \notag
\end{eqnarray}%
Here we note that the nonassociative star product and R-flux contributions
of order $\left\lceil \hbar ,\kappa \right\rceil $ are encoded into
generating functions, generating sources and nonlinear symmetries (\ref%
{nonlinsymr}) of s-metrics $\ ^{\shortmid }\mathbf{g}_{\alpha _{s}\beta
_{s}}(\tau ).$ Such $\tau $- and $\chi $-families of s-metrics can be found
as exact/parametric solutions (\ref{nonassocgeomflefcc}) with prescribed
effective running cosmological constants $\ _{s}^{\shortmid }\Lambda (\tau
). $ Finding a solution of the nonlinear system of PDEs (\ref%
{nonassocgeomflp1}), we can introduce it in the formula (\ref{nawfunctp1})
and compute the W-entropy of associated thermodynamic system. This
functional is important for a further analysis if such a nonassociative
model belongs, or not, to swampland.

Summarizing above considerations, for nonassociative geometric flows, we
formulate:

\vskip5pt \textbf{The canonical distance - W-functional conjecture and claim
for nonassociative geometric flows, CCL6:}

\begin{description}
\item[a)] \textsf{\ Conjecture:} \emph{For models of nonassociative
geometric flows determined by $\kappa $ -parametric functional $\
_{s}^{\shortmid }\widehat{\mathcal{W}}_{\kappa }^{\star }(\tau )$ (\ref%
{nawfunctp1}), the generalized distance $\ _{g}^{s}\aleph $ (\ref{mfdist})
can be defined in a form $\ _{W}^{s}\widehat{\aleph }$ determined by such a
canonical W-entropy functional defined and computed for corresponding
canonical data and solutions of nonassociative geometric flow equation (\ref%
{nonassocgeomflp}). For QG models at $\ _{s}^{\shortmid }\widehat{\mathcal{W}%
}_{\kappa }^{\star }=0,$ there exists a canonical infinite tower of phase
space states with zero effective masses.}

\item[b)] \textsf{\ Claim:} \emph{If certain nonassociative models of
geometric flows / Ricci solitons are constructed to posses nonlinear
symmetries (\ref{nonlinsymr}) to running cosmological constants $\
_{s}^{\shortmid }\Lambda (\tau )$, we can generate exact/ parametric
solutions of $\tau $-parametric running $\kappa $-linear modified Einstein
equations (\ref{nonassocgeomflefcc}) satisfying, or not, the conditions
stated by above conjecture. LC-configurati\-ons can be extracted for
additional nonholonomic constraints (\ref{lccondnonass}). The formulas for $%
\ _{W}^{s}\widehat{\aleph }$ can be expressed as functionals on $\
_{s}^{\shortmid }\Lambda (\tau )$ and respective volume forms and the
conclusion that such a nonassociative geometric and thermodynamic model
belongs, or not, to swampland depend on the type of prescribed generating
functions and running effective cosmological constants.}
\end{description}

In analogy to CCL4 and CCL5, we define and compute for CCL6 such a
generalized distance functional: 
\begin{equation}
\ _{W}^{s}\widehat{\aleph }\simeq \log (\ _{s}^{\shortmid }\widehat{\mathcal{%
W}}_{\kappa }^{\star }(\tau _{i})/\ _{s}^{\shortmid }\widehat{\mathcal{W}}%
_{\kappa }^{\star }(\tau _{f})).  \label{wdist}
\end{equation}%
If $\ _{s}^{\shortmid }\widehat{\mathcal{W}}_{\kappa }^{\star }(\tau _{f})$
is a fixed point of the nonassociative flow equations (\ref{nonassocgeomflp1}%
), the canonical tower of effective massless stated scale for $\ _{W}^{s}%
\widehat{\aleph }\rightarrow \infty $ as 
\begin{equation}
\ ^{\shortmid }m\sim M_{p}e^{-\hat{\alpha}|\ \ _{W}^{s}\widehat{\aleph }%
|}\simeq \left( \ _{s}^{\shortmid }\widehat{\mathcal{W}}_{\kappa }^{\star
}(\tau )\right) ^{\hat{\alpha}}.  \label{wmass}
\end{equation}

In \cite{kehagias19} (see formulas (88)-(95) in that work), there are
provided explicit computations for functionals (\ref{wdist}) and (\ref{wmass}%
) in some cases of associative and commutative geometric flows, for
instance, for high dimensional Einstein spaces and 2-d spacetime with $SO(2)$
symmetries. Another example was studied for modified gravity with F(R) Ricci
flows (section 4, in that work). Here we note that more general examples and
various applications for $R^{2}$-gravity with locally anisotropic
cosmological and BH solutions were studied in \cite{gheorghiu16}. We shall
provide explicit proofs of W-entropy and related formulas for generalized
distance and effective massless stated scales for explicit examples of
nonassociative BH solutions in next subsections.

\subsubsection{Generalized nonassociative phase space swampland conjectures
for effective (A) dS phase \newline space configurations and the Bekenstein--Hawking
entropy}

In \cite{bonnefoy19,lueben21}, there were studied possible physical
implications of the Black Hole Entropy Distance Conjecture, BHEDC. It was
postulated in the framework of a research elucidating a close connection
between the infinite distance conjecture in the context of swampland and
area law for the Bekenstein-Hawking entropy, $S_{BH}.$ That conjecture sates
that for BH spacetime configurations in QG, the formulas $\ _{g}^{s}\aleph $
(\ref{mfdist}) and $m$ (\ref{masst}) can be written, respectively, in the
form 
\begin{equation}
\ _{BH}^{s}\aleph \sim \log S_{BH}\mbox{ and }m\sim S_{BH}^{-\hat{\alpha}},%
\mbox{ where }\hat{\alpha}\sim \mathcal{O}(1).  \label{bhedc}
\end{equation}%
For such considerations, the limit $S_{BH}\rightarrow \infty $ is identified
with the Minkowski spacetime. The BHEDC was tested and generalized for
concrete string setups with AdS configurations; for metrics belonging the
family of RNdS solutions; multi-horizon spacetimes; for BH evaporation etc.

Nonassociative star product and R-flux deformations result in different
types of nonlinear systems of PDEs characterized by more general classes of
generic off-diagonal solutions, nonholonomic structures and (non) linear
connections. Such solutions are generated for corresponding nonlinear
symmetries and for coefficients of the fundamental geometric objects
depending on all phase space coordinates and flow evolution parameter.
Nevertheless, we can construct also some very special and physically
important classes of solutions with conventional horizons when the
nonassociative contributions are encoded into respective classes of
generating functions (for instance, rotoid deformations). Our strategy for
generalizing the BHEDC on nonassociative phase spaces $\ ^{\star }\mathcal{M}
$ is stated as follow:

\vskip5pt \textbf{The phase space black hole entropy distance conjecture and
claim for nonassociative geometric flows, CCL7:}

\begin{description}
\item[a)] \textsf{\ Conjecture:} \emph{\ Exact/ parametric solutions with
conventional phase space hyper-surface horizons, hh, defining models of
nonassociative geometric flows and/or gravity theories with $\kappa $%
-parametric modified R. Hamilton (\ref{nonassocgeomflp}) / Ricci soliton (%
\ref{naricsol}) equations, and describing $\tau $-families of star product
R-flux deformed BHs. The QG models for such quasi-stationary solutions are
characterized by a Bekenstein-Hawking entropy variable, $_{s}^{%
\shortmid}S_{hh}$ defined on corresponding phase spaces when the generalized
distance $\ _{hh}^{s}\aleph \sim \log (\ _{s}^{\shortmid}S_{hh}) $ and
effective mass scale for a respective tower of states is $\ _{s}^{\shortmid
}m\sim \ _{s}^{\shortmid }S_{hh}^{- \hat{\alpha}},$ for $\hat{\alpha}\sim
O(1).$}

\item[b)] \textsf{\ Claim:} \emph{For nonassociative geometric flow models
and gravitational configurations with nonlinear symmetries (\ref{nonlinsymr}%
) for running cosmological constants $\ _{s}^{\shortmid }\Lambda (\tau )$,
we can construct exact/parametric solutions of $\tau $-parametric running $%
\kappa $-linear modified Einstein equations (\ref{nonassocgeomflefcc})
satisfying, or not, the conditions stated by the phase space BHEDC. We can
extract BE and BH solutions for LC-configurations for additional
nonholonomic constraints (\ref{lccondnonass}). }
\end{description}

Additionally, in this subsection, we provide a list of four parametric
solutions for $\tau $-families of nonassociative BE and BH solutions which
are characterized by the conditions CCL7 and can be investigated as models
of nonassociative geometric flows for which the paradigm of
Bekenstein--Hawking thermodynamics holds true. {\small 
\begin{equation*}
\begin{tabular}{llll}
& \textbf{Type of solution} & 
\begin{tabular}{l}
{\ Nonlinear} \\ 
{\ quadratic} \\ 
{\ element}%
\end{tabular}
& {\ Generalized Bekenstein-Hawking entropy, }$_{s}^{\shortmid }S_{hh}$ \\ 
&  &  &  \\ 
1. & 
\begin{tabular}{l}
Nonassociative $\tau $ -deformed \\ 
double RNdS BHs, dissipation \\ 
to BEs \& Schwarzschild BHs,%
\end{tabular}
& 
\begin{tabular}{l}
${d\ }_{\shortmid }^{\flat }{s}_{[8d]}^{2}{(\tau ),}$ \\ 
{\ see (\ref{2rotsol});}%
\end{tabular}
& 
\begin{tabular}{l}
$_{s}^{\shortmid }S_{hh}=S_{0}(\tau ,r,\theta ,\varphi )+\ _{\shortmid
}S_{0}(\tau ,p,p_{\theta },p_{\varphi }),$ \\ 
$S_{0}=\frac{\Omega _{2}\times (r_{s})^{4}}{4}[1+\frac{4\kappa }{3}%
\underline{\chi }{(\tau ,r,\theta )}\sin {(\omega }_{0}{\varphi +\varphi }%
_{0})],$ \\ 
$\ _{\shortmid }S_{0}=\frac{\Omega _{2}\times (p_{s})^{4}}{4}[1+\frac{%
4\kappa }{3}\overline{\chi }{(\tau ,p,p}_{\theta })\sin (\ _{\shortmid }^{p}{%
\omega }_{0}{p}_{\varphi }+p_{\varphi }^{0})$ \\ 
{\ see (\ref{nabhthermod}).}%
\end{tabular}
\\ 
2. & 
\begin{tabular}{l}
Nonassociative $\tau $-running \\ 
couples of Schwarzschild-AdS \\ 
BHs and BEs deformations,%
\end{tabular}
& 
\begin{tabular}{l}
$d\ _{\shortmid }^{\epsilon }{s}_{[8d]}^{2}{(\tau ),}$ \\ 
see (\ref{pmrndsf2});%
\end{tabular}
& 
\begin{tabular}{l}
$\ _{s}^{\shortmid }S_{hh}=\ ^{\epsilon \chi }S(\tau ,r,\theta ,\varphi )+\
_{\shortmid }^{\epsilon \chi }S(\tau ,p,p_{\theta },p_{\varphi }),$ \\ 
$\ ^{\epsilon \chi }S=\ ^{\epsilon }S_{0}(\tau )[1+\frac{4\kappa }{3}%
\underline{\chi }(\tau ,r,\theta )\sin (\omega _{0}\varphi +\varphi _{0})],$
\\ 
$\ _{\shortmid }^{\epsilon \chi }S=\ _{\shortmid }^{\epsilon }S_{0}(\tau
)[1+4\kappa \overline{\chi }(\tau ,p,p_{\theta })\sin (\ _{\shortmid
}^{p}\omega _{0}\ p_{\varphi }+p_{\varphi }^{0})]$ \\ 
see (\ref{hbbhtherman}) and (\ref{hbbhtherm}).%
\end{tabular}
\\ 
3. & 
\begin{tabular}{l}
$d=5${\ \ RN AdS metric} \\ 
{\ \ embedded into a } \\ 
{\ 8-d phase space }$\ _{s}M,$%
\end{tabular}
& 
\begin{tabular}{l}
$d\ \breve{s}_{[5+3]}^{2},$ \\ 
{\ see (\ref{pm5d8d});}%
\end{tabular}
& $\ ^{0}\breve{S}=\frac{\ ^{0}\breve{A}}{4G_{[5]}}=\frac{\omega _{\lbrack
3]}\breve{r}_{h}}{4G_{[5]}},$ {see (\ref{bhth58}).} \\ 
4. & 
\begin{tabular}{l}
\\ 
$\tau $-running phase space \\ 
{\ RN-AdS BEs} \\ 
\end{tabular}
& 
\begin{tabular}{l}
${d\ }_{\shortmid }^{\chi }{s}_{[6\subset 8d]}^{2}{(\tau ),}$ \\ 
{\ see (\ref{sol4of});}%
\end{tabular}
& ${\breve{S}(\tau )=\ }^{0}{\breve{S}(1+}\frac{\kappa \ }{2}{\hat{\chi}}_{4}%
{(\tau ))},$ {see (\ref{rotbhbhthermv}).}%
\end{tabular}%
\ 
\end{equation*}%
}

There are possible different scenarios of nonassociative geometric flow
evolution defined by above outlined solutions. They depend on the type of
generating functions and prescribed $\tau $-running/fixed cosmological
constants, and respective nonlinear symmetries (\ref{nonlinsymr}). In the
case 1, the off-diagonal interactions and flow evolution may transform a
phase space double RNdS BH configuration into a phase system of BEs and
Schwarzschild BHs, when the thermodynamic variables are computed as rotoid
type locally anisotropic deformations of respective variables in
Bekenstein-Hawking thermodynamics. For certain classes of nonholonomic
constraints, the corresponding ellipsoidal deformations can be stable for a
fixed value $\tau _{0}$ (i.e. there are defined stable nonassociative phase
space Ricci solitons) and describe $\tau $-evolution preserving the prime
configuration. Such models are described by solutions of type 2. Similar
scenarios can be described by solutions of type 3 and 4 but with different
thermodynamic models and physical interpretations which is typical for the
phase space higher dimension BEs and BHs.

\subsection{Nonassociative geometric flows, swampland conjectures, and G.
Perelman thermodynamics}

\label{ss62}A class of solutions for nonassociative geometric flows and/or
nonassociative gravitational equations can be always characterized by an
associated G. Perelman thermodynamic model, when, for instance, the
conditions of CCL6 are satisfied but those stated in CCL7 are not
applicable. We have to investigate explicitly (for a class of $\tau $%
-quasi-stationary solutions) if and for which the conditions and statements
of CCL1-CCL6 are true, not true, or undetermined. In this subsection, we
study a certain important examples how the swampland conjectures have to be
extended to nonassociative Ricci flows. In explicit form, we show how such
constructions are related to CCL6 when the W-entropy is used for defining
and computing the generalized canonical distance functional and respective
tower of effective mass stated scaling, see respective formulas (\ref{wdist}%
) and (\ref{wmass}). Similar analysis on applicability and testing of
CCL1-CCL5 are not presented in this work because, the formulas with
canonical scalar curvature, F-functionals etc. are more cumbersome and not
related directly to computation of variables of G. Perelman thermodynamics.

\subsubsection{The canonical distance - W-functional conjecture and claim,
CCL6, for $\protect\tau $-running \newline nonassociative generating sources and
quasi-stationary solutions}

For any class of solutions of geometric flow equations (\ref%
{nonassocgeomflef}) with $\tau $-running effective sources $\
_{s}^{\shortmid }\Im (\tau )$ (\ref{cannonsymparamc2b}), represented in a
form (\ref{sol1rf}), or (\ref{sol3rf}), or (\ref{offdiagpolfr}), we can
compute the star product deformed W-functional $\ _{s}^{\shortmid }\widehat{%
\mathcal{W}}^{\star }(\tau )$ (\ref{nawfunct}). Let us consider a s-metric $%
\ _{s}^{\shortmid }\mathbf{g}[\hbar ,\kappa ,\tau ,\psi (\tau ),\ _{s}\Psi
(\tau ),\ _{s}^{\shortmid }\Im (\tau )]$ for quasi-stationary solutions of
type (\ref{sol1a}). We can compute the W-entropy and S-entropy functions as
in formulas (\ref{thvcann}) but for corresponding generating functions and
effective sources, and volume form $^{\shortmid }\delta \ ^{\shortmid }%
\mathcal{V}(\tau )$ (\ref{volume}), 
\begin{eqnarray}
\ _{s}^{\shortmid }\widehat{\mathcal{W}}_{\kappa }^{\star }(\tau )
&=&\int\nolimits_{\tau ^{\prime }}^{\tau }\frac{d\tau }{(4\pi \tau )^{4}}%
\int_{\ _{s}^{\shortmid }\widehat{\Xi }}\left( \tau \lbrack
\sum\nolimits_{s}\ _{s}^{\shortmid }\Im (\tau )]^{2}-8\right) \ ^{\shortmid
}\delta \ ^{\shortmid }\mathcal{V}(\tau ),  \label{wscc57} \\
\ _{s}^{\shortmid }\widehat{\mathcal{S}}_{\kappa }^{\star }(\tau )
&=&-\int\nolimits_{\tau ^{\prime }}^{\tau }\frac{d\tau }{(4\pi \tau )^{4}}%
\int_{\ _{s}^{\shortmid }\widehat{\Xi }}\left( \tau \lbrack
\sum\nolimits_{s}\ _{s}^{\shortmid }\Im (\tau )]-8\right) \ ^{\shortmid
}\delta \ ^{\shortmid }\mathcal{V}(\tau ).  \label{swcc57}
\end{eqnarray}%
So, nonassociative geometric flows can be characterized by two entropy type
variables: the "standard" statistical thermodynamics entropy (similar to
constructions in hydrodynamic models of moving media but in terms of
curvature scalars, metrics, connections) , $\ _{s}^{\shortmid }\widehat{%
\mathcal{S}}_{\kappa}^{\star }(\tau ),$ with $[\sum\nolimits_{s}\
_{s}^{\shortmid }\Im (\tau )];$ and the "minus" entropy, i.e. W-entropy, $\
_{s}^{\shortmid }\widehat{\mathcal{W}}_{\kappa }^{\star}(\tau ),$ with $%
[\sum\nolimits_{s}\ _{s}^{\shortmid }\Im (\tau )]^{2}.$

Using (\ref{swcc57}), we can modify the formulas (\ref{bhedc}) for BHEDC and
the conditions of CCL7, 
\begin{equation*}
\ _{\mathcal{S}_{\kappa }^{\star }}^{s}\widehat{\aleph }\ \sim \log (\
_{s}^{\shortmid }\widehat{\mathcal{S}}_{\kappa }^{\star }(\tau )) 
\mbox{ and
} \ _{s}^{\shortmid }m \sim (\ _{s}^{\shortmid }\widehat{\mathcal{S}}%
_{\kappa }^{\star }(\tau ))\ ^{-\hat{\alpha}\ }\mbox{ for } \ \hat{\alpha}%
\sim O(1).
\end{equation*}%
Alternatively, we can follow the CCL6 and compute (\ref{wdist}) \ and (\ref%
{wmass}) for (\ref{wscc57}). Respectively, we obtain a generalized distance
functional: 
\begin{equation}
\ _{W}^{s}\widehat{\aleph }\simeq \int\nolimits_{\tau _{f}}^{\tau _{i}}\frac{%
d\tau }{(4\pi \tau )^{4}}\int_{\ _{s}^{\shortmid }\widehat{\Xi }}\left( \tau
\lbrack \sum\nolimits_{s}\ _{s}^{\shortmid }\Im (\tau )]^{2}-8\right) \
^{\shortmid }\delta \ ^{\shortmid }\mathcal{V}(\tau ),  \label{wdistqs}
\end{equation}%
when the canonical tower of effective massless stated scale for $\ _{R}^{s}%
\widehat{\aleph }\rightarrow \infty $ as 
\begin{equation}
\ ^{\shortmid }m\sim M_{p}e^{-\hat{\alpha}|\ \ _{W}^{s}\widehat{\aleph }%
|}\simeq \left( \ \int\nolimits_{\tau _{i}}^{\tau _{f}}\frac{d\tau }{(4\pi
\tau )^{4}}\int_{\ _{s}^{\shortmid }\widehat{\Xi }}\left( \tau \lbrack
\sum\nolimits_{s}\ _{s}^{\shortmid }\Im (\tau )]^{2}-8\right) \ ^{\shortmid
}\delta \ ^{\shortmid }\mathcal{V}(\tau )\right) ^{\hat{\alpha}}.
\label{wmassqs}
\end{equation}%
In general, it is not clear if such nonassociative geometric flow models
belong, or not, to swampland. It depends of the data for generating
functions encoded in $\ ^{\shortmid }\delta \ ^{\shortmid }\mathcal{V}(\tau)$
and effective sources $\ _{s}^{\shortmid }\Im (\tau ).$ We have to analyze
such issues for explicit classes of solutions of (\ref{nonassocgeomflef}).

\subsubsection{CCL6 for $\protect\tau $-running effective shell cosmological
constants and quasi-stationary solutions}

Using nonlinear symmetries (\ref{nonlinsymr}), with $\ _{s}^{\shortmid }%
\mathbf{g}[\hbar ,\kappa ,\tau ,\psi (\tau ),\ _{s}\Psi (\tau ),\
_{s}^{\shortmid }\Im (\tau )]\rightarrow \ _{s}^{\shortmid }\mathbf{g}%
[\hbar,\kappa ,\tau ,\psi (\tau ), \ _{s}\Phi (\tau ),\ _{s}^{\shortmid
}\Lambda(\tau )],$ we can study the conditions of CCL5 involving
quasi-stationary solutions of type (\ref{sol1a}). This allows us to simplify
the formulas for explicit computation of W-entropy $\ _{s}^{\shortmid }%
\widehat{\mathcal{W}}_{\kappa }^{\star }(\tau )$ in terms of running/fixed
effective cosmological constants $\ _{s}^{\shortmid }\Lambda (\tau ),$ see (%
\ref{thvcann}) with $\ ^{\shortmid }\delta \ ^{\shortmid }\mathcal{V}(\tau )$
(\ref{volumf}). Introducing such values, respectively, in (\ref{wdist}) and (%
\ref{wmass}), we obtain: 
\begin{equation}
\ _{W}^{s}\widehat{\aleph }\simeq \int\nolimits_{\tau _{f}}^{\tau _{i}}\frac{%
d\tau }{(4\pi \tau )^{4}}\int_{\ _{s}^{\shortmid }\widehat{\Xi }}\left( \tau
\lbrack \sum\nolimits_{s}\ _{s}^{\shortmid }\Lambda (\tau )]^{2}-8\right) \
^{\shortmid }\delta \ ^{\shortmid }\mathcal{V}(\tau ),  \label{wdistqscc}
\end{equation}%
when the canonical tower of effective massless stated scale for $\ _{R}^{s}%
\widehat{\aleph }\rightarrow \infty $ is computed as 
\begin{equation}
\ ^{\shortmid }m\sim M_{p}e^{-\hat{\alpha}|\ \ _{W}^{s}\widehat{\aleph }%
|}\simeq \left( \ \int\nolimits_{\tau _{i}}^{\tau _{f}}\frac{d\tau }{(4\pi
\tau )^{4}}\int_{\ _{s}^{\shortmid }\widehat{\Xi }}\left( \tau \lbrack
\sum\nolimits_{s}\ \ _{s}^{\shortmid }\Lambda (\tau )]^{2}-8\right) \
^{\shortmid }\delta \ ^{\shortmid }\mathcal{V}(\tau )\right) ^{\hat{\alpha}}.
\label{wmassqscc}
\end{equation}%
Prescribing data for $\ _{s}^{\shortmid }\Lambda (\tau )$ and $\
^{\shortmid}\delta \ ^{\shortmid }\mathcal{V}(\tau ),$ we are able to
generate nonassociative geometric flow scenarios which belong, or not, two
swampland. Such an analysis is more simple than that for (\ref{wdistqs}) and
(\ref{wmassqs}) because respective formulas (\ref{wdistqscc}) and (\ref%
{wmassqscc}) contain effective cosmological constants which can be
prescribed in certain forms which allow generating viable and important
physical models.

Above formulas can be computed in explicit $\kappa $-parametric form when $\
_{s}^{\shortmid }\widehat{\mathcal{W}}_{\kappa }^{\star }(\tau )=\
_{s}^{\shortmid }\widehat{\mathcal{W}}_{0}+\kappa \ _{s}^{\shortmid }%
\widehat{\mathcal{W}}_{1}^{\star }(\tau ),\ $ for $\ ^{\shortmid }\delta \
^{\shortmid }\mathcal{V}=\ ^{\shortmid }\delta \ ^{\shortmid }\mathcal{V}%
_{0}+\kappa \ ^{\shortmid }\delta \ ^{\shortmid }\mathcal{V}_{1}$ (\ref%
{volumfd}). We omit in this work such cumbersome computations and
incremental formulas with $\kappa $--linear decomposition as in (\ref%
{klinthvar}), which can be considered for solutions of type (\ref{2rotsol}),
with $\chi $-polarization functions.

\subsubsection{CCL6 for nonassociative flows of phase space deformed
(double) RN-dS BHs}

\label{ss623} The W-entropy from (\ref{thvcannpd1}) was computed for target
s-metrics (\ref{dbhpspol}) with $\eta $-polarization functions defining
general quasi-stationary star-product R-flux deformations of double RN-dS
BHs in nonassociative phase spaces. Such configurations can be characterized
by effective cosmological constants $\ _{1}^{\shortmid }\Lambda _{0}=\
_{2}^{\shortmid }\Lambda _{0}=\check{\Lambda}\geq 0$ and $\
_{3}^{\shortmid}\Lambda _{0}= \ _{4}^{\shortmid }\Lambda _{0}=\ ^{\shortmid }%
\check{\Lambda}\geq 0.$ This class of solutions can be extended for running
cosmological constants $\ _{s}^{\shortmid }\Lambda (\tau )=[\check{\Lambda}%
(\tau ), \ ^{\shortmid }\check{\Lambda}(\tau )]\geq 0;$ when respective
effective sources $\ _{s}^{\shortmid }\Im (\tau )$ are related via nonlinear
symmetries (\ref{nonlinsymr}) to such $\ _{s}^{\shortmid }\Lambda (\tau ).$
In general, such nonassociative geometric flow deformations do not describe
BH configurations and their general physical interpretation is not clear.
For $\kappa $-parametric decompositions with $\chi $-generating functions (%
\ref{2rotsol}), we can model stable BE and BH configurations embedded
self-consistently into nonassociative phase space backgrounds.

Using (\ref{thvcannpd1}) and introducing $\ _{s}^{\shortmid }\Lambda (\tau)=
[\check{\Lambda}(\tau ),\ ^{\shortmid }\check{\Lambda}(\tau )]$ in formulas (%
\ref{wdistqscc}) and (\ref{wmassqscc}), we compute 
\begin{equation*}
\ _{W}^{s}\widehat{\aleph }\simeq \int\nolimits_{\tau _{f}}^{\tau _{i}}\frac{%
d\tau }{64(\pi \tau )^{4}} \frac{\tau \lbrack \check{\Lambda}(\tau )+\
^{\shortmid }\check{\Lambda}(\tau )]^{2}-2}{|\check{\Lambda}(\tau ) \
^{\shortmid }\check{\Lambda}(\tau )|}\ _{\eta }^{\shortmid \flat }\mathcal{V}%
(\tau ),
\end{equation*}%
when the canonical tower of effective massless stated scale for $\ _{R}^{s}%
\widehat{\aleph }\rightarrow \infty $ as 
\begin{equation*}
\ ^{\shortmid }m\sim M_{p}e^{-\hat{\alpha}|\ \ _{W}^{s}\widehat{\aleph }%
|}\simeq \left( \ \int\nolimits_{\tau _{i}}^{\tau _{f}}\frac{d\tau }{64(\pi
\tau )^{4}}\frac{\tau \lbrack \check{\Lambda}(\tau )+\ ^{\shortmid }\check{%
\Lambda}(\tau )]^{2}-2}{|\check{\Lambda}(\tau )\ ^{\shortmid }\check{\Lambda}%
(\tau )|}\ _{\eta }^{\shortmid \flat }\mathcal{V}(\tau )\right) ^{\hat{\alpha%
}}.
\end{equation*}%
Prescribing data for $\check{\Lambda}(\tau )$ and$\ ^{\shortmid }\check{%
\Lambda}(\tau )$ and $\ _{\eta }^{\shortmid \flat }\mathcal{V}(\tau ),$ we
are able to generate nonassociative geometric flow scenarios for deforming
double BH solutions in phase space, which belong, or not, to swampland. This
depends on behaviour of $\ _{W}^{s}\widehat{\aleph }$ (if there are limits
to $\inf$, or zero) for some types of chosen functions under the integral on 
$\tau$.

\subsubsection{CCL6 for nonassociative flows of phase space nonholonomic
deformed RN-AdS BHs}

\label{ss624} The conditions of CC6 can be tested for another class of
exact/parametric solutions (\ref{narnbhs}) and (\ref{sol4of}) with W-entropy
(\ref{thvcannpd2}) with volume form (\ref{aux7}). In such a target metric,
there is an effective cosmological constant $_{s}\Lambda (\tau )=\Lambda
_{\lbrack 5]}<0, $ for $s=1,2,3,4;$ when respective effective sources $\
_{s}^{\shortmid }\Im (\tau )$ are related via nonlinear symmetries (\ref%
{nonlinsymr}) to $\ \Lambda _{\lbrack 5]}.$ Using (\ref{thvcannpd2}), (\ref%
{wdistqscc}) \ and (\ref{wmassqscc}), we write: 
\begin{equation*}
\ _{W}^{s}\widehat{\aleph }\simeq \int\nolimits_{\tau _{f}}^{\tau _{i}}\frac{%
d\tau }{32(\pi \tau )^{4}}\frac{2\tau \Lambda _{\lbrack 5]}^{2}-1}{\Lambda
_{\lbrack 5]}^{2}}\ _{\eta }^{\shortmid }\mathcal{\breve{V}}(\tau ),
\end{equation*}%
when the canonical tower of effective massless is characterized by such a
scale behaviour for $\ _{R}^{s}\widehat{\aleph }\rightarrow \infty :$ 
\begin{equation*}
\ ^{\shortmid }m\sim M_{p}e^{-\hat{\alpha}|\ \ _{W}^{s}\widehat{\aleph }%
|}\simeq \left( \ \int\nolimits_{\tau _{i}}^{\tau _{f}}\ \frac{d\tau }{%
32(\pi \tau )^{4}}\frac{2\tau \Lambda _{\lbrack 5]}^{2}-1}{\Lambda _{\lbrack
5]}^{2}}\ _{\eta }^{\shortmid }\mathcal{\breve{V}}(\tau )\right) ^{\hat{%
\alpha}}.
\end{equation*}%
Such formulas can be generalized for a $\tau $-running cosmological constant 
$\Lambda _{\lbrack 5]}(\tau )<0$ with the same value for all shells.
Effective volume flows $\ _{\eta }^{\shortmid }\mathcal{\breve{V}}(\tau )$
can be adapted to nonholonomic s-distributions which allow to conclude if
such nonassociative geometric flow deformed phase space RN-AdS BHs belong,
or not, to swampland.

\section{Discussion and conclusions}

\label{sec7}

This is the fifth partner work in a series of articles devoted to the theory
of nonassociative geometric and information flows, modified gravity, and
applications in astrophysics and cosmology \cite%
{partner01,partner02,partner03,partner04}. Such a research program is
motivated by important results on nonassociative geometry and physics
involving theories with star product R-flux deformations in string and
M-theory \cite%
{mylonas13,kupriyanov15,gunaydin,szabo19,blumenhagen16,aschieri17}. In this
paper, we postulate the main functionals and derive the geometric evolution
equations for such a theory of nonassociative Ricci flows and gravity; state
the necessary conditions for a general decoupling and integrability of
necessary systems of nonlinear PDEs; and construct new classes of
exact/parametric solutions describing nonassociative flux evolution on
quasi-stationary and BH configurations. There are computed and analyzed
physical properties of nonassociative phase space thermodynamic variables
developing the paradigms defined (for subclasses of solutions with
conventional hyper-surface horizons) by the Bekenstein-Hawking entropy,
and/or following the G. Perelman W-entropy and geometric thermodynamic
approach which is applicable to general geometric evolution models and
generic off-diagonal solutions. Finally, we study key issues (formulated as
conjectures and claims) related to nonassociative extensions and revision of
the swampland program \cite%
{vaf05,oog06,pal19,gomez19,ooguri19,luest19,kehagias19,biasio20,biasio21,lueben21,biasio22}%
.

\subsection{Summary and discussion of main results}

Let us discuss how the results of this paper fit in with the Objectives
(Aims 1-5) stated in section \ref{ssint12}. We analyze the most important
ideas and innovative methods, consider possible interpretations, and present
the key points of novelty and originality:

\begin{enumerate}
\item Using nonassociative star product, $\star $, deformations, we define
in section \ref{sec2} (\textbf{first objective}) generalized G. Perelman F-
and W-functionals, $\ _{s}^{\shortmid }\widehat{\mathcal{F}}^{\star }(\tau )$
and $\ _{s}^{\shortmid}\widehat{\mathcal{W}}^{\star }(\tau ),$ see formulas (%
\ref{naffunct}) and (\ref{nawfunct}); and effective functionals $\
_{s}^{\shortmid }\widehat{\mathcal{F}}_{\kappa }^{\star }(\tau )$ (\ref%
{naffunctp}) and $\ _{s}^{\shortmid }\widehat{\mathcal{W}}_{\kappa
}^{\star}(\tau )$ (\ref{nawfunctp}), for computing $\kappa $-parametric
decompositions (on string constant $\kappa $, when $\tau$ is a temperature
like parameter). Such functionals are used for deriving nonassociative
versions of R. Hamilton equations describing nonassociative and
noncommutative geometric flow evolution models. For self-similar
configurations, they define nonassociative Ricci soliton geometries and, in
particular, nonassociative vacuum gravity theories with running cosmological
constants. We note that nonassociative geometric flow evolution theories and
applications have not been studied in modern mathematics and physics. There
were elaborated only approaches related to noncommutative geometry (a l\'{a}
A. Connes) \cite{v08}, geometric and quantum information flow theories and
modified gravity \cite{gheorghiu16,vacaru20,ib19,bubuianu19}; and, in
associative/commutative form, it is performed a recent research on swampland
conjectures and Ricci flows \cite{kehagias19,biasio20,lueben21,biasio22}.
Our nonassociative geometric models and solutions are self-consistent and
well-defined in a sense that they are nonholonomic deformations with
2+2+2+... and/or (3+1)+(3+1)+.... decompositions of dimensions, encoding
nonassociative geometric data, and generalizing the fundamental geometric
objects and evolution equations in standard Ricci flow theory \cite%
{hamilton82,friedan80,perelman1,perelman2,perelman3,kleiner06,morgan06,cao06}%
. Such geometric constructions can be performed in relativistic forms for
Lorentz manifolds and their (co) tangent bundles, when the nonassociativeand
noncommutative structures are determined by star product and R-flux
deformations in string and M-theory. \vskip3pt In \cite%
{blumenhagen16,aschieri17}, two similar models of nonassociative gravity
theories were elaborated up to defining and computing the nonassociative
Ricci tensor for the nonassociative star deformed Levi-Civita, LC,
connection, involving (non) symmetric metric structures. The originality of
nonassociative geometric methods elaborated for the nonholonomic shell
oriented structures in \cite{partner01,partner02} consists in developing a
formalism which allows us to apply the anholonomic frame and connection
method, AFCDM \cite{v00,lb18,gheorghiu16} for constructing exact and
parametric solutions in modified gravity theories, MGTs; see a brief review
in Appendix \ref{appendixb}. Using the AFCDM we can prove a general
decoupling and integration property of of nonassociative R. Hamilton
equations (\ref{nonassocgeomfl}), see section \ref{sec4} and \cite%
{partner03,partner04}, for related details for nonasscociative vacuum
Einstein equations. \vskip3pt Modern mathematics states explicit
limitations, in the sense of topology and geometric analysis methods, for
elaborating a general nonassociative and/or noncommutative Ricci flow
theory. This is because an infinite number of models of nonassociative/
noncommutative geometries can be elaborated (some of them present interest
in modern mathematical physics and information theory) and it was not
formulated at least an example of nonassociative/ noncommutative /
relativistic generalization of the Poincar\'{e}-Thurston conjecture.
Nevertheless, nonassociative geometric flow models with star product R-flux
deformations are well defined (at least in $\kappa $-parametric form, in the
framework of string/ M-theory, and corresponding phase space generalizations
of the Einstein gravity). Applying the AFCDM, we can construct physically
important solutions of respective nonlinear systems of PDEs.

\item Theories of nonassociative geometric flows and modified gravity are
characterized by respective statistical thermodynamic models defined by
W-entropy $\ _{s}^{\shortmid }\widehat{\mathcal{W}}^{\star }(\tau )$ and
related statistical generating functional $\ _{s}^{\shortmid }\widehat{%
\mathcal{Z}}^{\star }(\tau )$. The novelty of the \textbf{second objective }
(for section \ref{sec3}) consists in formulating a nonassociative geometric
thermodynamic theory determined by variables (\ref{nagthermodvalues})
encoding nonassociative geometric data in the $\kappa $-linear thermodynamic
variables $[\ _{s}^{\shortmid }\widehat{\mathcal{E}}_{\kappa }^{\star}(\tau
),\ _{s}^{\shortmid }\widehat{\mathcal{S}}_{\kappa }^{\star }(\tau),\
_{s}^{\shortmid }\widehat{\sigma }_{\kappa }^{\star }]$ (\ref%
{nagthermodvaluesp}). Such statistical thermodynamic variables are defined
using fundamental (non) associative/ commutative geometric objects (the
geometric measure, (non) linear derivatives, respective nonassociative Ricci
tensors and scalar curvatures). They can be computed, at least in $\kappa $%
-parametric forms, for all classes of solutions of corresponding geometric
flow/ gravitational field equations. Former results were for associative and
commutative theories \cite{kehagias19,biasio20,lueben21,biasio22,bonnefoy19}
and involving special classes of solutions with conventional horizons when
the concept of Bekenstein-Hawking entropy \cite{bek1,bek2,haw1,haw2} is
well-defined. Following a nonassociative generalized G. Perelman functional
formalism \cite{perelman1,kleiner06,morgan06,cao06,ib19}, we can elaborate
on a new statistical/ geometric thermodynamic paradigm for classical and
quantum gravity and information theories, and study nonlinear models of
quasi-stationary and locally anisotropic cosmological evolution in MGTs, see
details in \cite{gheorghiu16,vacaru20,ib19,bubuianu19,lb18,gheorghiu16}.

\item In our approach, the nonassociative geometric and gravity theories
with star product R-flux deformations are defined on nonholonomic phase
spaces modelled as cotangent Lorentz bundles, $\ _{s}^{\star }\mathcal{M}%
=T_{s}^{\ast }V,$ with nonholonomic shell, s, dyadic splitting of
dimensions, 2+2+2+2. Following the \textbf{third objective} of this work, in
section \ref{sec4}, we show how the AFCDM developed for nonassociative
gravity theories in \cite{partner02,partner03,partner04} can be applied for
generating $\tau $-evolving classes of quasi-stationary solutions of
nonassociative geometric flow equations written in $\kappa $-parametric form
(\ref{nonassocgeomflp}). An original and important result consists from the
proof that, for well-defined nonholonomic geometric and relativistic
physical assumptions, such equations can be written in two equivalent forms:

\begin{itemize}
\item 1] as $\tau $-running canonically s-deformed Einstein equations (\ref%
{nonassocgeomflef}), with effective sources $\ ^{\shortmid}\Im _{\alpha
_{s}\beta _{s}}(\tau )$ encoding nonassociative geometric data; and

\item 2] as equivalent families of modified vacuum gravitational equation (%
\ref{nonassocgeomflefcc}), with effective $\tau $-running shell cosmological
constants $\ _{s}^{\shortmid }\Lambda (\tau ).$
\end{itemize}

Such functional representations of $\kappa $-parametric nonassociative R.
Hamilton equations are important because they allow us to use directly the
AFCDM (as for the nonholonomic Einstein equations but with additional $\tau$%
-dependence) and generate various classes of generic off-diagonal solutions.
The method simplifies substantially if there are considered, for instance,
some general classes of quasi-stationary solutions involving nonlinear
symmetries of type (\ref{nonlinsymr}) (for generating locally anisotropic
cosmological solutions, we have to consider certain dual nonlinear
symmetries). \vskip3pt We constructed in general off-diagonal forms such
exact/parametric nonassociative geometric flow solutions for $\tau $%
-evolving quasi-stationary configurations: with effective sources, (\ref%
{sol1rf}); with running cosmological constants (\ref{sol2rf}); and for
various types of s-metric generating and polarizations functions, see (\ref%
{sol3rf}), or (\ref{offdiagpolfr}). The coefficients of such (non) symmetric
metrics and (non) linear connections depend, in general, on all conventional
space like, co-fiber coordinates, for certain Killing type phase space
symmetries, being determined by corresponding $\tau $-families of generating
and integration functions, generating effective sources and running
cosmological constants as in (\ref{integrfunctrf}). \vskip3pt The AFCDM is
an innovative geometric and analytic method for constructing
exact/parametric solutions of physically important systems of nonlinear PDEs
in (non) associative/ commutative MGTs and geometric flow generalizations.
It allows to generate vacuum and non-vacuum metrics for very general
off-diagonal ansatz not using special assumptions for diagonalizable metrics
resulting into some associated systems of nonlinear ODEs (such details, for
GR, are presented in \cite{misner,hawking73,wald82,kramer03}). Another
important difference from other analytic and numeric methods is that using
the AFCDM we work with certain canonical s-adapted connections which allow
to decouple necessary types of nonlinear systems of equations. We have to
impose additional nonholonomic constraints of type (\ref{lccondnonass}) in
order to extract LC-configurations and model their nonassociative geometric
flow evolution. It should be noted that the AFCDM can be applied for
constructing exact/parametric solutions of nonassociative vacuum
gravitational equations with coefficients proportional to the complex unity.
Such $\kappa $-parametric decompositions are for nonassociative Ricci
tensors are considered, for instance, in \cite{blumenhagen16,aschieri17},
when only the real coefficients are taken for analyzing possible
consequences for classical models. It is not clear what physical importance
may have such complex solutions in classical theories but, positively, they
can be used to study quasi-classical approximations in QG and string/
M-theory. In our approach, we can redefine all (non) associative/ geometric
constructions using almost complex/symplectic structures and related real
nonholonomic geometric models \cite{partner01,partner02,partner03}.

\item The \textbf{fourth objective} of this work, in section \ref{sec5}, is
twofold on its novelty of results and methods: \vskip3pt 1] There are
constructed and analyzed most important physical properties of two new
classes of $\tau $-running nonassociative phase space black hole, BH, black
ellipsoid, BE, and other types of more general quasi-stationary
deformations. Such generic off-diagonal solutions are classified with
respect to two different types of primary phase space s-metrics defined by
double Schwarzschild - AdS BHs and, in the second case, phase space
generalizations of Reisner-Norstr\"{o}m BHs, when the generated target
s-metrics have different interpretations. For certain types of nonholonomic
constraints and generating functions, there are generated $\tau $-evolving
BE type solutions with conventional phase space hyper-surfaces, when the
concept of Bekenstein--Hawking is applicable, and we compute respective
thermodynamic variables in sections \ref{ss523} and \ref{ss533}. \vskip3pt
2] Nevertheless, general models with nonassociative and generic off-diagonal
Ricci flow evolution of physically important quasi-stationary solutions do
not contain hyper-surfaces and we have to apply more general concepts of
nonassociative generalized G. Perelman thermodynamics. The formulas for
respective thermodynamic variables (\ref{thvcannpd}) are defined by
effective running shell cosmological constants, $\ _{s}^{\shortmid }\Lambda
(\tau )$ and certain volume forms depending on the prescribed nonholonomic
structure. Such values are computed in explicit forms for corresponding
classes of nonassociative Ricci flow deformed BHs solutions, see subsections %
\ref{ss524} and \ref{ss534}. They functionally depend on generating and
integration functions, which can be chosen in such forms when the
configurations are well-defined as effective relativistic thermodynamic
variables, preserving certain physical important properties under $\tau $%
-evolution, to belong, or not to swampland models etc.

\item We study nonassociative modifications of the swampland program \cite%
{vaf05,oog06,pal19,gomez19,ooguri19}, which consists the \textbf{fifth
objective} of this paper, see section \ref{sec6}. As motivated in modern
literature on high energy physics and gravity \cite%
{v08,vacaru20,lb18,ib19,bubuianu19,luest19,kehagias19,biasio20,biasio21,lueben21,biasio22}%
, the Ricci flow is closely related to the RG flow with respect to the
energy scale in the underlying two-dimensional non-linear sigma-models and
various modified gravity and string theories. This provides a connection
between RG, Ricci flows, and the swampland idea. \vskip3pt In this work, we
formulate a theory of nonassociative geometric flows using generalizations
of G. Perelman functionals and related thermodynamic models and develop the
AFCDM of constructing new classes of $\tau $-evolving nonassociative
quasi-stationary star-deformed BH and BE solutions. In general, the
thermodynamic physical properties of such nonassociative geometric flow and
gravity generic off-diagonal solutions can not be investigated following
only the Bekenstein-Hawking entropy paradigm. We analyze how the a series of
important swampland and Ricci flow conjectures can be extended to
nonassociative geometric flows and respective extensions to the modified G.
Perelman thermodynamic models. Then, we argue that we can prescribe certain
subclasses of nonholonomic constraints, generating and integration functions
when corresponding models satisfy the conditions of swampland conjectures
studied in \cite{kehagias19,biasio20,biasio21,lueben21,biasio22}.
Corresponding phase space generalizations and modifications involving
respective generalized distance functionals and effective mass scale of
towers of states are stated respectively in subsections \ref{ss61} and \ref%
{ss62} and numbered as a) Conjectures 1-7. We provide and analyze four
examples of $\tau $-running quasi-stationary phase space solutions for
nonassociative BH configurations with conventional hyper-surface horizons
when the concept of generalized Bekenstein--Hawking entropy is applicable,
see the end of subsection \ref{ss61}. This proves that for certain defined
geometric conditions the swampland conjectures can be extended to
conventional phase space and generalized for such nonassociative geometric
flow and gravitational configurations. \vskip3pt For more general
nonassociative $\kappa $-parametric off-diagonal solutions, characterized by
respective W-entropy functionals, we need a more tedious analysis in order
to conclude if a corresponding nonassociative geometric flow model belongs,
or not, to swampland following b) Claims 1-7. We formulate such criteria in
the form of Conjectures and Claims, CCL, 1-7. In subsection \ref{ss62}, we
provide explicit formulas for computing nonassociative and off--diagonal
modifications of general distances and towers of effective mass states for
various classes of $\tau $-running nonassociative quasi-stationary equations
following the CCL5 and effective cosmological constants $\ _{s}^{\shortmid
}\Lambda (\tau ).$
\end{enumerate}

\subsection{Conclusions, validity of claims and results, and perspectives}

\label{ssvalpersp}Discussing Objectives 1-5 for nonassociative Ricci flow
theory and applications, we have advocated for modifications of the
Swampland Program summarized in a series of Conjectures and Claims, CCL:

\begin{itemize}
\item \textbf{CCL1: The generalized distance conjecture and claim for
nonassociative phase space}. We conclude that for very special nonholonomic/
diagonal configurations and additional assumptions on the type of nonlinear
symmetries for the $\kappa$-parametric distance there are limits of the
general distance functional $\ _{g}^{s}\aleph $ (\ref{mfdist})$\rightarrow
\infty $ resulting in infinite towers of effective phase masses $\
^{\shortmid }m$ (\ref{masstf}). This extends for nonassociative phase space
the generalized distance conjecture studied in \cite{luest19,kehagias19} for
(associative and commutative) d-dimensional manifolds. In this work, we show
that using the AFCDM, we can construct new classes of exact/parametric
solutions when, for nonassociative configurations, when $\ _{g}^{s}\aleph $
and $\ ^{\shortmid }m$ may have a different behaviour because generic
off-diagonal interactions and a more rich nonholonomic structure of phase
spaces and nonassociative geometric flux evolution. This claims additional
investigations for any class of off-diagonal solutions in order to
understand if they result in effective models which belong, or not, to
swampland.

\item \textbf{CCL2: The AdS distance conjecture and claim for nonassociative
phase spaces}. Another important conclusion of this work is that combining
the conditions of CCL1 with the notion of (conformal) Weyl re-scaling of
nonassociative phase space metrics, we can extend the AdS Distance
conjecture in a form including a part a) as a conjecture for a subclass of
phase space quasi-stationary s-metrics with $\kappa$- and $\tau$ parametric
nonlinear symmetries (\ref{nonlinsymr}) connecting effective sources to
effective running cosmological constants. A part b) Claim in CCL2 is
motivated because the conformal transforms and symmetries are relevant only
to very special classes of solutions encoding nonassociative data and
generic off-diagonal geometric flow evolution scenarios and/or
nonassociative gravitational interactions. In general, we need an additional
analysis in order to determine if a class of parametric solutions for
nonassociative Ricci flows belong, or not, to swampland.

\item \textbf{CCL3: The fixed points of nonassociative geometric flows
conjecture and claim}. This concludes that in the nonassociative geometric
flow theory elaborated in this work there exist parametric solutions
defining infinite towers with effective zero masses points for flow
evolution toward a fixed point at infinite distance. The a) Conjecture in
CCL3 can be checked on nonassociative phase spaces, for instance, for
s-metrics (\ref{pm5d8d}) in a form similar to that for associative and
commutative Ricci flows in \cite{kehagias19}. For more general classes of
nonassociative quasi-stationary off-diagonal solutions, we need a more
rigorous analysis for chosen classes of generating functions and effective
cosmological constants. Different nonholonomic configurations may result, or
not, in well defined effective classical and QG models.

\item \textbf{CCL4: The canonical distance - scalar curvature conjecture and
claim for nonassociative geometric flows}. The AFCDM allows to decouple
(modified) Ricci flow and gravitational equations and construct
exact/parametric solutions if the canonical s-connection is used instead of
the LC-connection. An original result of the subsection \ref{ss613} consists
in deriving formulas for the generalized canonical distance and related
tower of effective mass states in terms of the canonical scalar curvatures
and running shell cosmological constants (see constructions related to (\ref%
{gdchccl4}) and (\ref{masstf1})). So, in principle, we can check if some
classes of parametric solutions belong or not to swampland if we formulate
all results in terms of the canonical s-connections. We can extract
LC-configurations imposing additional nonholonomic constraints but the
corresponding geometric constructions and analysis of possible physical
implications depend on the type of generated families of quasi-stationary
solutions. This imposes certain limitations for testing in general LC-form
the CCL4. Nevertheless, such formulas involving the canonical s-curvature
allow to express the necessary results in terms of $\ _{s}^{\shortmid
}\Lambda (\tau )$ which is important for testing, for instance, the CCL6 and
CCL7 (see below).

\item \textbf{CCL5: The canonical distance - F-functional conjecture and
claim for nonassociative geometric flows}. The idea to define the general
distance functional in terms of F- and W-functionals was elaborated for
associative and commutative geometric flows, see Conjectures B2 and B3 in 
\cite{kehagias19}. The subsection \ref{ss614} consists an original extension
of those constructions for nonassociative geometric flows determined by
respective $\kappa$-parametric generalizations of G. Perelman's functionals.
Using $\ _{s}^{\shortmid }\widehat{\mathcal{F}}_{\kappa }^{\star }(\tau )$ (%
\ref{naffunctp}), we can compute the generalized distance and state the
conditions of existence of canonical infinite tower of phase space states
with zero effective masse as in a) Conjecture of CCL5. The part b) Claim
states that in general, solving respective $\tau$- and $\kappa$-parametric
equations we can express necessary formulas in terms of respective
F-functionals and decide if a class of solutions belong or not two swampland
and speculate on how LC-configurations can be extracted. Nevertheless, we
conclude that the constructions with generalized F-functionals are quite
sophisticate when we study the statistic and geometric thermodynamic
properties of modified (nonassociative and/or other types) geometric flows
and Ricci soliton configurations and, in particular of quasi-stationary and
nonholonomic BH solutions.

\item \textbf{CCL6: The canonical distance - W-functional conjecture and
claim for nonassociative geometric flows}. The a) Conjecture and b) Claim
parts are similar to those stated for CCL5 but using the star deformed
W-functional. We derived the formulas $\ _{W}^{s}\widehat{\aleph }\simeq
\log (\ _{s}^{\shortmid }\widehat{\mathcal{W}}_{\kappa }^{\star }(\tau
_{i})/\ _{s}^{\shortmid }\widehat{\mathcal{W}}_{\kappa }^{\star }(\tau
_{f})) $ (\ref{wdist}) and $\ ^{\shortmid }m\sim M_{p}e^{-\hat{\alpha}|\ \
_{W}^{s}\widehat{\aleph } |}\simeq \left( \ _{s}^{\shortmid }\widehat{%
\mathcal{W}}_{\kappa }^{\star}(\tau )\right) ^{\hat{\alpha}}$ (\ref{wmass}),
for $\ _{W}^{s}\widehat{\aleph }\rightarrow \infty $, which are very
important and simplify the procedure for analyzing if certain classes of
exact/ parametric solutions belong, or not, to swampland. Using nonlinear
symmetries, we can express $\ _{s}^{\shortmid }\widehat{\mathcal{W}}_{\kappa
}^{\star }(\tau _{i})$ as a functional of running cosmological constants and
volume forms and, then, we can speculate on classes of generating functions
which drive a class of solutions to swampland, or inversely, keep it with
some well-defined properties. For associative and commutative Ricci flows,
such formulas were conjectured in \cite{kehagias19}.

\item \textbf{CCL7: The phase space black hole entropy distance conjecture
and claim for nonassociative geometric flows}. An original approach to
swampland and BH physics was elaborated in \cite{bonnefoy19,lueben21}
exploiting possible connections between the distance conjecture and area
low, involving Ricci flow conjectures, for the Bekenstein-Hawking entropy.
The statements of the Black Hole Entropy Distance Conjecture, BHEDC, can be
extended to certain classes of nonassociative phase solutions for geometric
flows/ Ricci solitons if the s-metrics contain certain hyper-surface
configurations, as we concluded in a) Conjecture of CCL7. Nevertheless, even
for such type of quasi-stationary $\kappa$- and $\tau$-running solutions
related via nonlinear symmetries to running cosmological constants, we may
need an additional analysis (the b) Claim of CCL7) in order to decide if
such a solution belong to swampland or not. We list a table with 4 examples
of such nonlinear quadratic elements and respective formulas for phase
generalized Bekenstein-Hawking entropy computed, for instance, for rotoid
configurations.
\end{itemize}

Summarizing CCL1-CCL7, we conclude that, in general, the solutions of
nonassociative geometric flow and Ricci soliton equations can be
characterized by a corresponding modified G. Perelman thermodynamic model,
when the conditions of CCL6 can be analyzed with respect to the purposes of
the Swampland Program, even the statements of CCL7 are not applicable. Here
we note that we can define and compute two types of entropy variables: the
"minus" entropy, i.e. W-entropy, $\ _{s}^{\shortmid }\widehat{\mathcal{W}}%
_{\kappa }^{\star }(\tau )$ and $\ _{s}^{\shortmid }\widehat{\mathcal{S}}%
_{\kappa }^{\star }(\tau )$, with effective shell sources $%
\sum\nolimits_{s}\ _{s}^{\shortmid }\Im (\tau )$ (\ref{swcc57}). Using
nonlinear symmetries / transforms (\ref{nonlinsymr}), such formulas allow us
to modify respectively the statements of CCL6 in certain forms which are
applicable to compute the generalized distance functional and the effective
mass stated scale in terms of effective running cosmological constants. We
present such important formulas for quasi-stationary solutions, see (\ref%
{wdistqscc}) and (\ref{wmassqscc}). Together with CCL6, the modified CCL7 is
tested by explicit examples of nonassociative quasi-stationary geometric
flow deformations of phase space double nonholonomic deformed RN-dS BHs and
nonholonomic deformed RN-AdS BHs, see respective formulas in subsections \ref%
{ss623} and \ref{ss624}.

\vskip5pt Let us discuss the proofs and validity of results and claims of
this article stated as Issues 1-5, \textbf{Is1-5}:

\begin{itemize}
\item \textbf{Is1:} \textit{Postulating nonassociative geometric flow and
vacuum gravitational equations.} It is not possible to formulate a general
theory of nonassociative or noncommutative geometric flows or MGTs involving
different types of non-Riemannian metric and (non) linear connection structures. For
a subclass of theories determined by R-flux deformations \cite%
{partner01,partner02,partner03,partner04}, it is possible to define star
variants of Riemannian and Ricci tensors and perform corresponding $\kappa$%
--linear decompositions of geometric objects (generalizing in nonholonomic
form the constructions from \cite{blumenhagen16,aschieri17}). The geometric
constructions are very different for other  models involving octonionic
or supersymmetric variables, spectral triples etc. \cite%
{v08,castro2,mylonas13,kupriyanov15,gunaydin,szabo19,vbv18}. Using $\
_{s}^{\shortmid }\widehat{\mathcal{R}}ic^{\star }$ and $\ _{s}^{\shortmid }%
\widehat{\mathcal{R}}sc^{\star }$ from (\ref{mafgeomobn}), we can postulate
the nonassociative geometric flow equations (\ref{nonassocheq}), or (\ref%
{nonassocgeomfl}), describing the evolution of $\tau $-families $\left( \
_{\star}^{\shortmid }\mathfrak{g}_{\alpha _{s}\beta _{s}}(\tau ), \
_{s}^{\shortmid }\widehat{\mathbf{D}}^{\star }(\tau )\right) $ (as we
considered in subsection \ref{ss231}). For some subclasses of nonassociative
Ricci solitons, we obtain nonassociative generalizations of the Einstein
equations (\ref{naricsol}), or (\ref{nonassocdeinst1}), which can be
formulated independently in abstract symbolic geometric form as in \cite{misner} but for nonassociative geometric s-objects.

\item \textbf{Is2: }\textit{Postulating nonassociative Perelman's
functionals and thermodynamic variables.} Considering $\ _{s}^{\shortmid }%
\widehat{\mathcal{R}}ic^{\star }$ and  $\ _{s}^{\shortmid }\widehat{\mathcal{%
R}}sc^{\star }$, and respective normalizing functions and volume forms, we
can define in abstract geometric form the F- and W-functionals (\ref%
{naffunct}) and (\ref{nawfunct}). Such Lyapunov type functionals consist
nonholonomic relativistic generalizations and star product deformations of
standard Perelman functionals (\ref{perelmfst}) for Ricci flows of
Riemannian metrics. The nonassociative W-functional (\ref{nawfunct}) is
similar to a "minus" nonassociative entropy, which allowed us to define
geometrically (generalizing the constructions from \cite{perelman1})
nonassociative versions of statistical/ geometric thermodynamic variables (%
\ref{nagthermodvalues}). Such nonassociative geometric thermodynamic models can be
defined in abstract symbolic geometric form and corresponding physical
motivations are provided at the end of subsection \ref{ss232}, see also
below the paragraphs Is3 - Is5.

\item \textbf{Is3:} \textit{Problems and a cure for formulating variational
procedures for deriving nonassociative geometric flow and nonassociative
gravitational field equations.} One of the fundamental G. Perelman's result 
\cite{perelman1} consisted in a proof that for Riemannian geometries the R.
Hamilton equations (\ref{hameqst}) can be derived using variational
procedures for F- or W-functionals (\ref{perelmfst}). Those geometric
constructions were used for elaborating rigorous proofs of the Poincar\'{e}%
-Thurston conjecture \cite{perelman1}, see details in monographs \cite%
{kleiner06,morgan06,cao06}. Unfortunately, it is not possible to formulate a
self-consistent variational principle for general nonassociative
deformations determined by a general twist product and in other types of
nonassociative and noncommutative theories. Here, we note that one can be
defined an infinite number of nonassociative and noncommutative differential
and integral calculi which is different from the commutative (pseudo)
Riemannian geometry. So, it is not possible to formulate some general forms
of nonassociative the Poincar\'{e}-Thurston conjecture and use "twisted"
variational constructions for corresponding proofs. Nevertheless, this does
not prohibit us to elaborate on physically important and well-defined models
of nonassociative geometric flows. We can postulate in symbolic geometric
form nonassociative geometric flow equations (\ref{nonassocheq}), or (\ref%
{nonassocgeomfl}), and then to study their thermodynamic properties using
associated nonassociative F-and W-functionals. For such nonassociative
geometric constructions, we can perform always parametric decompositions and
compute all terms for $\kappa $-deformed R. Hamilton equations (\ref%
{nonassocgeomflp}) and Perelman's functionals (\ref{naffunctp}) and (\ref%
{nawfunctp}). Working with $\kappa $--linear deformed theories, we can
always elaborate self-consistent variational procedures, when corresponding
nonassociative geometric flow/ gravitational equations are characterized by $%
\kappa $-linear thermodynamic variables (\ref{nagthermodvaluesp}).

\item \textbf{Is4:} \textit{Physically important nonvariational theories.}
There are various physical arguments (similarly to those for models of hydrodynamics
with parametric turbulence, or for thermodynamic models with parametric
phase transitions, multi-diffusion and branch decomposition and evolution)
to study nonassociative geometric flow theories. In such cases, the
definition of self-consistent variational procedures is not uniquely defined
but depends on some prescribed (non) linear / associative / commutative
symmetries, on the order of parametric decompositions and corresponding
nonholonomic s-distributions. We emphasize that the variational procedure
described in subsection \ref{ss232} can be performed in a quite general form
if we begin with a corresponding commutative canonical s-adapted variational
nonholonomic configuration and subject the constructions to $\kappa $%
-parametric deformations following the Convention 2 (\ref{conv2s}). Such a
variational procedure can be performed recurrently for higher orders on $%
\kappa $. Unfortunately, to prove some general convergence conditions
summarizing all terms on powers of $\kappa $ is an unsolved mathematical
problem, which is typical for nonlinear functional analysis. Theoretically,
we can consider an inverse situation and begin with general nonassociative
formulas for flow equations (\ref{nonassocheq}) and formal functionals (\ref%
{naffunct}) and (\ref{nawfunct}). Then, we can perform adapted $\kappa $%
-parametric decompositions involving  imaginary and real terms.
Unfortunately, in such cases, we are not able to elaborate a general method
for constructing exact/parametric solutions but following the inverse
procedure we can always apply the AFCDM as we proved in previous sections
(for quasi-stationary solutions) and in partner works \cite%
{partner02,partner03,partner04}. This is a generic property for any
nonassociative/ noncommutative differential and integral calculus.

\item \textbf{Is5: } \textit{Validity through exact and parametric solutions
of physically important nonlinear systems of PDE encoding nonassociative
data.} So, haven defined corresponding fundamental nonassociative geometric s-objects
like a star deformed metric, a canonical s-connection and respective
Ricci s-tensor and canonical scalar, we can always postulate geometrically
certain nonassociative geometric flow equations and F- and
W-functionals. Such values are well-defined in all orders on $\kappa $ but
for twisted theories there is not a general nonassociative variational proof
for nonassociative geometric flow/ gravitational equations from some
generalized Perelman/ action functionals. For applications in modern physics
and quantum information theory, we can consider only $\kappa $-linear
constructions, apply the AFCDM and construct physically important solutions
(star-deformed black holes, wormholes, locally anisotropic cosmological
s-metrics), and speculate on verifiable models encoding nonassociative data.
Recurrently, we can construct nonassociative and noncommutative solutions
with higher orders on $\kappa $ and  $\hbar $ but the technically the resulting formulas are much cumbersome.  For any such higher order
parametric configurations, we can define and compute effective F- and W-functionals,
and respective nonassociative thermodynamic variables. Re-defining the
measures (for respective normalizing s-functions and nonholonomic
s-distributions), we suppose that a corresponding variational procedure can
be formulated for any stated polynomial order on $\kappa $ and $\hbar $ (like we
considered in subsection \ref{ss232}). But this do not provide us a general
well-defined nonassociative variational theory with twist product. We can
construct and study physical properties of realistic nonassociative theories
considering physically important solutions encoding  $\kappa $-linear data.
\end{itemize}

Finally, with respect to above Is1-5, we note that (in certain similar forms) such
problems exist in all classical gravity and QG theories. For instance, the
classical Einstein equations are linear on $\varkappa =8\pi G/c^{4},$ were $G
$ is the Newtonian constant of gravitation and $c$ is the light velocity
constant. We do not have yet a well-defined theory of QG, but it is aways
possible to elaborate on physically important effects proportional to $%
\varkappa $ and $\hbar $ and consider higher orders on such constants
involving (square) curvature terms etc. Such effective models are
commutative or noncommutative. R-flux deformations from string theory result
in nonassociative geometric configurations determined by  twist products. This results in ambiguities for
constructing  general self-consistent principles, involves complex terms for
curvatures and Ricci tensors for some parametric decompositions on  $\kappa $ and $%
\hbar $   like in \cite{blumenhagen16,aschieri17,partner01}  etc.  This
reflects our not complete knowledge about string theory and QG. The priority
of nonholonomic s-adapted definition of star product   (\ref{starpn}) is
that using the  Convention 2 (\ref{conv2s}) we can define and compute
recurrently parametric  R-flux deformations of some (associative/commutative) variatonal equations and apply the AFCDM in order to construct physically important solutions of nonassociative geometric flows
and vacuum gravitational equations. In general form, such nonassociative
solutions are characterized by respective modified Perelman (thermodynamic)
functionals and variables. This states a new paradigm for formulating and investigating possible physical implications of nonassociative theories and provides a general computational method when 
nonassociative effects can be computed and the validity of certain claims can be 
verified at least (for simplicity)  in the linear approximation on  $\kappa $.

\vskip5pt Above conclusions and theoretical/ computational tests of
CCL1-CCL7 support the \textbf{The Main Hypothesis, MH, } of this work
(formulated at the end of subsection \ref{ssint12}) \emph{that the Swampland
Program has to be revised/ modified in order to elaborate explicit criteria
how to include nonassociative and noncommutative geometric flows, QM, QFTs,
and MGTs in elaborating QG theories. The corresponding Conjectures and
Claims allows us to select self-consistent nonassociative geometric and
physical models (defined by generic off-diagonal solutions) encoding at
least in parametric form nonassociative star product and R-flux data for
string / M-theory. In low-energy limits, such configurations can be
completed into QG in the UV forms and distinguished from another classes of
theories/models/ solutions which do not have such properties.}

\vskip5pt The MH modifies a series of future research purposes stated in our
partner works \cite{partner01,partner02,partner03,partner04}. We list and
speculate on five perspective directions of research involving
nonassociative geometric flow methods, classical and quantum gravity, and
quantum information models\footnote{%
for instance, Q1a is used for modifications involving nonassociative Ricci
flow of the query Q1 from \cite{partner04}; we also state a new query Q5.}:

\begin{itemize}
\item \textbf{Q1a}: \textit{Nonassociative Einstein-Yang-Mills-Higgs systems.%
} A full investigation of nonassociative geometric flow and gravity theories
should involve flux evolution and field equations with nontrivial matter
sources and other types of nonassociative/ noncommutative structures (for
instance, octonions etc.) which are not necessarily determined by R-flux
deformations but based on other types of star product and/or nonassociative
algebraic structures. One of the next steps is to formulate and study models
with nonassociative nonholonomic deformations and geometric flow evolution
of Einstein-Yang-Mills-Higgs systems resulting in nonassociative gravity and
matter field theories with nonsymmetric metrics, generalized connections and
nonzero sources. Such nonassociative models should generalize in
nonassociative form the Einstein-Eisenhart-Moffat theories \cite%
{einstein25,einstein45,eisenhart51,eisenhart51,moffat95}, see respective
nonholonomic and phase space Ricci flow constructions in \cite{sv08}.
References \cite{partner01,partner02,partner03,partner04} provide a series
of new ideas on further nonassociative/ nonsymmetric metric / nonholonomic
developments when the AFCDM can be applied for constructing various classes
of exact/parametric solutions (for instance, defining locally anisotropic
wormholes, BHs and BEs, solitonc hierarchies, nonassociative quasi-periodic
cosmological structures etc.).

\item \textbf{Q2a:} \textit{Nonassociative Finsler-Lagrange-Hamilton
geometric flows/gravity, their almost symplectic models, and deformation
quantization.  }Nonassociative geometric flow and gravity theories
determined by a star product of type (\ref{starpn}) on tensor products of
cotangent bundles consist of a class of nonassociative generalizations of
the Finsler-Lagrange-Hamilton geometry which have generalized
metrics/connections depending on velocity/ momentum-like coordinates. In
relativistic/ noncommutative and commutative/ supersymmetric / fractional
etc. variants, there are status reports and reviews of results \cite%
{vbv18,bubuianu19}. The importance of Finsler-like geometric objects and
nonholonomic variables (they can be defined even on Lorentz manifold enabled
with a nonolonomic fibered structure) is that they can be reformulated as
some equivalent almost Kaehler/ symplectic and almost complex variables.
This allows us to elaborate on realistic classical models for (non)
associative/ commutative geometric flow and gravity theories (to apply
complex variables for quantum models and almost complex variables for
classical models) and apply/develop methods of deformation and/or geometric
quantization. A more sophisticate program is to formulate and study models
of QG involving nonassociative/ noncommutative Lagrange-Hamilton structures
and/or almost symplectic variables.

\item \textbf{Q3a: }\textit{Elaborating Bekenstein-Hawking and G.\ Perelman
thermodynamic and locally anisotropic kinetic models for physically
important solutions in nonassociative classical and quantum theories.} The
sections \ref{sec2} and \ref{sec3} of this work provide a solution of query
Q3 stated at the end of \cite{partner04}. Here we remember that the
nonassociative gravity with star product and R-flux deformations was
formulated up to defining and computing the nonassociative Ricci tensor in 
\cite{blumenhagen16,aschieri17}. This was enough to develop models of
nonassociative/ noncommutative Ricci flows but those abstract geometric
and/or coordinate frame constructions based on LC-connection do not allow to
motivate respective generalizations of R. Hamilton equations, prove
decoupling and integration properties of such geometric flow equations and
find exact/ parametric solutions. To study possible physical implications we
had to elaborate a respective nonholonomic formalism with dyadic shell
structures, see details in \cite{partner01,partner02} and Appendices to this
work. Such constructions were motivated by finding physically important
exact/parametric solutions and computing respective Bekenstein-Hawking
and/or G. Perelman thermodynamic variables in section 4 of \cite{partner04}
and section \ref{sec5} in this paper.

\item \textbf{Q4a:} \textit{The program on nonassociative geometric and
quantum information flow theories. }We point again to perspectives and
importance of a research program to extend in nonassociative forms the
geometric flow information theory \cite{vacaru20,ib19} to theories with
nonassociative qubits and entanglement, conditional entropies etc. Such new
directions in modern QFT, strings and gravity, theory of quantum computers
have deep roots in nonassociative quantum mechanics and gauge models \cite%
{jordan32,jordan34,castro2,szabo19} and motivations from noncommutative
geometry, string and M-theory \cite%
{v08,mylonas13,kupriyanov15,gunaydin,blumenhagen16,aschieri17}.

\item \textbf{Q5:} \textit{Cosmological models and dark energy and dark
matter physics encodding quasi periodic structure and data for
nonassociative theories.  }In this work, we constructed and studied the main
properties of a series of new classes of nonassociative exact/
para\-met\-ric solutions for $\tau $-running quasi-stationary physically
important solutions defining star product and R-flux deformed BHs, BEs etc.
The AFCDM can be also developed and applied in certain dual forms when there
are generated locally anisotropic solutions \cite{sv18,lb18,ibubuianu21}.
Extending such geometric and analytic methods, we plan to elaborate on a
series of works on phase space and spacetime nonassociative quasi-crystal
configurations and related web filaments, nonassociative quasi-periodic
geometric flow evolution and pattern forming structures in accelerating and
inflationary cosmology and dark energy and dark matter physics. Such models
are generic off-diagonal (for certain classes of holonomic configurations,
the s-metrics can be diagonalized) encoding nonassociative star product
deformations. They are characterized by respective generalized G. Perelman
thermodynamic variables and put their imprints and requests for
modifications of the Swampland Program.
\end{itemize}

We shall develop on above directions and report on progress for queries
Q1a-Q4a,Q5 in future works.

\vskip6pt \textbf{Acknowledgments:} This is the second work together with 
\cite{partner04} supported by a Fulbright senior fellowship for SV and
hosted by the physics department at California State University at Fresno,
USA. Some results are related also to a research program proposed for the
PNRR-Initiative 8 of the Romanian Ministry of Research, Innovation and
Digitization and for a visiting program to be decided by CASLMU M\"{u}nchen.
It is also a partner work to \cite{partner01,partner02,partner03} extending
to nonassociative geometric flows former research programs and fellowships
at CERN (Geneva, Switzerland) and Max Planck Institut f\"{u}r Physik /
Werner Heisenberg Institut, M\"{u}nchen (Germany). SV is also grateful to
professors D. L\"{u}st, N. Mavromatos, J. Moffat, Yu. A. Seti, P. Stavrinos,
and M. V. Tkach for hosting visits and collaborations. Authors thank the
referee for very important critics and requests to provide substantial
modifications and proofs on s-adapted variational calculus for
nonassociative geometric flow equations performed in section \ref{ssngfeq}.

\appendix

\setcounter{equation}{0} \renewcommand{\theequation}
{A.\arabic{equation}} \setcounter{subsection}{0} 
\renewcommand{\thesubsection}
{A.\arabic{subsection}}

\section{The anholonomic frame and connection deformation method, AFCDM}

\label{appendixb}

We outline the AFCDM for constructing exact and parametric solutions for
nonlinear systems of PDEs (\ref{seinsta}). Details with proofs and methods
in commutative gravity theories are provided in \cite{ib19,bubuianu19} and,
for recent nonassociative generalizations, \cite%
{partner01,partner02,partner03,partner04}. In this section, we consider
effective sources parameterized in the form (\ref{cannonsymparamc2a})
encoding nonassociative star product and R-flux deformations. Such PDEs
can't be integrated in some general off-diagonal forms if we follow standard
methods from GR \cite{kramer03,misner,wald82,hawking73} when solutions are
found for some special diagonal ansatz of metrics transforming the
(modified) Einstein equations into certain systems of nonlinear ordinary
differential equations, ODE. For diagonalizable metrics, there are imposed
some higher order symmetries (spherical, cylindrical type etc.) which allow
to integrate the resulting ODEs in very general forms. The solutions are
classified by respective integrations constants. This is used, for instance,
for constructing black hole, BH, solutions. The AFCDM is a more genera
geometric method for generating exact and parametric solutions, when a
necessary type of auxiliary connection (for instance, the canonical
s-connection $\ _{s}^{\shortmid }\widehat{\mathbf{D}}$) is used. This allows
us to decouple and integrate systems of (modified) geometric flow and
gravitational field equations defined by generic off-diagonal metrics,
(non-) Riemannian connections etc. with coefficients depending, in
principle, on all spacetime and phase space coordinates. Different classes
of such exact/ parametric solutions are determined by respective classes of
generating functions, generating sources, integrating functions and
constants, and/or decompositions certain small physical parameters (for
instance, on string and Plank constants) etc.

To apply the AFCDM the quadratic linear element for generating
quasi-stationary solutions is parameterized by such an off-diagonal ansatz
for s-metrics: 
\begin{equation}
d\widehat{s}%
^{2}=g_{i_{1}}(x^{k_{1}})(dx^{i_{1}})^{2}+g_{a_{2}}(x^{i_{1}},y^{3})(\mathbf{%
e}^{a_{2}})^{2}+\ ^{\shortmid }g^{a_{3}}(x^{i_{2}},p_{5})(\ ^{\shortmid }%
\mathbf{e}_{a_{3}})^{2}+\ ^{\shortmid }g^{a_{4}}(\ ^{\shortmid
}x^{i_{3}},E)(\ ^{\shortmid }\mathbf{e}_{a_{4}})^{2},  \label{ans1dm}
\end{equation}%
for $%
i_{1},j_{1},k_{1}...=1,2;a_{2}=3,4;i_{2}=(i_{1},b_{2}),b_{2}=3,4;i_{3}=(i_{2},b_{3}),b_{3}=5,6;b_{4}=7,8 
$ etc.; and coordinates $%
x^{i_{1}}=(x^{1},x^{2});u^{i_{2}}=x^{i_{2}}=(x^{i_{1}},y^{b_{2}}),y^{4}=t;\
^{\shortmid }u^{i_{3}}=\ ^{\shortmid }x^{i_{3}}=(x^{i_{2}},p_{b_{3}});\
^{\shortmid }u^{i_{4}}=(\ ^{\shortmid }x^{i_{3}},p_{b_{4}}),p_{8}=E$. The
dual bases in (\ref{ans1dm}), 
\begin{equation*}
\mathbf{e}^{a_{2}}=dy^{a_{2}}+N_{k_{1}}^{a_{2}}(x^{i_{1}},y^{2})dx^{k_{1}},\
^{\shortmid }\mathbf{e}_{a_{3}}=dp_{a_{3}}+\ ^{\shortmid }N_{a_{3}k_{2}}^{\
}(x^{i_{2}},p_{6})dx^{k_{2}},\ ^{\shortmid }\mathbf{e}_{a_{4}}=dp_{a_{4}}+\
^{\shortmid }N_{a_{4}k_{3}}^{\ }(\ ^{\shortmid }x^{i_{3}},E)d\ ^{\shortmid
}x^{k_{3}},
\end{equation*}%
are determined by respective N-connection coefficients,%
\begin{eqnarray}
N_{k_{1}}^{3}
&=&w_{k_{1}}(x^{i_{1}},y^{3}),N_{k_{1}}^{4}=n_{k_{1}}(x^{i_{1}},y^{3});
\label{ans1n} \\
\ ^{\shortmid }N_{5k_{2}}^{\ } &=&n_{k_{2}}(x^{i_{2}},p_{5}),\ ^{\shortmid
}N_{6k_{2}}^{\ }=w_{k_{2}}(x^{i_{2}},p_{5});\ ^{\shortmid }N_{7k_{3}}^{\
}=n_{k_{3}}(x^{i_{3}},E),\ ^{\shortmid }N_{8k_{3}}^{\
}=w_{k_{3}}(x^{i_{3}},E).  \notag
\end{eqnarray}%
We emphasize that above coefficients defined with respect to s-adapted
frames do not depend on $u^{4}=y^{4}=x^{4}=t,$ i.e. there is a Killing
vector $\partial _{t}$ on shell $s=2;$ do not depend on $\ ^{\shortmid
}u^{6}=p_{6},$ i.e. there is a Killing symmetry on $\ ^{\shortmid }\partial
^{6}$ on shell $s=3;$ and do not depend on $\ ^{\shortmid }u^{7}=p_{7},$
i.e. there is a Killing symmetry on $\ ^{\shortmid }\partial ^{7}$ on shell $%
s=4.$ In similar forms, changing parameterizations of coordinates and
coefficients, we can consider ansatz with Killing symmetry on $\partial _{3}$
\ instead of $\partial _{4};$ on $\ ^{\shortmid }\partial ^{5}$ instead of $%
\ ^{\shortmid }\partial ^{6};$ on $\ ^{\shortmid }\partial ^{8}$ instead of $%
\ ^{\shortmid }\partial ^{7}.$ Such parameterizations allow to decouple and
integrate in explicit form various classes of vacuum and non-vacuum modified
gravitational equations.

\subsection{Off-diagonal quasi-stationary solutions with effective sources}

\label{sssol1}Tedious computations provided in \cite{partner02} (re-defined
in real momentum coordinates $\ ^{\shortmid }u^{\alpha }$) prove that
quasi-stationary solutions of the nonassociative parametric vacuum
gravitational equations (\ref{seinsta}) with effective sources (\ref%
{cannonsymparamc2a}) are defined by respective s-metric (\ref{ans1dm}) and
N-connection (\ref{ans1n}) coefficients: 
\begin{eqnarray}
g_{1}(x^{i_{1}}) &=&g_{2}(x^{i_{1}})=e^{\psi (\hbar ,\kappa ;x^{k_{1}})},
\label{sol1} \\
g_{3}(x^{i_{1}},y^{3}) &=&\frac{[\partial _{3}(\ _{2}\Psi )]^{2}}{%
4(~_{2}^{\shortmid }\mathcal{K})^{2}\{g_{4}^{[0]}-\int dy^{3}\frac{\partial
_{3}[(\ _{2}\Psi )^{2}]}{4(\ ~_{2}^{\shortmid }\mathcal{K})}\}},\
g_{4}(x^{i_{1}},y^{3})=g_{4}^{[0]}-\int dy^{3}\frac{\partial _{3}[(\
_{2}\Psi )^{2}]}{4(~_{2}^{\shortmid }\mathcal{K})},  \notag
\end{eqnarray}%
\begin{eqnarray*}
^{\shortmid }g^{5}(x^{i_{1}},y^{3},p_{5}) &=&\frac{[\ ^{\shortmid }\partial
^{5}(\ _{3}^{\shortmid }\Psi )]^{2}}{4(~_{3}^{\shortmid }\mathcal{K})^{2}\{\
^{\shortmid }g_{[0]}^{6}-\int dp_{5}\frac{\ ^{\shortmid }\partial ^{5}[(\
_{3}^{\shortmid }\Psi )^{2}]}{4(~_{3}^{\shortmid }\mathcal{K})}\}},\
^{\shortmid }g^{6}(x^{i_{1}},y^{3},p_{5})=\ ^{\shortmid }g_{[0]}^{6}-\int
dp_{5}\frac{\ ^{\shortmid }\partial ^{5}[(\ _{3}^{\shortmid }\Psi )^{2}]}{%
4(\ _{3}^{\shortmid }\mathcal{K})},\  \\
^{\shortmid }g^{8}(x^{i_{1}},y^{3},p_{5},p_{7}) &=&\frac{[\ ^{\shortmid
}\partial ^{7}(\ _{4}^{\shortmid }\Psi )]^{2}}{4(~_{4}^{\shortmid }\mathcal{K%
})^{2}\{\ ^{\shortmid }g_{[0]}^{8}-\int dp_{7}\frac{\ ^{\shortmid }\partial
^{7}[(\ _{4}^{\shortmid }\Psi )^{2}]}{4(~_{4}^{\shortmid }\mathcal{K})}\}},\
^{\shortmid }g^{8}(x^{i_{1}},y^{3},p_{5},p_{7})=\ ^{\shortmid
}g_{[0]}^{8}-\int dp_{7}\frac{\ ^{\shortmid }\partial ^{7}[(\
_{4}^{\shortmid }\Psi )^{2}]}{4(\ _{4}^{\shortmid }\mathcal{K})};\ 
\end{eqnarray*}%
\begin{eqnarray*}
\mbox{ and }N_{k_{1}}^{3} &=&w_{k_{1}}(x^{i_{1}},y^{3})=\frac{\partial
_{k_{1}}(\ _{2}\Psi )}{\partial _{3}(\ _{2}\Psi )}, \\
N_{k_{1}}^{4} &=&n_{k_{1}}(x^{i_{1}},y^{3})=\ _{1}n_{k_{1}}+\
_{2}n_{k_{1}}\int dy^{3}\frac{\partial _{3}[(\ _{2}\Psi )^{2}]}{4(\
~_{2}^{\shortmid }\mathcal{K})^{2}|g_{4}^{[0]}-\int dy^{3}\frac{\partial
_{3}[(\ _{2}\Psi )^{2}]}{4(~_{2}^{\shortmid }\mathcal{K})}|^{5/2}};
\end{eqnarray*}%
\begin{eqnarray*}
\ ^{\shortmid }N_{5k_{2}}^{\ } &=&w_{k_{2}}(x^{i_{2}},p_{5})=\frac{\partial
_{k_{2}}(\ _{3}^{\shortmid }\Psi )}{\ ^{\shortmid }\partial ^{5}(\
_{3}^{\shortmid }\Psi )}, \\
\ ^{\shortmid }N_{6k_{2}}^{\ } &=&n_{k_{2}}(x^{i_{2}},p_{5})=\
_{1}^{\shortmid }n_{k_{2}}+\ _{2}^{\shortmid }n_{k_{2}}\int dp_{5}\frac{\
^{\shortmid }\partial ^{5}[(\ _{3}^{\shortmid }\Psi )^{2}]}{4(\
_{3}^{\shortmid }\mathcal{K})^{2}|\ ^{\shortmid }g_{[0]}^{6}-\int dp_{5}%
\frac{\ ^{\shortmid }\partial ^{5}[(\ _{3}^{\shortmid }\Psi )^{2}]}{4(\
_{3}^{\shortmid }\mathcal{K})}|^{5/2}};
\end{eqnarray*}%
\begin{eqnarray*}
\ ^{\shortmid }N_{7k_{3}}^{\ } &=&w_{k_{3}}(x^{i_{2}},p_{5},p_{7})=\frac{\
^{\shortmid }\partial _{k_{3}}(\ _{4}^{\shortmid }\Psi )}{\ ^{\shortmid
}\partial ^{7}(\ _{4}^{\shortmid }\Psi )}, \\
\ ^{\shortmid }N_{8k_{3}}^{\ } &=&n_{k_{3}}(x^{i_{2}},p_{5},p_{7})=\
_{1}^{\shortmid }n_{k_{3}}+\ _{2}^{\shortmid }n_{k_{3}}\int dp_{7}\frac{\
^{\shortmid }\partial ^{7}[(\ _{4}^{\shortmid }\Psi )^{2}]}{4(\
_{4}^{\shortmid }\mathcal{K})^{2}|\ ^{\shortmid }g_{[0]}^{8}-\int dp_{7}%
\frac{\ ^{\shortmid }\partial ^{7}[(\ _{4}^{\shortmid }\Psi )^{2}]}{4(\
_{4}^{\shortmid }\mathcal{K})}|^{5/2}},
\end{eqnarray*}%
Above coefficients are functionals on such functions and parameters: 
\begin{eqnarray}
&&\mbox{generating functions: }\psi (\hbar ,\kappa ;x^{k_{1}});\ _{2}\Psi
(\hbar ,\kappa ;x^{k_{1}},y^{3});\ _{3}^{\shortmid }\Psi (\hbar ,\kappa
;x^{k_{2}},p_{5});\ _{4}^{\shortmid }\Psi (\hbar ,\kappa ;\ ^{\shortmid
}x^{k_{3}},p_{7});  \label{integrfunct} \\
&&\mbox{generating sources:}\ ~_{1}^{\shortmid }\mathcal{K}(\hbar ,\kappa
;x^{k_{1}});\ ~_{2}^{\shortmid }\mathcal{K}(\hbar ,\kappa
;x^{k_{1}},y^{3});\ ~_{3}^{\shortmid }\mathcal{K}(\hbar ,\kappa
;x^{k_{2}},p_{5});~_{4}^{\shortmid }\mathcal{K}(\hbar ,\kappa ;\ ^{\shortmid
}x^{k_{3}},p_{7});  \notag \\
&&\mbox{integrating functions: }g_{4}^{[0]}(\hbar ,\kappa ;x^{k_{1}}),\
_{1}n_{k_{1}}(\hbar ,\kappa ;x^{j_{1}}),\ _{2}n_{k_{1}}(\hbar ,\kappa
;x^{j_{1}});  \notag \\
&&\quad \ ^{\shortmid }g_{[0]}^{6}(\hbar ,\kappa ;x^{k_{2}}),\
_{1}n_{k_{2}}(\hbar ,\kappa ;x^{j_{2}}),\ _{2}n_{k_{2}}(\hbar ,\kappa
;x^{j_{2}});\ ^{\shortmid }g_{[0]}^{8}(\hbar ,\kappa ;\ ^{\shortmid
}x^{j_{3}}),\ _{1}^{\shortmid }n_{k_{3}}(\hbar ,\kappa ;\ ^{\shortmid
}x^{j_{3}}),\ _{2}^{\shortmid }n_{k_{3}}(\hbar ,\kappa ;\ ^{\shortmid
}x^{j_{3}}).  \notag
\end{eqnarray}%
We emphasize that $\psi (\hbar ,\kappa ;x^{k_{1}})$ is determined on shell $%
s=1$ as a solution of 2-d Poisson equation, $\partial _{11}^{2}\psi
+\partial _{22}^{2}\psi =2\ _{1}\mathcal{K}$, and may encode certain
nonassociative data if $\ _{1}\mathcal{K}$ contains nonholonomic
dependencies on such ones.

Any generic off-diagonal ansatz (\ref{sol1}) defines a class of exact
solutions determined by generating data (\ref{integrfunct}) and depend in
parametric form on $\hbar ,\kappa $ for any nonassociative star product and
R-flux data encoded in $_{s}^{\shortmid }\mathcal{K}$. Corresponding
quasi-stationary configurations are also characterized by nontrivial
coefficients of respective nonsymmetric metrics $\ _{\star }^{\shortmid }%
\mathfrak{a}_{\alpha _{s}\beta _{s}}$ computed by introducing in (\ref%
{aux40aa}) the s-metric and N-connection coefficients for (\ref{sol1}). Such
solutions are with nontrivial nonholonomic torsion but can be constrained to
subclasses of generating data which solve the conditions (\ref{lccondnonass}%
), see also (\ref{lccond}), and allow to extract LC-configurations, see
details in section 5.3.3 of \cite{partner02}.

\subsection{Nonlinear symmetries and solutions with effective cosmological
constants}

\label{bassnsym}Quasi-stationary solutions posses an important nonlinear
symmetry which allow to formulate them in different functional forms
emphasizing certain classes of effective sources and cosmological constants,
different types of generating functions and parametric decompositions, which
is important for finding other classes of solutions and investigating their
physical properties.

We can study nonassociative nonholonomic deformations of a \textbf{prime}
s-metric $\ _{s}^{\shortmid }\mathbf{\mathring{g}}$ (which may be, or not, a
solution of some (modified) gravitational equations) into a \textbf{target}
s-metric $\ _{s}^{\shortmid }\mathbf{g}$ defining a quasi-stationary
solution (\ref{sol1}) on $\ _{s}^{\star }\mathcal{M},$ 
\begin{equation}
\ _{s}^{\shortmid }\mathbf{\mathring{g}}\rightarrow \ _{s}^{\shortmid }%
\mathbf{g}=[\ ^{\shortmid }g_{\alpha _{s}}=\ ^{\shortmid }\eta _{\alpha
_{s}}\ ^{\shortmid }\mathring{g}_{\alpha _{s}},\ ^{\shortmid
}N_{i_{s-1}}^{a_{s}}=\ ^{\shortmid }\eta _{i_{s-1}}^{a_{s}}\ ^{\shortmid }%
\mathring{N}_{i_{s-1}}^{a_{s}}].  \label{offdiagdef}
\end{equation}%
Such nonholonomic s-deformations can be described in terms of gravitational
polarization functions \newline
($\eta $-polarizations), when the target s-metrics are parameterized 
\begin{equation}
\ ^{\shortmid }\eta _{\alpha _{s}}(\hbar ,\kappa ,x^{i_{s-1}},p_{a_{s}})\ %
\mbox{ and }\ ^{\shortmid }\eta _{i_{s-1}}^{a_{s}}(\hbar ,\kappa
,x^{i_{s-1}},p_{a_{s}}).  \label{dmpolariz}
\end{equation}%
For $\kappa $-linear s-deformations, we can introduce $\chi $-polarizations, 
\begin{eqnarray}
&&[\ ^{\shortmid }g_{\alpha _{s}}=\ ^{\shortmid }\ \zeta _{\alpha
_{s}}(1+\kappa \ ^{\shortmid }\chi _{\alpha _{s}})\ ^{\shortmid }\mathring{g}%
_{\alpha _{s}},\ ^{\shortmid }N_{i_{s}}^{a_{s}}=\ ^{\shortmid }\zeta
_{i_{s-1}}^{a_{s}}(1+\kappa \ ^{\shortmid }\chi _{i_{s-1}}^{a_{s}})\ \
^{\shortmid }\mathring{N}_{i_{s-1}}^{a_{s}}],\mbox{ when }  \label{espolariz}
\\
&&\ ^{\shortmid }\eta _{\alpha _{s}}=\ ^{\shortmid }\zeta _{\alpha
_{s}}(\hbar ,x^{i_{s-1}},p_{a_{s}})[1+\kappa \ \ ^{\shortmid }\chi _{\alpha
_{s}}(\hbar ,x^{i_{s-1}},p_{a_{s}})]\ \mbox{ and }  \notag \\
&&\ ^{\shortmid }\eta _{i_{s-1}}^{a_{s}}=\ ^{\shortmid }\zeta
_{i_{s-1}}^{a_{s}}(\hbar ,x^{i_{s-1}},p_{a_{s}})[1+\kappa \ ^{\shortmid
}\chi _{i_{s-1}}^{a_{s}}(\hbar ,x^{i_{s-1}},p_{a_{s}})].  \notag
\end{eqnarray}%
In detailed forms, general quasi-stationary deformations to solutions of
type (\ref{sol1}) determined by gravitational polarizations of type (\ref%
{dmpolariz}) and (\ref{espolariz}) are studied in section 2.3 of \cite%
{partner02} and appendix A.2 of \cite{partner04}.

Any target s-metric $\ _{s}^{\shortmid }\mathbf{g}$ (\ref{sol1}) satisfies
on shells $s=2,3,4$ certain nonlinear symmetries which allow to re-define
the generating functions and relate the effective sources to certain
effective shell cosmological constants, 
\begin{eqnarray}
(\ _{s}\Psi ,\ ~_{s}^{\shortmid }\mathcal{K}) &\leftrightarrow &(\
_{s}^{\shortmid }\mathbf{g},\ ~_{s}^{\shortmid }\mathcal{K})\leftrightarrow
(~_{s}^{\shortmid }\eta \ \ ^{\shortmid }\mathring{g}_{\alpha _{s}}\sim \
^{\shortmid }\zeta _{\alpha _{s}}(1+\kappa \ ^{\shortmid }\chi _{\alpha
_{s}})\ \ ^{\shortmid }\mathring{g}_{\alpha _{s}},\ ~_{s}^{\shortmid }%
\mathcal{K})\leftrightarrow  \label{nonlinsym} \\
(\ _{s}\Phi ,\ _{s}^{\shortmid }\Lambda _{0}) &\leftrightarrow &(\
_{s}^{\shortmid }\mathbf{g},\ \ _{s}^{\shortmid }\Lambda
_{0})\leftrightarrow (~_{s}^{\shortmid }\eta \ ^{\shortmid }\mathring{g}%
_{\alpha _{s}}\sim \ ^{\shortmid }\zeta _{\alpha _{s}}(1+\kappa \
^{\shortmid }\chi _{\alpha _{s}})\ \ ^{\shortmid }\mathring{g}_{\alpha
_{s}},\ \ _{s}^{\shortmid }\Lambda _{0}).  \notag
\end{eqnarray}%
In explicit form, such nonlinear transforms are defined by equations 
\begin{equation*}
\frac{\partial _{3}[(\ _{2}\Psi )^{2}]}{~_{2}^{\shortmid }\mathcal{K}}=\frac{%
\partial _{3}[(\ _{2}\Phi )^{2}]}{\ _{2}\Lambda _{0}},\frac{~^{\shortmid
}\partial ^{5}[(\ _{3}^{\shortmid }\Psi )^{2}]}{~_{3}^{\shortmid }\mathcal{K}%
}=\frac{~^{\shortmid }\partial ^{5}[(\ _{3}^{\shortmid }\Phi )^{2}]}{\
_{3}^{\shortmid }\Lambda _{0}},\frac{~^{\shortmid }\partial ^{8}[(\
_{4}^{\shortmid }\Psi )^{2}]}{~_{4}^{\shortmid }\mathcal{K}}=\frac{%
~^{\shortmid }\partial ^{8}[(\ _{4}^{\shortmid }\Phi )^{2}]}{\
_{4}^{\shortmid }\Lambda _{0}}.
\end{equation*}%
In integral forms, we obtain such transforms: 
\begin{eqnarray*}
(\ _{2}\Psi )^{2} &=&(\ _{2}\Lambda _{0})^{-1}\int dy^{3}(~_{2}^{\shortmid }%
\mathcal{K})\partial _{3}[(\ _{2}\Phi )^{2}]\ \mbox{ and/or }(\ _{2}\Phi
)^{2}=\ _{2}\Lambda _{0}\int dy^{3}(~_{2}^{\shortmid }\mathcal{K}%
)^{-1}\partial _{3}[(\ _{2}\Psi )^{2}], \\
(\ _{3}^{\shortmid }\Psi )^{2} &=&(\ _{3}^{\shortmid }\Lambda _{0})^{-1}\int
d~p_{5}(~_{3}^{\shortmid }\mathcal{K})[(\ _{3}^{\shortmid }\Phi )^{2}]%
\mbox{
and/or }\ (\ _{3}^{\shortmid }\Phi )^{2}=\ _{3}^{\shortmid }\Lambda _{0}\int
dp_{5}(~_{3}^{\shortmid }\mathcal{K})^{-1}[(\ _{3}^{\shortmid }\Psi )^{2}],
\\
(\ _{4}^{\shortmid }\Psi )^{2} &=&(\ _{4}^{\shortmid }\Lambda _{0})^{-1}\int
d~p_{7}(~_{4}^{\shortmid }\mathcal{K})[(\ _{4}^{\shortmid }\Phi )^{2}]%
\mbox{
and/or }\ (\ _{4}^{\shortmid }\Phi )^{2}=\ _{4}^{\shortmid }\Lambda _{0}\int
dp_{7}~(~_{4}^{\shortmid }\mathcal{K})^{-1}[(\ _{4}^{\shortmid }\Psi )^{2}].
\end{eqnarray*}%
The generating functions/ sources/ cosmological constants and gravitational
polarization functions (\ref{dmpolariz}) and (\ref{espolariz}) can be
re-defined for various geometric and analytic purposes when the nonlinear
symmetries are re-written in other equivalent forms:%
\begin{eqnarray*}
\partial _{3}[(\ _{2}\Psi )^{2}] &=&-\int dy^{3}(~_{2}^{\shortmid }\mathcal{K%
})\partial _{3}g_{4}\simeq -\int dy^{3}(~_{2}^{\shortmid }\mathcal{K}%
)\partial _{3}(\ ^{\shortmid }\eta _{4}\ \ \ \mathring{g}_{4})\simeq -\int
dy^{3}(~_{2}^{\shortmid }\mathcal{K})\partial _{3}[\ ^{\shortmid }\zeta
_{4}(1+\kappa \ ^{\shortmid }\chi _{4})\ \mathring{g}_{4}], \\
(\ _{2}\Phi )^{2} &=&-4\ _{2}\Lambda _{0}g_{4}\simeq -4\ _{2}\Lambda _{0}\
^{\shortmid }\eta _{4}\ \ \mathring{g}_{4}\simeq -4\ _{2}\Lambda _{0}\
^{\shortmid }\zeta _{4}(1+\kappa \ ^{\shortmid }\chi _{4})\ \mathring{g}_{4};
\end{eqnarray*}%
\begin{eqnarray*}
~\ ^{\shortmid }\partial ^{5}[(\ _{3}^{\shortmid }\Psi )^{2}] &=&-\int
dp_{5}(~_{3}^{\shortmid }\mathcal{K})\ ^{\shortmid }\partial ^{5}\
^{\shortmid }g^{6}\simeq -\int dp_{5}(~_{3}^{\shortmid }\mathcal{K})\
^{\shortmid }\partial ^{5}(\ ^{\shortmid }\eta ^{6}\ ^{\shortmid }\mathring{g%
}^{6})\simeq -\int dp_{5}(~_{3}^{\shortmid }\mathcal{K})\ ^{\shortmid
}\partial ^{5}[\ ^{\shortmid }\zeta ^{6}(1+\kappa \ ^{\shortmid }\chi ^{6})\ 
\mathring{g}^{6}], \\
(\ _{3}^{\shortmid }\Phi )^{2} &=&-4\ _{3}^{\shortmid }\Lambda _{0}\
^{\shortmid }g^{6}\simeq \ -4\ _{3}^{\shortmid }\Lambda _{0}\ ^{\shortmid
}\eta ^{6}\ ^{\shortmid }\mathring{g}^{6}\simeq -4\ _{3}^{\shortmid }\Lambda
_{0}\ ^{\shortmid }\zeta ^{6}(1+\kappa \ ^{\shortmid }\chi ^{6})\
^{\shortmid }\mathring{g}^{6};
\end{eqnarray*}%
\begin{eqnarray*}
~\ ^{\shortmid }\partial ^{7}[(\ _{4}^{\shortmid }\Psi )^{2}] &=&-\int
dp_{7}(~_{4}^{\shortmid }\mathcal{K})\ ^{\shortmid }\partial ^{7}\
^{\shortmid }g^{8}\simeq -\int dp_{7}(~_{4}^{\shortmid }\mathcal{K})\
^{\shortmid }\partial ^{7}(\ ^{\shortmid }\eta ^{8}\ ^{\shortmid }\mathring{g%
}^{8})\simeq -\int dp_{7}(~_{4}^{\shortmid }\mathcal{K})\ ^{\shortmid
}\partial ^{8}[\ ^{\shortmid }\zeta ^{8}(1+\kappa \ ^{\shortmid }\chi ^{8})\ 
\mathring{g}^{8}], \\
(\ _{4}^{\shortmid }\Phi )^{2} &=&-4\ _{4}^{\shortmid }\Lambda _{0}\
^{\shortmid }g^{8}\simeq \ -4\ _{4}^{\shortmid }\Lambda _{0}\ ^{\shortmid
}\eta ^{8}\ ^{\shortmid }\mathring{g}^{8}\simeq -4\ _{4}^{\shortmid }\Lambda
_{0}\ ^{\shortmid }\zeta ^{8}(1+\kappa \ ^{\shortmid }\chi ^{8})\
^{\shortmid }\mathring{g}^{8}.
\end{eqnarray*}

The formulas for nonlinear symmetries (\ref{nonlinsym}) allow to express
solutions (\ref{sol1}) in different equivalent forms and/or for different
approximations via corresponding functionals on generating sources,
effective cosmological constants, and generating functions: 
\begin{eqnarray}
\ \ _{s}^{\shortmid }\mathbf{g} &=&\ _{s}^{\shortmid }\mathbf{g}[\hbar
,\kappa ,\psi ,_{s}\Psi ,_{s}^{\shortmid }\mathcal{K}]=\ _{s}^{\shortmid }%
\mathbf{g}[\hbar ,\kappa ,\psi ,_{s}\Phi ,\ _{s}^{\shortmid }\Lambda _{0}]
\label{smetrfunctionals} \\
&\simeq &\ \ _{s}^{\shortmid }\mathbf{g}[\hbar ,\kappa ,\psi
,~_{s}^{\shortmid }\mathcal{K},g_{4},\ ^{\shortmid }g^{6},\ ^{\shortmid
}g^{8}]  \notag \\
&\simeq &\ \ _{s}^{\shortmid }\mathbf{g}[\hbar ,\kappa ,\psi ,\
_{s}^{\shortmid }\Lambda _{0},g_{4},\ ^{\shortmid }g^{6},\ ^{\shortmid
}g^{8}]  \notag \\
&\simeq &\ \ _{s}^{\shortmid }\mathbf{g}[\hbar ,\kappa ,\psi
,~_{s}^{\shortmid }\mathcal{K},\ ^{\shortmid }\eta _{4}\ \ \mathring{g}%
_{4},\ ^{\shortmid }\eta ^{6}\ ^{\shortmid }\mathring{g}^{6},\ ^{\shortmid
}\eta ^{8}\ ^{\shortmid }\mathring{g}^{8},\mathring{g}_{3},\ ^{\shortmid }%
\mathring{g}^{5},\ ^{\shortmid }\mathring{g}^{7}]  \notag \\
&\simeq &\ \ _{s}^{\shortmid }\mathbf{g}[\hbar ,\kappa ,\psi ,\
_{s}^{\shortmid }\Lambda _{0},~_{s}^{\shortmid }\mathcal{K},\ ^{\shortmid
}\eta _{4}\ \ \mathring{g}_{4},\ ^{\shortmid }\eta ^{6}\ ^{\shortmid }%
\mathring{g}^{6},\ ^{\shortmid }\eta ^{8}\ ^{\shortmid }\mathring{g}^{8},%
\mathring{g}_{3},\ ^{\shortmid }\mathring{g}^{5},\ ^{\shortmid }\mathring{g}%
^{7}]  \notag \\
&\simeq &\ \ _{s}^{\shortmid }\mathbf{g}[\hbar ,\kappa ,\psi
,~_{s}^{\shortmid }\mathcal{K},\ ^{\shortmid }\zeta _{4},\ ^{\shortmid }\chi
_{4},\ \mathring{g}_{4};\ \ ^{\shortmid }\zeta ^{6},\ ^{\shortmid }\chi
^{6},\ ^{\shortmid }\mathring{g}^{6};\ \ ^{\shortmid }\zeta ^{8},\
^{\shortmid }\chi ^{8},\ ^{\shortmid }\mathring{g}^{8},\mathring{g}_{3},\
^{\shortmid }\mathring{g}^{5},\ ^{\shortmid }\mathring{g}^{7}]  \notag \\
&\simeq &\ \ _{s}^{\shortmid }\mathbf{g}[\hbar ,\kappa ,\psi ,\
_{s}^{\shortmid }\Lambda _{0},~_{s}^{\shortmid }\mathcal{K},\ ^{\shortmid
}\zeta _{4},\ ^{\shortmid }\chi _{4},\ \mathring{g}_{4};\ \ ^{\shortmid
}\zeta ^{6},\ ^{\shortmid }\chi ^{6},\ ^{\shortmid }\mathring{g}^{6};\ \
^{\shortmid }\zeta ^{8},\ ^{\shortmid }\chi ^{8},\ ^{\shortmid }\mathring{g}%
^{8},\mathring{g}_{3},\ ^{\shortmid }\mathring{g}^{5},\ ^{\shortmid }%
\mathring{g}^{7}].  \notag
\end{eqnarray}

We conclude that using functional data we can construct a class of solutions
for certain prescribed effective sources $_{s}^{\shortmid }\mathcal{K}$ and
generating functions $_{s}\Psi $. To elaborate on solutions in classical and
quantum gravity and information theory, it can be useful to consider
equivalent (or almost equivalent, with "$\simeq $", for some decompositions
on a small parameter) representations of such solutions when they are
connected to certain effective cosmological constants $\ _{s}^{\shortmid
}\Lambda _{0}$ and generating functions $_{s}\Phi .$ We can consider
configurations when some coefficients of s-metrics are taken as explicit
generating functions, for instance, $g_{4},\ ^{\shortmid }g^{6},\
^{\shortmid }g^{8}$. In another case, some $\eta $-polarizations are
prescribed as generating functions, for instance, we can use $\ ^{\shortmid
}\eta _{4},\ ^{\shortmid }\eta ^{6},\ ^{\shortmid }\eta ^{8}.$ In \cite%
{partner03,partner04}, nonassociative star product and R-flux deformations
of black hole solutions into black ellipsoid configurations were studied
using $\kappa $-linear s-deformations determined by generating functions $\
^{\shortmid }\chi _{4},\ ^{\shortmid }\chi ^{6},\ ^{\shortmid }\chi ^{8}.$

\subsection{Solutions with effective cosmological constants}

We can transform the nonassociative vacuum s-metrics (\ref{sol1}), $\
_{s}^{\shortmid }\mathbf{g}[\hbar ,\kappa ,\psi ,_{s}\Psi ,_{s}^{\shortmid }%
\mathcal{K}]\rightarrow \ _{s}^{\shortmid }\mathbf{g}[\hbar ,\kappa ,\psi
,_{s}\Phi ,\ _{s}^{\shortmid }\Lambda _{0}]$ following the first line in the
functional representations (\ref{smetrfunctionals}):%
\begin{eqnarray}
g_{1}(x^{k_{1}}) &=&g_{2}(x^{k_{1}})=g_{1}[\psi ]=g_{2}[\psi ]=e^{\psi
(\hbar ,\kappa ;x^{k_{1}})},  \label{sol1a} \\
g_{3}(x^{k_{1}},y^{3}) &=&g_{3}[\ _{2}\Phi ]=-\frac{1}{g_{4}[\ _{2}\Phi ]}%
\frac{(\ _{2}\Phi )^{2}[\partial _{3}(\ _{2}\Phi )]^{2}}{|\ _{2}\Lambda
_{0}\int dy^{3}(~_{2}^{\shortmid }\mathcal{K})[\partial _{3}(\ _{2}\Phi
)^{2}]|},  \notag \\
g_{4}(x^{k_{1}},y^{3}) &=&g_{4}[\ _{2}\Phi ]=g_{4}^{[0]}-\frac{(\ _{2}\Phi
)^{2}}{4\ _{2}\Lambda _{0}};  \notag
\end{eqnarray}%
\begin{eqnarray*}
\ ^{\shortmid }g^{5}(x^{i_{2}},p_{5}) &=&\ ^{\shortmid }g^{5}[\
_{3}^{\shortmid }\Phi ]=-\frac{1}{\ ^{\shortmid }g^{6}[\ _{3}^{\shortmid
}\Phi ]}\frac{(\ _{3}^{\shortmid }\Phi )^{2}[\ ^{\shortmid }\partial ^{5}(\
_{3}^{\shortmid }\Phi )]^{2}}{|\ _{3}^{\shortmid }\Lambda _{0}\int
dp_{5}(~_{3}^{\shortmid }\mathcal{K})\ \ ^{\shortmid }\partial ^{5}[(\
_{3}^{\shortmid }\Phi )^{2}]|}, \\
\ ^{\shortmid }g^{6}(x^{i_{2}},p_{5}) &=&\ ^{\shortmid }g^{6}[\
_{3}^{\shortmid }\Phi ]=g_{[0]}^{6}-\frac{(\ _{3}^{\shortmid }\Phi )^{2}}{4\
_{3}^{\shortmid }\Lambda _{0}};
\end{eqnarray*}%
\begin{eqnarray*}
\ ^{\shortmid }g^{7}(x^{i_{2}},p_{5},p_{7}) &=&\ ^{\shortmid }g^{7}[\
_{4}^{\shortmid }\Phi ]=-\frac{1}{\ ^{\shortmid }g^{8}[\ _{4}^{\shortmid
}\Phi ]}\frac{(\ _{4}^{\shortmid }\Phi )^{2}[\ ^{\shortmid }\partial ^{7}(\
_{4}^{\shortmid }\Phi )]^{2}}{|\ _{4}^{\shortmid }\Lambda _{0}\int
dp_{7}(~_{4}^{\shortmid }\mathcal{K})\ \ ^{\shortmid }\partial ^{7}[(\
_{4}^{\shortmid }\Phi )^{2}]|}, \\
\ ^{\shortmid }g^{8}(x^{i_{2}},p_{5},p_{8}) &=&\ ^{\shortmid }g^{8}[\
_{4}^{\shortmid }\Phi ]=g_{[0]}^{8}-\frac{(\ _{4}^{\shortmid }\Phi )^{2}}{4\
_{4}^{\shortmid }\Lambda _{0}};
\end{eqnarray*}%
\begin{eqnarray*}
\ N_{3i_{1}}^{\ }(x^{k_{1}},y^{3}) &=&w_{i_{1}}[\ _{2}\Phi ]=\frac{\partial
_{i_{1}}\ \int dy^{3}(~_{2}^{\shortmid }\mathcal{K})\ \partial _{3}[(\
_{2}\Phi )^{2}]}{(~_{2}^{\shortmid }\mathcal{K})\ \partial _{3}[(\ _{2}\Phi
)^{2}]}, \\
N_{4k_{1}}^{\ }(x^{i_{1}},y^{3}) &=&n_{k_{1}}[\ _{2}\Phi ]=\ _{1}n_{k_{1}}+\
_{2}n_{k_{1}}\int dy^{3}\ \frac{g_{3}[\ _{2}\Phi ]}{|\ g_{4}[\ _{2}\Phi
]|^{3/2}} \\
&=&\ _{1}n_{k_{1}}+\ _{2}n_{k_{1}}\int dy^{3}\frac{(\ _{2}\Phi )^{2}[\
\partial _{3}(\ _{2}\Phi )]^{2}}{|\ _{2}\Lambda _{0}\int
dy^{3}(~_{2}^{\shortmid }\mathcal{K})[\ \partial _{3}(\ _{2}\Phi )^{2}]|}%
\left\vert g_{4}^{[0]}-\frac{(\ _{2}\Phi )^{2}}{4\ _{2}\Lambda _{0}}%
\right\vert ^{-5/2};
\end{eqnarray*}%
\begin{eqnarray*}
\ ^{\shortmid }N_{5k_{2}}^{\ }(x^{i_{2}},p_{5}) &=&w_{k_{2}}[\
_{3}^{\shortmid }\Phi ]=\frac{\partial _{k_{2}}\ \int
dp_{5}(~_{3}^{\shortmid }\mathcal{K})\ \ ^{\shortmid }\partial ^{5}[(\
_{3}^{\shortmid }\Phi )^{2}]}{(~_{3}^{\shortmid }\mathcal{K})\ \ \
^{\shortmid }\partial ^{5}[(\ _{3}^{\shortmid }\Phi )^{2}]}, \\
\ ^{\shortmid }N_{6k_{2}}^{\ }(x^{i_{2}},p_{5}) &=&n_{k_{2}}[\
_{3}^{\shortmid }\Phi ]=\ _{1}^{\shortmid }n_{k_{2}}+\ _{2}^{\shortmid
}n_{k_{2}}\int dp_{5}\ \frac{\ ^{\shortmid }g^{5}[\ _{3}^{\shortmid }\Phi ]}{%
|\ \ ^{\shortmid }g^{6}[\ _{3}^{\shortmid }\Phi ]|^{3/2}} \\
&=&\ _{1}^{\shortmid }n_{k_{2}}+\ _{2}^{\shortmid }n_{k_{2}}\int dp_{5}\frac{%
(\ _{3}^{\shortmid }\Phi )^{2}[\ ^{\shortmid }\partial ^{5}(\
_{3}^{\shortmid }\Phi )]^{2}}{|\ _{3}^{\shortmid }\Lambda _{0}\int
dp_{5}(~_{3}^{\shortmid }\mathcal{K})[\ ^{\shortmid }\partial ^{5}(\
_{3}^{\shortmid }\Phi )^{2}]|}\left\vert \ ^{\shortmid }g_{[0]}^{6}-\frac{(\
_{3}^{\shortmid }\Phi )^{2}}{4\ _{3}^{\shortmid }\Lambda _{0}}\right\vert
^{-5/2};
\end{eqnarray*}%
\begin{eqnarray*}
\ ^{\shortmid }N_{7k_{3}}^{\ }(x^{i_{2}},p_{5},p_{7}) &=&w_{k_{3}}[\
_{4}^{\shortmid }\Phi ]=\frac{\partial _{k_{3}}\ \int
dp_{7}(~_{4}^{\shortmid }\mathcal{K})\ \ ^{\shortmid }\partial ^{7}[(\
_{4}^{\shortmid }\Phi )^{2}]}{(~_{4}^{\shortmid }\mathcal{K})\ \ \
^{\shortmid }\partial ^{7}[(\ _{4}^{\shortmid }\Phi )^{2}]}. \\
\ ^{\shortmid }N_{8k_{3}}^{\ }(x^{i_{2}},p_{5},p_{7}) &=&n_{k_{3}}[\
_{4}^{\shortmid }\Phi ]=\ _{1}^{\shortmid }n_{k_{3}}+\ _{2}^{\shortmid
}n_{k_{3}}\int dp_{7}\ \frac{\ ^{\shortmid }g^{7}[\ _{4}^{\shortmid }\Phi ]}{%
|\ \ ^{\shortmid }g^{8}[\ _{4}^{\shortmid }\Phi ]|^{3/2}} \\
&=&\ _{1}^{\shortmid }n_{k_{3}}+\ _{2}^{\shortmid }n_{k_{3}}\int dp_{7}\frac{%
(\ _{4}^{\shortmid }\Phi )^{2}[\ ^{\shortmid }\partial ^{7}(\
_{4}^{\shortmid }\Phi )]^{2}}{|\ _{4}^{\shortmid }\Lambda _{0}\int
dp_{7}(~_{4}^{\shortmid }\mathcal{K})[\ ^{\shortmid }\partial ^{7}(\
_{4}^{\shortmid }\Phi )^{2}]|}\left\vert \ ^{\shortmid }g_{[0]}^{8}-\frac{(\
_{4}^{\shortmid }\Phi )^{2}}{4\ _{4}^{\shortmid }\Lambda _{0}}\right\vert
^{-5/2}.
\end{eqnarray*}%
In formulas for above coefficients, there are used such conventions: \newline
for indices: $%
i_{1},j_{1},k_{1},...=1,2;i_{2},j_{2},k_{2},...=1,2,3,4;i_{3},j_{3},k_{3},...=1,2,...6;x^{3}=\varphi ,y^{4}=t,~p_{8}=~E; 
$ and 
\begin{eqnarray}
&&\mbox{generating functions: }\psi (\hbar ,\kappa ,x^{k_{1}});\ _{2}\Phi
(\hbar ,\kappa ,x^{k_{1}}y^{3});\ _{3}^{\shortmid }\Phi (\hbar ,\kappa
,x^{k_{2}},p_{6});\ _{4}^{\shortmid }\Phi (\hbar ,\kappa ,~^{\shortmid
}x^{k_{3}},p_{7});  \notag \\
&&\mbox{generating sources:}~_{1}^{\shortmid }\mathcal{K}(\hbar ,\kappa
,x^{k_{1}});\ ~_{2}^{\shortmid }\mathcal{K}(\hbar ,\kappa
,x^{k_{1}},y^{3});\ ~_{3}^{\shortmid }\mathcal{K}(\hbar ,\kappa
,x^{k_{2}},p_{5});\ ~_{4}^{\shortmid }\mathcal{K}(\hbar ,\kappa
,~^{\shortmid }x^{k_{3}},p_{7});  \notag \\
&&\mbox{integration  functions: }g_{4}^{[0]}(\hbar ,\kappa ,x^{k_{1}}),\
_{1}n_{k_{1}}(\hbar ,\kappa ,x^{j_{1}}),\ _{2}n_{k_{1}}(\hbar ,\kappa
,x^{j_{1}});  \label{gensol1a} \\
&&\ ^{\shortmid }g_{[0]}^{6}(\hbar ,\kappa ,x^{k_{2}}),\ _{1}n_{k_{2}}(\hbar
,\kappa ,x^{j_{2}}),\ _{2}n_{k_{2}}(\hbar ,\kappa ,x^{j_{2}});\ ^{\shortmid
}g_{[0]}^{8}(\hbar ,\kappa ,\ ^{\shortmid }x^{j_{3}}),\ _{1}^{\shortmid
}n_{k_{3}}(\hbar ,\kappa ,~^{\shortmid }x^{j_{3}}),\ _{2}^{\shortmid
}n_{k_{3}}(\hbar ,\kappa ,~^{\shortmid }x^{j_{3}}).  \notag
\end{eqnarray}

Such functional representations of off-diagonal solutions allow to encode
possible contributions from effective cosmological constants when certain
dynamics of effective sources is re-distributed into off-diagonal terms of
s-metrics. Nevertheless, the contributions from $~_{s}^{\shortmid }\mathcal{K%
}$ are not completely excluded being present in integrals for certain
s-connection coefficients like $g_{3},\ ^{\shortmid }g^{5},\ ^{\shortmid
}g^{7}$ and all N-connection coefficients.

Changing the generating functions and generating sources/ cosmological
constants data, $\ [_{s}\Psi ,_{s}^{\shortmid }\mathcal{K}]\rightarrow
\lbrack _{s}\Phi ,\ _{s}^{\shortmid }\Lambda _{0}],$ we re-express the data
for quasi-stationary solutions of the nonassociative parametric vacuum
gravitational equations (\ref{seinsta}) with effective sources (\ref%
{cannonsymparamc2a}), defined by (\ref{sol1}) as solutions for $\
^{\shortmid }\widehat{\mathbf{R}}_{\ \ \gamma _{s}}^{\beta _{s}}[_{s}\Psi
,_{s}^{\shortmid }\mathcal{K},...]={\delta }_{\ \ \gamma _{s}}^{\beta _{s}}\
_{s}^{\shortmid }\mathcal{K}$ (\ref{modeinst}), into solutions of type (\ref%
{sol1a}) of $\ ^{\shortmid }\widehat{\mathbf{R}}_{\ \ \gamma _{s}}^{\beta
_{s}}[_{s}\Phi ,\ _{s}^{\shortmid }\Lambda _{0},_{s}^{\shortmid }\mathcal{K}%
,...]~=\ \delta _{\ \ \gamma _{s}}^{\beta _{s}}\ \ _{s}^{\shortmid }\Lambda
_{0}$ (\ref{nonassocrsol}). The functional structure of geometric objects is
subjected to certain (\ref{nonlinsym}) transforms when the data for
effective sources $_{s}^{\shortmid }\mathcal{K}$ are kept into off-diagonal
N-connection terms but the left side of modified Einstein equations is
stated with effective cosmological constants $\ _{s}^{\shortmid }\Lambda
_{0}.$

\subsection{Using some coefficients of s-metrics as generating functions}

\label{assmetrgf}We can consider 
\begin{eqnarray*}
g_{4}(x^{k_{1}},y^{3}) &=&g_{4}[\ _{2}\Psi ,~_{2}^{\shortmid }\mathcal{K}%
]=g_{4}[\ _{2}\Phi ,\ _{2}\Lambda _{0}];\ ^{\shortmid
}g^{6}(x^{i_{2}},p_{5})=\ ^{\shortmid }g^{6}[\ _{3}\Psi ,~_{3}^{\shortmid }%
\mathcal{K}]=\ ^{\shortmid }g^{6}[\ _{3}^{\shortmid }\Phi ,\ _{3}\Lambda
_{0}]; \\
\ ^{\shortmid }g^{8}(x^{i_{2}},p_{5},p_{7}) &=&\ ^{\shortmid }g^{8}[\
_{4}\Psi ,~_{4}^{\shortmid }\mathcal{K}]=\ ^{\shortmid }g^{8}[\
_{4}^{\shortmid }\Phi ,\ _{4}\Lambda _{0}],
\end{eqnarray*}%
from (\ref{sol1}) and (\ref{sol1a}) as generating functions for a s-metric (%
\ref{ans1dm}) and N-coefficients (\ref{ans1n}). In the first case,
expressing $\ _{s}\Psi =\ _{s}\Psi \lbrack ~_{s}^{\shortmid }\mathcal{K}%
,g_{4},\ ^{\shortmid }g^{6},\ ^{\shortmid }g^{8}],$ we obtain such
parameterizations of quasi-stationary solutions: 
\begin{eqnarray}
&&g_{1}(x^{k_{1}})=g_{2}(x^{k_{1}})=g_{1}[\psi ]=g_{2}[\psi ]=e^{\psi (\hbar
,\kappa ;x^{k_{1}})},  \label{sol1b} \\
&&g_{3}(x^{k_{1}},y^{3})=-\frac{(\partial _{3}g_{4})^{2}}{|\int
dy^{3}\partial _{3}[(~_{2}^{\shortmid }\mathcal{K})g_{4}]|\ g_{4}}%
,g_{4}(x^{k_{1}},y^{3})\mbox{  is a generating function on shell }s=2; 
\notag
\end{eqnarray}%
\begin{eqnarray*}
\ ^{\shortmid }g^{5}(x^{i_{2}},p_{5}) &=&-\frac{[\ ^{\shortmid }\partial
^{5}(\ ^{\shortmid }g^{6})]^{2}}{|\int dp_{5}\ ^{\shortmid }\partial
^{5}[(~_{3}^{\shortmid }\mathcal{K})\ ^{\shortmid }g^{6}]\ |\ ^{\shortmid
}g^{6}},\ ^{\shortmid }g^{6}(x^{i_{2}},p_{5})%
\mbox{ is a generating function on
shell }s=3, \\
\ ^{\shortmid }g^{7}(x^{i_{2}},p_{5},p_{7}) &=&-\frac{[~^{\shortmid
}\partial ^{7}(~^{\shortmid }g^{8})]^{2}}{|\int dp_{7}\ ~^{\shortmid
}\partial ^{7}[(~_{4}^{\shortmid }\mathcal{K})~^{\shortmid }g^{8}]\ |\
~^{\shortmid }g^{8}},\ ^{\shortmid }g^{8}(x^{i_{2}},p_{5},p_{7})%
\mbox{ is a generating function on
shell }s=4,;
\end{eqnarray*}%
\begin{eqnarray*}
\ N_{3i_{1}}^{\ }(x^{k_{1}},y^{3}) &=&w_{i_{1}}[g_{4}]=\frac{\partial
_{i_{1}}[\int dy^{3}(~_{2}^{\shortmid }\mathcal{K})\ \partial _{3}g_{4}]}{%
(~_{2}^{\shortmid }\mathcal{K})\ \partial _{3}g_{4}}, \\
N_{4k_{1}}^{\ }(x^{i_{1}},y^{3}) &=&n_{k_{1}}[\ g_{4}]=\ _{1}n_{k_{1}}+\
_{2}n_{k_{1}}\int dy^{3}\frac{(\partial _{3}g_{4})^{2}}{|\int dy^{3}\partial
_{3}[(~_{2}^{\shortmid }\mathcal{K})g_{4}]|\ [g_{4}]^{5/2}}\ ;
\end{eqnarray*}%
\begin{eqnarray*}
\ ^{\shortmid }N_{5k_{2}}^{\ }(x^{i_{2}},p_{5}) &=&w_{k_{2}}[\ ^{\shortmid
}g^{6}]=\frac{\partial _{k_{2}}[\int dp_{5}(~_{3}^{\shortmid }\mathcal{K})\
~^{\shortmid }\partial ^{5}(~^{\shortmid }g^{6})]}{(~_{3}^{\shortmid }%
\mathcal{K})~^{\shortmid }\partial ^{5}(~^{\shortmid }g^{6})}, \\
\ ^{\shortmid }N_{6k_{2}}^{\ }(x^{i_{2}},p_{5}) &=&n_{k_{2}}[\ ^{\shortmid
}g^{6}]=\ _{1}^{\shortmid }n_{k_{2}}+\ _{2}^{\shortmid }n_{k_{2}}\int dp_{5}%
\frac{[~^{\shortmid }\partial ^{5}(~^{\shortmid }g^{6})]^{2}}{|\int dp_{5}\
~^{\shortmid }\partial ^{5}[(~_{3}^{\shortmid }\mathcal{K})~^{\shortmid
}g^{6}]|\ [~^{\shortmid }g^{6}]^{5/2}};
\end{eqnarray*}%
\begin{eqnarray*}
\ ^{\shortmid }N_{7k_{3}}^{\ }(x^{i_{2}},p_{5},p_{7}) &=&w_{k_{3}}[\
^{\shortmid }g^{8}]=\frac{~^{\shortmid }\partial _{k_{3}}[\int dp_{7}(\
~_{4}^{\shortmid }\mathcal{K})\ ~^{\shortmid }\partial ^{7}(~^{\shortmid
}g^{8})]}{(~_{4}^{\shortmid }\mathcal{K})~^{\shortmid }\partial
^{7}(~^{\shortmid }g^{8})}, \\
\ ^{\shortmid }N_{7k_{3}}^{\ }(x^{i_{2}},p_{5},p_{7}) &=&n_{k_{3}}[\
^{\shortmid }g^{8}]=\ _{1}^{\shortmid }n_{k_{3}}+\ _{2}^{\shortmid
}n_{k_{3}}\int dp_{7}\frac{[~^{\shortmid }\partial ^{7}(~^{\shortmid
}g^{8})]^{2}}{|\int dp_{7}\ ~^{\shortmid }\partial ^{7}[(~_{4}^{\shortmid }%
\mathcal{K})~^{\shortmid }g^{8}]|\ [~^{\shortmid }g^{8}]^{5/2}}\ ,
\end{eqnarray*}%
The s-coefficients (\ref{sol1b}) define quasi-stationary solutions of type $%
\ _{s}^{\shortmid }\mathbf{g}[\hbar ,\kappa ,\psi ,~_{s}^{\shortmid }%
\mathcal{K},g_{4},\ ^{\shortmid }g^{6},\ ^{\shortmid }g^{8}]$ from (\ref%
{smetrfunctionals}).

Above coefficients can be re-defined to include functional dependencies on
effective cosmological constants $\ _{s}^{\shortmid }\Lambda $ if we begin
with (\ref{sol1a}) and express $\ _{s}\Phi =\ _{s}\Phi \lbrack \
_{s}^{\shortmid }\Lambda _{0},~_{s}^{\shortmid }\mathcal{K},g_{4},\
^{\shortmid }g^{6},\ ^{\shortmid }g^{8}].$ This way we generate a solution
of type $\ _{s}^{\shortmid }\mathbf{g}[\hbar ,\kappa ,\psi ,\
_{s}^{\shortmid }\Lambda _{0},~_{s}^{\shortmid }\mathcal{K},g_{4},\
^{\shortmid }g^{6},\ ^{\shortmid }g^{8}]$ from (\ref{smetrfunctionals}).

\subsection{Parametric quasi-stationary gravitational polarizations and $%
\protect\kappa $-linear flows}

\label{assgrpolfl}To model off-diagonal deformations of a prescribed prime
metric into target ones, $\ _{s}^{\shortmid }\mathbf{\mathring{g}}=[\
^{\shortmid }\mathring{g}_{\alpha _{s}},\ ^{\shortmid }\mathring{N}%
_{i_{s-1}}^{a_{s}}]\rightarrow \ _{s}^{\shortmid }\mathbf{g}$ (\ref%
{offdiagdef}) by $\eta $-polarizations (\ref{dmpolariz}), we can consider as
generating functions such values: 
\begin{equation*}
\psi (\hbar ,\kappa ;x^{k_{1}}),\ ^{\shortmid }\eta _{4}(x^{k_{1}},y^{3}),\
^{\shortmid }\eta ^{6}(x^{i_{2}},p_{5}),\ ^{\shortmid }\eta
^{8}(x^{i_{2}},p_{5},p_{7}).
\end{equation*}%
This allows us to compute all polarization functions $\ ^{\shortmid }\eta
_{\alpha _{s}}(\hbar ,\kappa ,x^{i_{s-1}},p_{a_{s}})\ $and $\ ^{\shortmid
}\eta _{i_{s-1}}^{a_{s}}(\hbar ,\kappa ,x^{i_{s-1}},p_{a_{s}})$ following
the standard AFCDM. The explicit form of $\eta $-polarizations depend, for
instance, on the type of target quasi-stationary solutions we search. If the
target is of type (\ref{sol1}), we generate s-metrics with functional
dependence $\ _{s}^{\shortmid }\mathbf{g}[\hbar ,\kappa ,\psi
,~_{s}^{\shortmid }\mathcal{K},\ ^{\shortmid }\eta _{4}\ \ \mathring{g}%
_{4},\ ^{\shortmid }\eta ^{6}\ ^{\shortmid }\mathring{g}^{6},\ ^{\shortmid
}\eta ^{8}\ ^{\shortmid }\mathring{g}^{8},\mathring{g}_{3},\ ^{\shortmid }%
\mathring{g}^{5},\ ^{\shortmid }\mathring{g}^{7}]$ stated by classifications
(\ref{smetrfunctionals}). In explicit form, with complex variables, such a
s-metric was constructed in appendix B.2 to \cite{partner02}, see formulas
(B.4) in that work. We can introduce effective cosmological constants $\
_{s}^{\shortmid }\Lambda _{0}$ and generate s-metrics with quasi-stationary
dependence $\ _{s}^{\shortmid }\mathbf{g}[\hbar ,\kappa ,\psi ,\
_{s}^{\shortmid }\Lambda _{0},~_{s}^{\shortmid }\mathcal{K},\ ^{\shortmid
}\eta _{4}\ \ \mathring{g}_{4},\ ^{\shortmid }\eta ^{6}\ ^{\shortmid }%
\mathring{g}^{6},\ ^{\shortmid }\eta ^{8}\ ^{\shortmid }\mathring{g}^{8},%
\mathring{g}_{3},\ ^{\shortmid }\mathring{g}^{5},\ ^{\shortmid }\mathring{g}%
^{7}]$ if such a target metric is of type (\ref{sol1a}). More general
classes of solutons describing nonassociative geometric flows of
quasi-stationary metrics with $\eta $-polarizations, effective sources $\
_{s}^{\shortmid }\Im (\tau )$ (\ref{cannonsymparamc2b}) and running
cosmological constants $\ _{s}^{\shortmid }\Lambda (\tau )$ are studied in
section \ref{ssqsgrpol}.

In \cite{partner02,partner03,partner04}, we constructed quasi-stationary 4-d
and 8-d solutions of nonassociative vacuum gravitational equations when the
prime metrics are certain black hole, BH, metrics and the target s-metrics
for black ellipsoid, BE, s-metrics are generated as $\kappa $-linear
s-deformations determined by generating functions $\ ^{\shortmid }\chi
_{4},\ ^{\shortmid }\chi ^{6},\ ^{\shortmid }\chi ^{8},$ see (\ref{espolariz}%
). The coefficients of such s-metrics, with functional dependence of type 
\newline
$\ _{s}^{\shortmid }\mathbf{g}[\hbar ,\kappa ,\psi ,~_{s}^{\shortmid }%
\mathcal{K},\ ^{\shortmid }\zeta _{4},\ ^{\shortmid }\chi _{4},\ \mathring{g}%
_{4};\ \ ^{\shortmid }\zeta ^{6},\ ^{\shortmid }\chi ^{6},\ ^{\shortmid }%
\mathring{g}^{6};\ \ ^{\shortmid }\zeta ^{8},\ ^{\shortmid }\chi ^{8},\
^{\shortmid }\mathring{g}^{8},\mathring{g}_{3},\ ^{\shortmid }\mathring{g}%
^{5},\ ^{\shortmid }\mathring{g}^{7}]$ from (\ref{smetrfunctionals}, were
computed in a general form involving complex coordinates in appendix B.3,
formulas (B.7) of \cite{partner02}. This way, we generate a class of $\kappa 
$-parametric target quasi-stationary s-metrics of type (\ref{sol1}). In a
similar form, we can construct $\kappa $-parametric target quasi-stationary
s-metrics of type (\ref{sol1a}) with dependencies on $\ _{s}^{\shortmid
}\Lambda _{0}.$ Such generic off-diagonal configurations are described by
functionals of type \newline
$\ _{s}^{\shortmid }\mathbf{g}[\hbar ,\kappa ,\psi ,\ _{s}^{\shortmid
}\Lambda _{0},~_{s}^{\shortmid }\mathcal{K},\ ^{\shortmid }\zeta _{4},\
^{\shortmid }\chi _{4},\ \mathring{g}_{4};\ \ ^{\shortmid }\zeta ^{6},\
^{\shortmid }\chi ^{6},\ ^{\shortmid }\mathring{g}^{6};\ \ ^{\shortmid
}\zeta ^{8},\ ^{\shortmid }\chi ^{8},\ ^{\shortmid }\mathring{g}^{8},%
\mathring{g}_{3},\ ^{\shortmid }\mathring{g}^{5},\ ^{\shortmid }\mathring{g}%
^{7}]$ from (\ref{smetrfunctionals}). We note also that nonassociative BE
target solutions are generated for special rotoid polarizations of $\
^{\shortmid }\chi _{4},\ ^{\shortmid }\chi ^{6},\ ^{\shortmid }\chi ^{8}$ as
we studied in \cite{partner03,partner04}. The goal of this appendix is to
generalize such solutions in real variables for nonassociative geometric
flow with small parametric deformations depending on $\tau $-parameter.

In a $\tau $-running family of quasi-stationary s-metrics of type (\ref%
{offdiagpolfr}), the $\eta $-polarizations are expressed as $\kappa $-linear
functions when s-metric and N-connection coefficients of families of prime
s-metrics are transformed into respective families of target ones, 
\begin{equation*}
\ _{s}^{\shortmid }\mathbf{\mathring{g}}(\tau )\rightarrow \ _{s}^{\shortmid
\kappa }\mathbf{g}(\tau )=[\ ^{\shortmid }g_{\alpha _{s}}(\tau )=\
^{\shortmid }\zeta _{\alpha _{s}}(\tau )(1+\kappa \ ^{\shortmid }\chi
_{\alpha _{s}}(\tau ))\ ^{\shortmid }\mathring{g}_{\alpha _{s}}(\tau ),\
^{\shortmid }N_{i_{s}}^{a_{s}}(\tau )=\ ^{\shortmid }\zeta
_{i_{s-1}}^{a_{s}}(\tau )(1+\kappa \ ^{\shortmid }\chi
_{i_{s-1}}^{a_{s}}(\tau ))\ ^{\shortmid }\mathring{N}_{i_{s-1}}^{a_{s}}(\tau
)].
\end{equation*}%
The $\zeta $- and $\chi $-coefficients for deformations the $\eta $%
-polarization generating functions (\ref{etapolgen}) are respectively $%
\kappa $-linearized as data 
\begin{eqnarray}
\psi (\tau ) &\simeq &\psi (\hbar ,\kappa ;\tau ,x^{k_{1}})\simeq \psi
_{0}(\hbar ,\tau ,x^{k_{1}})(1+\kappa \ _{\psi }\chi (\hbar ,\tau
,x^{k_{1}})),\mbox{ for }\   \label{epsilongenfdecomp} \\
\ \eta _{2}(\tau ) &\simeq &\eta _{2}(\hbar ,\kappa ;\tau ,x^{k_{1}})\simeq
\zeta _{2}(\hbar ,\tau ,x^{k_{1}})(1+\kappa \chi _{2}(\hbar ,\tau
,x^{k_{1}})),\mbox{ we can consider }\ \eta _{2}(\tau )=\ \eta _{1}(\tau ); 
\notag \\
\ \ ^{\shortmid }\eta _{4}(\tau ) &\simeq &\ ^{\shortmid }\eta _{4}(\hbar
,\kappa ;\tau ,x^{k_{1}},y^{3})\simeq \ ^{\shortmid }\zeta _{4}(\hbar ,\tau
,x^{k_{1}},y^{3})(1+\kappa \ ^{\shortmid }\chi _{4}(\hbar ,\tau
,x^{k_{1}},y^{3})),  \notag \\
\ \ \ ^{\shortmid }\eta ^{6}(\tau ) &\simeq &\ ^{\shortmid }\eta ^{6}(\hbar
,\kappa ;\tau ,x^{i_{2}},p_{5})\simeq \ ^{\shortmid }\zeta ^{6}(\hbar
,\kappa ;\tau ,x^{i_{2}},p_{5})(1+\kappa \ ^{\shortmid }\chi ^{6}(\hbar
,\kappa ;\tau ,x^{i_{2}},p_{5})),\   \notag \\
\ \ ^{\shortmid }\eta ^{8}(\tau ) &\simeq &\ ^{\shortmid }\eta ^{8}(\hbar
,\kappa ;\tau ,x^{i_{2}},p_{5},p_{7})\simeq \ ^{\shortmid }\zeta ^{8}(\hbar
,\kappa ;\tau ,x^{i_{2}},p_{5},p_{7})(1+\kappa \ ^{\shortmid }\chi
^{8}(\hbar ,\kappa ;\tau ,x^{i_{2}},p_{5},p_{7})).  \notag
\end{eqnarray}%
In above formulas, $\psi (\tau )$ and $\eta _{2}(\tau )=\ \eta _{1}(\tau )$
are related to define a $\tau $-family of solutions 2-d Poisson equation $%
\partial _{11}^{2}\psi (\tau )+\partial _{22}^{2}\psi (\tau )=2\ _{1}\Im
(\tau ).$

For parameterizations (\ref{epsilongenfdecomp}), we can re-write the
geometric evolution of quasi-stationary s-metrics in a small $\kappa $%
-parametric form with $\chi $-generating functions (for simplicity, we do
not write in this formula the phase space coordinate and parametric $\tau $%
-dependence of coefficients):

\begin{eqnarray*}
d\ ^{\shortmid }\widehat{s}^{2}(\tau ) &=&\ ^{\shortmid }\widehat{g}_{\alpha
_{s}\beta _{s}}(\hbar ,\kappa ;\tau
,x^{k},y^{3},p_{a_{3}},p_{a_{4}};g_{4}(\tau ),\ ^{\shortmid }g^{6}(\tau ),\
^{\shortmid }g^{8}(\tau ),~_{s}^{\shortmid }\Im (\tau ))d\ ^{\shortmid
}u^{\alpha _{s}}d\ ^{\shortmid }u^{\beta _{s}} \\
&=&e^{\psi _{0}}(1+\kappa \ ^{\psi }\ ^{\shortmid }\chi
)[(dx^{1})^{2}+(dx^{2})^{2}]
\end{eqnarray*}%
\begin{eqnarray*}
&&-\{\frac{4[\partial _{3}(|\zeta _{4}\mathring{g}_{4}|^{1/2})]^{2}}{%
\mathring{g}_{3}|\int dy^{3}\{\ _{2}\Im \partial _{3}(\zeta _{4}\mathring{g}%
_{4})\}|}-\kappa \lbrack \frac{\partial _{3}(\chi _{4}|\zeta _{4}\mathring{g}%
_{4}|^{1/2})}{4\partial _{3}(|\zeta _{4}\mathring{g}_{4}|^{1/2})}-\frac{\int
dy^{3}\{\ _{2}\Im \partial _{3}[(\zeta _{4}\mathring{g}_{4})\chi _{4}]\}}{%
\int dy^{3}\{\ _{2}\Im \partial _{3}(\zeta _{4}\mathring{g}_{4})\}}]\}%
\mathring{g}_{3} \\
&&\{dy^{3}+[\frac{\partial _{i_{1}}\ \int dy^{3}\ _{2}\Im \ \partial
_{3}\zeta _{4}}{(\mathring{N}_{i_{1}}^{3})\ _{2}\Im \partial _{3}\zeta _{4}}%
+\kappa (\frac{\partial _{i_{1}}[\int dy^{3}\ _{2}\Im \partial _{3}(\zeta
_{4}\chi _{4})]}{\partial _{i_{1}}\ [\int dy^{3}\ _{2}\Im \partial _{3}\zeta
_{4}]}-\frac{\partial _{3}(\zeta _{4}\chi _{4})}{\partial _{3}\zeta _{4}})]%
\mathring{N}_{i_{1}}^{3}dx^{i_{1}}\}^{2}
\end{eqnarray*}%
\begin{eqnarray}
&&+\zeta _{4}(1+\kappa \ \chi _{4})\ \mathring{g}_{4}\{dt+[(\mathring{N}%
_{k_{1}}^{4})^{-1}[\ _{1}n_{k_{1}}+16\ _{2}n_{k_{1}}[\int dy^{3}\frac{\left(
\partial _{3}[(\zeta _{4}\mathring{g}_{4})^{-1/4}]\right) ^{2}}{|\int
dy^{3}\partial _{3}[\ _{2}\Im (\zeta _{4}\mathring{g}_{4})]|}]
\label{offdncelepsilon} \\
&&+\kappa \frac{16\ _{2}n_{k_{1}}\int dy^{3}\frac{\left( \partial
_{3}[(\zeta _{4}\mathring{g}_{4})^{-1/4}]\right) ^{2}}{|\int dy^{3}\partial
_{3}[\ _{2}\Im (\zeta _{4}\mathring{g}_{4})]|}(\frac{\partial _{3}[(\zeta
_{4}\mathring{g}_{4})^{-1/4}\chi _{4})]}{2\partial _{3}[(\zeta _{4}\mathring{%
g}_{4})^{-1/4}]}+\frac{\int dy^{3}\partial _{3}[\ _{2}\Im (\zeta _{4}\chi
_{4}\mathring{g}_{4})]}{\int dy^{3}\partial _{3}[\ _{2}\Im (\zeta _{4}%
\mathring{g}_{4})]})}{\ _{1}n_{k_{1}}+16\ _{2}n_{k_{1}}[\int dy^{3}\frac{%
\left( \partial _{3}[(\zeta _{4}\mathring{g}_{4})^{-1/4}]\right) ^{2}}{|\int
dy^{3}\partial _{3}[\ _{2}\Im (\zeta _{4}\mathring{g}_{4})]|}]}]\mathring{N}%
_{k_{1}}^{4}dx^{\acute{k}_{1}}\}^{2}+  \notag
\end{eqnarray}%
\begin{eqnarray*}
&&-\{\frac{4[\ ^{\shortmid }\partial ^{5}(|\ ^{\shortmid }\zeta ^{6}~\
^{\shortmid }\mathring{g}^{6}|^{1/2})]^{2}}{\ ^{\shortmid }\mathring{g}%
^{5}|\int dp_{5}\{\ _{3}^{\shortmid }\Im ~\ ^{\shortmid }\partial ^{5}[(\
^{\shortmid }\zeta ^{6}\ ^{\shortmid }\mathring{g}^{6})]\}|}-\kappa \lbrack 
\frac{\ ^{\shortmid }\partial ^{5}(\ ^{\shortmid }\zeta ^{6}\ ^{\shortmid }%
\mathring{g}^{6})}{\ ^{\shortmid }\partial ^{5}(\ ^{\shortmid }\zeta ^{5})}-%
\frac{\partial _{i_{2}}[\int dp_{5}\ _{3}^{\shortmid }\Im ~\ ^{\shortmid
}\partial ^{5}(\ ^{\shortmid }\zeta ^{6}~\ ^{\shortmid }\mathring{g}^{6})]}{%
\partial _{i_{2}}\ [\int dp_{5}\ _{3}^{\shortmid }\Im \ ^{\shortmid
}\partial ^{5}(\ ^{\shortmid }\zeta ^{6})]}]\}\ ^{\shortmid }\mathring{g}^{5}
\\
&&\{dp_{5}+[\frac{\partial _{i_{2}}\ \int dp_{5}\ _{3}^{\shortmid }\Im ~\
^{\shortmid }\partial ^{5}(\ ^{\shortmid }\zeta ^{6})}{(\ ^{\shortmid }%
\mathring{N}_{i_{2}5})\ _{3}^{\shortmid }\Im ~\ ^{\shortmid }\partial ^{5}(\
^{\shortmid }\zeta ^{6})}+\kappa (\frac{\partial _{i_{2}}[\int dp_{5}\
_{3}^{\shortmid }\Im \ ^{\shortmid }\partial ^{5}(\ ^{\shortmid }\zeta ^{6}\
^{\shortmid }\mathring{g}^{6})]}{\partial _{i_{2}}\ [\int dp_{5}\
_{3}^{\shortmid }\Im \ ^{\shortmid }\partial ^{5}(\ ^{\shortmid }\zeta ^{6})]%
}-\frac{\ ^{\shortmid }\partial ^{5}(\ ^{\shortmid }\zeta ^{6}\ ^{\shortmid }%
\mathring{g}^{6})}{\ ^{\shortmid }\partial ^{5}(\ ^{\shortmid }\zeta ^{5})}%
)](\ ^{\shortmid }\mathring{N}_{i_{2}5})dx^{i_{2}}\}
\end{eqnarray*}%
\begin{eqnarray*}
&&+\ ^{\shortmid }\zeta ^{6}(1+\kappa \ ^{\shortmid }\chi ^{6})~\
^{\shortmid }\mathring{g}^{6}\{dp_{5}+[\ (\ ^{\shortmid }\mathring{N}%
_{i_{2}6})^{-1}[\ _{1}^{\shortmid }n_{i_{2}}+16\ _{2}^{\shortmid
}n_{i_{2}}[\int dp_{5}\{\frac{\left( \ ^{\shortmid }\partial ^{5}[(\
^{\shortmid }\zeta ^{6}\ ^{\shortmid }\mathring{g}^{6})^{-1/4}]\right) ^{2}}{%
|\int dp_{5}~\ ^{\shortmid }\partial ^{5}[\ _{3}^{\shortmid }\Im (\
^{\shortmid }\zeta ^{6}\ ^{\shortmid }\mathring{g}^{6})]|}]+ \\
&&+\kappa \frac{16\ _{2}^{\shortmid }n_{i_{2}}\int dp_{5}\frac{\left( \
^{\shortmid }\partial ^{5}[(\ ^{\shortmid }\zeta ^{6}\ ^{\shortmid }%
\mathring{g}^{6})^{-1/4}]\right) ^{2}}{|\int dp_{5}\ _{3}^{\shortmid }\Im \
^{\shortmid }\partial ^{5}[(\ ^{\shortmid }\zeta ^{6}\ ^{\shortmid }%
\mathring{g}^{6})]|}(\frac{\ ^{\shortmid }\partial ^{5}[(\ ^{\shortmid
}\zeta ^{6}\ ^{\shortmid }\mathring{g}^{6})^{-1/4}\ ^{\shortmid }\chi ^{6})]%
}{2\ \ ^{\shortmid }\partial ^{5}[(\ ^{\shortmid }\zeta ^{6}\ ^{\shortmid }%
\mathring{g}^{6})^{-1/4}]}+\frac{\int dp_{5}\ \ _{3}^{\shortmid }\Im \
^{\shortmid }\partial ^{5}[(\ ^{\shortmid }\zeta ^{6}\ ^{\shortmid }%
\mathring{g}^{6})\ ^{\shortmid }\chi ^{6}]}{\int dp_{5}\ _{3}^{\shortmid
}\Im \ ^{\shortmid }\partial ^{5}[(\ ^{\shortmid }\zeta ^{6}\ ^{\shortmid }%
\mathring{g}^{6})]})}{\ _{1}^{\shortmid }n_{i_{2}}+16\ _{2}^{\shortmid
}n_{i_{2}}[\int dp_{5}\frac{\left( \ ^{\shortmid }\partial ^{5}[(\
^{\shortmid }\zeta ^{6}\ ^{\shortmid }\mathring{g}^{6})^{-1/4}]\right) ^{2}}{%
|\int dp_{5}\ \ _{3}^{\shortmid }\Im \ ^{\shortmid }\partial ^{5}[(\
^{\shortmid }\zeta ^{6}\ ^{\shortmid }\mathring{g}^{6})]|}]}]](\ ^{\shortmid
}\mathring{N}_{i_{2}6})dx^{i_{2}}\}^{2}
\end{eqnarray*}%
\begin{eqnarray*}
&&-\{\frac{4[\ ^{\shortmid }\partial ^{7}(|\ ^{\shortmid }\zeta ^{8}\
^{\shortmid }\mathring{g}^{8}|^{1/2})]^{2}}{~\ ^{\shortmid }\mathring{g}%
^{7}|\int dp_{7}\{\ _{4}^{\shortmid }\Im \ ^{\shortmid }\partial ^{7}(\
^{\shortmid }\zeta ^{8}~\ ^{\shortmid }\mathring{g}^{8})\}|}-\kappa \lbrack 
\frac{\ ^{\shortmid }\partial ^{7}(\ ^{\shortmid }\chi ^{8}|\ ^{\shortmid
}\zeta ^{8}~\ ^{\shortmid }\mathring{g}^{8}|^{1/2})}{4\ ^{\shortmid
}\partial ^{7}(|\ ^{\shortmid }\zeta ^{8}\ ^{\shortmid }\mathring{g}%
^{8}|^{1/2})}-\frac{\int dp_{7}\{\ _{4}^{\shortmid }\Im \ ^{\shortmid
}\partial ^{7}[(\ ^{\shortmid }\zeta ^{8}\ ^{\shortmid }\mathring{g}^{8})\
^{\shortmid }\chi ^{8}]\}}{\int dp_{7}\{\ _{4}^{\shortmid }\Im \ ^{\shortmid
}\partial ^{7}[(\ ^{\shortmid }\zeta ^{8}\ ^{\shortmid }\mathring{g}^{8})]\}}%
]\}\ ^{\shortmid }\mathring{g}^{7} \\
&&\{dp_{7}+[\frac{\ ^{\shortmid }\partial _{i_{3}}\ \int dp_{7}\
_{4}^{\shortmid }\Im \ \ ^{\shortmid }\partial ^{7}(\ ^{\shortmid }\zeta
^{8})}{(\ ^{\shortmid }\mathring{N}_{i_{3}7})\ _{4}^{\shortmid }\Im \
^{\shortmid }\partial ^{7}(\ ^{\shortmid }\zeta ^{8})}+\kappa \lbrack \frac{%
\ ^{\shortmid }\partial _{i_{3}}[\int dp_{7}\ _{4}^{\shortmid }\Im \
^{\shortmid }\partial ^{7}(\ ^{\shortmid }\zeta ^{8}\ ^{\shortmid }\mathring{%
g}^{8})]}{\ ^{\shortmid }\partial _{i_{3}}\ [\int dp_{7}\ _{4}^{\shortmid
}\Im \ ^{\shortmid }\partial ^{7}(\ ^{\shortmid }\zeta ^{8})]}-\frac{\
^{\shortmid }\partial ^{7}(\ ^{\shortmid }\zeta ^{8}\ ^{\shortmid }\mathring{%
g}^{8})}{\ ^{\shortmid }\partial ^{7}(\ ^{\shortmid }\zeta ^{8})}]](\
^{\shortmid }\mathring{N}_{i_{3}7})d\ ^{\shortmid }x^{i_{3}}\}^{2}
\end{eqnarray*}%
\begin{eqnarray*}
&&+\ ^{\shortmid }\zeta ^{8}(1+\kappa \ ^{\shortmid }\chi ^{8})\ ^{\shortmid
}\mathring{g}^{8}\{dp_{7}+[(\ ^{\shortmid }\mathring{N}_{i_{3}8})^{-1}[\
_{1}^{\shortmid }n_{i_{3}}+16\ _{2}^{\shortmid }n_{i_{3}}[\int dp_{7}\{\frac{%
\left( \ ^{\shortmid }\partial ^{7}[(\ ^{\shortmid }\zeta ^{8}\ ^{\shortmid }%
\mathring{g}^{8})^{-1/4}]\right) ^{2}}{|\int dp_{7}\ _{4}^{\shortmid }\Im ~\
^{\shortmid }\partial ^{7}(\ ^{\shortmid }\zeta ^{8}\ ^{\shortmid }\mathring{%
g}^{8})|}] \\
&&+\kappa \frac{16\ _{2}^{\shortmid }n_{i_{3}}\int dp_{7}\frac{\left( \
^{\shortmid }\partial ^{7}[(\ ^{\shortmid }\zeta ^{8}\ ^{\shortmid }%
\mathring{g}^{8})^{-1/4}]\right) ^{2}}{|\int dp_{7}\ \ _{4}^{\shortmid }\Im
\ ^{\shortmid }\partial ^{7}[(\ ^{\shortmid }\zeta ^{8}\ ^{\shortmid }%
\mathring{g}^{8})]|}(\frac{\ ^{\shortmid }\partial ^{7}[(\ ^{\shortmid
}\zeta ^{8}\ ^{\shortmid }\mathring{g}^{8})^{-1/4}\ ^{\shortmid }\chi ^{8})]%
}{2\ \ ^{\shortmid }\partial ^{7}[(\ ^{\shortmid }\zeta ^{8}\ ^{\shortmid }%
\mathring{g}^{8})^{-1/4}]}+\frac{\int dp_{7}\ \ _{4}^{\shortmid }\Im \
^{\shortmid }\partial ^{7}[(\ ^{\shortmid }\zeta ^{8}\ ^{\shortmid }%
\mathring{g}^{8})\ ^{\shortmid }\chi ^{8}]}{\int dp_{7}\ \ _{4}^{\shortmid
}\Im \ ^{\shortmid }\partial ^{7}(\ ^{\shortmid }\zeta ^{8}\ ^{\shortmid }%
\mathring{g}^{8})})}{\ _{1}^{\shortmid }n_{i_{3}}+16\ _{2}^{\shortmid
}n_{i_{3}}[\int dp_{7}\frac{\left( \ ^{\shortmid }\partial ^{7}\ [(\
^{\shortmid }\zeta ^{8}\ ^{\shortmid }\mathring{g}^{8})^{-1/4}]\right) ^{2}}{%
|\int dp_{7}\ \ _{4}^{\shortmid }\Im \ ^{\shortmid }\partial ^{7}(\
^{\shortmid }\zeta ^{8}\ ^{\shortmid }\mathring{g}^{8})|}]}]]\ (\
^{\shortmid }\mathring{N}_{i_{3}8})dx^{i_{3}}\}^{2}.
\end{eqnarray*}

Quasi-stationary solutions of type (\ref{offdncelepsilon}) can be
constructed for $\kappa $-parametric decompositions of $\tau $-running
gravitational polarization functions beginning with quadratic linear
elements (\ref{sol2rf}) with nonholonomic frames (\ref{sol2nrf}) and
respective nonlinear transforms (\ref{nonlinsymr}) and (\ref{nonlinsymrex}),
when 
\begin{eqnarray*}
\lbrack \ _{s}^{\shortmid }\mathbf{g}(\tau ),\ _{s}\Psi (\tau ),\
~_{s}^{\shortmid }\Im (\tau )] &\leftrightarrow &[\ _{s}^{\shortmid }\mathbf{%
g}(\tau ),\ \ _{s}\Phi (\tau ),\ ~_{s}^{\shortmid }\Im (\tau ),\
_{s}^{\shortmid }\Lambda (\tau )]\leftrightarrow \lbrack \ _{s}^{\shortmid }%
\mathbf{g}(\tau ),\ ~_{s}^{\shortmid }\eta (\tau )\ ^{\shortmid }\mathring{g}%
_{\alpha _{s}}(\tau ),\ ~_{s}^{\shortmid }\Im (\tau ),\ _{s}^{\shortmid
}\Lambda (\tau )] \\
&\leftrightarrow &[\ _{s}^{\shortmid }\mathbf{g}(\tau ),\ ^{\shortmid }\zeta
_{\alpha _{s}}(\tau )(1+\kappa \ ^{\shortmid }\chi _{\alpha _{s}}(\tau ))\ \
^{\shortmid }\mathring{g}_{\alpha _{s}}(\tau ),\ ~_{s}^{\shortmid }\Im (\tau
),\ _{s}^{\shortmid }\Lambda (\tau )].
\end{eqnarray*}%
We omit details on such technical constructions but present examples in next
appendix section and some applications, for instance, for double BE and BHs
phase space flow solutions in section (\ref{ssgfevdbh}).

If we fix $\tau =\tau _{0}$ for self-similar configurations of s-metrics (%
\ref{offdncelepsilon}), we generate $\kappa $-parametric solutions for
nonassociative Ricci solitons (\ref{naricsolp}). Such solutions were
constructed for nonassociative vacuum Einstein equations with effective
sources$~_{s}^{\shortparallel }\mathcal{K}$ and complex variables in \cite%
{partner02} (see appendix B.3 with formulas (B.7) in that work).

\end{document}